\documentclass[journal = chreay, manuscript=review]{achemso}
\usepackage{graphicx}
\graphicspath{{./figures/}}
\usepackage{color}
\usepackage{amsmath,amsfonts,amsthm,amssymb}
\usepackage{xcolor}
\usepackage[pdftex,colorlinks=true,linkcolor=black,citecolor=black,urlcolor=black]{hyperref}
\usepackage[capitalise]{cleveref}
\usepackage{siunitx}
\DeclareSIUnit\bar{bar}
\usepackage{bm}
\usepackage{nicefrac}

\usepackage{notes2bib}
\usepackage{hyperref}
\let\footnote=\bibnote
\newcommand*{\citen}[1]{%
  \begingroup
    \romannumeral-`\x % remove space at the beginning of \setcitestyle
    \setcitestyle{numbers}%
    \cite{#1}%
  \endgroup   
}
\newcommand{\redlork}[1]{{#1}}
\newcommand{\bluelork}[1]{{#1}}

\newcommand{\kbT}{k_\textnormal{B} T}
\newcommand{\etc}{etc}
\newcommand{\ie}{i.e}
\newcommand{\eg}{e.g}
\newcommand{\cf}{\textit{cf}}
\newcommand{\etal}{et al}

\def\rmd{\mathrm{d}}

\title[Transport in Nanoporous Materials]{Theory and Modeling of Transport for Simple Fluids in Nanoporous Materials: From Microscopic to Coarse-Grained Descriptions}

\author{Alexander Schlaich}
\affiliation{ Univ. Grenoble Alpes, CNRS, LIPhy, 38000 Grenoble, France}
\alsoaffiliation{Institute for Atomistic Modeling of Materials in Aqueous Media,\\
  Hamburg University of Technology,
  21073 Hamburg, Germany}
\alsoaffiliation{ Stuttgart Center for Simulation Science,
University of Stuttgart, 70569 Stuttgart, Germany}
\alsoaffiliation{ Institute for Computational Physics, University of Stuttgart,
70569 Stuttgart, Germany}

\author{Jean-Louis Barrat}
\affiliation{ Univ. Grenoble Alpes, CNRS, LIPhy, 38000 Grenoble, France}
\author{Benoit Coasne}
\affiliation{ Univ. Grenoble Alpes, CNRS, LIPhy, 38000 Grenoble, France}
\email{benoit.coasne@univ-grenoble-alpes.fr}

\begin{document}

%\maketitle

%%%%%%%%%%%%%%%%%%%%%%%%%%%%%%%%%%%%%%%%%%%%%%%%%%%%%%%%%%%%%%%%%%%%%%%%%%%%%%%%%%%%%%%%
%  Abstract

\begin{abstract} 
  We present the state-of-the-art theoretical modeling, molecular simulation, and coarse-graining strategies for the transport of gases and liquids in nanoporous materials (pore size 1--100 nm). Special emphasis is placed on the transport of small molecules in zeolites, active carbons, metal-organic frameworks, but also in nanoporous materials with larger pores such as ordered and disordered mesoporous oxides. We present different atomistic and mesoscopic methods as well as the theoretical formalisms. Attention is given to the investigation of different molecular transport coefficients---including the self, collective and transport diffusivities---but also to the determination of free energy barriers and their role in overall adsorption/separation process rates. We also introduce other available approaches such as hierarchical simulations and upscaling strategies. 
  This review focuses on simple fluids in prototypical nanoporous materials. While the phenomena covered here capture the main physical mechanisms in such systems, complex molecules will exhibit additional specific features. For the sake of clarity and brevity, we also omit multicomponent systems (e.g. fluid mixtures, electrolytes, etc.) and electrokinetic effects arising when charged systems are considered (ionic species, charged surfaces, etc.), both of which add to the complexity.
\end{abstract}

{
  \hypersetup{linkcolor=black}
  \tableofcontents
}

%%%%%%%%%%%%%%%%%%%%%%%%%%%%%%%%%%%%%%%%%%%%%%%%%%%%%%%%%%%%%%%%%%%%%%%%%%%%%%%%%%%%%%%%
%  INTRODUCTION

%!TeX spellcheck = en_US
%!TeX encoding = utf8
%!TeX program = pdflatex
%!TeX root = ../manuscript.tex

\section{Introduction}

Transport in nanoporous environments (pore size in the range  1--100 nm\cite{thommes_physisorption_2015}) are ubiquitous in chemistry, biology and physics.\cite{weitkamp_catalysis_1999, schuth_handbook_2002,voort_introduction_2019, davis_ordered_2002,karger_diffusion_2012}
In particular, such dynamical phenomena are key aspects to rationalize the role of nanoconfinement and surface forces on the behavior of fluids adsorbed in nanoporous solids.\cite{bhatia_molecular_2011, coasne_multiscale_2016,
smit_molecular_2008, bouchaud_anomalous_1990,karger_diffusion_2016,
schneider_transport_2016, huber_soft_2015,
karger_transport_2015,bukowski_connecting_2021, bocquet_nanofluidics_2010,
faucher_critical_2019, aluru_fluids_2023}
Such situations are also relevant to important applications related to the energy and environment fields:\cite{voort_introduction_2019, schuth_handbook_2002,
weitkamp_catalysis_1999} chemistry and chemical engineering (zeolites and mesoporous materials are used
for phase separation and catalysis), energy (supercapacitors and fuel cells), environment (water remediation, nuclear waste storage, and desalination), etc. From a fundamental viewpoint, the dynamics of fluids in the vicinity of surfaces or confined in nanoporous materials remain mysterious in many aspects. 
Indeed, the subtlety of the surface molecular interactions in such environments has not been fully embraced yet, and new phenomena still get uncovered experimentally --
therefore making this topic more active than ever.\cite{kavokine_fluids_2021,
bocquet_nanofluidics_2020,coasne_multiscale_2016}
In particular, beyond known adsorption and confinement effects that affect the chemistry and physics of fluid transport, there is a large amount of experimental and theoretical works highlighting the role of morphological/topological pore disorder\cite{karger_diffusion_2012,
schneider_transport_2016,karger_transport_2015, karger_mass_2013} 
 and specific surface interactions
(hydrophilic/hydrophobic,\cite{joseph_why_2008, suk_water_2010, huber_soft_2015,
bukowski_connecting_2021} 
insulating/metallic,\cite{kavokine_fluctuation-induced_2022,
schlaich_electronic_2022} etc.).
Moreover, there is also an increasing number of reports providing evidence for
the complex multiscale behavior of fluid transport in nanoporous media (pore
scale \textit{versus} material network scale).\cite{bhatia_molecular_2011,
galarneau_probing_2016, dutta_interfacial_2018, dutta_molecular_2023}
Such behavior, which challenges existing theoretical frameworks, points to the
needs for fundamental developments in the theoretical description of multiscale transport in nanoporous materials.

As a generic feature, nanoporous materials possess a large intrinsic specific surface which scales as the surface to volume ratio $S_{\textrm{sp}} \sim S/\rho V$, where
$S$, $V$, and $\rho$ are the surface area, volume and
density.\cite{thommes_physisorption_2015, voort_introduction_2019}
Typically, regardless of its geometry, the specific surface area of a pore of a
size $D_\mathrm{p}$ scales as $S_{\textrm{sp}} \sim 1/D_\mathrm{p}$; 
using the density of silica, $\rho \sim 2.65$ g/cm$^3$, the specific surface
area is of the order of $100$ m$^2$/g for $D_\mathrm{p} \sim  1$ nm, $10$
m$^2$/g for $D_\mathrm{p} \sim 10$ nm, 1 m$^2$/g for $D_\mathrm{p} \sim 1$
$\mu$m, and so on. The realm of nanoporous solids includes different classes of materials such as zeolites, metal organic frameworks, oxides (\eg.\ alumina, titania,
and silica), active porous carbons, \etc.\cite{thommes_physical_2010,
schuth_handbook_2002, auerbach_handbook_2003}
Optimizing processes involving nanoporous materials requires better understanding
adsorption and transport in the vicinity of surfaces and in confining
environments~\cite{coppens_nature-inspired_2012, sahimi_statistical_1990,
gheorghiu_optimal_2004}. 
The specific surface area and surface chemistry are obviously key parameters
that control the efficiency of a given chemical engineering process involving
nanoconfined fluids.
However, other important ingredients have to be considered to unravel
and characterize  the role of nanoconfinement in nanoporous materials.
These additional ingredients include the pore size $D_\mathrm{p}$ which, in
addition to being a property that controls the specific surface area
developed by the nanoporous material (see above), also governs the
thermodynamics and dynamics of the molecules within their
porosity~\cite{smit_molecular_2008,coasne_adsorption_2013,bhatia_molecular_2011,
coasne_multiscale_2016}.
Typically, depending on pore size and thermodynamic conditions (such as
temperature, pressure, \etc.), the nanoporous material can be either completely
filled by a liquid containing the fluid or incompletely filled with an adsorbed
film at the pore surface while the pore center remains filled by a gas phase. 
Correspondingly, as will be discussed in the present review, different transport
types can be observed depending on pore size $D_\mathrm{p}$, kinetic diameter of
the molecule $\sigma$, fluid/solid interactions and thermodynamic conditions
such as temperature $T$ and pressure $P$.
These transport mechanisms include but are not limited to molecular sieving,
surface diffusion, molecular diffusion, Knudsen diffusion,
\etc.\cite{karger_diffusion_2012, bukowski_connecting_2021}
Other properties such as pore morphology (pore shape, geometrical defects,
\etc.) and network topology (pore connectivity and accessibility) strongly 
affect the efficiency of an adsorption, separation or catalytic process by modifying the  
transport properties of the different interacting/reacting 
species~\cite{torquato_random_2005,coasne_multiscale_2016,coppens_nature-inspired_2012}.

\begin{figure}[!htbp]
\centering
\includegraphics[width=1.1\linewidth]{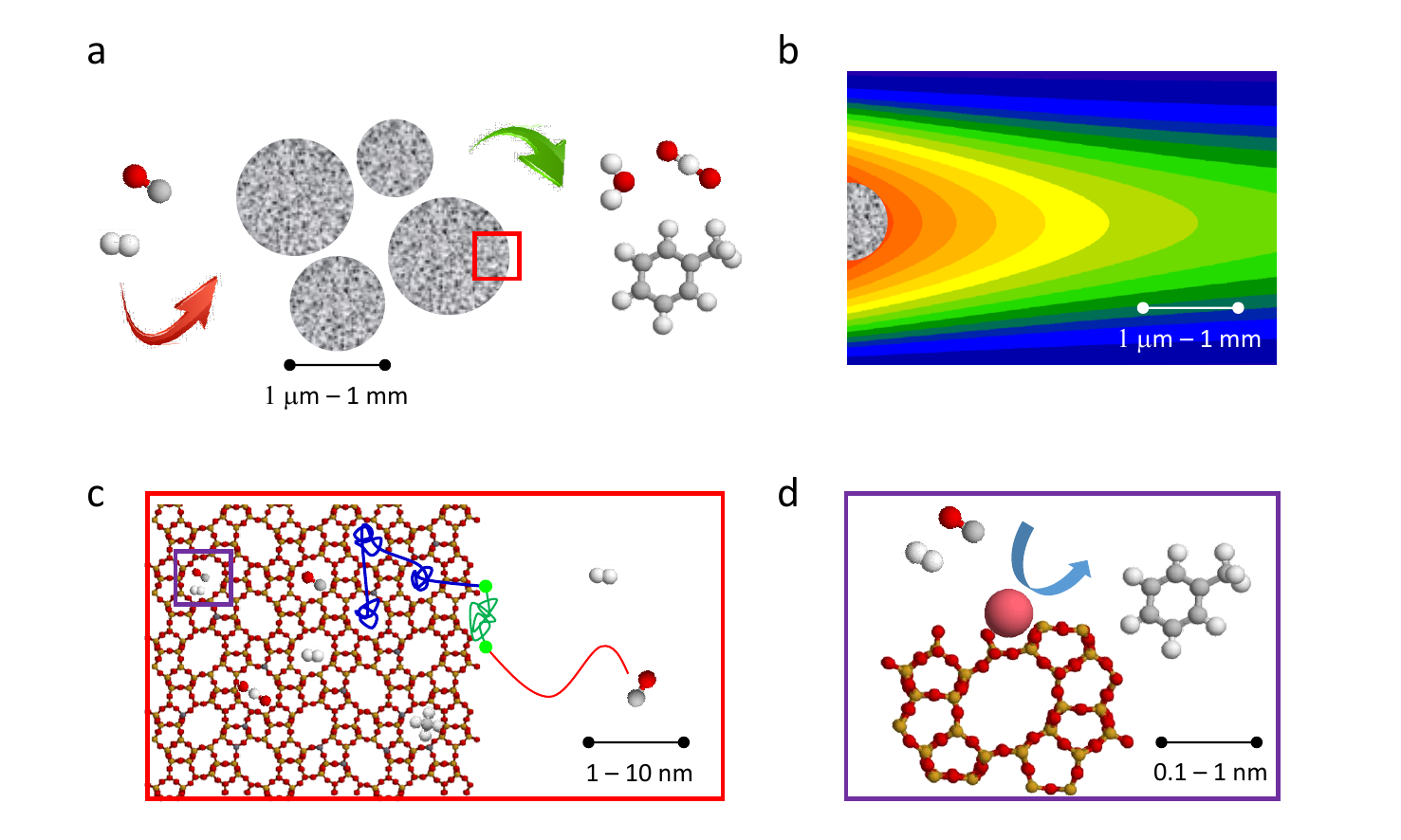}
\caption{\textbf{Multiscale theoretical approach to an engineering process in porous materials.} 
\textbf{(a)} Schematic view of the use of nanoporous materials for heterogeneous
adsorption, separation or catalysis.
Here, a zeolite is considered as an example but the scheme presented is valid 
for any nanoporous material types including mesoporous alumina, silica, active
carbons, \etc.
The competitive adsorption and/or reaction between different molecules occur at
active sites on the material internal surface.
Typically, the nanoporous material is shaped into pellets which form a granular
material with dimensions that should optimize the transport of  molecules within
the material but also to optimize access to the porosity.
The description of adsorption, separation or catalysis requires to develop a
multiscale approach which includes at least the three characteristic time and
length scales schematically depicted in b, c, and d. 
\textbf{(b)} Transport at the macroscopic scale ($\sim$1 $\mu$m -- 1 mm) is
usually described using hydrodynamics with methods known as computational fluid
dynamics tools. 
\textbf{(c)} At the mesoscopic scale ($\sim$1--100 nm), molecular dynamics and
other molecular or mesoscopic simulation tools such as Monte Carlo techniques, 
dissipative particle dynamics, free energy calculations, \etc.\ are used to 
describe adsorption and dynamics of molecules at the surface of the 
porous material and within its porosity. 
\textbf{(d)} At the microscopic/molecular scale  ($\sim$ 0.1--1 nm), quantum
mechanical approaches, which include \textit{ab-initio} techniques and density
functional theory calculations, allow describing at the atomic scale the
molecular reactions that correspond to the elementary adsorption, separation
and/or catalytic steps.}
\label{fig:FIG1.1}
\end{figure}

\bluelork{In order to set the frame of the present review, let us consider the transport of a fluid in a multiscale porous material as depicted in Fig. 1. While we consider here the adsorption, separation and/or catalysis of fluid molecules in such systems as an illustrative example, we stress that the depicted situation pertains to all fields where the transport of a fluid in a porous material containing nanoporosity is considered. This includes  situations such as transport in geological porous media and porous membranes (where many phenomena couple to lead to complex non-reactive or reactive transport). As a result, while the following introductory discussion is focused on a specific class of systems, we believe that the concepts described and phenomena covered are ubiquitous in all the fields listed above. The complexity of describing and predicting transport in a multiscale porous material stems from two main difficulties (\cref{fig:FIG1.1}) 
\cite{hansen_multiscale_2012}.}
First, such processes involve a variety of complex phenomena that pertain to
chemistry (chemical reaction or adsorption at active sites), chemical physics
(local thermodynamic equilibria and molecular separation in the vicinity of the
active sites), physics (diffusion and possible coupling between transport and
adsorption properties), and chemical engineering (transport properties and heat
transfer at the scale of the adsorption device/catalytic reactor).
Second, while these phenomena occur at different scales -- typically, from the
molecular to the engineering scales -- they are coupled so that describing the
overall process requires the development of robust upscaling strategies.
From a theoretical viewpoint, available approaches to adsorption and reaction
in a nanoporous material rely on the decomposition of the problem 
into the three following steps. 
(1) \textit{adsorption or chemical reaction at the surface of the nanoporous
material.}
These processes are usually investigated using the so-called \textit{ab-initio}
methods, which solve the electronic structure problem based on the
underlying principles of quantum mechanics.
Such approaches --- which in principle can be employed to provide an accurate
prediction of the system --- are not discussed within this work as they fall out
of its scope and have been thoroughly reviewed
recently.\cite{santiso_multi-scale_2004, xie_insight_2019, li_electronic_2021}
(2) \textit{Thermodynamic equilibrium and transport properties  in the
nanoporous material}.
These phenomena can be investigated using the large panel of numerical methods 
that are coined as molecular simulation.
Such molecular modeling strategies, which are at the heart this review, allow simulating both 
adsorption equilibrium and transport of pure molecular components and their 
mixtures in a porous material. 
(3) \textit{Process efficiency at the macroscopic scale.}
The process yield under given thermodynamic and hydrodynamic conditions can be
estimated using
mesoscopic, coarse-grained or continuum models such as computational fluid
dynamics, as we will discuss in this review both for equilibrium and non-equilibrium transport conditions.

From a modeling point of view, significant efforts are usually devoted to
understanding chemical processes (reactive or non-reactive) at the molecular
scale and the process efficiency at the macroscopic
scale.\cite{keil_molecular_2018, detmann_modeling_2021, ladd_reactive_2021}
On the other hand, in practice, the thermodynamics and dynamics of the molecular
species in the nanoporous solids are often treated only in an empirical manner
despite the availability of fairly simple molecular simulation approaches.
This is in part due to the lack of simple robust approaches to upscale
the parameters obtained at a smaller scale into coarse-grained calculations at a
larger scale.
Yet, molecular simulation techniques are very efficient at providing, for a
reasonable computational cost, adsorption equilibrium and 
transport coefficients in nanoporous
materials.\cite{smit_molecular_2008,bukowski_connecting_2021}
Such methods, which include but are not limited to Molecular Dynamics and Monte Carlo
techniques,\cite{frenkel_understanding_2001,allen_computer_2017} are much less
fraught with theoretical difficulties than quantum
mechanics\cite{santiso_multi-scale_2004, balbuena_molecular_1999} since
electron-level details are not considered and the set of atoms
or molecules are treated as classical objects obeying classical statistical 
physics.\cite{barrat_basic_2003,chaikin_principles_1995,hansen_theory_2013,
mcquarrie_statistical_2000}
Thanks to this simplification, molecular simulations involving thousands or
even hundreds of thousands of atoms are possible and timescales of up to
hundreds of nanoseconds and even beyond can be reached. 
Molecular simulation methods also include mesoscopic coarse-grained methods such
as Brownian Dynamics and Dissipative Particle Dynamics as well as simple random
walk models which allow probing even larger systems and on much larger
timescales.
Finally, other strategies that will be also discussed in this review
include free energy techniques, lattice gas models, and upscaling strategies.\\

\bluelork{
This review aims to be a comprehensive, critical, and accessible document of general interest to the chemistry, physical chemistry, chemical physics, and geologists communities. In particular, researchers working in domains including but not limited to adsorption, separation, catalysis, batteries, micro/nanofluidics are among the targeted audience. We also believe that engineering communities (e.g.\ chemical engineering, soil engineering, petrochemical engineering) would benefit from such an up-to-date review on transport in nanoporous materials.} The specificities of this review can be listed as follows. 
(1) We present the state of the art on the fundamentals of self-diffusion and transport in nanoporous materials. As an important feature, the present review
adopts a multiscale view of the problem under consideration: both fundamental
aspects and effective strategies are considered. 
(2) Considering the time/length scales involved in diffusion and transport in
porous materials (from in pore dynamics to dynamics within the pore network), we
treat all relevant approaches from molecular simulation (e.g.\ molecular
dynamics, free energy calculations) to mesoscopic approaches (e.g.\ effective
theories, random walk techniques, pore network models) but also upscaling
strategies.
(3) We discuss the determination of transport barriers, including
surface/geometrical barriers and free energy barriers as encountered when phase
transitions occur. 
Their role in overall transfer and process rates is also discussed.
Finally, we also cover available approaches to include these aspects
in a mesoscopic fashion such as hierarchical simulations.
(4) For each approach, we provide the basic underlying concepts in a way that
helps the reader to get familiar with such fundamental aspects without being an
expert.
While the different strategies considered in this review all rely on
similar physical formalisms (statistical mechanics ground), we feel that no
review encompassing these methods in a multiscale picture is available to date. 
(5) This contribution covers both self-diffusion and transport aspects and is not restricted to a given type of materials but deals with the broad class of
nanoporous materials including -- but not limited to -- zeolites, active carbons, Metal Organic Frameworks, ordered and disordered nanoporous oxides.
In this context, the present review considers both small and large nanoporous
materials with pore sizes ranging from the subnanometer scale to tens of
nanometers.
(6) Last but not least, by considering both ordered and disordered materials, we
feel that this review connects two important parts of the literature on fluids in
nanoporous solids: 
    ($i$) dynamics at the pore scale, which is mostly investigated using simple
    pore geometries (e.g.\ cages, channels, slits, etc.)
and ($ii$) transfer through porous structures where morphological and
    topological features of the host network are known to impact the overall,
    macroscopic dynamics.

The present review covers important insights as well as methodological developments
in the very broad field of transport of simple fluids in the vicinity or
confined within nanoporous materials.
It is thus  complementary to the existing literature as it covers
additional aspects compared to available
reviews.\cite{bhatia_molecular_2011, smit_molecular_2008, karger_diffusion_2016,
schneider_transport_2016, huber_soft_2015, karger_transport_2015,
bukowski_connecting_2021, bocquet_nanofluidics_2010, gubbins_role_2010}
First, we cover all transport modes (diffusion, advection, and convection). In particular, we wish to connect the molecular mechanisms observed in
nanoporous media to classical approaches such as Poiseuille law, Darcy law,
Darcy-Forchheimer law, etc. 
Second, we include additional aspects connected to transport in nanoporous media
such as liquid imbibition and activated
transport across interfaces. 
Third, while the impact of disorder at the pore scale and of pore network
morphology/topology are discussed in available reviews, we here cover
additional aspects by discussing coarse-grained descriptions such as
Lattice--Boltzmann calculations and dynamic mean-field DFT for instance.
Finally, we also treat physical models such as free volume theories,
Intermittent Brownian Motion, De Gennes narrowing, etc., which so far have not
been discussed in available reviews on similar topics. 
As a final remark on this review and its content, to allow interested readers to
deepen their understanding of the underlying scientific and practical aspects,
we decided to provide a limited set of references only for every covered
concept.
However, while we omit many relevant works in our reference list for
the sake of clarity and readability, we acknowledge that a large number of
references on each given effect/phenomenon are available (such missing
references can be found in other available reviews cited in this introduction). 

To provide a comprehensive overview, the remainder of this work is organized as follows. First, we consider the self-diffusion and tracer diffusion of molecules in systems at equilibrium, where transport is driven by local density/concentration gradients as induced by thermal fluctuations. While the two concepts are often invoked in different contexts, they refer to the same phenomenon corresponding to the diffusion of a tagged, i.e.\ identified, molecule in a given medium. Second, we consider all transport mechanisms as observed when applying a thermodynamic gradient to the system. In general, to avoid any confusion, transport will be used as a generic term to encompass all transport mechanisms (diffusion, advection, etc.). In Section 2, we discuss the fundamentals of transport in porous media. We introduce the different transport coefficients relevant to the dynamics of simple fluids confined in nanoporous materials but also important concepts such as tortuosity, diffusion factor, surface energy barriers, etc. In Section 3 we review the state of the art of molecular simulation and other statistical mechanics approaches applied to the problem of self-diffusion and tracer diffusion in nanoporous materials. In this section, we only consider diffusion at the pore scale. In contrast, Section 4 treats the problem of diffusion at the porous network scale. In both Sections 3 and 4, in addition to molecular simulation approaches, we also discuss other available methods such as hierarchical simulations (in which one probes the dynamics in a precalculated  free energy landscape), random walk approaches, etc., but also  theoretical formalisms based on statistical mechanics, such as Intermittent  Brownian Dynamics, to tackle the problem of dynamics in heterogeneous media. In Sections 5 and 6, we review the application of molecular simulation and theoretical approaches to the problem of transport in porous media as induced by a thermodynamic gradient. After reviewing the different mechanisms that pertain to transport in porous media, we consider both stationary transport under a constant driving force and transient transport such as in situations dealing with mass uptakes, adsorption/desorption kinetics, \etc. Section 5  first deals with gradient-driven transport at the pore scale while Section 6 considers the problem of gradient-driven transport at a much larger scale with techniques such as lattice and pore network models but also Lattice Boltzmann calculations.  In Section 7, we provide  some concluding remarks and perspectives for future work.

%%%%%%%%%%%%%%%%%%%%%%%%%%%%%%%%%%%%%%%%%%%%%%%%%%%%%%%%%%%%%%%%%%%%%%%%%%%%%%%%%%%%%%%%
% Fundamentals of diffusion and transport in Nanoporous Materials

%!TeX spellcheck = en_US
%!TeX encoding = utf8
%!TeX program = pdflatex
%!TeX root = ../manuscript.tex

\section{General principles of diffusion and transport}
\label{chap:fundamentals}

\subsection{Different frameworks}
\label{sec:fundamentals_frameworks}

\subsubsection{Fluid dynamics}
\label{sec:fundamentals_frameworks_hydrodynamics}

Fluid transport in general as well as in porous media can be described using
the Navier--Stokes equation for momentum transfer, which writes
for an incompressible fluid (constant density $\rho$)~\cite{landau_fluid_2013}:
\begin{equation}
    \rho \Big ( \frac{\partial \textbf{u}}{\partial t} + \textbf{u} \cdot
    \nabla \textbf{u} \Big ) =
    - \nabla P + \eta \nabla^2 \textbf{u} + \textbf{f}
\label{EQ3.1}
\end{equation}
\bluelork{This equation,  which is valid when the system can be described as a continuous medium, corresponds to  Newton’s second law. It relies on key assumptions that define its applicability. In particular, in its form given in \cref{EQ3.1}, it assumes that the fluid is incompressible (i.e.\ its density is constant) and Newtonian (i.e.\ its viscosity is independent of the applied shear rate) and that the fluid can be treated as homogeneous fluid particles (therefore omitting strong structuring of the fluid especially near interfaces). Moreover, as will be discussed later in this review, the use of Eq. (1) assumes that the momentum relaxation timescale is much larger than the molecular relaxation timescale.} The left
hand side term in the equation above is the fluid particle acceleration $\textbf{a} =
\textrm{d}\textbf{u}/\textrm{d}t$ at a given time $t$ multiplied by the fluid
mass density $\rho$, while the right hand side term corresponds to the sum
of the forces per unit volume applied to the fluid particle ($\textbf{u}$ is
the fluid particle velocity). In more detail, the left hand side term is the total
time derivative of the fluid particle momentum which gives rise to a linear and
a non-linear term:
\begin{equation}
 \rho \frac{\textrm{d}\textbf{u}}{\textrm{d}t} =  \rho \Big ( \frac{\partial
 \textbf{u}}{\partial t} + \textbf{u} \cdot \nabla \textbf{u} \Big )
\label{EQ3.2}
\end{equation}
While the linear term simply is a time-dependent acceleration describing
the change in the fluid velocity with time, the second term is a spatial
contribution which describes the effect of flow field spatial heterogeneity on
the fluid velocity. The forces applied to the fluid particle
include the viscous forces that arise because of velocity heterogeneity in the
fluid ($\eta$ is the shear viscosity), any pressure gradient $\nabla
P$, and any other body forces $\textbf{f}$, such as gravity, that could be applied to
the fluid. In the context of fluid transport in nanoporous materials, the gravity
term is usually omitted as the corresponding force is negligible compared to
the force induced by the pressure gradient and the intermolecular forces
between the fluid molecules and those with the porous material. The
Navier--Stokes equation is very general so that it can be used to describe
a broad range of phenomena including laminar, convective and turbulent flows
for both compressible and incompressible flows.
In particular, as will be discussed in \cref{sec:selfColTransportD},  an
incompressible fluid at a small Reynolds numbers obeys Poiseuille's law\cite{bird_transport_2002} 
(equivalent to Darcy's law\cite{osullivan_perspective_2022}), where the flow is purely viscous.

There are several dimensionless numbers that  characterize fluid
transport in general and in porous materials in
particular.\cite{civan_porous_2011, bird_transport_2002}
The Reynolds number $\textrm{Re}$ is a dimensionless quantity that describes
the ratio of the inertial (convective) contribution to the viscous force term
in the Navier--Stokes equation given in \cref{EQ3.1}:
$\textrm{Re} = \rho u L/ \eta = u L/\nu$ where $L$ is the characteristic length
in the system and $\nu = \eta/\rho$ is the kinematic viscosity of the fluid
(with $\rho$ the fluid density).
\bluelork{For small Reynolds numbers, the flow is laminar which means that liquids flow in parallel layers with no strong mixing between adjacent layers having different velocities. On the other hand, for large Reynolds numbers, the flow is turbulent as a result of the strong inertia contributions which arise when the viscous force is not sufficient to overcome the convection flow between adjacent fluid particles. For intermediate Re, the system is in the laminar-turbulent transition regime where convection makes the laminar flow rough and then unstable (and eventually turbulent). The critical Reynolds numbers describing the boundaries between the different regimes depend on the geometry in which the fluid flows. Typically, as examples, the flow is laminar for $\textrm{Re} < 20$ and fully turbulent for $\textrm{Re} > 1500$ for a falling fluid film \cite{bird_transport_2002} and laminar for $\textrm{Re} < 10$ and fully turbulent for $\textrm{Re} < 2000$ for a fluid flowing in a packed bed \cite{Rhodes1989}.} In porous media, depending on the typical pore size $D_\mathrm{p}$, the flow can
be laminar or turbulent as predicted using the Reynolds number expression
with the characteristic length $L$ taken equal to $D_\mathrm{p}$.
However, in practice, for nanoporous materials typically used in adsorption,
separation and catalysis, it is easy to show that
$\textrm{Re} \ll 1$.
For instance, in the case of water involved in processes at room temperature,
$\rho \sim 1 \,\si{\gram/\centi\meter^3}$ and 
$\eta = 1 \,\mathrm{cp} = 1\, \si{\milli\pascal.\second}$,
confined in pores of a size $D_\mathrm{p}$ in the range  1--100 nm, we find 
$\textrm{Re} < 10^{-4}$.\footnote{
	\label{footnoteRe}
	Let us consider the Poiseuille flow in a slit pore induced by a pressure gradient
	$\nabla P$ corresponding to a pressure drop
	$\Delta P = 100\,\si{\bar} = 10^7\,\si{\pascal}$ over a length $L = 0.01\,\si{\meter}$.
	For such a viscous flow, as described later in this review,	the average fluid flow velocity follows as $u = D_\mathrm{p}^2 \Delta
	P/(32 L \eta)$.	Using $\eta = 1$ $\si{\milli\pascal.\second}$ for water at
	room temperature, this yields $\textrm{Re} = \rho u D_\mathrm{p}/\eta \sim
	10^{-4}$ for $D_\mathrm{p} = 100$ $\si{\nano\meter}$.
}
As a result, for all examples considered in this review, we will always assume
that the flow in nanoporous materials is laminar unless stated otherwise.

There are other dimensionless numbers that are of utmost importance when
investigating transport in porous materials.  
\bluelork{Of particular relevance for the
present review, the Mach number describes the ratio of the flow velocity $u$ and sound velocity $c$, \ie.\ $\textrm{Ma} = u/c$ (typically, the sound velocity is $c \sim 350$ m/s in air and 3500 m/s in water at room temperature).} This dimensionless number
indicates whether the fluid flow can be treated as
incompressible or compressible (even when a compressible fluid such as a gas is
considered). For small $\textrm{Ma}$, typically $\textrm{Ma} < 1$, the flow
velocity is smaller than the sound velocity, \ie.\ the velocity of a
compression wave, so that a fluid element can be treated as incompressible
because it is not affected by the compression wave that its displacement
generates.
In contrast, for large $\textrm{Ma}$, typically $\textrm{Ma} \gg 1$, the 
compression wave generated by the fluid particle motion affects its own motion
so that the fluid must be treated as compressible.
In practice, even when transport of a compressible fluid (gas or vapor for instance) in nanoporous
materials is considered, the fluid flow is usually slow compared to the velocity of
sound such that $\textrm{Ma} \ll 1$.
This is an important result as this implies that the Navier--Stokes equation
given in \cref{EQ3.1} --- which was written for an incompressible fluid ---
remains
valid in many cases. For instance, taking N$_2$ flow at room temperature, the
sound velocity is $c \sim 350\,\si{\meter/\second}$,\footnote{
	\label{footnoteMa}
	For a fluid, the sound velocity $c$ is given by the Newton--Laplace equation
	which involves the stiffness coefficient $K_s$ (\ie.\ the bulk elastic
	modulus for fluids) and the mass density $\rho$, $c = \sqrt{K_s/ \rho}$.
	In the case of an ideal gas at a given pressure $P$, $K_s = \gamma_a P$ so that
	$c = \sqrt{\gamma_a P/\rho}$ where $\gamma_a = C_p/C_v$, the adiabatic index,
	is equal to the ratio of the heat capacities at constant pressure and constant volume
	($\gamma_a = 1.4$ for a diatomic ideal gas such as N$_2$). Using the ideal gas law, 
	$P = \rho/M RT$ where $M$ is the molar mass, $R$ the ideal gas constant, and $T$
	the temperature, we obtain $c = \sqrt{\gamma R T/M} \sim 350$ m/s for N$_2$ 
	at room temperature.}
which leads to $\textrm{Ma} \sim
10^{-3}$.
Finally, another important dimensionless number that should be considered when dealing
with fluid transport in nanoporous media is the capillary number $\textrm{Ca}$.
The latter describes the ratio of the viscous forces with
respect to the force corresponding to the Laplace pressure across a gas/liquid
interface (or any fluid/fluid interface in general such as when fluid mixtures
are considered). In more detail, $\textrm{Ca} \sim \eta u / \gamma$ where we
recall that $\eta$ is the fluid dynamical viscosity, $u$ its velocity,
and $\gamma$ the fluid surface tension. While the flow is dominated by
capillary forces for small Ca, the flow is dominated by viscous forces for
large Ca.

\subsubsection{Diffusion, advection, convection, reaction}
\label{sec:diffusion-advection}

Let us consider the transport of tracers (molecules) in any porous material --- by
tracers, we refer either to solute molecules in a liquid or to a single tagged
molecule that is of the same nature as the liquid but we distinguish this
single molecule to probe its motion. Here, we restrict ourselves to the case of
passive tracers in a prescribed flow field $\textbf{u}(\textbf{r})$.
In a very general way, the transport of
such tracers can be described using the reaction-diffusion-advection equation
which describes the change in the tracer density over time~\cite{bird_transport_2002}:
\begin{equation}
 \frac{\partial c}{\partial t} =
 \nabla \cdot (D_\mathrm{s} \nabla c) - \nabla \cdot (\textbf{u}c) + R
\label{EQ3.3}
\end{equation}
where $c$ is the tracer concentration at a time $t$, $D_\mathrm{s}$ is the
self-diffusion coefficient, $\textbf{u}$ is the fluid velocity and $R$ is a
source/sink term that adds or removes tracers from the system ($R$ can be
either positive or negative).
In more detail, this expression is a mass balance
equation where the change in the concentration $c$ over an infinitesimal time
step is the sum of (1) the diffusion contribution,
$\nabla \cdot \left(D_\mathrm{s}\nabla c\right)$, which disperses the tracers in the
fluid, (2) the change because of the fluid motion carrying away the
concentration with a velocity $u$, $\nabla \cdot (\textbf{u}c)$, and (3) a
sink/source term, $R$, such as a chemical reaction
in the case of catalysis which increases or decreases locally the
concentration of tracers.

\bluelork{Here, it must be made clear that the various dispersion contributions to \cref{EQ3.3} arise from very different mechanisms. Acknowledging that conventions and definitions vary from one scientific field to another, we introduce the following concepts that will be used and discussed in detail in this review. The reaction term in the above equation describes either a chemical reaction that would consume or create tracer
molecules or a physical adsorption/desorption contribution which can be seen as a non-chemical reaction that changes locally the concentration of molecules. The diffusion term --- as will be discussed in \cref{sec:onsager} --- is due to
diffusion which occurs both under uniform chemical potential (such as the self-diffusion or tracer diffusion probed using pulsed field gradient nuclear magnetic resonance or quasi-elastic neutron scattering) and non-uniform chemical potential (such as
transport diffusion measured when the system is subjected to a thermodynamic gradient).
Advection refers to the different mechanisms where the fluid is set in motion by thermodynamic gradients (temperature, pressure, concentration, \etc.), whereas convection here refers to the transport induced by the non-linear acceleration term (inertial effects), which was introduced when discussing the Navier--Stokes equation above. In this context, we note that the term convection often also refers to the collective motion of fluid regardless of the flow mechanism or nature.} 

Assuming for now that the Fickian diffusion term is known
(see \cref{sec:selfColTransportD}), the derivation of the
advection-diffusion-reaction equation given in \cref{EQ3.3} is
straightforward. A simple mass balance equation at any spatial position implies
that the change over time in the local concentration is related to the sum
of the incoming/outgoing tracer fluxes $J$ and the sink/source term:
\begin{equation}
 \frac{\partial c}{\partial t} = - \nabla \cdot J + R
\label{EQ3.4}
\end{equation}
$J$ is the sum of the diffusive flux, \ie.\ $J_\textrm{d}= - D_\mathrm{s}
\nabla c$,
and the advective flux, \ie.\ $J_\textrm{a}=  \textbf{u}c$ and $R$ is the
rate of creation/removal of tracers ($\nabla \cdot$ is the divergence
operator). \bluelork{\Cref{EQ3.4} as presented here is identical to \cref{EQ3.3}, but written in a more compact form to gather the diffusion and advection term together.} 
In another common case of self-diffusion coefficient being homogeneous in space, the advection-diffusion equation given in
\cref{EQ3.3} can be recast in a simpler form:
\begin{equation}
\frac{\partial c}{\partial t} =
D_\mathrm{s} \nabla^2 c - \nabla \cdot (\textbf{u}c) + R
\label{EQ3.5}
\end{equation}
However, in general, the self-diffusion coefficient depends on the local environment (for instance the local pore size in a
disordered porous medium, the crowding by other constituents or the ``self-crowding'' by other, non tagged molecules of the same nature as the tracer) so that the most general form of the
advection-diffusion equation given in \cref{EQ3.3} must be kept.
It is interesting to note the
analogy between the advection-diffusion equation in \cref{EQ3.5} and
the Navier--Stokes equation for the momentum transfer in \cref{EQ3.1}.
While the former corresponds to a mass balance condition for the tracer
diffusion, the latter describes the momentum conservation condition within the
flowing fluid.
In principle, the reaction-advection-diffusion equation could be
used to model reaction and transport in porous media. However, in practice,
the resolution of such equations remains only at the qualitative level with
input parameters that cannot be derived from molecular thermodynamic and
dynamical coefficients. This is due to the fact that the different terms ---
\ie.\ reactive, diffusive, and advective contributions --- in this equation are
strongly coupled at the molecular scale so that any set of effective parameters
will fail to describe the complexity and richness of the phenomena occurring in
the nanoscale porosity (even when non-reactive transport is considered).

The advection-diffusion equation emphasizes the competition/combination of
diffusive and advective transport in porous media.
This competition is also characterized by the Peclet number.
This dimensionless number, $\textrm{Pe}$, describes the ratio
of the advective fluxes, $J_a$, to the diffusive fluxes, $J_d$,
$\textrm{Pe} = J_a / J_d = u D_\textrm{p}/D_\mathrm{s}$ where $u$ is the fluid
velocity, $D_\mathrm{s}$ is the self-diffusion coefficient, and $D_\textrm{p}$
is the characteristic length scale taken equal to the pore size.
$\textrm{Pe}$ therefore describes the relative efficiency of diffusion and
advection to disperse tracers within the porous medium.
%As already discussed in the introduction, the
%height equivalent to a theoretical plate (HETP) typically employed in chromatography %is similar to the Peclet number as it is corresponds to the column element size (a %so-called plate of a thickness HETP) where the adsorption/diffusion time in the %direction perpendicular to the flow, $t_d \sim \mathrm{HETP}^2/D_\mathrm{s}$, is %comparable to the time needed to advectively transport the liquid over the same size %HETP, $t_a \sim \textrm{HETP}/u$~\cite{martin_new_1941,holcapek_handbook_2017}.
%Going back to the Peclet number,
For very small $\textrm{Pe}$, \ie.\ when the
advection contribution can be neglected (such as when no thermodynamic gradient
is applied to induce transport or when the flow induced by the thermodynamic
gradient is negligible compared to the diffusive flow), the advection-diffusion
equation is equivalent to the well-known ``equation of porous media'' for a
purely diffusive regime and without any chemical reaction ($R = 0$)~\cite{civan_porous_2011}:
\begin{equation}
 \frac{\partial c}{\partial t} =
 \nabla \cdot \big \lbrack D_\mathrm{s}(\vec{r}) \nabla c \big \rbrack
\label{EQ3.6}
\end{equation}
where $D_\mathrm{s}(\vec{r})$ is the local, environment-dependent self-diffusion
coefficient.
In the context of nanopores, owing to the very small pore size considered,
transport mostly occurs through diffusion, \ie.\ $\textrm{Pe} \ll 1$ ---
typically, both the advective flow rate  $u \sim \nabla P$ and the
self-diffusion coefficient $D_\mathrm{s}$ decrease with the pore size
$D_\textrm{p}$ but
there is always a critical pressure gradient $\nabla P_c$ below which
$D_\textrm{p} u \ll D_\mathrm{s}$ (\ie.\ $\textrm{Pe} \ll 1$).
%For very small Pe numbers, \ie.\ negligible advection, diffusion/reaction
%efficiency in the porous catalyst can be described using the
%Thiele modulus as introduced in the introduction. More in detail, the Thiele
%modulus $\chi = \Lambda  [k_v/D_{\textrm{eff}}]^{1/2}$ is a dimensionless
%number which compares the intrinsic rate of the chemical reaction, $k_v$, and
%the effective diffusion coefficient of the reacting species in the porous
%catalyst $D_{\textrm{eff}}$ ($\Lambda$ is the characteristic length which is
%taken equal to the catalyst particle size)~\cite{sahimi_statistical_1990,gheorghiu_optimal_2004}.

\subsection{Onsager theory of transport}
\label{sec:onsager}

\subsubsection{Transport coefficients}

We now briefly discuss Onsager's phenomenological theory of transport as it allows  introducing key concepts
for the transport of molecules in nanoporous media -- namely the self, collective and
transport diffusivities~\cite{chaikin_principles_1995,groot_non-equilibrium_1962}.
Let us consider a system which is characterized at the
macroscopic scale by a set of thermodynamic extensive quantities $X_i$ (energy,
volume, number of molecules). 
The thermodynamics of the system is governed by the entropy-state function of
these extensive variables which is maximum at equilibrium:
$S = S(X_i)$~\cite{callen_thermodynamics_1985,pottier_nonequilibrium_2010}.
The differential of the entropy, known as the Gibbs relation, introduces the
conjugated intensive variable $F_i$, the so-called affinity, of each extensive
quantity $X_i$:
\begin{equation}
 \textrm{d}S = \sum_i \frac{\partial S (X_i, X_j, \dots)}{\partial X_i} \bigg
 \rvert_{j \neq i}  \textrm{d}X_i = \sum_i F_i \hspace{1mm} \textrm{d}X_i
\label{EQ3.7}
\end{equation}
where the bar ``$\rvert$'' in the partial derivative indicates that all other
quantities $j \neq i$ are kept constant.
For most practical situations, the extensive quantities $X_i$ are the internal
energy $E$, the volume $V$ and the number of molecules $N_k$ for each chemical
component $k$ in the system, which leads to $S = S(E, V, N_1, ..., N_k)$ and the
following Gibbs relation:
\begin{equation}
 \textrm{d}S = \frac{1}{T} \textrm{d}E + \frac{P}{T} \textrm{d}V -
 \sum_{i = 1}^k \frac{\mu_i}{T} \textrm{d}N_i
\label{EQ3.8}
\end{equation}
where we used that $\partial S/\partial E = 1/T$, $\partial S/\partial V = P/T$,
and $\partial S/\partial N_i = -\mu_i/T$ for $i = \lbrace 1, ..., k \rbrace$ 
to define the different affinities $F_i$. 

Transport phenomena induced by thermodynamic gradients such as $\nabla T$,
$\nabla P$, $\nabla \mu$, \etc.\ in a given thermodynamic system can be
described using the second law of thermodynamics under the local thermodynamic
equilibrium  approximation. The  latter assumption states that, even under
non-equilibrium conditions, the system can be subdivided into mesoscopic volume
elements which are (1) small enough to assume that local thermodynamic
properties do not vary within these elements but
(2) large enough to be treated like thermodynamic subsystems.
Typically, the size $\lambda$ of these mesoscopic elements should be such that
the change $\Delta \chi$ in a given thermodynamic property $\chi$ due to the
gradient $\nabla \chi = \Delta \chi/\lambda$ is smaller than the 
thermodynamic fluctuations at equilibrium $\delta \chi$ \ie.\ $\Delta \chi /\chi < \delta \chi /\chi \ll 1$
(the last part of the inequality states that the equilibrium fluctuations must
be small enough to assume that the system can be treated using macroscopic
thermodynamics)~\cite{pottier_nonequilibrium_2010}. Under such local
thermodynamic equilibrium, the change in any thermodynamic quantity $\chi$ can
be described using a variable that depends on spatial coordinates $\textbf{r}$
and time $t$, \ie.\ $\chi(\textbf{r},t)$ (typically, $\chi$ is the temperature,
fluid density, pressure, \etc.).
As discussed in Ref.~\citen{barrat_basic_2003}, the
local thermodynamic equilibrium condition requires a time scale separation with 
a much shorter molecular relaxation time (typically, $\tau_m \sim 1\,\si{\pico\second}$)
compared to the macroscopic evolution time governed by the local thermodynamic gradient
(typically, $\tau_M \gg 1\,\si{\pico\second}$).
Under thermodynamic equilibrium, the maximization of the entropy implies that
the affinities are homogeneous throughout the system, \ie.\ $\nabla F_i = 0 \,
\forall i$.\footnote{
	\label{maxentropy}
	 Let us consider a thermodynamic system subdivided into two subsystems 1
	 and 2. For each conserved quantity $X$ such as the energy $E$ or number of
	 molecules $N$, the sum of their contribution in each subsystem is
	 constant, \ie.\ $X_1 + X_2 = 0$, which implies that $\partial X_1 = -
	 \partial X_2$.
	 Moreover, the equilibrium condition is given by the entropy maximization
	 principle, \ie.\ $\partial S/\partial X_1 = \partial S_1/\partial X_1 -
	 \partial S_2/\partial X_1 = 0$ which leads to $F_1 = F_2$ since $\partial
	 S/\partial X = F$.
}
\bluelork{We recall that the different affinities $F_i$, which were introduced right after \cref{EQ3.8}, correspond to partial derivatives of the entropy with respect to the extensive variables $N$, $V$, $E$, \etc. We also indicate here that $i$ simply corresponds to an index that lists the different affinity gradients that can be applied to the system. In contrast to the situation without any affinity gradients, a non-zero affinity gradient $\nabla F_i$ will induce a flux in the conjugated extensive variable $X_i$, $J_i = \textrm{d}X_i/\textrm{d}t$. For small affinity gradients $\nabla F_i$, the flux
$J_i$ in the extensive quantities $X_i$ varies linearly with $\nabla F_i$. Note that an affinity gradient $\nabla F_i$ involves a direct flux of its conjugated
extensive variable $X_i$ but also an indirect flux of all other conjugated
variables $X_j$ with $j \neq i$, \cf. \cref{EQ3.7} and
Refs.~\citen{pottier_nonequilibrium_2010,barrat_basic_2003}):
\begin{equation}
 J_i = \sum_j L_{ij} \nabla F_j
\label{EQ3.9}
\end{equation}
where the coefficients $L_{ij}$ are called Onsager transport
coefficients. An important property of these coefficients --- arising from the
time reversibility of the equations governing the motion of atoms and molecules
at the microscopic scale --- is known as Onsager reciprocal relations: $L_{ij}
=  L_{ji}\,\forall i,j$.} For the system defined above (internal energy $E$,
volume $V$ and  number of molecules $N_k$ for each chemical component $k$),
\cref{EQ3.9} leads to the following equations:
\begin{align}
 J_E = L_{EE} \nabla F_E + L_{EV} \nabla F_V + \sum_{j = 1}^k L_{EN_j} \nabla F_{N_j}  \label{EQ3.10} \\
%  J_V = L_{VE} \nabla F_E + L_{VV} \nabla F_V + \sum_{j = 1}^k L_{VN_j} \nabla F_{N_j} \label{EQ3.11} \\
   J_{N_i} = L_{N_iE} \nabla F_E + L_{N_iV} \nabla F_V + \sum_{j = 1}^k L_{N_i N_j} \nabla F_{N_j} 
 \label{EQ3.12}
\end{align}
Under specific conditions (\ie.\ when the system is subjected to a single
thermodynamic/affinity gradient), the general linear equations given in
\cref{EQ3.10,EQ3.12} simplify to a number of well-known
equations. They include Fourier law when the chemical potential/pressure are
uniform: $J_E = - \lambda_T \nabla T$ with $\lambda_T = L_{EE}/T^2$. Another
important example, particularly relevant in the context of the present review,
is the case of isothermal/isobaric system which leads to the following linear
relationship for the molecule flux:
\begin{equation}
J_N  = - L_{NN} \nabla \Big ( \frac{\mu}{T} \Big ) =  - \frac{L_{NN}}{T} \nabla \mu
\label{EQ3.13}
\end{equation}
This equation, which governs diffusion under static pressure/temperature
conditions in local thermodynamic equilibrium, is a cornerstone of diffusion in
porous media as it allows defining the different diffusivities that can be
measured in typical transport experiments: the self-diffusivity $D_\mathrm{s}$,
the collective diffusivity $D_0$, and the transport diffusivity $D_\mathrm{T}$.
Before discussing in details below each diffusion mechanism, a few comments are
in order (see also Refs.~\citen{karger_diffusion_2012,
karger_transport_2015, karger_diffusion_2016}).
On the one hand, the self-diffusion $D_\mathrm{s}$ pertains to the diffusion of
a single molecule so that it corresponds either to a very diluted solute
molecule in a liquid or to a single molecule (tracer) in a liquid, that would be
tagged to follow its trajectory. On the other hand, the
collective diffusivity $D_0$ refers to the collective displacement of the fluid
in response to a chemical potential gradient. As will be shown below, the
collective diffusivity is a very important parameter as it is formally linked
to the so-called permeability . Finally,  the transport
diffusivity $D_\mathrm{T}$ is similar  to the collective diffusivity $D_0$, but  with
the induced transport  written as a response to a concentration/density
gradient instead of a chemical potential gradient.

While these three diffusion coefficients are identical in the limit of very
dilute systems such as in gas transport or for ultraconfined molecules, they
strongly depart from each other when the molecular density becomes
non-negligible. In particular, as far as the difference between $D_\mathrm{s}$
and $D_0$ is concerned, both direct molecular interactions between molecules
and the so-called hydrodynamic interactions --- \ie.\ when the
velocity field created by a moving molecule affects the trajectory of the
others --- are responsible for the marked differences seen between these two
diffusion coefficients and their dependence on density, temperature,\etc.
 From a practical viewpoint, the three diffusion coefficients $D_\mathrm{s}$,
 $D_0$ and $D_\mathrm{T}$ are probed using different experimental techniques,
 see \eg.\ Ref.~\citen{karger_diffusion_2012}.
Typically, the self-diffusivity $D_\mathrm{s}$ can be determined using Pulsed
Field
Gradient Nuclear Magnetic Resonance (PFG-NMR) and Quasi Elastic Neutron
Scattering (QENS). PFG-NMR probes the dynamics corresponding to displacements over
microns, while QENS probes dynamics at the nm scale. As a result, in a real
material with defects (pore collapse/amorphization, impurities,
vacancies, \etc.), PFG-NMR is sensitive to these defects since it probes
displacements over lengths that are comparable to the typical distance between
them --- therefore, PFG-NMR leads to transport coefficients that can be more
than an order of magnitude lower than those measured using QENS. Yet, other
NMR methods can be used to probe dynamics over much shorter distances and
times. Finally, while QENS techniques typically probe the self-diffusivity because
hydrogen is a strongly incoherent scatterer  (large incoherent scattering
length), deuteration allows probing with the same technique the collective
diffusivity because deuterium atoms are strong coherent scatterers. As for the
transport diffusivity $D_\mathrm{T}$, macroscopic transport experiments are
needed to probe this effective transport coefficient. Similarly, as will be
illustrated in the rest of this section, different theoretical and numerical
methods can be used to probe $D_\mathrm{s}$, $D_0$ ad $D_\mathrm{T}$.
Typically, both equilibrium and non-equilibrium molecular dynamics can be used
to determine $D_\mathrm{s}$ and $D_0$. Assessing
$D_\mathrm{T}$ is more complex as it usually requires to determine $D_0$ and
the correction factor to account for the change when replacing the chemical
potential gradient by a concentration/density gradient (see discussion below).

\subsubsection{Self, collective and transport diffusivities}
\label{sec:selfColTransportD}

\bluelork{In the framework of the Onsager phenomenological theory of transport introduced above,  transport coefficients relevant to very different experimental situations can be described. In the spirit of \cref{EQ3.10,EQ3.12}, several Onsager transport coefficients can be defined from the corresponding thermodynamic driving forces directly related to the derivatives of entropy to their entropy derivatives. By extension, any other transport coefficients can be defined in the context of linear response theory by considering different driving forces. This includes important examples such as Ohm's law for electrical conductivity, Fourier's law for thermal conductivity, etc.  As far as transport in nanoporous materials is concerned, the following important   transport coefficients, which characterize the mass flow $\mathbf{J} = \rho \mathbf{v}$ -- defined as the fluid density $\rho$ multiplied by the flow velocity $\mathbf{v}$, can be introduced. They describe experimental situations corresponding to (1) the tracer or self-diffusion of particles governed by concentration gradients $\nabla c$, (2) the flow induced by a pressure gradient $\nabla P$,  (3) the flow induced by a density gradient $\nabla   \rho$, and (4) the flow induced by a chemical potential gradient $\nabla \mu$:\cite{smit_molecular_2008}
\begin{equation}\label{eq2}
\mathbf{J} = - D_\textrm{s} \nabla c
\textrm{,} \hspace {3mm}
\mathbf{J} = - \rho K \nabla P
\textrm{,} \hspace {3mm} 
\mathbf{J} = - D_\textrm{T} \nabla \rho
\textrm{,} \hspace {3mm} \textrm{and} \hspace {3mm} 
\mathbf{J} = - \rho \frac{D_0}{k_\textrm{B}T} \nabla \mu
\end{equation}
In these equations, $D_\textrm{s}$ is the self-diffusivity,  $K$ the permeability, $D_\textrm{T}$ the transport diffusivity, and $D_0$ the collective diffusivity. As will be seen below and in the rest of this document, these coefficients correspond to different physical situations as encountered in applications but they are linked through simple thermodynamic quantities obtained from simple experiments. For instance, $D_\textrm{T} \sim D_0 (\partial \mu / \partial \rho)_T$ where the proportionality factor can be inferred from the adsorption isotherm $\rho(\mu,T)$ at constant temperature $T$ as measured under static conditions \cite{smit_molecular_2008}. In what follows, we discuss in detail the different transport coefficients above with special emphasis on their connections through thermodynamic factors.}
\bluelork{While the notations above will be used throughout this review, we acknowledge that the conventions for the different diffusion coefficients differ from one field to another. In this context, we wish to highlight the forthcoming Technical Report on ``Diffusion in Nanoporous Materials with special Consideration of the Measurement of Determining Parameters'' in the framework of the IUPAC project ``Diffusion in nanoporous solids'' \cite{IUPAC2024}.}

\noindent \textbf{Self-diffusivity $D_\mathrm{s}$}. For a very dilute system,
the chemical potential $\mu$ can be expressed by making use of the ideal-gas law as
$\mu \sim k_{\textrm{B}}T \ln c$,
where $k_{\textrm{B}}$ is the Boltzmann constant, $T$ is the temperature, and
$c$ is the concentration. Upon inserting this expression in the
isothermal/isobaric version of the Onsager linear relationship given in
\cref{EQ3.13}, one obtains Fick’s first law:
\begin{equation}
J_N  = - D_\mathrm{s}(c) \nabla c
\label{EQ3.14}
\end{equation}
with $D_\mathrm{s}(c) =  L_{NN}(c)/k_{\textrm{B}} c$.
Generally, $L_{NN}$ is concentration-dependent with the dilute limit $L_{NN}(c)
\propto D_\mathrm{s} c$ where $D_\mathrm{s}$ is a molecular property
independent of $c$ ($D_\mathrm{s}$ depends on the molecular mass, i.e.\ the thermal velocity).
It is important to reckon that, as already mentioned above, diffusion occurs
even under equilibrium conditions, \ie.\ when
there is no net molecule flux that would be induced by a thermodynamic gradient
applied to the system. In many textbooks, the concentration $c$
is replaced by the density $\rho$ as it allows encompassing the situation where
the self-diffusivity refers to the diffusion of a single molecule in a pure
fluid at equilibrium under static conditions.
%In that case, the self-diffusion is induced by the density fluctuations which
%occur at any finite (i.e.\ non-zero) temperature.

Inserting Fick’s first law into the mass conservation equation, \ie.\ $\partial
c/\partial t + \nabla \cdot J = 0$, leads to Fick’s second law:
\begin{equation}
\frac{\partial c}{\partial t} = \nabla \cdot \big \lbrack D_\mathrm{s} \nabla  c \big \rbrack
\label{EQ3.15}
\end{equation}
For uniform (or weakly position dependent) self-diffusivities, \ie.\
$D_\mathrm{s}\sim$ constant, the latter equation can be recast in a simpler form:
\begin{equation}
\frac{\partial c}{\partial t} = D_\mathrm{s} \Delta c
\label{EQ3.16}
\end{equation}
As expected, this equation is strictly equivalent to the
advection-diffusion-reaction equation derived in \cref{sec:diffusion-advection}
if no reaction and advection contributions are considered. Starting from an
initial configuration at $t = 0$ where all particles are located at $\textbf{r}
= 0$, the spatial and time dependent solution of Fick’s second law is known as
the Gaussian propagator
$P(\textbf{r},t)$~\cite{klafter_probability_1994,klafter_first_2011}:
\begin{equation}
c(\textbf{r},t) =  N P(\textbf{r},t) = \frac{N
\exp{(-\textbf{r}^2}/4D_\mathrm{s} t)}{(4 \pi D_\mathrm{s} t)^{3/2}}
\label{EQ3.17}
\end{equation}
where $N$ is the total number of diffusing species (this constant is needed here
since the propagator is a quantity defined for a single molecule while the
concentration is for the $N$ molecules). The Gaussian propagator is the
probability density that a molecule moves by a vector $\textbf{r}$ over a time
$t$. The equation above states that, starting from the initial configuration,
the concentration in a position $\textbf{r}$ at a time $t$ is simply given by
the probability that molecules move by a vector $\textbf{r}$.
The average mean-square displacement
$\left<r(t)^2\right>$ of the $N$ molecules over a time $t$ can be estimated from the Gaussian propagator:
\begin{equation}
\left<r^2(t)\right> = \lvert \textbf{r}(t) - \textbf{r}(0) \rvert^2 =
\int r^2 P(\textbf{r},t) \textrm{d}\textbf{r}  = 6 D_\mathrm{s} t
\label{EQ3.18}
\end{equation}
The last equation is an important cornerstone of diffusion as it allows one, using
the Gaussian propagator formalism, to recover Einstein's formula which
relates the mean square displacement to the self-diffusivity in Brownian
motion, \ie.\ $ \left<r^2(t)\right> = 2d D_\mathrm{s} t$ where $d$ is the space
dimensionality of the system ($d$ = 3 in the example treated above). 

\begin{figure}[htbp]
	\centering
	\includegraphics[width=0.95\linewidth]{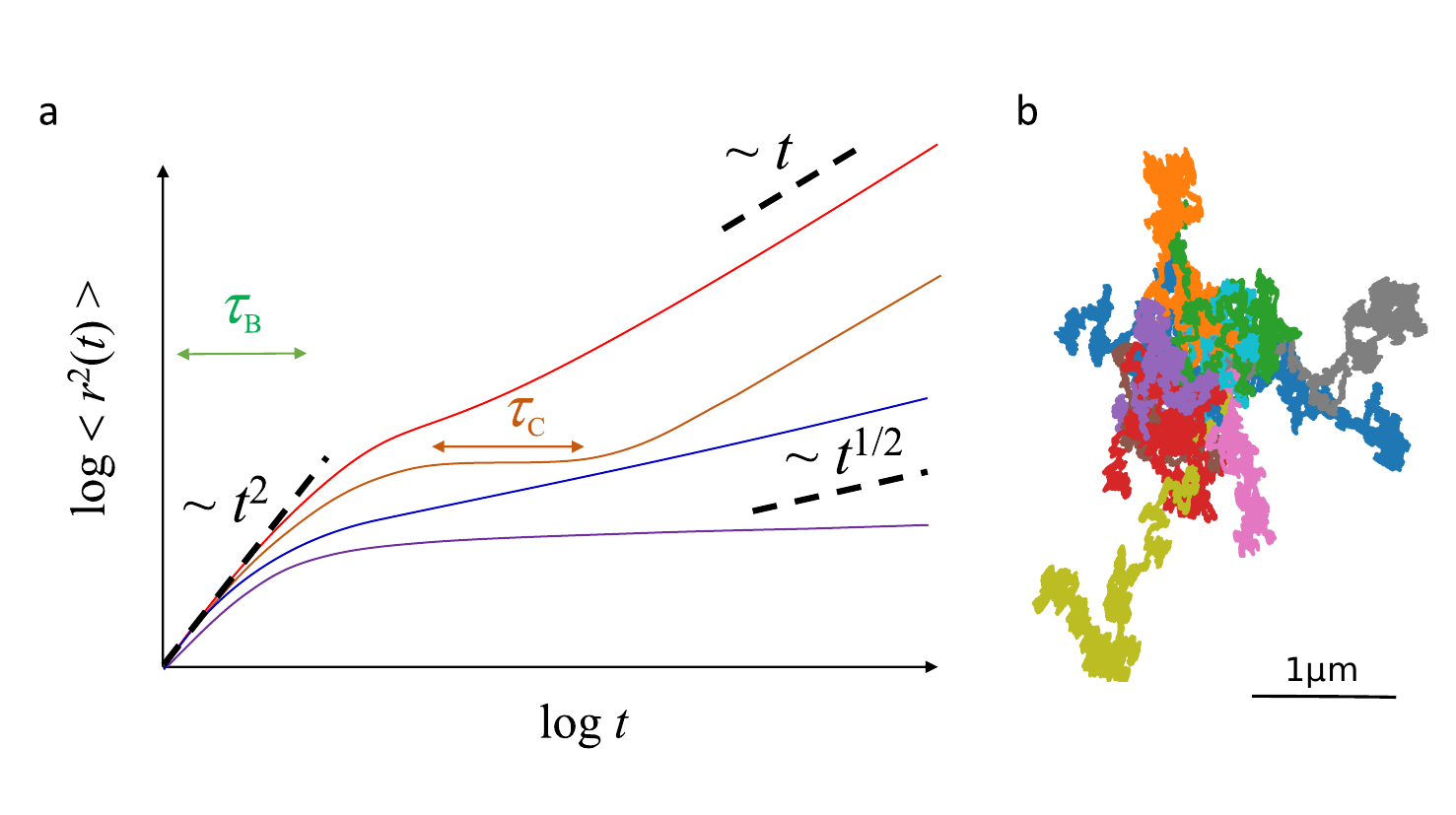}
	\caption{\textbf{Mean square displacements $\left<r(t)^2\right>$
	observed as a function of time $t$ in a log-log scale.}
		(a) In the short time limit, the regime is said to be ballistic with
		$\left<r(t)^2\right> \sim t^2$. This regime is observed until an
		average time $\tau_B$, which corresponds to the characteristic time
		before the molecule collides with other molecules. At longer times, in
		general, the molecule follows the normal diffusion regime also known as
		Fickian regime where $\left<r(t)^2\right> \sim t$. This general
		behavior is shown with the red line. In complex systems such as
		glasses, $\left<r(t)^2\right>$ can display cage effects where the
		molecule remains trapped for a time $\tau_C$ inside cages formed by
		other molecules before escaping and entering the Fickian regime (brown
		line). In case of diffusion in extremely confining porous materials,
		\ie.\ when molecules are trapped for times much larger than the
		observation  time $\tau_{obs}$, the cage
		effect appears over timescales such as $\tau_C \gg \tau_{obs}$. As a
		result, in this regime, $\left<r(t)^2\right> \sim$~constant (purple
		line). In  unidimensional confining channels, diffusion  can obey a
		non-Fickian regime known as single-file diffusion with
		$\left<r(t)^2\right> \sim t^{1/2}$ (blue line). (b) Typical
		trajectories observed over $100\,\mathrm{ns}$ for a set of 10 molecules
		in bulk liquid nitrogen at $77\,\si{\kelvin}$ (each molecule trajectory
		is shown with a different color code).}
	\label{fig:3_1}
\end{figure}

In practice, \cref{EQ3.18} is routinely used in molecular dynamics studies to
determine the self-diffusivity $D_\mathrm{s}$ from the mean square displacement
$\left<r(t)^2\right>$ as a function of time $t$. \Cref{fig:3_1} shows schematic
yet characteristic examples of mean square displacements in different physical
situations~\cite{klafter_probability_1994}. $\left<r(t)^2\right>$ scales
quadratically with time $t$, \ie.\ $\left<r(t)^2\right> \sim t^2$ in the short
time limit $t<\tau_B$; this regime corresponds to the ballistic regime where the
molecule obeys Newton's equation of motion until it collides with other molecules
at a typical time $\tau_B$. At longer times, the molecule follows the normal
diffusion regime, also known as Fickian regime, where $\left<r(t)^2\right> \sim
t$ as expected from \cref{EQ3.18}. These two asymptotic regimes are very
general but, as discussed in Ref.~\citen{klafter_probability_1994}, other
regimes can be observed. From a very general standpoint, the time dependence of
the mean square displacement can be described as a power law,
$\left<r(t)^2\right> \sim t^\alpha$, where $\alpha$ is a real number smaller or
larger than 1.
While $\alpha = 1$ corresponds to normal or Fickian diffusion, other regimes
are coined as anomalous diffusion [note that only the normal regime has an
underlying Gaussian propagator as defined in \cref{EQ3.17}]. In more detail,
$\alpha = 0$ corresponds to a localized state, $\alpha < 1$ to a subdiffusive
regime, $\alpha > 1$ to a superdiffusive regime,
$\alpha = 2$ to the ballistic regime, and $\alpha = 3$ to a fully developed
turbulent regime. Two typical examples of subdiffusive regimes, \ie.\
single-file diffusion $\alpha=1$ and confined diffusion $\alpha=0$, are shown
in \cref{fig:3_1}. 

%In the general case, the mobility $M$, which describes the proportionality
%factor between the mean square displacement $\left<r(t)^2\right>$ and the time
%$t$, can be used, $\left<r(t)^2\right> = 2d M t$ where $d$ is the
%dimensionality of the system as discussed above (note that $M = D_\mathrm{s}$
%in the
%Fickian regime).
As a specific example of the subdiffusive regime, we consider
single-file diffusion which can apply to ultraconfining materials where the
porosity consists of narrow cylindrical pores.\cite{hahn_molecular_1996,
chen_transition_2010}
Single-file diffusion is just given as an example here so that we
provide the main ingredients/results without further explanation (also, we note
that the typical time-dependence of the corresponding mean-square displacement
cannot be described in the framework of Fickian diffusion).
In this regime, owing to the very small pore size, molecules cannot pass each
other so that their dynamics is severely hindered.
The typical scaling for this regime is $\left<r(t)^2\right> \sim
t^{1/2}$, which indeed falls in the category of subdiffusion. 
As shown by Kärger and coworkers, except when the repulsive fluid/fluid
potential is infinite (hard core potential), there is always a time beyond which
molecules eventually pass each other and the Fickian regime is recovered.
The typical time for this crossover is given by $\tau_c \sim
1/D_\mathrm{s}^2$~\cite{hahn_molecular_1996}.
However, for practical use, in the framework of molecular dynamics simulations,
considering reasonable timescales that can be reached, it is likely that the
crossover time cannot be accessed so that only single-file diffusion is reached
(in contrast to experiments where much longer times are often probed)
\cite{hahn_molecular_1996}.

%As another important aspect, 
It should be emphasized that diffusion in complex,
\ie.\ heterogeneous, materials 
%such as many nanoporous media 
can also lead to
different Fickian diffusion regimes. For instance, depending on pore size and
temperature, the mean square displacement as a function of time in materials
made of different domains can show different regimes where $\left<r(t)^2\right>
\sim t$. While the diffusivity at short times corresponds to diffusion within a
given domain, the diffusivity at long times corresponds to a diffusion time for
molecules that explore the entire material with significant exchanges between
the different porosity domains (see the discussion by Roosen-Runge \etal.\ in
Ref.~\citen{roosen-runge_analytical_2016}). When diffusion is probed on short
time scales, the Gaussian propagator is characteristic for different
subpopulations that do not exchange and rather explore the domain they belong
to. On the other hand, as the typical time scale becomes much larger than the
exchange time between domains, the diffusion coefficient being probed is a
single effective diffusivity which depends on the diffusivity in the different
domains. From an experimental viewpoint, this can also be used in PFG-NMR for
instance to probe interconnectivity in multiscale porous media such as
hierarchical zeolites (see Ref.~\citen{galarneau_probing_2016} for instance); while diffusion
corresponds to the superimposition of diffusion in microporous and mesoporous
domains when short timescales are considered (\ie.\ both slow and fast
diffusions are observed), a single intermediate effective diffusivity is
obtained when long time scales are considered.

\begin{figure}[tbhp]
	\centering
	\includegraphics[width=0.95\linewidth]{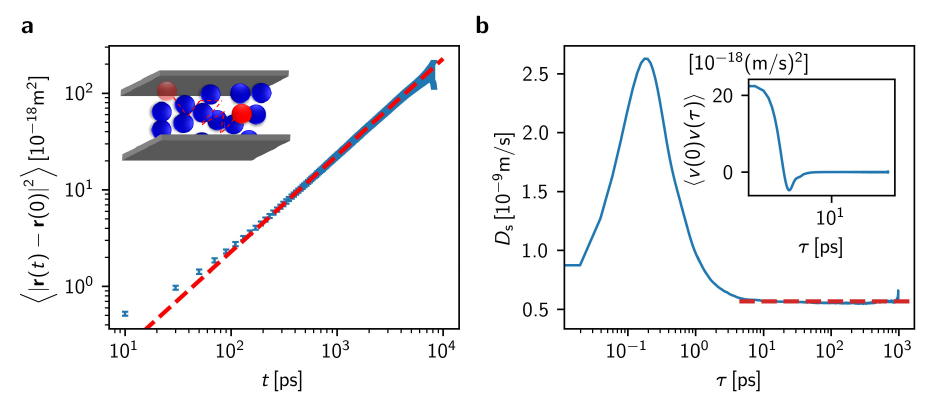}
	\caption{\textbf{Self-diffusivity $D_\mathrm{s}$ of $N_2$ in a silica
	mesopore of width $D_\mathrm{p}=2\,\mathrm{nm}$.}
		(a) Mean square displacement (symbols) and fit according to \cref{EQ3.18}.
			Note that the uncertainity increases with $t$ as fewer initial
			positions $\mathbf{r}(0)$ are averaged over.
		(b) Self diffusivity obtained from the velocity autocorrelation
			according to \cref{EQ3.20} (solid line).
			Dashed line shows the fitting results from (a).
			The inset shows the velocity autocorrelation before integration.
			Note that for a slit pore, the dimensionality is $d=2$ and only the components parallel to
			the pore surface for displacement and velocity vectors are
			considered.
			Data shown are obtained from a total simulation time of
			$100\,\si{\nano\second}$.}
	\label{fig:Ds_msd_velcor}
\end{figure}

Determining the self-diffusion coefficient using molecular dynamics 
from the average mean square displacements can fail as the typical time scale
attainable using computers might be insufficient to reach the Fickian regime.
This is particularly true for ultraconfined nanoporous materials or when low
temperatures or strong fluid/solid interactions are considered. Another
strategy consists of calculating the self-diffusivity $D_\mathrm{s}$ from the
velocity
autocorrelation function using the Green--Kubo
formalism~\cite{hansen_theory_2013}. The velocity autocorrelation function
$\left<\textbf{v}(0) \cdot  \textbf{v}(t)\right>$ is a time autocorrelation
function which describes the correlation between the velocity $\textbf{v}(0)$
of an atom at a time $t = 0$ and the velocity $\textbf{v}(t)$ of the same atom
at a time $t$ later:

\begin{equation}
\left<\textbf{v}(0) \cdot  \textbf{v}(t)\right> = \frac{1}{N} \sum_{i = 1}^N
\textbf{v}_i(0) \cdot  \textbf{v}_i(t)
\label{Eq3.19}
\end{equation}
where the brackets denote a statistical average over the $N$ atoms forming the
system. Once $\left<\textbf{v}(0) \cdot  \textbf{v}(t)\right>$ has been
determined using a molecular dynamics simulation under static equilibrium
conditions, the self-diffusivity can be obtained readily as:\footnote[MSDvsGK]{
	We here provide a formal proof of \cref{EQ3.20} which follows the
	derivation proposed by Hansen and McDonald~\cite{hansen_theory_2013}.
	The displacement of a particle over a time $t$ can be written as:
	\begin{equation*}
	\textbf{r}(t) - \textbf{r}(0) = \int_0^t \textbf{v}(t') \textrm{d}t'
	\end{equation*}
	Upon taking the square and averaging the latter expression, the mean square
	displacement over a time $t$ can be expressed as:
	\begin{equation*}
	\left<  \big| \textbf{r}(t) - \textbf{r}(0) \big|^2  \right>
	= \left< \int_0^t \textbf{v}(t') \textrm{d}t' \cdot \int_0^t \textbf{v}(t'') \textrm{d}t''   \right>
	\end{equation*}
	Using the symmetry properties upon time reversal and invariance with respect
	to time translation of the  velocity autocorrelation function, the latter
	equation can be recast as:
	\begin{align*}
	\left<  \big| \textbf{r}(t) - \textbf{r}(0) \big|^2  \right>
	= 2  \int_0^t \textrm{d}t' \int_0^{t'} \textrm{d}t''
	 \left< \textbf{v}(t') \cdot \textbf{v}(t'')    \right>
	= 2  \int_0^t \textrm{d}t' \int_0^{t'} \textrm{d}t''
	 \left< \textbf{v}(t'-t'') \cdot \textbf{v}(0)    \right>
	\end{align*}
	This equation can be simplified using the variable change $t^{\prime\prime}$ to
	$s = t^\prime - t^{\prime\prime}$ and an integration by parts with respect to
	$t'$:
	\begin{align*}
	\left<  \left| \textbf{r}(t) - \textbf{r}(0) \right|^2  \right>
	= 2  \int_0^t \textrm{d}t' \int_0^{t'} \textrm{d}t''
	 \left< \textbf{v}(t') \cdot \textbf{v}(t'')  \right>
	= 2t  \int_0^t \Big (1 - \frac{s}{t} \Big)
	\left< \textbf{v}(s) \cdot \textbf{v}(0)    \right>
	\textrm{d}s
	\end{align*}
	By comparing the latter to the definition of the self-diffusivity
	$D_\mathrm{s}$ in a system
	of a dimension $d$ according to \cref{EQ3.18},
	\begin{equation*}
	D_\mathrm{s} = \lim_{t \longrightarrow \infty} \frac{\Big <  \big|
	\textbf{r}(t) - \textbf{r}(0) \big|^2  \Big >}{2dt},
	\end{equation*}
	we arrive at \cref{EQ3.20}.
}
\begin{equation}
	D_\mathrm{s} = \frac{1}{d} \int_0^\infty \left<\textbf{v}(0) \cdot
	\textbf{v}(t)\right> \textrm{d}t
	\label{EQ3.20}
	\end{equation}
where we recall that $d$ is the dimensionality of the system.
The latter expression reveals the equivalence between the
self-diffusion coefficients obtained from the mean square displacement,
$D_\mathrm{s}^{\textrm{MSD}} 
= \lim_{t \longrightarrow \infty} \left<  \left|
\textbf{r}(t) - \textbf{r}(0) \right|^2  \right>/2dt$,
and from the velocity autocorrelation function,
$D_\mathrm{s}^{\textrm{VACF}} = 
1/d \times \int \left< \textbf{v}(t) \cdot \textbf{v}(0)  \right>
\textrm{d}t$, \ie.\ $D_\mathrm{s}^{\textrm{MSD}} =
D_\mathrm{s}^{\textrm{VACF}} = D_\mathrm{s}$.
Note that the computational cost of evaluating the correlation in
\cref{EQ3.20} can be significantly reduced using the convolution
theorem.\footnote{
Let $G(f)$ and $H(f)$ be the Fourier transform of two functions $g(t)$ and
$h(t)$ in the time domain. The convolution theorem states that
\begin{equation*}
G(f)H(f) = \int\limits_{-\infty}^\infty g(\tau) h(t-\tau) \mathrm{d}\tau.
\end{equation*}
For real functions $g$ and $h$, we can use that $H(-f) =
H^\star(f)$, where the $\star$ denotes the complex conjugate.
This allows to write the correlation theorem,
\begin{equation*}
	G(f) H^\star(f) = \int\limits_{-\infty}^\infty
	g(\tau+t)h(\tau) \mathrm{d}\tau,
\end{equation*}
which after backwards Fourier transform yields the correlation between
$g$ and $h$. If $g$ and $h$ are discrete and finite ($t_\mathrm{max}<\infty$)
observables defined only for $t\geq 0$ as in \cref{EQ3.20}, one can make use of
the cyclic property of the discrete time Fourier transform and average
$g(\tau+t)h(\tau)$ efficiently over all initial times $\tau$ using the Fast
Fourier Transform (FFT),
\begin{equation*}
\left<g(\tau+t)h(\tau)\right>_\tau =
\frac{1}{N_\tau(t)}\mathsf{iFFT}\left(\mathsf{FFT}(g)
\cdot
\left[\mathsf{FFT}(h)\right]^\star\right),
\end{equation*}
where the normalizing factor $N_\tau(t)$ counts the number of samples (\ie.\
the number of initial times $\tau$) that are considered in the data series for
each lag time $t$.
For the special case of an autocorrelation function, $g=h$, one obtains
$G(f)\cdot G^\star(f)=\left|G(f)\right|^2$.
Also note, that because of the cyclic property of the FFT one usually adds zeros
to the signal for $-t_\mathrm{max}<t<0$ (zero padding).
The advantage of employing the FFT for correlation functions is the scaling of
$\mathcal{O}(N\log N)$ as compared to $\mathcal{O}(N^2)$ for a direct
calculation.
}$^,$\cite{arfken_mathematical_2005}
From a practical standpoint, the comparison between these two techniques
constitutes an important consistency check to validate the robustness of a
given molecular dynamics study. \Cref{fig:Ds_msd_velcor} compares the
self-diffusivity $D_\mathrm{s}$ obtained using these two techniques for $N_2$
at 77~K in mesoporous silica. As can be seen, provided long enough
simulations are carried out, the two techniques give the same value within
statistical accuracy.

\noindent \textbf{Collective diffusivity $D_0$}.
In the previous section, Fick’s first law was derived from the
isothermal/isobaric linear response law given in \cref{EQ3.13} by considering a
very diluted system. As discussed earlier, this assumption is fully justified
for a solute molecule dispersed in a liquid at very low concentration or for a
tracer diffusion (since a tagged particle can indeed be considered as
infinitely diluted in the liquid formed by all other molecules). However, in
most general situations, this assumption breaks down and the general form given
in \cref{EQ3.13} must be kept:
\begin{equation}
J_N  =  - \frac{L'_{NN}}{ k_\textrm{B} T} \nabla \mu
\label{EQ3.29}
\end{equation}
where we replaced $L_{NN}$ by $L'_{NN} = k_\textrm{B} L_{NN}$
in order to be consistent with the units of energy given in the following
	derivation.
In the framework of the linear response theory, the transport coefficient $L'_{NN}$
can be determined from the Green--Kubo formula involving the time correlation
function of the fluid velocity~\cite{yoshida_molecular_2014}.
In what follows, we present the standard derivation based on the linear response
theory in which the Hamiltonian $\mathcal{H}_0$ of the system at thermodynamic
equilibrium is perturbed using a time-dependent perturbation
$\Delta\mathcal{H}(t)$~\cite{hansen_theory_2013,maginn_transport_1993}:
\begin{equation}
\mathcal{H} = \mathcal{H}_0 + \Delta\mathcal{H}(t)
\label{EQ3.30}
\end{equation}
with $\Delta \mathcal{H}(t)$ taken to be a simple time-oscillating function:
\begin{equation}
\Delta\mathcal{H}(t) = - \mathcal{A}(\textbf{r}^N)
\mathcal{F}_0 \exp(-i \omega t)
\label{EQ3.31}
\end{equation}
$\omega$ is the frequency of the perturbation field while
$\mathcal{A}(\textbf{r}^N)$ is the conjugated quantity of the perturbation
field (the product of the field with its conjugated variable must be
homogeneous to an energy). In practice, to derive a microscopic expression for
$L'_{NN}$, we choose $F_0 = - \nabla \mu_x$, \ie.\ a homogeneous force field
corresponding to a chemical potential gradient $\nabla \mu_x$ along $x$ which
applies to all particles in the fluid. The corresponding conjugated variable
then is $\mathcal{A}(\textbf{r}^N) = \sum_{i = 1}^N x_i$.
In the linear response theory, the change $\left< \Delta \mathcal{B} \right>$
induced by the perturbation $\Delta\mathcal{H}(t)$ in an observable
$\mathcal{B}$ is given by~\cite{zwanzig_time-correlation_1965}:
\begin{equation}
\big < \Delta \mathcal{B} \big > =
M_{BA}(\omega) \mathcal{F}_0 \exp(-i \omega t)
\label{EQ3.32}
\end{equation}
with
\begin{equation}
M_{BA}(\omega) =
\frac{1}{k_\textrm{B}T} \int_0^\infty
\big < \mathcal{B}(t) \dot{\mathcal{A}}
\big > \exp(i \omega t) \textrm{d}t
\label{EQ3.33}
\end{equation}
where $\dot{\mathcal{A}} = \textrm{d}\mathcal{A}/\textrm{d}t$ is the time
derivative of the observed quantity $\mathcal{A}$.
On the one hand, the observed quantity is the molecule flux defined in
\cref{EQ3.12}: $\left< \Delta \mathcal{B}\right> = J_N = \rho v_x = 1/V \sum_{i
= 1}^N v_{x,i}$  ($\rho$ is the fluid density of the $N$ molecules while $v_x$
and $v_{x,i}$ are the average flow velocity and velocity of molecule $i$ along
$x$, respectively). On the other hand, we recall that $\mathcal{A} = \sum_{i =
1}^N x_i$ so that $\dot{\mathcal{A}} = \sum_{i = 1}^N v_{x,i}$ . With these
definitions, \cref{EQ3.32,EQ3.33} lead to:
\begin{equation}
M_{BA}(\omega) =
\frac{1}{Vk_\textrm{B}T} \int_0^\infty
\Big < \sum_{i,j} v_{x,i}(t) v_{x,j}(0) \Big > \exp(- i \omega t) \textrm{d}t
\label{EQ3.34}
\end{equation}
For an isotropic medium, $\sum_{i,j} v_{x,i}(t) v_{x,j}(0) = 1/3 \sum_{i,j}
\textbf{v}_i(t) \cdot \textbf{v}_j(0)$ which allows recasting \cref{EQ3.34} as:
\begin{equation}
M_{BA}(\omega) =
\frac{1}{3Vk_\textrm{B}T} \int_0^\infty
\Big < \sum_{i,j} \textbf{v}_i(t) \cdot \textbf{v}_j(0) \Big > \exp(- i \omega t) \textrm{d}t
\label{EQ3.35}
\end{equation}
For an anisotropic medium, \cref{EQ3.34} has to be used but, in that case,
transport coefficients $L_{\alpha\beta}$ are also anisotropic and must be
replaced by a vector $L_{\alpha\beta}^k$ with $k = x$, $y$, $z$.

In conventional transport experiments, one is often interested in the static response
to a stationary perturbation, \ie.\ a thermodynamic gradient, which is
obtained by considering the limit $\omega \to 0$ in \cref{EQ3.32},
\ie.\ $J_N = - M_{BA}(0) \nabla \mu$ with:
\begin{equation}
M_{BA}(0) =
\frac{1}{3Vk_\textrm{B}T} \int_0^\infty
\Big < \sum_{i,j} \textbf{v}_i(t) \cdot \textbf{v}_j(0) \Big >  \textrm{d}t
\label{EQ3.36}
\end{equation}
Comparison between Eq. (\ref{EQ3.29}) and $J_N = - M_{BA}(0) \nabla \mu$ leads to:
\begin{equation}
L'_{NN} = k_\textrm{B}T M_{BA}(0) =
\frac{1}{3V} \int_0^\infty
\Big < \sum_{i,j} \textbf{v}_i(t) \cdot \textbf{v}_j(0) \Big >  \textrm{d}t
\label{EQ3.37}
\end{equation}
Note that these definitions and formalism are only applicable to a fluid that
can exchange momentum with a reservoir --- in our case the porous solid. If the
total momentum was conserved, $L'_{NN}$ would vanish identically in the frame of
the center of mass --- and in fact application of a force would lead to a constant
acceleration, rather than a finite velocity and current.

Let us now introduce the collective --- sometimes referred to as corrected ---
diffusivity, $D_0$:
\begin{equation}
D_0 =
\frac{1}{3N} \int_0^\infty
\Big < \sum_{i,j} \textbf{v}_i(t) \cdot \textbf{v}_j(0) \Big >  \textrm{d}t
\label{EQ3.38}
\end{equation}
Comparison between \cref{EQ3.29,EQ3.36,EQ3.37} leads to:
\begin{equation}
J_N  =  -  \frac{\rho D_0}{ k_\textrm{B} T} \nabla \mu
\label{EQ3.39}
\end{equation}
with $D_0 = {L'_{NN}}/{\rho}$.

\begin{figure}[htbp]
	\centering
	\includegraphics[width=\linewidth]{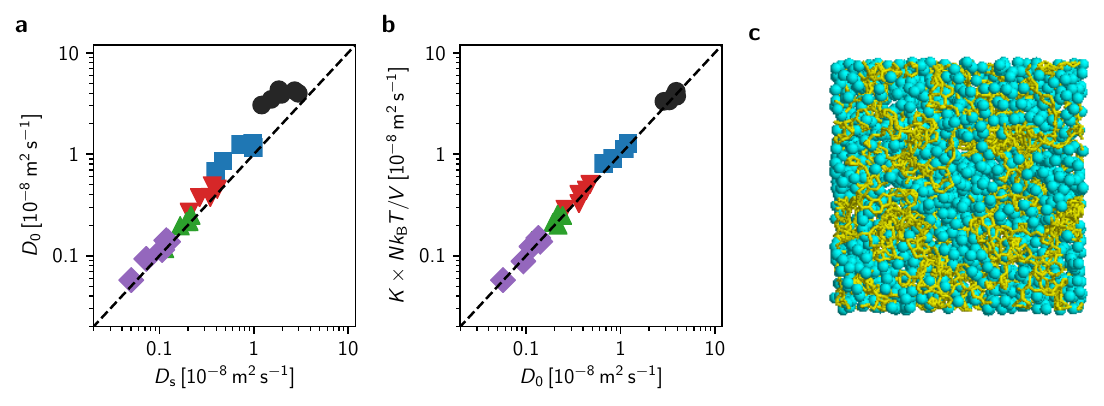}
	\caption{\textbf{Self/collective diffusivities and permeance.}
		(a) Collective diffusivity $D_0$ as a function of the self-diffusivity
		$D_\mathrm{s}$ for different hydrocarbons confined at $T = 450$ K in a
		disordered porous carbon with subnanometric pores. The black, blue,
		green, orange, and red circles are for methane, propane, hexane, nonane
		and dodecane, respectively. The dashed line indicates the bisector $D_0
		= D_\mathrm{s}$.
		(b) Equivalence between the permeance $K$ and the collective diffusivity
		$D_0 = KNk_\textrm{B}T/V$ for different hydrocarbons confined at $T =
		450$ K in a disordered porous carbon with subnanometric pores.
		The color code is the same as in (a). (c) Typical molecular
		configuration corresponding to the data shown in (a) and (b) with
		methane molecules diffusing in a disordered porous carbon. 
		The cyan spheres are the methane molecules while the yellow segments are
		the chemical bonds between carbon atoms in the host porous material.
		Adapted with permission from Ref.~\protect\citen{falk_subcontinuum_2015}.
		Copyright 2015 Macmillan Publishers Limited under Creative Commons
		Attribution 4.0 International License
		(http://creativecommons.org/licenses/by/4.0/).
		}
	\label{fig:3_3}
\end{figure}

The microscopic expression given in \cref{EQ3.38} for the collective diffusivity
$D_0$ can be separated into velocity time correlations for the same molecule $i
= j$ and cross-correlations between different molecules $i \neq j$:
\begin{equation}
D_0 =
\frac{1}{3N} \int_0^\infty
\Big < \sum_{i = 1}^N \textbf{v}_i(t) \cdot \textbf{v}_i(0) \Big >  \textrm{d}t
+
\frac{1}{3N} \int_0^\infty
\Big < \sum_{i = 1}^N \sum_{j \neq i}^N \textbf{v}_i(t) \cdot \textbf{v}_j(0) \Big >  \textrm{d}t
\label{EQ3.40}
\end{equation}
where the first term is identical to the microscopic expression for the
self-diffusivity $D_\mathrm{s}$ given in \cref{EQ3.20}.
This important result shows that
the collective diffusivity, which describes the fluid response to any chemical
potential gradient applied to the system, is the sum of an individual response
characterized by the self-diffusivity $D_\mathrm{s}$ and a collective
contribution that arises from the direct collective interactions between fluid molecules. There is a
number of situations where the velocities between different molecules are
uncorrelated so that $D_\mathrm{s} \sim D_0$. This includes very dilute
systems for which velocity cross-correlations are negligible. Other situations
include adsorbed fluids in the limit of very small adsorbate loading or in
ultra-confined environments where the velocity cross-correlations are
negligible compared to the individual (\ie.\ self) velocity correlations
induced by the interactions with the pore walls. As an illustration,
\cref{fig:3_3}~(a) shows the self and collective diffusivities for different
alkanes (methane, propane, hexane, nonane and dodecane) confined in the
porosity of a host porous carbon mimicking kerogen in gas
shales~\cite{falk_subcontinuum_2015}. For a given molecule type, while
$D_\mathrm{s} \sim D_0$ at low densities, $D_0 > D_\mathrm{s}$ at larger
densities as the velocity cross-correlations become non-negligible. The fact
that $D_0 > D_\mathrm{s}$ indicates that collective molecular interactions, \ie.\ the
fact that a given molecule motion creates a velocity field affecting its neighbors,
make overall diffusion faster.
Comparison between different fluid molecules in \cref{fig:3_3}(a) shows that
differences between $D_\mathrm{s}$ and $D_0$ are more pronounced upon
decreasing the molecule size.
This result can be explained by crowding effects; for a given
cavity size, a larger number of small molecules can be confined compared to
large molecules so that the role of velocity cross-correlations is more
pronounced for small molecules.

\bluelork{Like \cref{EQ3.20} for the self-diffusivity $D_\mathrm{s}$, the microscopic expression given in \cref{EQ3.38} for the collective diffusivity $D_0$ belongs to the family of Green–Kubo relations which relate a transport coefficient to microscopic fluctuations\cite{alfazazi_interpreting_2024}. 
The latter can serve as the basis of a theoretical model of transport based on the analysis of the fluctuations of microscopic variables. 
Such microscopic relations are particular cases of the general fluctuation dissipation theorem (FDT)~\cite{barrat_basic_2003,pottier_nonequilibrium_2010,
bedeaux_fluctuation-dissipation_2022, galteland_local_2022}. The latter considers  the response of an observable (e.g.\ position) to a driving force at finite frequency. In particular, its imaginary part that describes the out of phase response and therefore the dissipation, to the power spectrum of the time correlations in a system at equilibrium. The physical significance of this theorem, sometimes described as Onsager's regression principle, is that the fluctuations that exist in a system at equilibrium (without driving force) follow a  dynamics that is identical to the time evolution of a disturbance induced by an external driving.}

In the spirit of the connection between the Green--Kubo  framework and Einstein formula for the self-diffusivity $D_\mathrm{s}$
discussed above,\bibnotemark[MSDvsGK]
a direct relationship can be established between the microscopic
fluctuation-based formula given in \cref{EQ3.38} and the mean-square
displacements of the fluid particles (more exactly, the mean square
displacement of the center of mass of the fluid):
\begin{equation}
D_0 = \lim_{t \longrightarrow \infty} \frac{1}{2Ndt}\left\langle
\sum_{i,j} \lbrack \textbf{r}_i(t) - \textbf{r}_i(0) \rbrack \cdot
\lbrack \textbf{r}_j(t) - \textbf{r}_j(0) \rbrack \right\rangle
\label{EQ3.41}
\end{equation}
where we recall that $d$ is the dimensionality of the system.
For the sake of brevity, the formal derivation of this equation will not be
provided here as it follows very closely the step-by-step derivation provided
for the self-diffusivity to arrive at 
\cref{EQ3.20}.\bibnotemark[MSDvsGK]
\redlork{As already mentioned, when considering a non isolated system (such as a system in equilibrium with a reservoir or a nanoconfined fluid in contact with a solid surface), momentum exchange between the system and its surrounding allows defining a collective diffusion coefficient from the position of the center of mass (mean squared displacements) or its velocity fluctuations (Green--Kubo formula). In contrast, for a perfectly isolated system (such as a bulk fluid), the total momentum conservation makes that the molecule current is zero at all times so that it is not possible to assess the  collective diffusivity of the system. 
In that case, as exemplified in Ref. \citen{kellouai_gennes_2024}, it is possible to measure the collective diffusivity from the coherent scattering functions $F_\textrm{coh}(q,t)$ taken for different non zero wave vectors $q$. In practice, by taking the characteristic relaxation times $\tau(q)$ as a function of $q$, one can estimate the collective diffusivity from $\tau(q) \sim 1/[D_0(q) q ^2]$ in the limit of $q$ approaching zero. 
As another solution, it was recently proposed to use a ``counterscope’’ in which one explores number/density fluctuations in real space using virtual boxes of variable size $L$.~\cite{mackay_countoscope_2024} Using this approach, one defines a scale-dependent transport coefficient whose limit in $L \to \infty$ provides the macroscopic collective diffusion. In spirit, this ``counterscope'' technique is similar to deriving the collective diffusivity $D_0(q)$ by taking the limit when $q \sim 1/L \to 0$.}

The different expressions for $D_\mathrm{s}$ and $D_0$ show that the latter
is a collective property that depends on the total
motion of all molecules while the former only depends on individual motions.
This trivial statement has important practical implications when undertaking a
molecular simulation study. Independently of the formalism used (\ie.\
Green--Kubo \textit{versus} mean square displacement calculations), all molecule velocities
contribute to calculating $D_\mathrm{s}$ therefore leading to improved
statistics.
In contrast, $D_0$ is a collective property where all molecule motions are
combined into a single quantity which results in much poorer statistics.
Finally, for both $D_\mathrm{s}$ and $D_0$, the Green--Kubo or mean square
displacement formulas allow one to  evaluate their value from equilibrium molecular
dynamics simulations, \ie.\ when no thermodynamic gradient is applied to induce
transport.
However, while this approach should always be successful in principle, long
time tails are difficult to assess using typical molecular dynamics simulations
and can thus lead to erroneous estimates for the different transport
coefficients. As a result, non-equilibrium molecular dynamics simulations in the linear response limit,
where transport is induced using an applied thermodynamic gradient, are
sometimes preferred as they provide a direct measurement of the fluxes induced
by a chemical potential, pressure and/or
temperature gradient (see \eg.\ Ref.~\citen{yoshida_generic_2014} for a
comparison between equilibrium and non-equilibrium molecular dynamics applied
to electrokinetic effects in nanopores).

As mentioned earlier, the collective diffusivity is formally linked to the
so-called permeance or permeability through the fluctuation-dissipation
theorem. Let us consider experiments on liquid transport where the fluid can be
assumed to be incompressible such as water at room temperature. In that case,
fluid transport is usually induced through a pressure drop which triggers a
viscous flow described using Darcy equation.\cite{osullivan_perspective_2022}  In more detail, as will be shown
below, the molecular flow $J_N = \rho v$ induced by a pressure gradient $\nabla
P$ corresponds to a flow rate (velocity) $v = - k/\eta \nabla P = - K \nabla
P$  where $\eta$ is the fluid viscosity, $k$ the permeability and $K = k/\eta$
the permeance. \bluelork{Exact definitions for the permeability and permeance vary from a
document to another in the literature. In the context of membranes and membrane processes, we note that IUPAC has established strict recommendations regarding the definitions of permeance, permeability, etc.\cite{koros1996terminology}}
For an incompressible liquid, using the Gibbs--Duhem equation $\rho
\,\textrm{d}\mu = \textrm{d}P$, we have $J_N = \rho v = - K \rho^2 \nabla \mu$.
Comparison between this expression and \cref{EQ3.39} for the collective
diffusivity  leads to: $K = D_0/\rho k_\textrm{B}T$. This is a very important
result of the fluctuation-dissipation theorem  which can be easily verified
using molecular simulation for incompressible fluids if the fluid response $v$
is linear in the driving force $\nabla P$, \ie.\ for properly chosen values
$\nabla P$. \Cref{fig:3_3}(b) compares the permeance $K$ and collective
diffusivity $D_0$ for alkanes confined in the same nanoporous carbon as that
considered in \cref{fig:3_3}(a). In agreement with the prediction of the
fluctuation-dissipation theorem, $K = D_0/\rho k_\textrm{B}T$ for all fluids
and densities considered~\cite{falk_subcontinuum_2015}.

\noindent \textbf{Transport diffusivity $D_\mathrm{T}$}.
As introduced in \cref{sec:onsager}, because the number of molecules $N$ and
the chemical potential $\mu$ are conjugated variables, the direct driving force for
molecular fluxes $J_N$ is a chemical potential gradient $\nabla \mu$. This is
also clear from \cref{EQ3.39}, where the molecular flux $J_N$ is expressed as a
linear response to $\nabla \mu$ with a transport coefficient directly
proportional to the collective diffusivity $D_0$. Yet, in most practical
experiments, transport is induced using a concentration, density or pressure
gradient which raises the question of the comparison between the different
transport
coefficients~\cite{smit_molecular_2008,karger_diffusion_2012,karger_diffusion_2016}.
While the concentration/density gradient $\nabla \rho$ is not \textit{stricto
sensu} an affinity as defined in  Onsager's theory of transport, it is always
possible to assume a linear response between a molecular flux $J_N$ and $\nabla
\rho$:
\begin{equation}
J_N = - D_\mathrm{T} \nabla \rho
\label{EQ3.42}
\end{equation}
$D_\mathrm{T}$ is the so-called transport diffusivity that can be
measured using experiments where transport is induced by applying a
concentration gradient across the sample. 
As discussed earlier, in many textbook notations, the number density $\rho$ (in
molecule per unit volume) is replaced by the concentration $c$ as they are
strictly equivalent (except for the fact that the concentration explicitly
refers to mixtures).
%Such experiments include transport in columns experiments where the column contains a porous monolithic solid or a packed granular material or mass uptake experiments where an initially empty material is set in contact with a
Comparison between \cref{EQ3.42,EQ3.39} provides a straightforward link between
$D_\mathrm{T}$ and $D_0$:
\begin{equation}
D_\mathrm{T} = \frac{\rho D_0}{k_\textrm{B}T} \frac{\partial \mu}{\partial
\rho}\bigg \lvert_T
\label{EQ3.43}
\end{equation}
where $\nabla \mu/\nabla \rho \sim \partial \mu / \partial \rho$ was used and
``$\lvert_T$'' recalls that the derivation only applies to a system at constant
temperature $T$. For any fluid, the chemical potential $\mu$ is related to the
fugacity $f$ which corresponds to the pressure that the fluid would have if it
were an ideal gas (\ie., $f = P$ for an ideal gas):
$\mu = k_{\textrm{B}}T \ln (f\Lambda^3/k_{\textrm{B}}T)$ where $\Lambda$ is the
de Broglie thermal wavelength. Using $\partial \mu = k_{\textrm{B}}T \partial
\ln f$ at constant temperature $T$ and  $\partial \rho / \rho = \partial \ln
\rho$, \cref{EQ3.43} can be recast as:
\begin{equation}
D_\mathrm{T} =  D_0\frac{\partial \ln f}{\partial \ln \rho}\bigg \lvert_T
\label{EQ3.44}
\end{equation}
The term $\partial \ln f / \partial \ln \rho \lvert_T$ is known as the Darken
or thermodynamic factor in the literature. Physically, it describes how a
change in the local concentration converts into a chemical potential (recalling
that, in the context of Onsager's theory of transport, the chemical
potential is the driving force for molecular diffusion). In the limit of small
gas chemical potentials/densities, the gas fugacity is equal to the gas
pressure and the density scales linearly with pressure, \ie.\ $\rho
\sim P$. This is known as Henry's law which introduces the Henry constant
$K_\mathrm{H}$ as a proportionality factor, \ie.\
$\rho = K_\mathrm{H} P$ --- this relation is valid both for gas solubility in
liquids and in porous solids (\ie\ adsorption).\footnote{
	Henry’s law can be derived as follows by starting from the equality of the
	fugacity $f$ (or, equivalently, the gas chemical potential $\mu$) in the
	gas phase and in the liquid phase: $f_g = f_l$. Under the assumption that
	the gas behaves as an ideal gas, $f_g = P$. As for $f_l$, we can write $f_l
	= \gamma x f_l^{(0)}$ where $\gamma$  is the activity coefficient, $x$ is
	the mole fraction of gas molecules in the liquid, and $ f_l^0$ is the
	fugacity of the liquid phase for a pure fluid, \ie.\ $x = 1$.
	Using a polynomial expansion for
	$\ln \gamma \sim a_1 (1 - x) + a_2 (1 - x)^2 + a_3 (1 - x)^3 + \ldots$,
	one recovers the two important asymptotic limits:
	(a) $\ln \gamma = 0$ for $x = 1$ as expected for a pure liquid for which
	$\gamma \sim  1$ and (b) $\ln \gamma \sim$ constant for $x \ll 1$.
	As a result, for $x \ll 1$, and noting that $\rho \sim x$, one obtains
	Henry’s law with $K_\mathrm{H} \sim  \left[ \gamma f_l^{(0)} \right]^{-1}$
	as a good approximation that describes the fugacity equality between the
	gas and liquid phases. In practice, the simple scaling between gas
	concentration and pressure in Henry’s law holds provided that the gas
	partial pressure remains low enough and that the host solid or liquid does
	not chemically react with the solubilized gas.
}$^,$\cite{henry_iii_1997}
In such an asymptotic regime where $\rho \to 0$, the thermodynamic factor goes to
$\partial \ln \mu / \partial \ln \rho \to  1$, which leads to $D_\mathrm{T}  =
D_0$. Moreover, since $D_\mathrm{s}  = D_0$ at very low fluid densities, we
have $D_\mathrm{T} = D_0 = D_\mathrm{s}$ in the Henry regime. In contrast, as
the density increases, $D_\mathrm{T}  \neq D_0 \neq D_\mathrm{s}$.
\Cref{EQ3.44} is an important result as it shows that the effective transport
diffusivity $D_\mathrm{T}$ can be easily determined by assessing independently
the collective diffusivity $D_0$ and the thermodynamic factor $\partial \ln \mu
/ \partial \ln \rho$. In practice, using molecular simulation tools, the former
is usually determined from either at equilibrium or non-equilibrium molecular
dynamics simulations while the latter can be evaluated using Monte Carlo simulation
in the Grand Canonical ensemble to estimate the adsorption isotherm $\rho(\mu)$.
%(see \cref{chap:molsim} for a presentation of these different techniques).
In \cref{chap:diffusion_pore}, simple yet representative examples will be
provided to illustrate how $D_\mathrm{T}$ can be assessed using such a two-step
strategy.

\section{Self and tracer diffusion at the pore scale}
\label{chap:diffusion_pore}

\subsection{Diffusion mechanisms}
\label{sec:diffusion_mechanisms}

Fluid transport in porous media occurs
through diffusion and advection whenever a thermodynamic gradient such as a
pressure or concentration gradient is applied to the system. Diffusion
occurs both in systems at equilibrium, \ie.\ in the absence of any
thermodynamic gradient, and in non-equilibrium situations, \ie.\ in
the presence of a thermodynamic gradient. In the latter case, an important
question arises regarding the transport resulting from the combination of
advection and diffusion in nanopores. In this section, in a first step, the
different diffusion mechanisms that pertain to fluid transport in nanoporous
media are presented with special attention to their dependence on pore size
$D_\textrm{p}$, fluid molecule size $\sigma$ and thermodynamic variables
such as fluid density $\rho$, pressure $P$, temperature $T$, \etc.
Then, in a second step, combination rules which allow one to describe the transport resulting
 from different diffusion mechanisms are introduced. Advection
mechanisms will be discussed in
\Cref{chap:transport_pore,chap:transport_network} together with underlying
theoretical models such as Poiseuille’s flow and Darcy’s law.

\begin{figure}[htbp]
	\centering
	\includegraphics[width=0.95\linewidth]{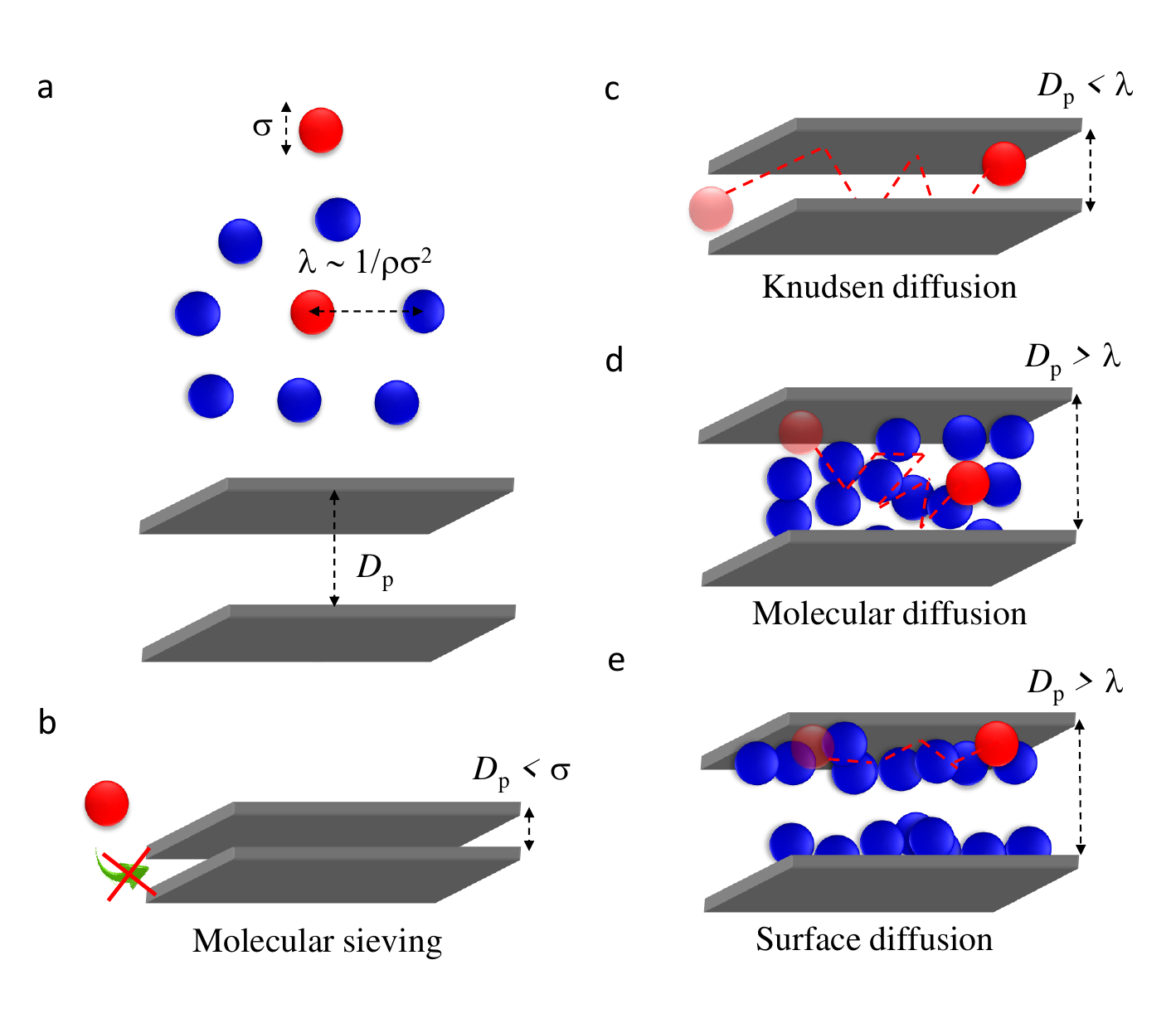}
	\caption{\textbf{Diffusion mechanisms in porous materials}.
		(a) Let us consider a fluid composed of molecules having a molecular size
		$\sigma$.
		The mean free path $\lambda$ in the gas or liquid scales as
		$1/\rho\sigma^2$ where $\rho$ is the number density. (b) For very
		small pore sizes $D_\textrm{p}$, typically $D_\textrm{p} < \sigma$,
		molecular sieving is observed as only molecules smaller than the pore
		size can enter and diffuse through the porosity. (c) For $D_\textrm{p}
		< \lambda$, diffusion occurs through Knudsen diffusion which results
		from the fluid mean free path affected by collisions with the pore
		surface. (d) For $D_\textrm{p} > \lambda$, diffusion in the porosity is
		coined as molecular diffusion which resembles molecular diffusion in
		the bulk fluid. (e) For strong wall fluid interactions and pores large
		enough, the confined fluid forms an adsorbed film at the pore surface
		which coexists with the gas	phase in the pore center. If the density
		difference between the adsorbed and gas phases is large, diffusion in
		the porosity predominantly occurs through surface diffusion within the
		adsorbed film.}
	\label{fig:4_1}
\end{figure}

Let us consider a fluid made up of molecules having a molecular size $\sigma$
which diffuses in a pore of size $D_\textrm{p}$. The fluid is taken at
thermodynamic conditions $T$, $P$ such that its density is $\rho$ and the
corresponding mean free path $\lambda \sim 1/\rho\sigma^2$ as shown in
\cref{fig:4_1}(a).
The main diffusion mechanisms that can be encountered in a
nanoporous material are presented in \cref{fig:4_1}(b)-(e) for increasing
pore diameters $D_\textrm{p}$. In the following, we discuss
diffusion in nanopores not as a function of the absolute pore size
$D_\textrm{p}$ but its ratio to the molecular size $\sigma$ and mean free path
$\lambda$ (see also Ref.~\citen{borah_transport_2012} for a discussion on the
relation between fluid size, entropy, and diffusivity).

\subsubsection{Molecular sieving}

For $D_\textrm{p} < \sigma$, molecular sieving is observed as the typical pore
size is not large enough to accommodate any fluid molecule
[\cref{fig:4_1}(b)].
Obviously, for this regime, no transport --- either through
diffusion or advection --- can be observed. While molecular sieving for a pure
fluid is trivial, separation processes in the case of mixtures can rely on such
effects when fluid components with different molecular sizes are considered.
%Osmotic flow such as in reverse osmosis membranes is another situation where molecular sieving effects are particularly relevant. In this case, which will be treated in \Cref{chap:transport_pore} as it pertains to transport induced by a thermodynamic gradient and not diffusion, one considers a nanoporous membrane with very small pores that are permeable to a liquid --- \eg.\ water --- but not permeable to solute molecules --- like  ions for example. Such membranes are typically used to separate water containing a large concentration of ions upstream from ion-free water downstream.

\subsubsection{Knudsen diffusion}

\textbf{Principles.}
Knudsen diffusion occurs in rarefied gases \ie.\ for gases
at very low pressures and, hence, very low densities.\cite{gruener_knudsen_2008}
In practice, in this
regime, collisions between gas molecules can be neglected and molecular
displacements can be assumed to occur only through collisions with the surface
of the host nanoporous material [\cref{fig:4_1}(c)]. The dimensionless
Knudsen number Kn is the quantity that characterizes the importance of
Knudsen diffusion in transport mechanisms as a function of pore size
$D_\textrm{p}$ and fluid mean free path $\lambda$:
\begin{equation}
\textrm{Kn} = \frac{\lambda}{D_\textrm{p}}
\label{EQ4.1}
\end{equation}
The mean free path $\lambda \sim 1/\rho\sigma^2$ follows from the
pressure $P$, temperature $T$ and molecular size $\sigma$:\footnote{
	Let us consider a single fluid molecule diffusing through a medium
	containing a density $\rho$ of frozen objects with which the molecule can
	collide. Introducing the cross-section area for collision
	$s \sim \pi \sigma^2$, the mean free path can be written as
	$\lambda \sim 1/\rho \pi \sigma^2$.
	For a gas, since the objects are other gas molecules, they cannot be
	considered  as immobile so that the relative velocity of the gas molecules
	must be taken into account which leads to a slightly different expression:
	$\lambda \sim 1/\sqrt{2} \rho \pi \sigma^2$.  Assuming that the gas can be
	treated as ideal, \ie.\ $\rho = P/k_\textrm{B}T$, leads to:
	$\lambda = k_\textrm{B}T /\sqrt{2 } P \pi \sigma^2$.}
\begin{equation}
\lambda = \frac{k_\textrm{B}T}{\sqrt{2} P \pi \sigma^2}
\label{EQ4.2}
\end{equation}
$\sigma$ can be obtained from the covolume $b$ in the van der Waals equation
describing the gas, or using for example the Lennard--Jones parameter
$\sigma_{\textrm{LJ}}$ from typical force-fields when the gas molecule is
treated as a single-sphere particle. $\sigma$ can also be taken as the kinetic diameter estimated from
bulk self-diffusivity/viscosity data using Stokes Einstein relation.
Equivalently, using the kinetic theory of gases to estimate the viscosity 
$\eta = m/3\sqrt{2} \pi \sigma^2 \times \sqrt{8k_{\textrm{B}}T/ \pi m}$,
\cref{EQ4.2} can be recast as:

\begin{equation}
\lambda = \frac{3\eta}{2P} \sqrt{\frac{\pi k_{\textrm{B}}T}{2 m}}
\label{EQ4.3}
\end{equation}
For $\textrm{Kn} > 10$, collisions between gas molecules and the surface of
the host porous material prevail and molecular diffusion can be neglected. For
$\textrm{Kn} \ll 0.1$, interactions between gas molecules prevail and Knudsen
diffusion can be neglected. As will be discussed below, for values in the range
$0.1 < \textrm{Kn} < 10$, the different mechanisms must be considered.

To derive the expression for the Knudsen diffusivity, we consider a
cylindrical pore of diameter $D_\textrm{p}$ and length $L_\textrm{p}$.
This pore is set in contact with an ideal gas at pressure $P$ and
temperature $T$ so that its density is $\rho = P/k_{\textrm{B}}T$. Let
$J_{\textrm{K}}$ be the flux \ie.\ the number of molecules per unit of
time and surface area that pass through the cylindrical pore. In the Knudsen
regime, $\textrm{Kn} \gg 1$, this flux can be written as
$J_{\textrm{K}} = - \omega v_{T} \rho$, \bluelork{where
$\omega$ is the probability that a molecule entering one pore cross section area crosses another pore cross section area} and $v_{T} = \sqrt{8k_{\textrm{B}}T/\pi m}$ is the mean thermal velocity (the latter expression is directly obtained from the Maxwell
distribution function) and $\omega = D_{\textrm{p}} /3L_{\textrm{p}}$ for a
cylindrical pore with $D_\textrm{p} \ll L_\textrm{p}$.
On average, for a pore or a porous material in contact with the same gas
density downstream and upstream, the total flow $J_\textrm{K}$ passing through
Knudsen diffusion is equal to 0 since there are as many molecules traveling in
one direction than in the other.
On the other hand, if a gas density gradient
$\Delta \rho$ is applied along the pore, the net flow is
$J_\textrm{K} = -\omega v_{T} \Delta \rho$.
This relationship allows one to estimate the Knudsen permeability
$\Pi_{\textrm{K}}$ and the Knudsen diffusivity $D_{\textrm{K}}$ which are
defined as
$J_{\textrm{K}} = - \Pi_{\textrm{K}} \Delta P = - D_{\textrm{K}} \nabla \rho$.
Invoking the ideal gas law, $\Delta \rho = \Delta P/k_{\textrm{B}}T$ and
$\nabla \rho = \Delta \rho/ L_{\textrm{p}}$, one obtains: $\Pi_\textrm{K} =
D_\textrm{p}/3L_\textrm{p} \times \sqrt{8/\pi m k_\textrm{B} T}$ and
$D_\textrm{K} = D_\textrm{p}/3 \times \sqrt{8 k_\textrm{B} T /\pi m }$.
These expressions are valid for a cylindrical pore but they can be generalized
to any porous material by introducing its porosity $\Phi$ and tortuosity
$\tau$ to correct the permeability and diffusivity.\footnote{
	Tortuosity will be introduced later in detail ---
	here, it should be simply considered as the ratio of the pore
	physical length to the pore height.}
This yields the corresponding quantities $X_\textrm{K} = \Phi/\tau
X_\textrm{K,O}$ where $ X_\textrm{K,O}$ is the permeability or diffusivity for
a cylindrical pore having the same pore size $D_\textrm{p}$. This leads to the
following general expressions:
\begin{equation}
\Pi_\textrm{K } = \frac{D_\textrm{p}\Phi}{3L_\textrm{p}\tau} \times
\sqrt{\frac{8}{\pi m k_\textrm{B}T}}
\label{EQ4.4}
\end{equation}
\begin{equation}
D_\textrm{K } = \frac{D_\textrm{p}\Phi}{3 \tau} \times \sqrt{\frac{8
		k_\textrm{B}T }{\pi m }}
\label{EQ4.5}
\end{equation}
In practice, $\tau$ is often treated as a fitting parameter. It is interesting
to note that the Knudsen diffusivity has a temperature dependence
$D_\textrm{K} \sim \sqrt{T}$ which drastically differs from the dependence for
other regimes.
This allows in principle to quantify the Knudsen contribution with respect
to molecular diffusion since the latter is expected to follow an
Arrhenius-type temperature dependence, $D_\mathrm{s} \sim \exp [- \Delta E /
k_\textrm{B}T]$ (where $\Delta E$ is the activation energy). Similarly,
according to the kinetic theory of gases, the viscous gas flow --- as will be
seen below --- follows a temperature dependence $\sim 1/\eta$ where $\eta \sim
\sqrt{T}$  which differs from the temperature dependence expected for the
Knudsen diffusivity. Finally, it can be noted that the Knudsen diffusivity is
sensitive to the mass of the gas molecules diffusing within the porous
material, $D_\textrm{K} \sim \sqrt{m}$. 

\noindent \textbf{Corrections to the Knudsen regime.}
In the derivation above, we implicitly assumed that in the Knudsen regime
molecules  colliding with the surface of the porous solid undergo diffuse
scattering. This corresponds to an important approximation as rigorously only
a fraction $f$ of the colliding molecules undergoes this mechanism while the
complementary fraction $1-f$ undergoes specular reflections (elastic
collisions).
$f$ is known as the {tangential momentum accommodation coefficient}. In
this context, diffuse scattering indicates that, upon colliding with the
surface of the host porous medium, some molecules become adsorbed at the
surface, get thermalized and are then released into the porous volume with a
velocity that is independent of their initial velocity at the collision time.
\Cref{EQ4.5} can be generalized to any value of $f$:
\begin{equation}
D_\textrm{K } = \frac{D_\textrm{p}\Phi}{3 \tau} \times \frac{2-f}{f}
\sqrt{\frac{8 k_\textrm{B}T }{\pi m }}
\label{EQ4.6}
\end{equation}
Specular reflections of the fraction $1-f$  of the molecules colliding with
the surface lead to a surface slippage phenomenon, where the term $(2-f)/f$ in
\cref{EQ4.6} must be seen as a slippage coefficient.
In \Cref{chap:transport_pore}, it will be shown that
this correction is inversely proportional to the pressure of the gas.
Due to this effect, the Knudsen permeability defined in \cref{EQ4.4}
underestimates Knudsen permeabilities measured in experimental set-ups. Thus,
the permeability must be corrected to obtain a permeability $\Pi_\textrm{K}^*$
accounting for this slippage effect which adds up to the Knudsen permeability
$\Pi_\textrm{K}$. This correction is known in the literature as the
Klinkenberg effect~\cite{ziarani_knudsens_2012,bear_modeling_2018}:
\begin{equation}
\Pi_\textrm{K }^* = \Pi_\textrm{K }(1 + a/P)
\label{EQ4.7}
\end{equation}
where $a$ is the correction term in pressure units. In practice, this
correction is negligible for Kn $< 0.1$, very small for Kn $\sim 1$ and
important or even very important for Kn $> 10$.

\begin{figure}[htbp]
	\centering
	\includegraphics[width=0.95\linewidth]{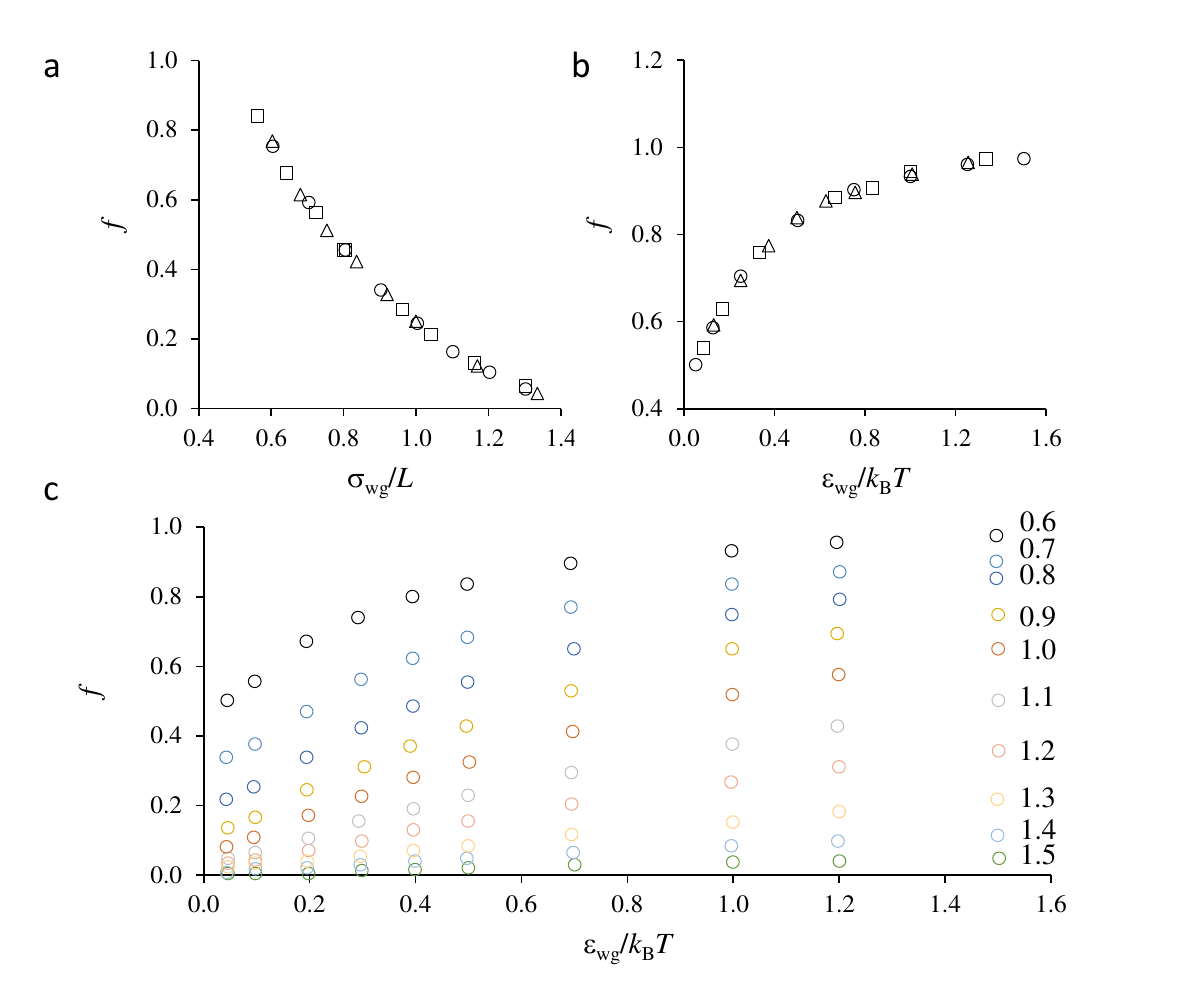}
	\caption{\textbf{Knudsen diffusion}. Effect of the  different molecular
		parameters on the {tangential momentum accommodation 
			coefficient} $f$.
		(a) Effect of the wall/gas interaction parameter $\sigma_\mathrm{wg}$
		normalized to the characteristic
		distance $L$ between substrate atoms. In this case, all calculations
		are taken at constant $\varepsilon_\mathrm{wg}/k_\textrm{B}T$ = 100 with
		$T$ = 300 K.
		The circles, squares and triangles correspond to $L$ = 0.4, 0.5 and 0.6
		nm, respectively. (b) Effect of the solid/fluid interaction
		$\varepsilon_\mathrm{wg}$ normalized to the thermal energy
		$k_\textrm{B}T$. In this case, all calculations are for
		$\sigma_\mathrm{wg}$ = 0.24 nm and $L = 0.4$ nm.
		The circles, squares and triangles correspond to $T =$ 200, 300 and 400
		K, respectively. (c) Effect of the solid/fluid interaction
		$\varepsilon_\mathrm{wg}$ normalized to the thermal energy $k_\textrm{B}T$
		for different ratios $\sigma_\mathrm{wg}/L$ indicated in the figure.
		Data adapted with permission from Arya~\etal.~\cite{arya_molecular_2003}.
		Copyright 2003 Taylor \& Francis.}
	\label{fig:4_2}
\end{figure}

\noindent \textbf{Molecular simulation.}
In spite of the abundant literature on Knudsen diffusion in porous media, only
a few studies have examined the role of molecular parameters such as the size
of the diffusing gas molecule or the strength of the molecular interactions
between the gas and the solid surface. In this context, the molecular
simulation work carried out by Maginn and coworkers
is particularly relevant.\cite{arya_molecular_2003}
These authors have investigated the role of
different parameters such as the temperature $T$, the size $\sigma$  of the gas
molecule and the solid/fluid molecular interaction energy
$\varepsilon_\mathrm{wg}$.
Let us consider a Lennard--Jones fluid whose parameters are $\sigma_\mathrm{g}$
and $\varepsilon_\mathrm{g}$ (\ie.\
interaction parameters between fluid molecules).
These parameters can be estimated for any gas.
Indeed, the phase diagram of the Lennard--Jones fluid is universal;
this means that for any pair of parameters $\sigma_\mathrm{g}$ and
$\varepsilon_\mathrm{g}$, certain thermodynamic points
such as the temperature of the triple point $T_\mathrm{tr}$,
the temperature of the critical point
$T_\mathrm{c}$, the pressure of the critical point $P_\mathrm{c}$,
the melting temperature $T_\mathrm{m}$ are uniquely defined.
For example, the critical point of
Lennard--Jones fluids in reduced units is given by
$T_\mathrm{c} \sim 1.32 \varepsilon_g$~\cite{thol_equation_2016}.
Other thermodynamic properties such as the critical point density
$\rho_c$ also take unique values when expressed in Lennard-Jones reduced units. For a given
gas, the corresponding parameters $\sigma_\mathrm{g}$ and $\varepsilon_\mathrm{g}$
can also be estimated from the second coefficient of the virial $B(T)$. These
values as well as the parameters $\sigma_\mathrm{g}$ and $\varepsilon_\mathrm{g}$
can be found in the book
by Hirschfelder \etal.~\cite{hirschfelder_molecular_1954}.

The parameters describing the interactions within the solid phase
$\sigma_\mathrm{ww}$ and $\varepsilon_\mathrm{ww}$ can
be defined in a similar way. Typical parameters are as follows:
$\varepsilon_\mathrm{ww}/k_\textrm{B} = 230\,\mathrm{K}$
and $\sigma_\mathrm{ww} = 0.27\,\mathrm{nm}$ for oxygen atoms
in oxide materials and $\varepsilon_\mathrm{ww}/k_\textrm{B}$ = 28 K and
$\sigma_\mathrm{ww} = 0.34\,\mathrm{nm}$ for carbon atoms in carbonaceous materials.
From these parameter sets, using the Lorentz--Berthelot combination
rules,\cite{allen_computer_2017,hansen_theory_2013}
we can estimate the cross-terms governing the solid/gas interactions,
$\sigma_\mathrm{wg} = (\sigma_\mathrm{ww}+\sigma_\mathrm{gg})/2$ and
$\varepsilon_\mathrm{wg} = \sqrt{\varepsilon_\mathrm{ww}\epsilon_\mathrm{gg}}$.
In what follows, all quantities are normalized;
energies are normalized to the thermal energy $k_\textrm{B}T$,
while the different $\sigma$ and lengths are normalized to the solid lattice
parameter $L$ ($L \sim 0.25\,\mathrm{nm}$ for graphite/carbon and $L \sim
0.28\,\mathrm{nm}$ for siliceous materials).
The use of normalized data is very convenient because it
allows one to extend the results of Maginn and coworkers to any gas/solid systems.
These authors have studied for a large number of parameters $T$,
$\sigma_\mathrm{wg}$
and $\varepsilon_\mathrm{wg}$ the values reached for $f$ in the Knudsen regime
(\cref{fig:4_2}).
\Cref{fig:4_2}(a) shows that, for a given interaction energy
$\varepsilon_\mathrm{wg}$, $f$ does not depend on $L$ when plotted as a function of $\sigma_\mathrm{wg}/L$.
Similarly, in \cref{fig:4_2}(b), for a ratio $\sigma_\mathrm{wg}/L$, if $f$
is plotted as a function of $\varepsilon_\mathrm{wg}/k_\textrm{B}T$, we obtain a
master curve. The set of results obtained by means of molecular dynamics
simulation are shown in \cref{fig:4_2}(c) where $f$ is plotted as a function
of $\varepsilon_\mathrm{wg}/k_\textrm{B}T$ for different $\sigma_\mathrm{wg}/L$
ratios.	Indeed, \cref{EQ4.6} suggests a diverging transport coefficient for
	$f\to 0$, \ie.\ in the limit of purely specular reflections.
This unphysical behavior would correspond to perfectly smooth and athermal surfaces, which does not correspond to a real system, therefore a finite $f$ should always be considered. 

%\redlork{JL - I looked at the reference and find it  rather obscure, so I suggest to remobve the following : This non-physical asymptotic regime is due to the fact that molecular interactions are completely neglected in the Knudsen description above.	This can be corrected for by taking into account the length 	variation of a particle traveling through the pore of length $L_\mathrm{p}$ and getting scattered at an angle $\Theta$. 	This length variation can be estimated as	$\Delta L_\mathrm{p} = D_\mathrm{p} \tan \Theta$, \ie.\ each time a particle is scattered its effective length for traversing 	the pore is increased. Accounting for this effect via the transmission probability $\alpha_\mathrm{K}$, one obtains a generalized expression for the Knudsen diffusion by replacing the slippage factor appearing in \cref{EQ4.6}:\cite{qian_generalized_2023}
%	\begin{equation}
%		\frac{2-f}{f} \to \frac{1}{f+\alpha_\mathrm{K}-f\alpha_\mathrm{{K}}}
%		\label{eq:knudsen_generalized}
%	\end{equation}
%	For $f\to 1$ the regular Knudsen diffusion is recovered in both cases,	however, as shown in Ref.~\citen{qian_generalized_2023}, only the generalized expression \cref{eq:knudsen_generalized} correctly captures the behavior for $f\to 0$.}

\subsubsection{Molecular diffusion}
\label{sec:molecular_diffusion}

\begin{figure}[htbp]
	\centering
	\includegraphics[width=0.95\linewidth]{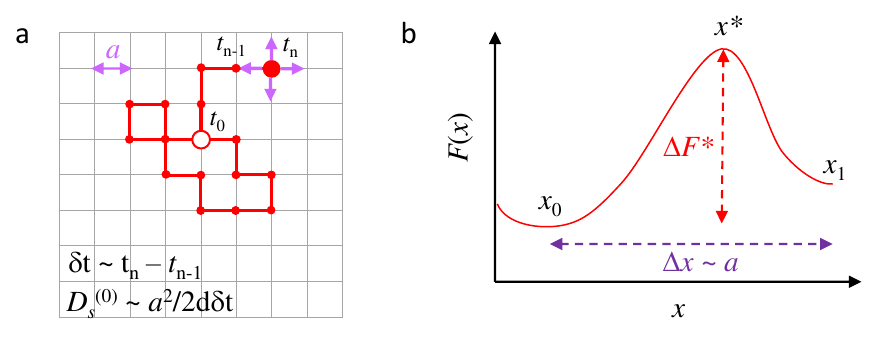}
	\caption{\textbf{Random walk and transition state theory}. (a) In the
		random walk model, the molecule moves by a quantity $a$ in a random
		direction over a characteristic time $\delta t$. Starting from an
		initial position at $t = t_0$, the mean square displacement at a time
		$t_n$, \ie.\ after $n$ time steps, is $r^2(t) \sim 2d
		D_\mathrm{s}^{(0)} n \delta t$ where $D_\mathrm{s}^{(0)}$ is the
		self-diffusivity and $d$
		the dimensionality of the system. (b) The random walk model assumes
		that there is no free energy barrier involved in the displacement from
		one site to another.
		In contrast, at the molecular scale, displacements are stochastic
		processes with an acceptance rate that depends on the free energy
		barrier $\Delta F^*$ to move from a position $x_0$ to a position $x_1$.
		In such cases, the hopping rate $k \sim \exp (-\Delta
		F^*/k_{\textrm{B}}T)$ as described in detail in the text.}
	\label{fig:4_3}
\end{figure}

In contrast to Knudsen diffusion, which relies on a clear underlying mechanism
(gas transport driven by a mean free path affected by collisions with the pore
surface), there is no general prediction for molecular diffusion in nanopores
($\lambda < D_{\textrm{p}}$) as it might correspond to different mechanisms.
Yet, several robust theoretical predictions for the self-diffusivity in pores
can be established, which remain valid in most physical situations encountered
when dealing with diffusion in porous media. Self-diffusion can be discussed
by introducing the random walk model where a molecule, the walker, jumps from a
site to a random neighboring site on a lattice within a time $\delta t$
[\cref{fig:4_3}(a)].
This
%which corresponds to the solution of Brownian motion,
leads to an average mean-square displacement of the particle at a
time $t$ given by $ \left\langle r^2(t) \right\rangle \sim 2d D_\mathrm{s} t  $
where $a$ is the lattice spacing and $d$ is the dimensionality of the lattice
(1D, 2D, \etc.).
Comparison between this expression and the mean-square displacement predicted
using the Gaussian propagator expected in the Fickian regime leads to:
$D_\mathrm{s}^{(0)} = a^2/6\delta t = k_0 a^2/6$
where $k_0 = 1/\delta t$ is the hopping rate on the lattice model. In
practice, $k_0 \sim v_0/a$ is related to the mean thermal velocity in one direction $v_0 =
\sqrt{2 k_\textrm{B} T/\pi m}$ and the lattice parameter $a$ (the latter is of the
order of the intermolecular distance in the liquid).

In the derivation above, by relying on the random walk model, we implicitly assumed that there is no free energy barrier involved in the displacement from one site to another. This assumption holds in a coarse-grained picture when considering displacement over a supramolecular size (in which case, any free energy barrier is encompassed in the value of the self-diffusivity $D_\mathrm{s}$).
However, at the molecular scale, it is clear that displacements from
one site to another site are stochastic processes with an acceptance rate that depends on the local free energy barrier $\Delta F^*$ (the sign $*$ is used to indicate that we refer to a free energy barrier and not a free energy difference between two stable states).
When such free energy barriers exist, the hopping rate $k$ is not simply $k_0 =
1/\delta t$ but $k \sim k_0 \exp (-\Delta F^*/k_{\textrm{B}}T)$. If we assume
that the free energy barrier $\Delta F^*$ is temperature independent in the
temperature range considered, the above ingredients lead to the following
expression:
\begin{equation}
D_\mathrm{s} = D_\mathrm{s}^{(0)} \exp{\Big (-\frac{\Delta F^*}{k_\textrm{B}T}
\Big )}
\label{EQ4.8}
\end{equation}
The latter equation, which has a characteristic temperature dependence $\ln D_\mathrm{s}
\sim 1/T$, is known as Arrhenius law.\cite{kramers_brownian_1940,
hanggi_reaction-rate_1990}

The transition state theory provides a more rigorous
basis for \cref{EQ4.8}~\cite{camp_transition_2016,voter_transition_1984}.
Let us consider a molecule in a bulk or
confined liquid where the typical free energy profile as a function of the
molecule position is shown in \cref{fig:4_3}(b). The molecule, initially
located at a position $x_0$, will move to an available site $x_1$ by crossing
the free energy barrier $\Delta F^*$.
Typically, mapping this free energy approach to the
random walk model shown in \cref{fig:4_3}(a) implies that $a \sim x_1-x_0$.
In the transition state theory, the rate $k_\mathrm{TST}$, which is defined as
the number of molecules located in $x_0$  that cross the free energy barrier
per unit of time, is given by the probability that a molecule is located at the
free energy barrier multiplied by the frequency at which it crosses the barrier.
By virtue of the definition of the free energy profile $F(x)$
in the region $\lbrack 0, x^\star \rbrack$, the probability
$p(x^*)\textrm{d}x$ to have a molecule located in a region $\textrm{d}x$ at
the free energy barrier $\Delta F^\star$ located in $x^*$ is:
\begin{equation}
{\int_0^{x^*} \exp\left[-\beta  \Delta F(x)\right] \,\rmd x}
\label{EQ4.9}
\end{equation}
where we used the reciprocal temperature $\beta = 1/k_\textrm{B}T$.
The frequency with which a molecule crosses the barrier
once reaching the top of the barrier can be defined as $1/\textrm{d}t \sim
v_0/\textrm{d}x$, where  the average velocity $v_0$ of the molecule at the top
of the barrier is taken equal to the mean thermal velocity
$v_0 = \sqrt{2 k_\textrm{B} T/\pi m}$. From these estimates, one predicts
the rate $k_\mathrm{TST} \sim \nicefrac{1}{2} \, p(x^*)\,\textrm{d}x
\,/\,\textrm{d}t$ according to:
\begin{equation}
k_\mathrm{TST} = \sqrt{\frac{k_\textrm{B} T}{2\pi m}} \times
\frac{\exp\left(-\beta \Delta F^*\right)}{\int_0^{x^*} \exp\left[-\beta \Delta
F(x)\right]\,\rmd x}
\label{EQ4.10}
\end{equation}
where the factor 1/2 accounts for the fact that only molecules going in the
direction from $x_0$ to $x_1$ will cross the barrier (hence not those going in
the direction from $x_1$ to $x_0$). In practice, the last expression
overestimates the crossing rate as a non-negligible number of molecules
recross the energy barrier in the other direction.
In order to correct \cref{EQ4.10}
for this effect, one needs to compute the transmission
coefficient, $\kappa \in \lbrack 0, 1\rbrack$, known as the Bennett--Chandler
dynamic correction factor~\cite{chandler_statistical_1978}.
Such computations require to run independent molecular dynamics simulations in
which molecules are initially positioned at the top of the energy barrier with
a velocity selected randomly from a Maxwell--Boltzmann distribution at a
temperature $T$. With these simulations, $\kappa$ is  readily obtained by counting the number of
molecules that do cross the free energy barrier and the corrected
transition state expression writes:
\begin{equation}
k_\mathrm{TST} = \kappa \sqrt{\frac{k_\textrm{B} T}{2\pi m}} \times
\frac{\exp\left(-\beta \Delta F^*\right)}{\int_0^{x^*} \exp\left[-\beta \Delta
	F(x)\right]\,\rmd x}
\label{EQ4.11}
\end{equation}
By using the latter expression for the hopping rate $k_\mathrm{TST}$, we can
establish a simple equation for the self-diffusivity as
$D_\mathrm{s}  = z_0 k_\mathrm{TST} a^2 / 2 d$
where $z_0$ is the number of sites accessible by a single jump (e.g. 
 $z_0 = 6$ for a cubic lattice). In practice, $z_0 = 2d$ so that
$D_\mathrm{s} = k_\mathrm{TST} a^2$.

\subsubsection{Surface diffusion}

Let us come back to the situation described in the previous paragraph where intermediate adsorbed amounts $\Gamma$ are considered (\ie.\ incomplete pore fillings).
\redlork{While surface diffusion within an adsorbed film is always possible, it often becomes very small or even negligible when strongly adsorbed molecules are considered. In fact, when the adsorption energy is just above or a few times the thermal energy $k_\textrm{B}T$, jump diffusion through the adsorbed film necessarily occurs. Let us consider an adsorbed layer coexisting with a gas phase having a very low density (this system is typically obtained at low temperatures where the gas density can be very low). In that case, diffusion only occurs through surface diffusion within the adsorbed film since diffusion in the gas phase is negligible -- not because the gas diffusivity is low but because the very low fraction of molecules within the gas phase makes their contribution to overall diffusion very small or negligible. Another illustrative example of surface diffusion corresponds to transport in an adsorbed layer where the free energy landscape near the solid surface is smooth. In that case, the molecules can be strongly adsorbed because of a large binding energy but the absence  of significant surface corrugation leads to significant diffusivity within the adsorbed film.}

Provided that the temperature is low enough, the pore surface is covered with an adsorbed film while the pore center is filled by the gas. If the density ratio between the adsorbed and gas phases sufficiently differs from 1, molecular diffusion mostly occurs through  diffusion within the surface phase (situations where self-diffusion occurs through a combination of diffusion within the surface and through the gas phase will be treated in \Cref{sec:diffusion_in_nanopores}).
Such diffusion restricted to the adsorbed phase, which is
referred to as surface diffusion in the literature, is illustrated in \cref{fig:4_1}(e).  Like molecular diffusion in pores, the underlying mechanisms that lead to surface diffusion can be complex and of different
nature. Yet, the Reed--Ehrlich model~\cite{reed_surface_1981} described in the rest of this subsection provides a robust and simple theoretical framework to describe such surface diffusion. As discussed below, in its asymptotic limit of negligible lateral interactions between adsorbed molecules, this model leads to a very simple expression for the adsorbed amount dependence of the collective diffusivity $D_0$ known as the surface diffusion model where $D_0 \sim 1 - \theta$ ($0 \leq \theta \leq 1$ is the site occupancy in the adsorbed phase).

As discussed in \Cref{chap:fundamentals}, upon fluid transport in a porous
material with a flow $J$, the transport diffusivity $D_\mathrm{T}$ defined as
$J = - D_\mathrm{T} \nabla c$ and the collective diffusivity $D_0$ defined
as $J = - c D_0/k_{\textrm{B}}T \times \nabla \mu$
are related through the so-called Darken or thermodynamic factor:
\begin{equation}
D_\mathrm{T}(\theta) =  D_0(\theta) \frac{\partial
(\mu/k_\textrm{B}T)}{\partial \ln c}\bigg\lvert_T
\label{EQ4.12}
\end{equation}
where $c$ is the fluid concentration in the porous material. Formally, this
equation is strictly equivalent to \cref{EQ3.44} since the fugacity is $f =
k_{\textrm{B}}T/\Lambda^3 \exp[\mu/k_\textrm{B} T]$, where $\Lambda$ is
de Broglie's thermal wavelength. Let us describe the adsorbed phase as a 2D
lattice gas where the occupancy $\theta$ is defined from the number of
adsorbed molecules $N$ normalized  to the number of available sites for
adsorption $N_0$, \ie.\ $\theta = N/N_0$. The jump rate $k(\theta)$ at
a given occupancy $\theta$ defines the local diffusivity $D_0(\theta) \sim
k(\theta) a^2$ where $a$ is the lattice parameter which scales with the
molecular size of the fluid (more precisely, if $c_{s}$ is the maximum density
at the pore surface, $a \sim c_s^{1/2}$). \Cref{EQ4.12} allows one to write:
\begin{equation}
D_\mathrm{T}(\theta) =  k(\theta) a^2 \frac{\partial
(\mu/k_\textrm{B}T)}{\partial \ln
	\theta}\bigg\lvert_T
\label{EQ4.13}
\end{equation}
where we used that $\partial \ln c = \partial \ln \theta$ since $\theta =
c/c_s$.

The last equation shows that $D_\mathrm{T}$ is the product of the jump rate
$k(\theta)$
and the thermodynamic factor $\partial(\mu/k_\textrm{B}T)/\partial \ln \theta$
which can be evaluated independently of each other. The jump rate $k(\theta)$
is proportional to the probability $P^{(z)}$ that $z$ of the $z_0$ neighboring
sites of an adsorbed molecule are occupied by other molecules --- typically, in
a 2D square lattice, $z_0 =  4$ --- multiplied by the jump rate in such a
configurational environment $k^{(z)}$:
\begin{equation}
k(\theta) = \sum_{z = 0}^{z_0} \frac{z_0 - z}{z_0} \times P^{(z)} k^{(z)}
\label{EQ4.14}
\end{equation}
where the contribution $(z_0 - z)/z_0$ simply accounts for the number of sites
in which the molecule can jump. Assuming additive lateral interactions between
the adsorbed molecules and its neighbors, \ie.\ $E \sim w z$ with $w$
the interaction with a single neighbor, $P{(z)}$ can be estimated in the
quasi-chemical approximation --- an extension of the Bragg-Williams
approximation proposed by Fowler and Guggenheim:\cite{hill_introduction_1986,
zaafouri_cooperative_2020}
\begin{equation}
P^{(z)} = \frac{z_0 !}{z ! (z_0-z)!} \frac{(\eta \epsilon)^z}{(1+ \eta
	\epsilon)^{z_0}}
\label{EQ4.15}
\end{equation}
with $\eta = \exp(-w/k_\textrm{B}T)$ and $\epsilon = (\beta-1 +
2\theta)/\lbrack 2\theta(1-\theta) \rbrack$ with $\beta = \lbrack 1-4\theta
(1-\theta)(1-\eta)\rbrack^{1/2}$. As for the jump rate $k^{(z)}$, Reed and
Ehrlich assumed that it scales as a power law with an exponent corresponding to
the number of nearest neighbors: $k^{(z)} = k^{(0)} \eta^{-z}$.\cite{reed_surface_1981}
Gathering all these results into \cref{EQ4.14} leads to:
\begin{equation}
k^{(z)} = k^{(0)}  \frac{(1+\epsilon)^{z_0-1}}{(1+ \eta \epsilon)^{z_0}}
\label{EQ4.16}
\end{equation}
In the quasi-chemical approximation, the chemical potential of the
adsorbed phase writes:
\begin{equation}
\mu = \mu_0 + k_\textrm{B}T \ln \left[ \frac{\theta}{1-\theta} \right]
+ \frac{z_0}{2}  k_\textrm{B}T \ln \left[ \frac{(\beta-1 + 2\theta)
	(1-\theta)}{(\beta + 1-2 \theta) \theta} \right]
\label{EQ4.17}
\end{equation}
where the first, second, and third terms on the right hand side correspond
respectively to (1) the chemical potential at the bulk saturating vapor
pressure where $\theta \rightarrow 1$, (2) the occupancy term for the lattice
gas model in the absence of lateral interactions between adsorbed molecules
and (3) a correction term that accounts for the lateral interactions between
adsorbed molecules. By using \cref{EQ4.17} to derive the thermodynamic
factor from \cref{EQ4.12}, we obtain:
\begin{equation}
\left. \frac{\partial(\mu/k_\textrm{B}T)}{\partial \ln \theta} \right\vert_T =
\frac{1}{1-\theta}
\left[ 1 + \frac{z_0(1-\beta)}{2\beta} \right]
\label{EQ4.18}
\end{equation}
By inserting \cref{EQ4.16,EQ4.18} into \cref{EQ4.13}, we can
establish a simple expression for the transport diffusivity:
\begin{equation}
D_\mathrm{T}(\theta) = \frac{k(0) a^2 (1+
\epsilon)^3}{(1-\theta)(1+\eta\epsilon)^4}
\left\lbrack 1 + \frac{2(1-\beta)}{\beta} \right\rbrack
\label{EQ4.19}
\end{equation}
As an important remark, note that the model by Reed--Ehrlich in the limit of negligible
adsorbate/adsorbate interactions ($w \sim 0$)  allows one to
recover the well-known asymptotic limit of the Langmuir model. Briefly, for
$w \sim 0$, we have $\eta \sim 1$, $\beta \sim 1$ and $\epsilon \sim
\theta/(1-\theta)$ which leads to:
\begin{equation}
\mu = \mu_0 + k_\textrm{B}T \ln \left[ \frac{\theta}{1-\theta} \right]
\label{EQ4.20}
\end{equation}
and
\begin{equation}
D_\mathrm{T}(\theta) = k(0) a^2 \hspace{3mm} \textrm{and} \hspace{3mm}
D_0(\theta) =
k(0) a^2 (1 - \theta)
\label{EQ4.21}
\end{equation}
as expected in the Langmuir regime.
In particular, by using the chemical
potential/pressure for an ideal gas, $\mu \sim k_\textrm{B}T \ln P$,
\cref{EQ4.20} can be recast into the  Langmuir adsorption isotherm:
$\theta = \alpha P/(1 + \alpha P)$ where $\alpha$ is an affinity parameter
that depends on the fluid/solid interaction
strength.\cite{langmuir_adsorption_1918}

\begin{figure}[htbp]
	\centering
	\includegraphics[width=0.95\linewidth]{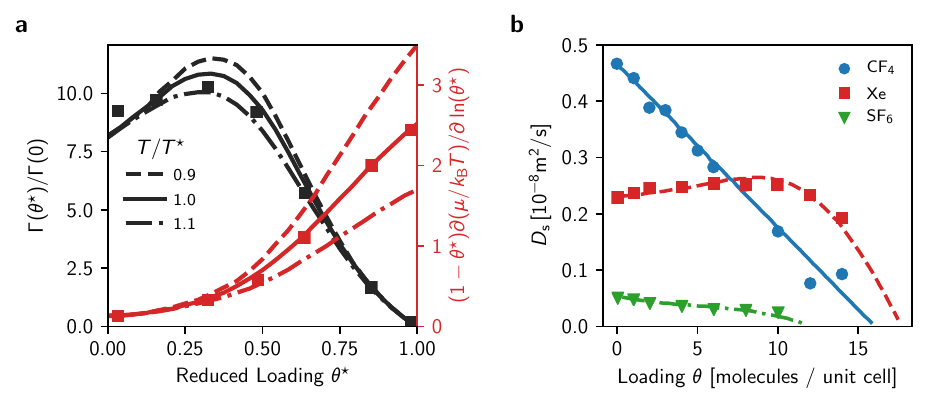}
	\caption{\textbf{Surface diffusion model}.
		(a) Effect of loading at different temperatures on the jump rate
		$\Gamma(\theta)$ and the thermodynamic factor
		$\partial(\mu/k_\textrm{B}T)/\partial \ln \theta \big \vert_T$ as
		derived within  the quasi-chemical approximation for repulsive
		interactions ($w = k_\textrm{B}T^* > 0$). Note that the jump rate is
		normalized to $\Gamma(0)$, \ie.\
		the jump rate in the zero loading limit, and the thermodynamic factor
		is multiplied by $(1-\theta^*)$ for clarity. The lines are the
		predictions of the theoretical model while the symbols are the results
		from Monte Carlo simulations for $T = T^*$. Adapted with permission from
		Ref.~\protect\citen{reed_surface_1981}.
		Copyright 1982 North-Holland Publishing Company.
		(b) Effect of loading $\theta$ on the diffusivity of CF$_4$, Xe, and
		SF$_6$ in  silicalite zeolite (MFI). The lines are the results from the
		surface	diffusion model by Reed and Ehrlich. The open symbols are 
		Molecular Dynamics simulation data from Skoulidas and
		Sholl~\cite{skoulidas_transport_2002,skoulidas_molecular_2003} and
		Chempath~\etal.~\cite{chempath_nonequilibrium_2004}. Adapted with permission
		from Ref.~\protect\citen{krishna_modeling_2004}.
		Copyright 2004 Elsevier Inc.}
	\label{fig:4_4}
\end{figure}

In their seminal paper,\cite{reed_surface_1981}
Reed and Ehrlich discussed the effect of surface
coverage $\theta$ on the jump rate $k(\theta)$ as well as on the thermodynamic
factor $\partial(\mu/k_\textrm{B}T)/\partial \ln \theta \big \vert_T$. In more
detail, these authors showed that the predictions of their model based on the
quasi-chemical approximation are in good agreement with
results from Monte Carlo simulations (\cref{fig:4_4}). Upon considering
repulsive lateral interactions between adsorbed molecules, $w = k_\textrm{B}T^*
> 0$, the jump rate was found to  be strongly affected by the loading $\theta$
with important deviations observed with respect to the prediction for the
Langmuir regime.
	It  was also recently shown that, indeed, upon increasing the coverage $\theta$
	there is a continuous transition to classical hydrodynamics for adsorbed
	water films.\cite{gravelle_transport_2022}

\subsection{Diffusion in nanopores}
\label{sec:diffusion_in_nanopores}

In \Cref{sec:diffusion_mechanisms}, the different diffusion mechanisms that
can occur in a pore have been discussed and identified as a function of simple
parameters --- namely, the pore size $D_\textrm{p}$, fluid molecule size
$\sigma$, and mean free path $\lambda$.
Despite the rather exhaustive description given of these phenomena,
the emerging picture often remains insufficient to describe transport in a
nanoporous medium. Even when a single pore of a simple geometry is considered,
the description provided in the previous section is incomplete for the two
following reasons. First, except in well defined asymptotic regimes (vanishing
confined fluid density, incompressible confined liquid, \etc.),  the
physical diffusion mechanism in a given pore can show a non trivial dependence
on fluid pressure and/or density.
As will be discussed below, even though some hints at the
effect of pore loading on diffusivity have been already provided in
\Cref{sec:diffusion_mechanisms}, the theoretical description of such density
and/or pressure dependence  often varies from one example
to another. Second, even for a single pore, in most situations, diffusion
arises from a combination of different mechanisms rather than a single
mechanism. Such intrinsic complexity raises the question of the formalism
required to describe such combined mechanisms and, more practically, of the
existence of simple combining rules.
In what follows, these different points are examined in detail. In particular,
we will see under what specific conditions the usual approximations consisting
of assuming that Knudsen and molecular diffusions occur in series while
surface diffusion occurs in parallel are valid.

\begin{figure}[htbp]
	\centering
	\includegraphics[width=0.95\linewidth]{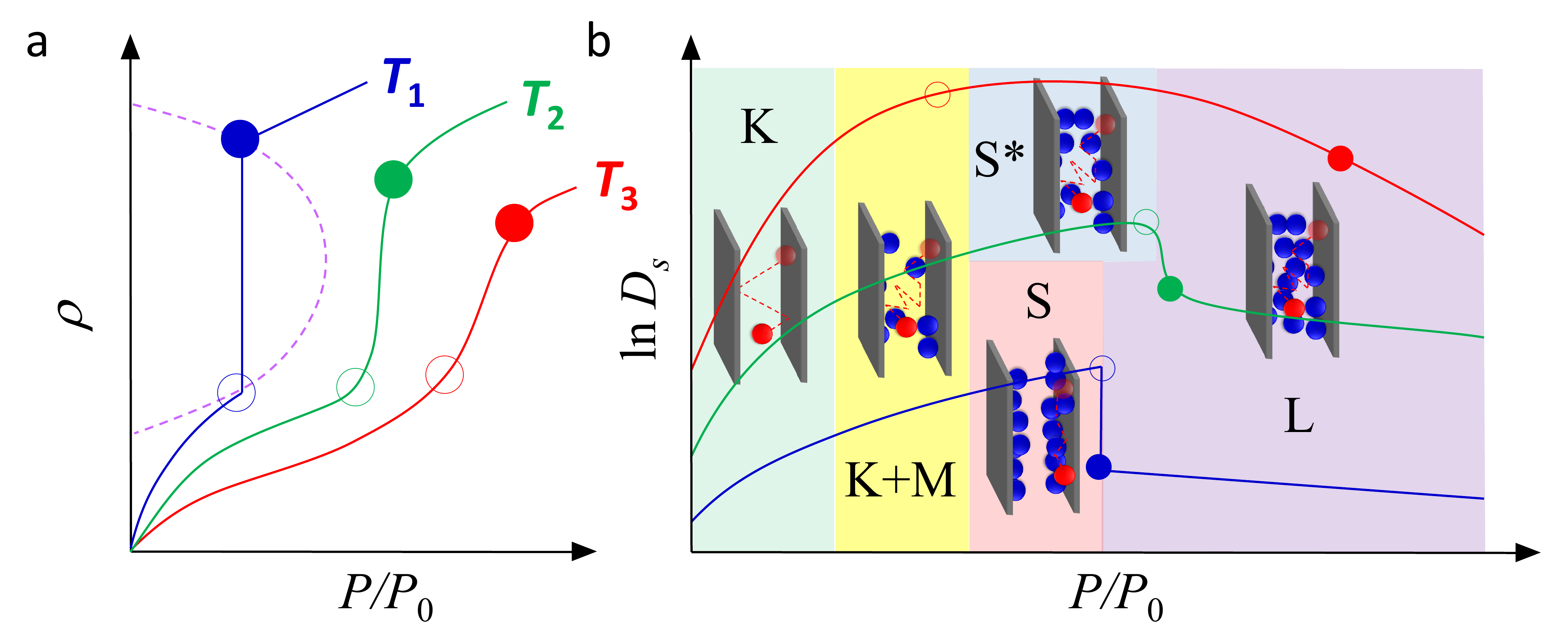}
	\caption{\textbf{Diffusion at the pore scale.} (a) Typical adsorption
		isotherms for a fluid in a single nanopore of a size $D_{\textrm{p}}$
		at different temperatures $T_1 < T_2 < T_3$. The $y$-axis shows the
		fluid density $\rho$ as a function of the reduced gas pressure $P/P_0$
		shown in the $x$-axis (where $P_0$ is the saturating vapor pressure for
		fluids below their critical point $T_c$ and
		its extrapolated value for $T > T_c$). At low temperature, the fluid
		is below its capillary critical point $T_1 < T_{cc}(D_{\textrm{p}})$
		and pore filling proceeds through the formation of an adsorbed film
		coexisting with the bulk gas phase in the pore center followed by
		discontinuous capillary condensation at a pressure $P < P_0$\cite{coasne_adsorption_2013}. At higher
		temperatures $T_3 > T_2 > T_{cc}$, the fluid is in a pseudo-critical
		state and capillary condensation is replaced by a continuous and
		reversible pore filling.\cite{coasne_temperature_2005}
		(b) Typical self-diffusivities for a fluid in
		a single nanopore of a size $D_{\textrm{p}}$ at
		different temperatures $T_1 < T_2 < T_3$ as a function of the fluid
		pressure $P/P_0$. Independently of temperature, the diffusivity first
		increases upon increasing the pressure (and therefore the confined
		fluid density) as fluid molecules first occupy strongly adsorbing sites
		(large negative adsorption energies) before filling weaker adsorbing
		sites (less negative adsorption energies). In all cases, in the very
		low density/pressure range, molecules first obey the Knudsen diffusion
		regime (K) before entering a hybrid regime combining Knudsen and
		molecular diffusion (K + M). Upon further increasing the pressure or
		confined fluid density, surface diffusion (S) is observed with two
		possible scenarios depending on the density difference $\Delta\rho =
		\rho_\mathrm{s} - \rho_\mathrm{b}$ between the adsorbed fluid phase at
		the pore surface and the bulk-like fluid phase in the pore center. At
		high temperature, $\lvert \Delta\rho \lvert  \ll \rho_\mathrm{s}$ so
		that significant exchange is observed between these two phases (S$^*$).
		At low temperature, $\lvert \Delta\rho \lvert \sim \rho_\mathrm{s}$ so
		that exchange between the two phases is limited (S).}
	\label{fig:4_5}
\end{figure}

\Cref{fig:4_5} shows the self-diffusivity $D_\mathrm{s}$ as a function of fluid
pressure $P$ at different temperatures for a simple fluid confined in a slit
pore of a width $D_\textrm{p}$ of about a few $\sigma$.
At all temperatures, the self-diffusivity shows a non-monotonic behavior with
$D_\mathrm{s}$ that first increases with pressure $P$ and then decreases upon
further increasing $P$.
At low temperature, \ie.\ below
the so-called critical capillary temperature $T_\textrm{cc}$,\cite{evans_theory_1984,
evans_fluids_1986,evans_capillary_1986,nakanishi_multicriticality_1982,
nakanishi_critical_1983, dillmann_monte_2001, binder_monte_2003,
thommes_pore_1994, morishige_adsorption_1998}
the self-diffusivity exhibits a marked discontinuous change in $D_\mathrm{s}$ as
capillary condensation occurs.
In contrast, at higher temperature, \ie.\ above
$T_\textrm{cc}$, the self-diffusivity decreases in a continuous and reversible
fashion as pore filling becomes also continuous and reversible (for a detailed
discussion on the effect of confinement on the fluid critical temperature and
the so-called critical capillary temperature, the reader is invited to read
Refs.~\citen{coasne_adsorption_2013,schlaich_dispersion_2019}). There is a
number of physical situations encountered in the self-diffusivity data shown
in \cref{fig:4_5} that were already examined in the previous section: namely,
the Knudsen regime, the surface diffusion regime and the molecular diffusion
regime.
In contrast, intermediate regimes obtained for fluid densities or pressures
where these different regimes coexist remain to be addressed. Such transition
regimes, which are discussed in the rest of this section, include the
combination of Knudsen and molecular diffusion through the pore center, the
combination of surface diffusion and molecular diffusion through the pore
center but also the change in the molecular diffusion upon increasing the
fluid density within the pore.

\subsubsection{Combined Knudsen/molecular diffusion}

In the limit of very low confined fluid density, the diffusion of gases within
a porous material occurs through Knudsen diffusion characterized by a Knudsen
diffusivity $D_\mathrm{K}$ and a corresponding permeance $\Pi_\mathrm{K}$ given
in \cref{EQ4.4,EQ4.5} in \Cref{sec:diffusion_mechanisms}.
However, in practice, even for very low densities, depending on
temperature, diffusion in a porous material can show non negligible departures
from such equations because diffusion through collisions between fluid
molecules cannot be completely ignored. Such molecular diffusion needs to be
accounted for to derive a simple expression for the effective self-diffusivity
arising from the combination of Knudsen and molecular diffusion. This problem
is analogous to the treatment of electron diffusion in solids which occurs
through the combination of two different diffusion
mechanisms~\cite{kasap_springer_2017,ashcroft_solid_1976}:
(1) diffusion through interactions with lattice vibrations, \ie.\
phonons, and (2) diffusion through collisions with impurities. Such a derivation
leads to the so-called Matthiessen’s rule which is useful to describe electrical
conductivity and resistivity in real metals and semiconductors. In what
follows, the Matthiessen’s rule  in the context of diffusion of fluid
molecules in the combined Knudsen/molecular diffusion regime is derived.

Let us consider an ensemble of $N(0)$ fluid molecules diffusing in a porous
material at a density such that the mean free path is $\lambda$. The number of
molecules that have not undergone a collision with another molecule or with
the pore surface at a time $t$ is $N(t)$. Let us now define the characteristic
time $\tau_\mathrm{M}$ such that $\textrm{d}t/\tau_\mathrm{M}$ is the
probability that a particle  undergoes a collision with another fluid molecule
over the time $\textrm{d}t$.
Similarly, we define the characteristic time $\tau_\mathrm{K}$ such that
$\textrm{d}t/\tau_\mathrm{K}$ is the probability that a particle undergoes a
collision with the pore surface over the time $\textrm{d}t$.
These definitions imply that the number of molecules $N(t + \textrm{d}t)$ that
have not undergone a collision at a time $t + \textrm{d}t$  is given by:
\begin{equation}
N(t + \textrm{d}t)
= N(t) - N(t)\frac{\textrm{d}t}{\tau_\mathrm{M}} -
N(t)\frac{\textrm{d}t}{\tau_\mathrm{K}}
= N(t) \lbrack 1 - \textrm{d}t/\tau\rbrack
\label{EQ4.22}
\end{equation}
where we have introduced $1/\tau = 1/\tau_\mathrm{K} + 1/\tau_\mathrm{M}$.
Note that the additivity of the different processes
(collisions with the wall or with another fluid molecule)
is equivalent to assuming that they are independent of each other while
diffusion in real porous media might show departure from this simple
condition.
Upon integrating the latter equation, one arrives at the distribution
$N(t) = N(0) \exp(-t/\tau)$. Let us now consider the 3D average velocity
autocorrelation function
$C_{vv}(t) = \left\langle \textbf{v}(0) \cdot \textbf{v}(t)\right\rangle$ for
the $N(t)$ molecules that have not undergone a collision at a time $t$ (where
$\left\langle\textbf{v}(0) \cdot \textbf{v}(t)\right\rangle = 1/N \times
\Sigma_{i = 1}^{N(t)} \left\langle\textbf{v}_i(0) \cdot
\textbf{v}_i(t)\right\rangle$
is averaged over each fluid molecule $i = \{1,N(t)\}$).
In the spirit of \cref{EQ4.22}, at time $t + \textrm{d}t$, the average
velocity autocorrelation function is given by:
\begin{equation}
C_{vv}(t + \textrm{d}t) = C_{vv}(t) \lbrack 1 - \textrm{d}t/\tau\rbrack
\label{EQ4.23}
\end{equation}
which leads to $C_{vv}(t) = v_\textrm{T}^2 \exp(-t/\tau)$ (where $v_\textrm{T}$
is the average constant velocity of the fluid). This expression relies on the
assumption that the contribution $C_{vv}(t)$ from molecules colliding between
$t$ and $t + \textrm{d}t$ averages to zero (uncorrelated velocities before and
after collision).
The self-diffusivity $D_\mathrm{s}$ is then readily obtained from the
Green--Kubo expression based on the velocity time autocorrelation function
$C_{vv}(t)$ of the fluid molecules:
\begin{equation}
D_\mathrm{s} = \frac{1}{3} \int_0^\infty \left\langle\textbf{v}(0) \cdot
\textbf{v}(t) \right\rangle \textrm{d}t
= \frac{1}{3} \int_0^\infty C_{vv}(t) \textrm{d}t = v_\mathrm{T}^2 \tau/3
\label{EQ4.24}
\end{equation}

\begin{figure}[htbp]
	\centering
	\includegraphics[width=0.95\linewidth]{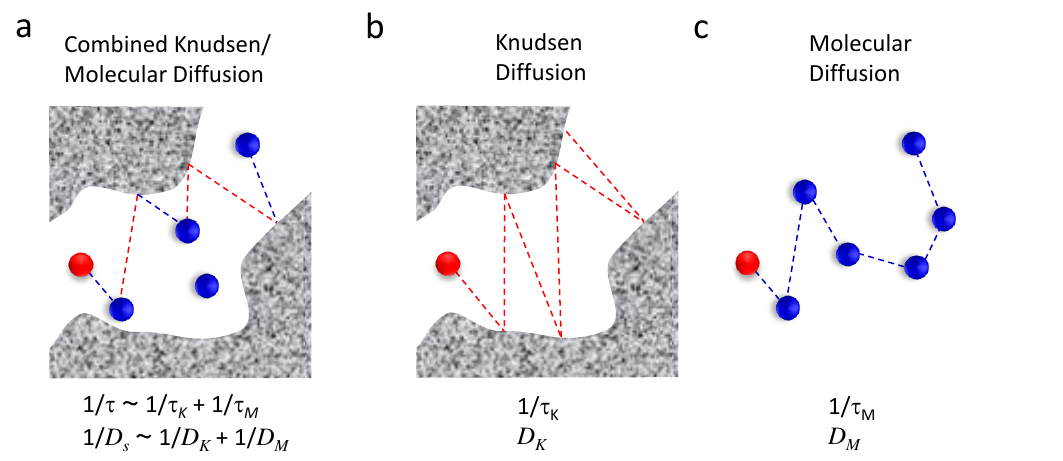}
	\caption{\textbf{Combined Knudsen/molecular diffusion.} (a) Schematic
		illustration of the diffusion of a tagged molecule (red sphere) in a
		porous material delimited by pore walls represented as grey areas. In
		the combined Knudsen/molecular diffusion regimes, the tagged molecule
		undergoes collisions with both the pore surface (dashed red segments)
		and with other fluid molecules (dashed blue segments with other fluid
		molecules shown	as blue spheres). This transition regime between the
		Knudsen and molecular diffusion regimes can be described using an
		effective model where the two mechanisms --- Knudsen diffusion as shown
		in (b) and molecular diffusion as shown in (c) --- are assumed to be
		independent of each other.}
	\label{fig:4_6}
\end{figure}

With such an effective approach, the reciprocal of the characteristic
diffusion time, $1/\tau$,  is the sum of the inverse of the two characteristic
times, \ie.\ $1/\tau \sim 1/\tau_\mathrm{K} + 1/\tau_\mathrm{M}$.
This result, known as Matthiessen’s rule,
and the underlying derivation above show that the self-diffusivity for a
confined fluid under thermodynamic conditions of combined Knudsen/molecular
diffusion regime should display an effective diffusivity $D_\mathrm{s}$
such that $1/D_\mathrm{s} = 1/D_\mathrm{K} + 1/D_\mathrm{M}$ where
$D_\mathrm{K}$ is the Knudsen diffusivity and $D_{s}$ is the
molecular diffusion (defined in conditions where only one
mechanism --- \ie.\ Knudsen versus molecular diffusion --- pertains,
cf.\ \cref{fig:4_6}).
It is important to keep in mind that the derivation above relies on the
assumption  that the two mechanisms are independent of each other (while, in
practice, the probability for a given fluid molecule of colliding with the pore
surface or with another fluid  molecule can be linked as conditional properties).

\subsubsection{Surface versus volume diffusion}
\label{sec:surface_versus_volume_diffusion}

Knudsen diffusion and combined Knudsen/molecular diffusion are limited to very
low gas densities where gas collisions with the pore surface are dominant
(mean free path $\lambda$ larger or comparable with the pore size
$D_\textrm{p}$).
However, in most practical situations, the gas density is such that this limit
is not relevant as the porosity gets filled with a gas or liquid phase whose
mean free path $\lambda$ is smaller than the pore size $D_\textrm{p}$. In this
case, as depicted in \cref{fig:4_5}, there are typically two situations that
can be encountered.
Depending on the gas pressure/temperature, either the pore is completely
filled with the liquid/fluid phase or the pore surface is covered by an
adsorbed phase having a density close to that of the liquid phase while the
pore center is occupied by a gas phase having a much lower density.
In both cases, the dynamics of the confined fluid is heterogeneous with strong
differences between diffusion of the molecules at the pore surface and
molecules in the pore center.
This raises the question of the ratio of the  surface and volume diffusion
contributions and how they should be combined to predict the overall (effective) diffusivity.

Like in a homogeneous medium, the heterogeneous dynamics in a pore either
completely or partially filled with a fluid phase is described through the
self correlation function $G_\mathrm{s}(\textbf{r},t)$. This function
corresponds to the probability that a molecule moves of a distance $\textbf{r}$
over a time $t$:
\begin{equation}
G_\mathrm{s}(\textbf{r},t) = \left\langle  \delta(\textbf{R}(t) - \textbf{R}(0)
- \textbf{r}(t) ) \right\rangle
\label{EQ4.25}
\end{equation}
where $\textbf{R}(t)$ and $\textbf{R}(0)$ are the molecule position at times
$t$ and $0$, respectively. The symbols $\langle \cdots \rangle$ denote average
over all molecules and times.
$G_\mathrm{s}(r,t)$ is related to the mean propagator which was
introduced in \Cref{chap:fundamentals} when discussing the solution of the
diffusion equation (Fick's second law).
In the Fickian regime, the propagator is a Gaussian
function as given in \cref{EQ3.17}. At this stage, it is convenient to
introduce the spatial Fourier transform $I_\mathrm{s}(\textbf{q},t)$ of the
self-correlation function $G_\mathrm{s}(\textbf{r},t)$,
$I_\mathrm{s}(\textbf{q},t) = \int
G_\mathrm{s}(\textbf{r},t) \textrm{e}^{i\textbf{q}\cdot\textbf{r}}
\textrm{d}\textbf{r}$.
The functions $G_\mathrm{s}(\textbf{r},t)$ and $I_\mathrm{s}(\textbf{q},t)$,
which can be probed experimentally through techniques such as PFG-NMR or neutron
scattering experiments, contain all the information on the self-diffusion
dynamics.
However, in general, for a heterogeneous medium such as fluid diffusing in a nanoporous material,
these functions are complex to interpret as they include the detailed dynamics
such as the crossing dynamics between different domains for which there is no
simple general model~\cite{roosen-runge_analytical_2016}.
In particular, as discussed in more detail in \Cref{chap:diffusion_network}
on diffusion at the porous network scale, these functions include the effects
of possible surface barriers at the border between different domain types.
Even at the pore scale, such crossing dynamics between the adsorbed phase and
the fluid phase manifests itself in the functions $G_\mathrm{s}(\textbf{r},t)$
and $I_\mathrm{s}(\textbf{q},t)$ in a fashion that is difficult to capture and
is system-dependent.
However, there are two asymptotic limits that can be considered for
heterogeneous systems depending on the ratio of the diffusion times within
each domain type and the crossing time between different domains. Typically,
for diffusion at the pore scale, one has to consider two domains
$\alpha = \lbrack 1, 2\rbrack$:
(1) the region in the vicinity of the pore surface where the molecules
experience the interaction potential generated by the host porous solid and
(2) the pore center where the molecules behave in a bulk-like fashion as they
do not feel the interaction potential generated by the host porous solid.
On the one hand, for a typical displacement on a distance $r \sim 2\pi/q$, the
diffusion time is $\tau_D^\alpha \sim (D_\mathrm{s}^\alpha q^2)^{-1}$ where
$D_\mathrm{s}^\alpha$ is the self-diffusion coefficient for molecules located
in domains $\alpha$.
On the other hand, the typical crossing time $\tau_C^\alpha$ can be
estimated as the mean first-passage time which corresponds to the time for a
molecule in a domain $\alpha$ to reach the boundary domain.
In the limit $\tau_C^\alpha \gg \tau_D^\alpha \hspace{1mm} \forall \alpha$,
known as the slow-switching regime, molecule exchange between the different
domains is limited and the correlation function $I_\mathrm{s}(\textbf{q},t)$ is
dominated by diffusion in each of the domains. In this slow-switching limit,
the correlation function $I_\mathrm{s}(\textbf{q},t)$ is given by the sum of
all intra-domain diffusion contributions $D_\mathrm{s}^\alpha$ pondered by the
average fraction $x^\alpha$ of molecules in these domains:
\begin{equation}
I_\mathrm{s}(\textbf{q},t) = \sum_\alpha x_\alpha \exp(-D_\mathrm{s}^\alpha q^2
t)
\label{EQ4.26}
\end{equation}
where $q = \lvert \textbf{q} \lvert$.
In contrast, in the limit $\tau_C^\alpha \ll \tau_D^\alpha \hspace{1mm} \forall
\alpha$, known as the fast-switching regime, molecule exchange between the
different domains is significant over the typical time scale needed to diffuse
through the different domains.
In other words, in this limit, for  displacements over
a length scale $r \sim 2\pi/q$, each molecule explores the different domains so
that the observed diffusivity is an effective diffusivity that reflects  the
diffusivities in all the different domains.
In this fast-switching limit, the correlation function
$I_\mathrm{s}(\textbf{q},t)$
is given by a single contribution with a characteristic diffusion coefficient
$\overline{D_{s}}$:
\begin{equation}
I_\mathrm{s}(\textbf{q},t) = \exp \left( - \sum_\alpha  x_\alpha
D_\mathrm{s}^\alpha q^2 t \right) = \exp \left( - \overline{D_{s}} q^2 t
\right)
\label{EQ4.27}
\end{equation}

The two asymptotic situations considered above are limiting cases which can
show strong departure with experimental measurements. Before addressing in
more detail this problem using the formalism of Intermittent Brownian
Motion, it is instructive to consider the following reasoning which illustrates
the breakdown of the fast and slow switching limits. Let us consider the
average displacement $\left\langle\Delta r^2(\tau)\right\rangle$ of molecules
in a single pore over a typical time $\tau$.
As introduced above, the fluid molecules will diffuse through two
domains $\alpha = \lbrack 1, 2\rbrack$ (vicinity of the pore surface and pore
center) as depicted in \cref{fig:4_7}(a). Each molecule trajectory
$\textbf{r}(\tau) - \textbf{r}(0)$ from its initial position at time $t = 0$ to its position at time $\tau$ can be split into the different
displacements $\Delta \textbf{r}_k^\alpha$ which correspond to the
$k^\mathrm{th}$ trajectory segment in the domain $\alpha$. With such notations,
the mean square displacement  over a time $\tau$ writes:
\begin{equation}
\Delta r^2(\tau)
= \left\langle (\textbf{r}(\tau) - \textbf{r}(0))^2 \right\rangle
= \left\langle \left(\sum_{\alpha = 1}^{M} \sum_k \Delta \textbf{r}_\alpha^k
\right)^2 \right\rangle
\label{EQ4.28}
\end{equation}
where  $M = 2$ is the number of domain types (adsorbed region and pore
center).
It can reasonably be assumed that displacements in the same domain type
$\alpha$ are uncorrelated, \ie.\ $\left\langle \Delta
\textbf{r}_\alpha^k \Delta \textbf{r}_\alpha^{k'}\right\rangle = 0$ when
$k \neq k'$ (because these two trajectory segments are separated by a trajectory segment in the other domain type $\alpha' \neq \alpha$).
Similarly, when considering trajectory segments in different domains $\alpha'
\neq \alpha$, it can be assumed that only two consecutive segments are
correlated so that $\left\langle \Delta \textbf{r}_\alpha^k \Delta
\textbf{r}_{\alpha'}^{k'}\right\rangle = 0$ when $k \neq k'$. With these
assumptions, the mean square displacement in \cref{EQ4.28} can be rewritten
as:
\begin{equation}
\Delta r^2(\tau)
\sim \sum_{\alpha = 1}^{M} \sum_k \left\langle  \left(\Delta
\textbf{r}_\alpha^k \right)^2 \right\rangle +
\mathop{\sum_{\alpha, \alpha' = 1}}_{\alpha \neq \alpha'}^{M} \sum_k
\left\langle \Delta \textbf{r}_\alpha^k \Delta \textbf{r}_{\alpha'}^k
\right\rangle
\label{EQ4.29}
\end{equation}
Assuming each trajectory segment $\Delta \textbf{r}_\alpha^k$ obeys Fickian
diffusion, it can be written that $\sum_k (\Delta \textbf{r}_\alpha^k)^2 \sim
6 D_\mathrm{s}^\alpha \tau_\alpha$ where $D_\mathrm{s}^\alpha$ is the
self-diffusion coefficient
of the fluid molecules in domain $\alpha$ while $\tau_\alpha$ is the  time
spent by the molecules in domains $\alpha$ in the course of the trajectory of a
duration $\tau$. \Cref{EQ4.29} can be used to estimate the effective
diffusivity $\overline{D_{s}}$ defined as $\overline{D_{s}} = \Delta
r^2(\tau)/6\tau$:
\begin{equation}
\overline{D_{s}} =
\sum_{\alpha = 1}^{M} \frac{\tau_\alpha}{\tau} D^\alpha
+ \frac{1}{6\tau} \left\langle \Delta \textbf{r}_\alpha^k \Delta
\textbf{r}_{\alpha'}^k \right\rangle
= \sum_{\alpha = 1}^{M} x_\alpha D^\alpha
+ \frac{1}{6\tau} \left\langle \Delta \textbf{r}_\alpha^k \Delta
\textbf{r}_{\alpha'}^k \right\rangle
\label{EQ4.30}
\end{equation}
The principle of ergodicity --- which implies that the average time fraction spent by a
molecule in domains $\alpha$, $\tau_\alpha/\tau$, is equal to the fraction of
molecules located in these domains, $x_\alpha$ --- was used to obtain the
second equality.
Upon further assuming that consecutive trajectory segments  are uncorrelated,
$\left\langle \Delta \textbf{r}_\alpha^k \Delta \textbf{r}_{\alpha'}^k
\right\rangle = 0$, we arrive at the simple combination rule for the
effective diffusivity:
$\overline{D_{s}} = \sum_\alpha x_\alpha D_\mathrm{s}^\alpha$.
The latter condition is known as the fast-exchange model where, independently
of the geometrical distribution of domains in the porous solid, the effective
diffusivity is given by a combination rule where the domains are assumed to be
visited in parallel. In the case of  partially filled pores, the conditions
under which the fast-exchange model is expected to be valid was discussed by
Kaerger, Valiullin, and
coworkers~\cite{zeigermann_diffusion_2012,karger_mass_2013}.
In practice, this approximated expression should be considered with caution as
\cref{EQ4.30} shows that correlations between two consecutive trajectory
segments in different domains $\alpha \neq \alpha'$ strongly affect the
effective diffusivity in heterogeneous media. In particular, when significant
recrossing between domains is expected, cross correlations
$\left\langle \Delta \textbf{r}_\alpha^k \Delta \textbf{r}_{\alpha'}^k
\right\rangle $ are expected to be negative on average so that they should
significantly decrease the effective diffusivity $\overline{D_{s}}$.
Moreover, as will be established in \Cref{chap:diffusion_network} on diffusion
at the porous network scale, from simple arguments based on the analogy with
electrical transport in resistance networks, intermittent
adsorption/relocation
should be considered as a model of domains visited in series so that the
effective diffusivity should be given by a combination rule:
$1/\overline{D_{s}} = \sum_\alpha x_\alpha/D_\mathrm{s}^\alpha$.

From a very general viewpoint, there is no simple model to predict the
cross-correlation terms in \cref{EQ4.30} as they are system-dependent.
However, theoretical frameworks such as
Langevin and Fokker--Planck equations are invaluable to describe such
correlations and the dynamics of molecules in heterogeneous environments. In
particular, using either a path integral approach or a transition matrix
approach, Roose-Runge \etal.\ derived simple expressions which allow one to
probe the dynamics in complex heterogeneous media from data available using
PFG-NMR, neutron scattering, or dynamic light
scattering~\cite{roosen-runge_analytical_2016}.
This  powerful formalism will be introduced in \Cref{chap:diffusion_network}
where diffusion at the porous network scale will be considered (we will also
consider the different simple combination rules such as model in series, in
parallel and the effective medium theory). As far as dynamics at the pore
scale is concerned, another formalism, known as ``Intermittent Brownian
Motion'', is presented in the following subsection as it provides a simple
framework to describe adsorption and diffusion in a pore partially or
completely filled with a fluid phase.

\subsubsection{Intermittent Brownian Motion}
\label{sec:intermittent_brownian_motion}

\begin{figure}[htbp]
	\centering
	\includegraphics[width=0.95\linewidth]{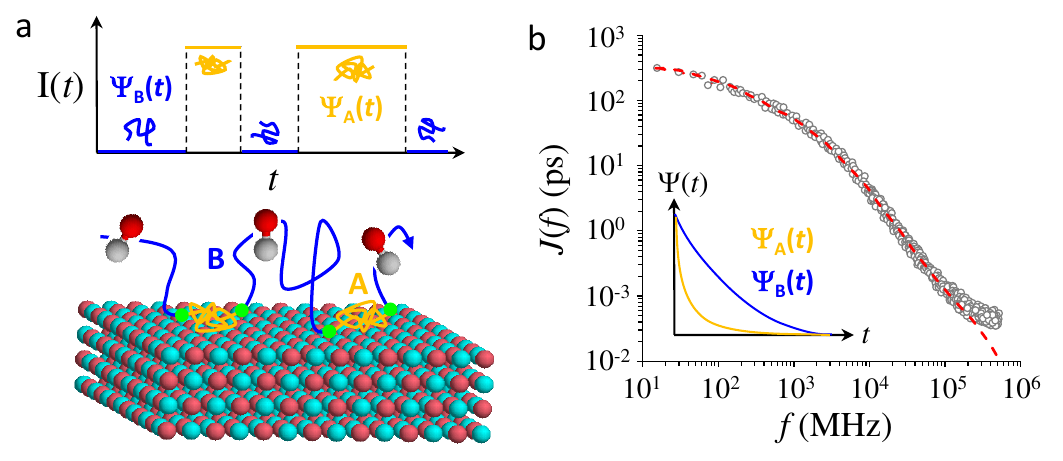}
	\caption{\textbf{Intermittent Brownian Motion.} (a) Principle of the
		Intermittent Brownian Motion formalism applied to fluid adsorption and
		diffusion in a porous material. Each molecule trajectory subdivides
		into segments during which the molecule diffuses at the pore surface
		--- orange subtrajectories --- and segments during which the molecule
		is relocated/diffuses through the bulk part of the porosity --- blue
		subtrajectories. Each type of segments is characterized by a time
		distribution $\Psi_\alpha(t)$ with $\alpha =$ \textbf{A} or
		\textbf{B} which is related to the mean first passage time within each
		domain.
		For a given molecule, the function $I(t)$ is defined as $I(t) \neq 0$
		if the molecule is adsorbed at the surface and $I(t) = 0$ when it
		relocates through the non-adsorbed fluid phase in the pore center. (b)
		From the statistical distributions, $\Psi_\alpha(t)$ ($\alpha =$
		\textbf{A}, \textbf{B}), the spectral density $J(f)$ of the function
		$I(t)$ can be predicted. The example shown here corresponds to the
		specific situation of water adsorbed in a hydrophilic silica slit pore
		of a width $D_{\textrm{p}}$.
		The symbols are data obtained using molecular dynamics simulations
		while the line is the prediction from the Intermittent Brownian Motion
		formalism (note that the statistical distributions shown in the insert
		are schematic examples). Data adapted with permission from
		Ref.~\protect\citen{levitz_molecular_2013}.
		Copyright 2013 Royal Society of Chemistry.}
	\label{fig:4_7}
\end{figure}

In order to introduce the framework of Intermittent Brownian Motion, we
follow the approach proposed by Levitz as derived in
Ref.~\citen{levitz_random_2005}. Let us consider a fluid confined in a porous
medium with pores large enough to assume that molecules are located either in
the vicinity of the pore surface or in the bulk-like region of the pore center.
In practice, a molecule can be considered located in the pore surface region if
its interaction energy with the solid atoms forming the porous medium is
larger than the thermal energy $k_\textrm{B}T$. Every molecular trajectory can
be described as an alternating  series of adsorption steps at the
pore surface, followed by relocation steps within the pore center
[\cref{fig:4_7}(a)].
Let us define the function $I(t)$ which is equal to 1 if the molecule is
adsorbed at time $t$ and 0 if the molecule is in the pore center
[\cref{fig:4_7}(a)].
The formalism of Intermittent Brownian Motion allows one to describe  the long-time
dynamics of such a confined fluid from the time autocorrelation $C(t) =
\left\langle I(t)I(0) \right\rangle/\eta_A$ where $\eta_A = \tau_A/(\tau_A+\tau_B)$ is the time fraction spent in adsorption steps ($\tau_A$ and $\tau_B$ are the average adsorption and relocation times). At this stage, let us also introduce the spectral density of the time correlation function $C(t)$: $J(\omega) =
\int_{-\infty}^{\infty} C(t) \exp(i\omega t) \textrm{d}t$. $C(t)$ is defined
as a time average over the variable $\tau$:
\begin{equation}
C(t) = \frac{\left\langle I(t)I(0) \right\rangle}{\eta_A}
= \frac{1}{\eta_A T}\int_{0}^{T}
I(\tau)I(t+\tau) \textrm{d}\tau
\label{EQ4.31}
\end{equation}

Microscopic information on the residence time within the adsorbed phase, bulk
relocation time, and time exchange between these different domains is all
encoded in the function $L(t) = 1/T \times \int_{0}^{T} I(\tau)I(t+\tau)
\textrm{d}\tau$ --- and equivalently its Fourier transform $\tilde{L}(\omega)$.
While there is no direct measurement of such time correlation functions or their
spectral density, experimental techniques such as NMR relaxometry provide data
from which they can be inferred.
In this technique, the characteristic relaxation time for the different
populations is used as it is a direct signature of the molecule environment.
Typically, by performing such relaxation measurements separated by a time lapse
$\Delta$, it is possible to probe molecules that move from one environment to
another or that remain within the same environment over this specific time $\Delta$.

For reasons that will become clearer below, let us introduce the second time
derivative of the function $L(t)$:\footnote{
	A formal derivation of this relationship is as follows.
	$L(t) \sim \int_{0}^{T} I(\tau)I(t+\tau) \textrm{d}\tau$
	so that $L'(t) \sim \int_{0}^{T} I(\tau)I'(t+\tau) \textrm{d}\tau
	\sim \int_{0}^{T} I(\tau - t)I'(\tau) \textrm{d}\tau$
	where the last equality is obtained by using the invariance of the
	function $L'(t)$ upon time shift \ie.\ $A(\tau)B(t+\tau) = A(\tau
	-t)B(\tau)$.
	Starting from the last expression, we derive a second time to obtain
	$L''(t) \sim - \int_{0}^{T} I'(\tau-t)I'(\tau) \textrm{d}\tau \sim
	- \int_{0}^{T} I'(\tau)I'(t+\tau) \textrm{d}\tau$.
}
\begin{equation}
L''(t) =  -\frac{1}{T} \times \int_{0}^{T}
I'(\tau)I'(t+\tau) \textrm{d}\tau
\label{EQ4.32}
\end{equation}
Typically, as shown in \cref{fig:4_7}(a), $I(t)$ can be described as a
series of Heaviside functions $H(t-t_i)$:
$I(t) = \sum_i (-1)^i H(t-t_i)$. Inserting this expression into
\cref{EQ4.32} and using $H'(t-t_i) = \delta(t-t_i)$ lead to:
\begin{equation}
L''(t) = - \frac{1}{T} \times \sum_{i,j} (-1)^{i+j} \int_{0}^{T}
\delta(\tau-t_j)\delta(t+\tau-t_i) \textrm{d}\tau = - \frac{1}{T} \times \sum_{i,j} (-1)^{i+j}
\delta(t-[t_i - t_j])
\label{EQ4.33}
\end{equation}

In the Intermittent Brownian Motion, each molecular trajectory is a succession
of adsorption steps ($2n \rightarrow 2n + 1$) and relocation steps through
the  porosity in the pore center ($2n+1 \rightarrow 2n+2$). The last equality
in \cref{EQ4.33} shows that $L''(t)$ is the sum of Dirac functions with (1)
terms $i = j$, (2) terms $j = i+1$ and $i = 2n$ which correspond to an
adsorption step, (3) terms $j = i+1$ and $i = 2n+1$ which correspond to a
relocation step, (4) terms $j = i+2$ and $i = 2n$ which correspond to an
adsorption step followed by a relocation step, (5) terms $j = i+3$ and $i =
2n+1$ which correspond to an adsorption step followed by a relocation step
followed by an adsorption step, (6) terms $j = i+3$ and $i = 2n$ which
correspond to a relocation step followed by an adsorption step followed by a
relocation step, and so on. Such a hierarchy can be described using the
following averages:
\begin{eqnarray}
\label{EQ4.34}
\langle \delta(t-(t_{2n+1} - t_{2n})) \rangle  &=& \psi_A(t)\\
\label{EQ4.35}
\langle \delta(t-(t_{2n+2} - t_{2n+1})) \rangle  &=& \psi_B(t)\\
\label{EQ4.36}
\langle \delta(t-(t_{2n+2} - t_{2n})) \rangle  &=& \psi_A(t) \ast \psi_B(t)\\
\label{EQ4.37}
\langle \delta(t-(t_{2n+3} - t_{2n+1})) \rangle  &=& \psi_B(t) \ast \psi_A(t)\\
\label{EQ4.38}
\langle \delta(t-(t_{2n+4} - t_{2n})) \rangle  &=& \psi_A(t) \ast \psi_B(t)
\ast \psi_A(t)\\
\label{EQ4.39}
\langle \delta(t-(t_{2n+5} - t_{2n+1})) \rangle  &=& \psi_B(t) \ast \psi_A(t)
\ast \psi_B(t) \\
&\dots& \nonumber
\end{eqnarray}
where the symbol $\ast$ denotes convolution, \ie.\ $f(t)\ast g(t) =
\int_{-\infty}^{\infty} f(\tau) g(t+\tau) \textrm{d}\tau$. Based on these
definitions, $\psi_A(t)$ [or $\psi_B(t)$, respectively] is the probability that
an adsorption step [or relocation step, respectively] lasts a time $t$.
These two functions, which are schematically illustrated in
\cref{fig:4_7}(b), contain the fingerprint of the adsorption properties of the
confined fluid as well as of the porous medium (specific surface area, mean
pore size, tortuosity, \etc.).  Using the statistical distributions
above and considering that $L''(t)$ is an even function, $L''(t)$ in \cref{EQ4.33} can be expressed as:
\begin{multline*}
L''(t) = \frac{N}{T} [-2\delta(t) + \psi_A(t) + \psi_B(t) - 2\psi_A(t) \ast \psi_B(t) \\
+ \psi_A(t) \ast \psi_B(t) \ast \psi_A(t) + \psi_B(t) \ast \psi_A(t) \ast
\psi_B(t) - \dots ] \hspace{1mm} (t \geq 0)
\label{EQ4.40}
\end{multline*}
\begin{multline}
L''(t) = \frac{N}{T} [-2\delta(t) + \psi_A(-t) + \psi_B(-t) - 2\psi_A(-t) \ast \psi_B(-t) \\
+ \psi_A(-t) \ast \psi_B(-t) \ast \psi_A(-t) + \psi_B(-t) \ast \psi_A(-t) \ast
\psi_B(-t) - \dots ] \hspace{1mm} (t \leq 0)
\end{multline}
where $N$ is the number of adsorption/relocation steps over the total duration
$T$. 
From \cref{EQ4.40}, the Fourier transform $\tilde{L}''(\omega) =
\int_{-\infty}^{\infty} L''(t) \exp(i\omega t) \textrm{d}t$ of the function
$L''(t)$ reads:\footnote{
Here, we use that $\psi_A(t)$ and $\psi_B(t) = 0$ for $t < 0$ so that
$\tilde{\psi}_A(\omega) = \int_{-\infty}^\infty \psi_A(t) \exp(i\omega t)
\textrm{d}t = \int_{0}^\infty \psi_A(t) \exp(i\omega t) \textrm{d}t$ and
$\tilde{\psi}_B(\omega) = \int_{-\infty}^\infty \psi_B(t) \exp(i\omega t)
\textrm{d}t = \int_{0}^\infty \psi_B(t) \exp(i\omega t) \textrm{d}t$. 
As a result, upon calculating $\tilde{L}''(\omega)$, we use $\tilde{L}''(\omega)
\sim -2 + \int_0^\infty \psi_A(t) \exp(i \omega t) \textrm{d}t +
\int_{-\infty}^0 \psi_A(-t) \exp(i \omega t) \textrm{d}t + \cdots \sim -2 +
\operatorname{Re} [2 \times \tilde{\psi}_A(\omega) + \cdots]$
}
\begin{multline}
\tilde{L}''(\omega) = \frac{N}{T} \operatorname{Re} \Big [-2 + 2 \times   \Big
(\tilde{\psi}_A(\omega) + \tilde{\psi}_B(\omega) -
2\tilde{\psi}_A(\omega)\tilde{\psi}_B(\omega) \\
+\tilde{\psi}_A(\omega)\tilde{\psi}_B(\omega)\tilde{\psi}_A(\omega) +
\tilde{\psi}_B(\omega)\tilde{\psi}_A(\omega)\tilde{\psi}_B(\omega) -  \dots \Big
) \Big]
\label{EQ4.41}
\end{multline}
The latter equation can be rearranged as:
\begin{multline}
\tilde{L}''(\omega) \times \tilde{\psi}_A(\omega)\tilde{\psi}_B(\omega) =
\frac{N}{T} \operatorname{Re} \Big [-2\tilde{\psi}_A(\omega)\tilde{\psi}_B(\omega)
+ 2\tilde{\psi}_A^2(\omega)\tilde{\psi}_B(\omega) +
2\tilde{\psi}_A(\omega)\tilde{\psi}_B^2(\omega) \\
- 4\tilde{\psi}_A^2(\omega)\tilde{\psi}_B^2(\omega) +
2\tilde{\psi}_A^3(\omega)\tilde{\psi}_B(\omega) +
2\tilde{\psi}_A(\omega)\tilde{\psi}_B^3(\omega) -  \dots  \Big ]
\label{EQ4.42}
\end{multline}
By subtracting \cref{EQ4.41,EQ4.42}, we obtain
$\tilde{L}''(\omega) \times(1-\tilde{\psi}_A(\omega)\tilde{\psi}_B(\omega)) = N/T \times
\operatorname{Re}[-2 + 2\tilde{\psi}_A(\omega) + 2\tilde{\psi}_A(\omega) -
2\tilde{\psi}_A(\omega)\tilde{\psi}_B(\omega)]$  which can be recast as:
\begin{equation}
\tilde{L}''(\omega)  =
- \frac{2N}{T}\times \operatorname{Re} \left\lbrack
\frac{(1-\tilde{\psi}_A(\omega))(1-\tilde{\psi}_B(\omega))}{1-\tilde{\psi}_A(\omega)\tilde{\psi}_B(\omega)}
\right\rbrack
\label{EQ4.43}
\end{equation}
Finally, by virtue of the Fourier transform properties, $\tilde{L}''(\omega) =
-\omega^2 \tilde{L}(\omega)$, we obtain:
\begin{equation}
\tilde{L}(\omega)  =
\frac{2N}{T\omega^2}\operatorname{Re} \left\lbrack
\frac{(1-\tilde{\psi}_A(\omega))(1-\tilde{\psi}_B(\omega))}{1-\tilde{\psi}_A(\omega)\tilde{\psi}_B(\omega)}
\right\rbrack
\label{EQ4.44}
\end{equation}
Equivalently, recalling that $C(t) = L(t)/\eta_A$ with $\eta = \tau_A/[\tau_A + \tau_B]$ and noting that $N/T = 1/[\tau_A+\tau_B]$ (\ie.\ the number of adsorption/relocation segments per time unit), we obtain $J(\omega) = 2/\tau_A \omega^2 \times 
\operatorname{Re}[(1-\tilde{\psi}_A(\omega))(1-\tilde{\psi}_B(\omega))]/[1-\tilde{\psi}_A(\omega)\tilde{\psi}_B(\omega)]$.
This equation is a very important expression as it shows that  multiscale
diffusion in a porous medium can be described from two simple
statistical distributions: the adsorption time distribution $\psi_A(t)$ and
relocation time distribution $\psi_B(t)$. Such functions can be estimated from
molecular simulation and fitted against simple mathematical functions. In a
second step, by inserting these molecular time distributions into
\cref{EQ4.44}, the long-time dynamics of the confined fluid can be described
up to time and length scales that go well beyond those accessible using
molecular simulation.  \Cref{fig:4_7}(b) illustrates the application of the
Intermittent Brownian Motion model to molecular simulation data for water in a
hydrophilic silica pore having a width $D_{\textrm{p}} = 2$
nm~\cite{levitz_molecular_2013}.
At high frequencies $f = \omega/2\pi$, some expected discrepancy is observed
between the molecular simulation data and the predictions based on the
Intermittent Brownian Motion model. This is due to the coarse-grained,
\ie.\ mesoscopic, level employed in the theory which does not capture
the details of the molecular dynamics explicitly treated in molecular
simulation. On the other hand, for frequencies smaller than $10^5$ MHz, the
Intermittent Brownian Motion model is found to capture very accurately the
molecular simulation data.
In particular, while molecular simulation cannot probe timescales much longer
than 10-100 ns ($f < 10^2-10^3$ MHz), the data in \cref{fig:4_7}(b) suggests
that the Intermittent Brownian Motion framework allows accurately upscaling
molecular dynamics to the long time, \ie.\ macroscopic, limit. In
practice, while being a robust and efficient theory, in order to calculate the
time distributions $\psi_A(t)$ and $\psi_B(t)$,
the Intermittent Brownian Motion requires to subdivide a given confined fluid
into an adsorbed phase in the vicinity of the surface and a bulk-like phase in
the pore center. For disordered porous materials, this task can be complex
even if descriptors such as the minimum distance to the pore surface and/or the
value of the fluid/wall energy at a given position $\textbf{r}$ can be used.
Even for a simple pore geometry, such as cylindrical and slit pores considered
in Ref.~\citen{levitz_molecular_2013}, $\psi_A(t)$ and $\psi_B(t)$ must
be calculated by using a minimum distance criterion to the pore surface
(typically, $r < R - r_c$ where $r$ is the distance to the pore surface, $R$
the pore radius or half-width, and $r_c$ the cutoff). While this appears as an
arbitrary choice at first, it was shown in Ref.~\citen{levitz_molecular_2013}
that using $r_c \sim 2\sigma$, where $\sigma$ is the fluid molecule size,
leads to accurate predictions such as those illustrated in \cref{fig:4_7}.
Such a choice $r_c \sim 2\sigma$ can be rationalized by the fact that, at
temperatures under which adsorption occurs, the range of  fluid--wall
intermolecular forces is about $2\sigma$ (as can be inferred from typical
density profiles which show that the adsorbate displays important positional
ordering --- with marked density oscillations --- that extends up to two layers from
the pore surface). Recently, the intermittent brownian formalism was extended to
disordered nanoporous materials by obtaining residence and relocation distributions and associated characteristic times through a mapping with molecular dynamics simulations.\cite{bousige_bridging_2021}

\subsubsection{Density-dependent diffusion}

Until now,  we have focused on the different diffusion
mechanisms that can occur in a single pore (Knudsen diffusion, surface
diffusion, molecular diffusion, \etc.). For each mechanism, the
theoretical foundations were provided together with a description of transition regimes between them. In the present subsection, we turn
to a different aspect which is the description of density, \ie.\
loading, effects on the self-diffusivity in a single pore. The schematic
behavior summarized in \cref{fig:4_5} can be considered representative of any
fluid confined below or above their capillary critical point $T_\mathrm{cc}$.
The exact shape of the adsorption isotherms and self-diffusivity as a function
of pressure plots will depend on pore size, temperature, strength of the
fluid/wall interaction, \etc.
However, irrespective of the adsorbate/adsorbent couple considered, the
self-diffusivity usually varies in a non-monotonic fashion with pressure (and,
therefore, with density since the density increases monotonically with
pressure). Such  a typical diffusivity/density dependence with the existence
of a maximum is seen in many experiments on fluids, e.g.  in
activated carbons~\cite{karger_molecular_1989} or in 
zeolites~\cite{brandani_concentration_1995}.
However, in other experiments, this maximum diffusivity
regime might not be observable because it is located in the very low
pressure/density range.
Before going into more details, the non-monotonic behavior observed when
plotting the self-diffusivity as a function of fluid pressure or density can
be understood as follows.\cite{coasne_adsorption_2006}
At low densities,  the self-diffusivity $D_\mathrm{s}$ increases
with density as the fluid molecules get adsorbed in adsorption
sites of  decreasing energies. In other words, while the first adsorbed
molecules are located in strongly adsorbing sites leading to a very slow
self-diffusivity, further adsorption takes place  in less energetic sites so that
the average diffusivity increases.  
%At low fluid densities, the
%self-diffusivity $D_\mathrm{s}$ of the confined fluid is small as most fluid
%molecules are adsorbed in sites of very low (more attractive) energy.
%Upon increasing the density, the adsorption sites of low energy are occupied and
%further adsorption takes place in sites where the molecules
%have a larger self-diffusivity. 
As the pore gets filled,
the self-diffusivity of the fluid molecules decreases upon increasing the
density as steric repulsion/crowding  becomes dominant. In what follows, we
first discuss the regime corresponding to the range where the self-diffusivity
increases with density --- as will be shown below, this regime can be simply
described using a spatially averaged diffusivity. We then discuss the second
regime where the decrease in the self-diffusivity upon increasing the density
is governed by steric repulsion. This second regime  can be accounted for by
considering the change in the volume accessible to the diffusing molecules
through a simple free volume theory.

\noindent \textbf{Spatially-averaged diffusivity.}
Before addressing the increase in the diffusivity upon increasing
the fluid loading, it is instructive to consider the following simple model.
Consider a pore of any arbitrary geometry filled with a fluid phase of
$N$ molecules at a given temperature $T$ and chemical potential $\mu$. If the
pore is completely filled by the fluid phase, it can reasonably be assumed that
fluid molecules explore on a reasonable time scale the whole porosity so that
the problem can be treated in the fast switching limit. As was done in the
previous section, within this approximation, we can assume that the mean
square displacements of each fluid molecule is a succession of mean square
displacements in different pore regions so that the effective diffusivity
writes:
\begin{equation}
\overline{D_\mathrm{s}} = \frac{1}{N} \int D_\mathrm{s}(\textbf{r})
\rho(\textbf{r})
\textrm{d}^3\textbf{r}
=  \frac{1}{\pi R_\mathrm{p}^2 \overline{\rho}}    \int_0^{R_{p}}
D_\mathrm{s}(r)
\rho(r)
\times  2\pi r \textrm{d}r
\label{EQ4.45}
\end{equation}
where $D_\mathrm{s}(\textbf{r})$ and $\rho(\textbf{r})$ are the
self-diffusivity and
density of the fluid at the position $\textbf{r}$ in the pore. In
the second equality, the specific case of an infinite cylindrical pore of
radius $R_\mathrm{p}$ is considered and the average density $\overline{\rho}$
is
introduced so that $N = \pi R_\mathrm{p}^2 \overline{\rho}$.
In the framework of the transition state theory, it
was shown in \Cref{sec:molecular_diffusion} that self-diffusion can be
described as an activated process involving an activation free energy
$\Delta F^0$, \ie.\ $D_\mathrm{s} \sim \exp \lbrack -\Delta F^0 / \kbT\rbrack$.
For a confined fluid, a simple physical assumption is 
that the activation energy for diffusion is that of the bulk augmented by the
fluid/wall potential $\phi(\textbf{r})$, \ie.\
$\Delta F = \Delta F^0 + \phi(\textbf{r})$.
In other words, it can be assumed that the attractive interaction potential
generated by the pore surface slows down the confined/adsorbed molecules which
therefore display a slower self-diffusivity with respect to their bulk
counterpart.
With these arguments, one simply obtains that $D_\mathrm{s}(r) = D_\mathrm{s}^0
\exp\lbrack-\phi(r)/k_{\textrm{B}}T\rbrack$ where $D_\mathrm{s}^0$ is the bulk
self-diffusivity~\cite{bhatia_modeling_2008}.
If we further assume that the density is constant within the
pore $\rho(r) \sim \overline{\rho}$ and that the fluid/wall potential can be
described as a square well interaction potential such that $\phi(r) = -
\epsilon$ for $R_\mathrm{p}-r < 2\sigma$ where $\sigma$ is the size of the
fluid
molecule (such a fluid/wall interaction range is consistent with the fact that
most adsorbed fluids show properties that depart from their bulk counterpart
when located at a distance smaller than $2\sigma$ from the pore surface while
the bulk fluid properties are recovered beyond this value). Upon inserting
these different approximations in \cref{EQ4.45}, one arrives at:
\begin{equation}
\overline{D_\mathrm{s}} = \frac{D_\mathrm{s}^0}{R_\mathrm{p}^2} \left\lbrack
(R_\mathrm{p} -
2\sigma)^2 +
\frac{D_\mathrm{s}^s}{D_\mathrm{s}^0}\left(R_\mathrm{p}^2 - (R_\mathrm{p} -
2\sigma)^2\right)
\right\rbrack
\sim D_\mathrm{s}^0 \left\lbrack 1 - \frac{4\sigma}{R_\mathrm{p}} \left(1 -
\frac{D_\mathrm{s}^s}{D_\mathrm{s}^0}
\right) \right\rbrack
\label{EQ4.46}
\end{equation}
where the second equality was obtained by expanding each term while keeping
only leading order terms in $\sigma$ (\ie.\ assuming
$\sigma^2 \ll R_\mathrm{p}^2$) and $D_\mathrm{s}^s$ is the self-diffusivity in
the surface layer of width $2\sigma$.

\Cref{EQ4.46} is a general expression that applies to cylindrical pores filled
by a liquid
with a homogeneous density profile. Using simple yet realistic approximations,
more specific situations including partially filled pores can be described as
follows. We write that the self-diffusivity $D_\mathrm{s}$ is a function of the
distance $r$ to the pore surface that decays exponentially  with a
characteristic molecular length $r_0$ ($r_0 < R_\mathrm{p}$): $D_\mathrm{s}(r)
= D_\mathrm{s}^0 + (D_\mathrm{s}^s - D_\mathrm{s}^0)\exp\lbrack -(R_\mathrm{p}
- r)/r_0\rbrack$.  The latter function ensures that the
local self-diffusivity $D_\mathrm{s}(r)$ varies from the surface
self-diffusivity $D_\mathrm{s}^s$
to the bulk surface diffusivity  $D_\mathrm{s}^0$  as the distance to the
surface increases.

\begin{itemize}
\item{For a totally filled pore,
upon inserting the expression for the local self-diffusivity in \cref{EQ4.45}, the
effective diffusivity $\overline{D_\mathrm{s}}$ writes:
\begin{align}
\overline{D_\mathrm{s}} &= D_\mathrm{s}^0 + \frac{2}{R_\mathrm{p}^2}
\int_0^{R_\mathrm{p}}
(D_\mathrm{s}^s - D_\mathrm{s}^0) \times r
\exp \left\lbrack -\frac{(R_\mathrm{p} - r)}{r_0} \right \rbrack \textrm{d}r
\label{EQ4.47}\\
&= D_\mathrm{s}^0 + 2(D_\mathrm{s}^s - D_\mathrm{s}^0) \left\lbrack
\frac{r_0}{R_\mathrm{p}} - \frac{r_0^2}{R_\mathrm{p}^2}
+ \frac{r_0^2}{R_\mathrm{p}^2} \exp(-R_\mathrm{p}/r_0) \right\rbrack
\sim D_\mathrm{s}^0 + \frac{2r_0}{R_\mathrm{p}}(D_\mathrm{s}^s -
D_\mathrm{s}^0)
\nonumber
\end{align}
In the third equality, only the leading order terms $r_0/R_\mathrm{p}$
were kept
(\ie.\ $r_0^2/R_\mathrm{p}^2$ and $r_0^2/R_\mathrm{p}^2 \times
\exp(-R_\mathrm{p}/r_0) \sim 0)$~\cite{chemmi_noninvasive_2016}.
Interestingly, the second expression in \cref{EQ4.47} is strictly
equivalent to that in \cref{EQ4.46} if $r_0$ is taken equal to $2\sigma$,
%In agreement with experiments on fluid diffusion in nanopores (see for
%instance the recent work by Petit~\etal.~\cite{chemmi_noninvasive_2016}),
%\cref{EQ4.47} suggests that the effective self-diffusivity
%$\overline{D_\mathrm{s}}$ in nanopores decreases upon decreasing the pore size
%$R_\mathrm{p}$.
}

\item{For a partially filled pore at low temperature, the density profile can be assumed to be a step function with $\rho(r) =
\overline{\rho}$ for $R_\mathrm{p}-R' = t$ where $t$ is the film thickness and $\rho =
\rho_g \sim  0$ otherwise ($\rho_g$, which is the gas density under the
considered temperature and pressure, is taken equal to zero as the system is at low temperature).
With such a profile, \cref{EQ4.45} leads to\cite{bhatia_hydrodynamic_2003}:
\begin{multline}
\overline{D_\mathrm{s}} = \frac{2}{R_\mathrm{p}^2}    \int_{R_\mathrm{p}'}^{R_{p}}  D_\mathrm{s}(r) r
\textrm{d}r
= \frac{(R_\mathrm{p}-R')^2}{R_\mathrm{p}^2} D_\mathrm{s}^0 + 2(D_\mathrm{s} -
D_\mathrm{s}^0) \times \\
\left[
\frac{r_0}{R_\mathrm{p}} - \frac{r_0^2}{R_\mathrm{p}^2}
- \left( \frac{R'r_0}{R_\mathrm{p}^2} - \frac{r_0^2}{R_\mathrm{p}^2} \right)
\exp(-(R_\mathrm{p}-R')/r_0)
\right]
\label{EQ4.48}
\end{multline}
For $R_\mathrm{p} >
r_0$, the terms in $r_0/R_\mathrm{p}$ in \cref{EQ4.48} can be neglected so that
the dependence over the film thickness $t = R_\mathrm{p} - R'$ is given by the
first term only.
This predicts that $\overline{D_\mathrm{s}}$ increases with $t$ and, therefore,
with pore loading/pressure.\footnote{
	This simple treatment is important as it allows capturing the overall
	effect of fluid density and pressure on the effective diffusivity in
	nanopores in the low pressure range (see \cref{fig:4_5}). The above
	treatment can be recovered easily from the simple scaling behavior derived
	earlier $\overline{D_\mathrm{s}} = \sum_{\alpha = 1,2} x_\alpha
	D_\mathrm{s}^\alpha$ with $\alpha = 1$ and $\alpha = 2$ for the adsorbed
	phase, respectively.  Within the approximation $\rho_g \sim 0$, such a
	linear combination rule leads to $\overline{D_\mathrm{s}} \sim
	D_\mathrm{s}^1$ since the fraction of molecules is $x_g \sim \rho_g \sim 0$.
	In other words, at temperatures low enough, molecular diffusion through the
	gas phase can be neglected so that, independently of the diffusion
	combination rules considered, diffusion only occurs through the adsorbed
	phase with a diffusion coefficient equal to the surface diffusion.}
}

\item{For a partially filled pore at high temperature,
diffusion through the pore center becomes an important contribution as the gas
density is not negligible.  In such a regime, $\rho(r)$ can be taken as a
decaying function $\rho(r) = \rho_0 \exp \lbrack -(R_\mathrm{p}-r)/r_1
\rbrack$ with a
characteristic length $r_1$ while the function $D_\mathrm{s} = D_\mathrm{s}^0 +
(D_\mathrm{s}^s - D_\mathrm{s}^0)
\exp \lbrack -(R_\mathrm{p}-r)/r_0  \rbrack$ is kept (in practice, $r_0$ and
$r_1$ do
not have to be identical but are similar since they are both affected by the same surface molecular interactions). Direct integration of the
density $\rho(r)$ allows estimating the average density as $\overline{\rho} =
(\pi R_\mathrm{p}^2)^{-1} \int_0^{R_\mathrm{p}} 2 \pi r \rho_0
\exp\lbrack-(R_\mathrm{p} - r)/r_1 \rbrack
\textrm{d}r = 2r_1^2\rho_0/R_\mathrm{p}^2 \times \left\lbrack R/r_1 - 1 +
\exp(-R_\mathrm{p}/r)
\right\rbrack$. Finally, using the position-dependent density
$\rho(r)$ and self-diffusivity $D_\mathrm{s}(r)$ in \cref{EQ4.47} leads to
(with
$\tilde{r}^{-1} = r_0^{-1} + r_1^{-1}$):
\begin{align}
\overline{D_\mathrm{s}} &= \frac{1}{\overline{\rho} \pi R_\mathrm{p}^2}
\int_0^{R_\mathrm{p}} 2
\pi r
D_\mathrm{s}(r) \rho(r) \textrm{d}r
\nonumber \\
&= \frac{2}{\overline{\rho} R_\mathrm{p}^2} \int_0^{R_\mathrm{p}} r
D_\mathrm{s}^0 \rho(r)
\textrm{d}r
+ \frac{2}{\overline{\rho} R_\mathrm{p}^2} \int_0^{R_\mathrm{p}} r
(D_\mathrm{s}^s -
D_\mathrm{s}^0) \rho(r)
\textrm{d}r
\nonumber \\
&= D_\mathrm{s}^0 +
\frac{2\rho_0}{\overline{\rho}R_\mathrm{p}^2}(D_\mathrm{s}^s
- D_\mathrm{s}^0)
\int_0^{R_\mathrm{p}} \exp \lbrack - (R_\mathrm{p} - r)/\tilde{r} \textrm{d}r
\nonumber \\
&= D_\mathrm{s}^0 + (D_\mathrm{s}^s - D_\mathrm{s}^0) \frac{\tilde{r}^2}{r_1
R_\mathrm{p}}
\left \lbrack
\frac{R_\mathrm{p}}{\tilde{r}} - 1 +  \exp \lbrack -R_\mathrm{p}/\tilde{r}
\rbrack
\right \rbrack
\label{EQ4.49}
\end{align}%
}%
\end{itemize}

\begin{figure}[htbp]
	\centering
	\includegraphics[width=0.95\linewidth]{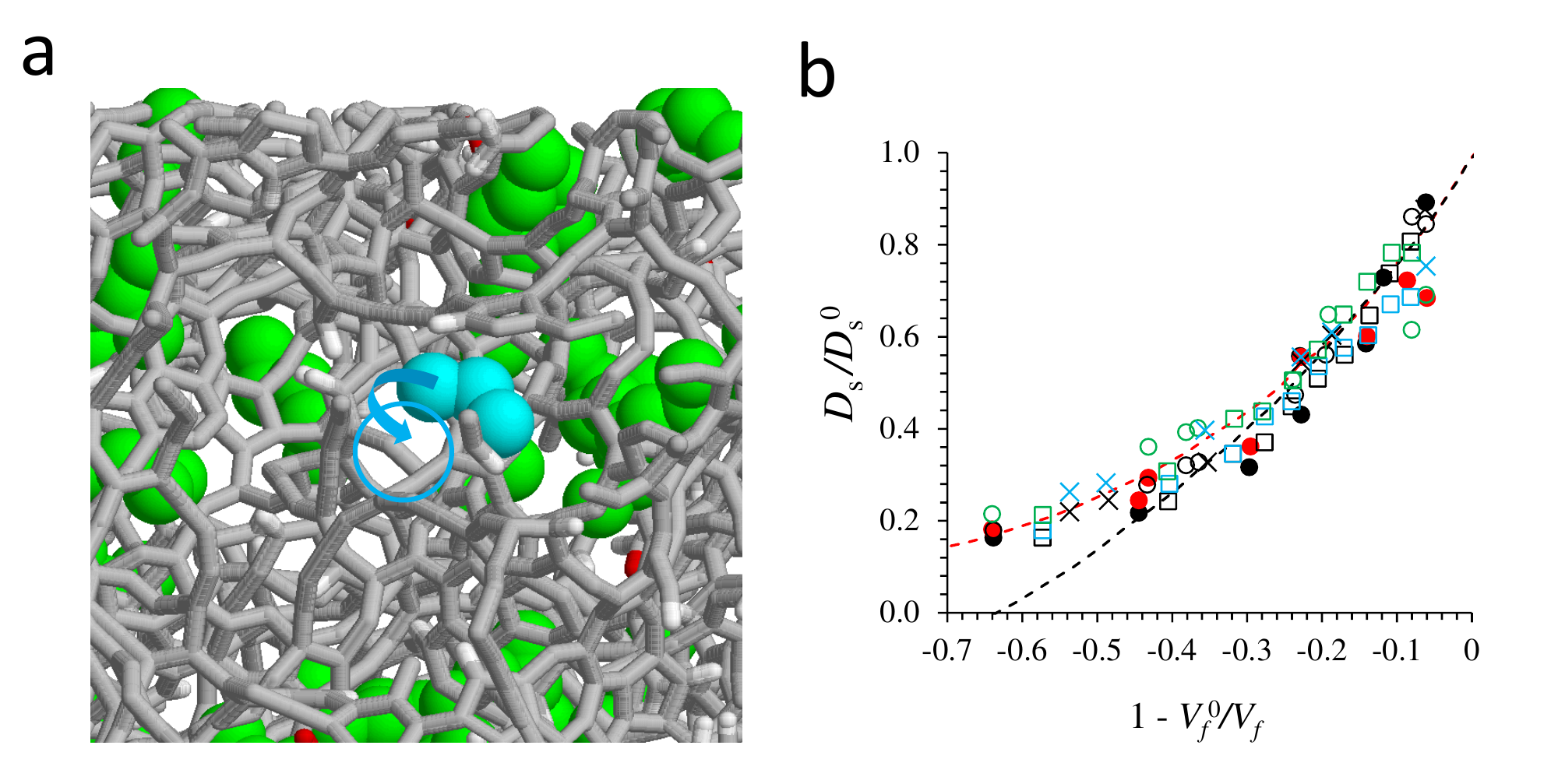}
	\caption{\textbf{Free volume theory.} (a) Principle of the free volume
		theory applied to diffusion in a porous material. A tagged molecule,
		here a propane molecule depicted in blue, will diffuse provided a
		cavity (denoted by the blue circle) is available around it. In the
		context of diffusion in a porous material, the free volume refers here
		to all cavities --- irrespective of their size --- available to the
		diffusing molecules \ie.\ regions that do not intersect with a host
		matrix atom, another fluid molecule, or their covolume.
		(b) Self-diffusivities $D_\mathrm{s}$
		normalized to the bulk self-diffusivity $D_\mathrm{s}^0$ as a function
		of the free volume fraction $V_\mathrm{f}^0/V_\mathrm{f}$ for alkane
		mixtures in the disordered porous carbon shown in (a).
		$V_\mathrm{f}^0$ and $V_\mathrm{f}$ are the free
		volume of the empty and filled porous matrix, respectively. The symbols
		denote different mixtures:
		methane/dodecane (filled circles), methane/hexane (empty circles),
		methane/propane (crosses), and methane/propane/hexane (squares). The
		color code is methane (black), propane (blue), hexane (green), and
		dodecane (red). The red line is the prediction of the free volume
		theory $D_\mathrm{s} \propto \exp \lbrack 1 -
		V_\mathrm{f}^0/V_\mathrm{f} \rbrack$
		obtained from
		the single-component fluids. The black dashed line corresponds to the
		prediction of the surface diffusion model. Figure adapted with permission from
		Ref.~\protect\citen{obliger_free_2016}.
		Copyright 2016 American Chemical Society.}
	\label{fig:4_8}
\end{figure}

\noindent \textbf{Free volume theory.}
In the previous subsection, we considered a spatially-averaged diffusivity
which captures the different regimes observed upon varying the average
confined fluid density $\overline{\rho}$. However, as schematically shown in
\cref{fig:4_5}, upon further increasing $\overline{\rho}$
or --- equivalently --- the pressure $P$, most experimental and simulation
data show that the effective self-diffusivity $\overline{D_\mathrm{s}}$
decreases.
Such a pressure or density-driven
behavior is accurately described using a simple free volume theory (see for
instance Refs.~\citen{falk_subcontinuum_2015,obliger_free_2016}).
In this model, density effects are accounted for by writing that a fluid
molecule diffuses provided  a free cavity is available around it as
illustrated in \cref{fig:4_8}.
In what follows, we first derive the free volume theory initially
proposed by Cohen and Turnbull~\cite{cohen_molecular_1959}
before illustrating its quantitative application to self-diffusion in porous
media. 
Let us define the free volume $v$ of a cavity available in the confined fluid
as the volume of the cavity minus the volume occupied by a fluid molecule. If
we define the  diameter $a(v)$ of this cavity having a free volume $v$, the
contribution from this cavity to the average fluid self-diffusivity is assumed
to be $D_\mathrm{s}(v) = g a(v) u$ where $u$ is the average fluid velocity
(taken equal
to the thermal velocity, $u = \sqrt{3k_\textrm{B}T/m}$), and $g$ is an
effective geometry factor. The average diffusivity $D_\mathrm{s}$ is simply
defined as
an integral over every possible cavity sizes:
\begin{equation}
D_\mathrm{s} = \int_{v^*}^\infty p(v) D_\mathrm{s}(v) \textrm{d}v
\label{EQ4.50}
\end{equation}
where $p(v)$ is the probability to find a cavity having a free volume $v$ and
$v^*$ is the minimum cavity volume in which the molecule can diffuse. There
are two important properties that must be verified by the probability
distribution $p(v)$:
\begin{equation}
\int_{0}^\infty p(v) \textrm{d}v = 1 \hspace{2mm} \textrm{and} \hspace{2mm}
\int_{0}^\infty \gamma v p(v) \textrm{d}v = V_\mathrm{f}
\label{EQ4.51}
\end{equation}
While the first condition simply ensures normalization, the second condition
ensures that the total cavity volume corresponds to the free volume
$V_\mathrm{f}$. The numerical factor $\gamma$ is a factor which accounts for
the fact that several molecules can share part of the same free volume.
If one assumes that the distribution $p(v)$ is exponential, 
$p(v) = \gamma N/V_\mathrm{f} \exp(-\gamma N v/V_\mathrm{f})$,  inserting
this expression into \cref{EQ4.50} together with the
assumption $D_\mathrm{s}(v) = D_\mathrm{s}(v^*)$ for all $v$ leads to
\begin{equation}
D_\mathrm{s} = D_\mathrm{s}(v^*) \exp \left \lbrack - \frac{\gamma v^*
N}{V_\mathrm{f}} \right \rbrack
= g a(v^*) u \exp \left[ - \frac{\gamma v^* N}{V_\mathrm{f}} \right]
\label{EQ4.52}
\end{equation}

As shown in Refs.~\citen{falk_subcontinuum_2015,obliger_free_2016}, the latter 
equation is a robust framework to describe diffusion
in nanoporous media. In particular, it was shown to be  more general than the surface diffusion model (as discussed in what follows, the surface diffusion model can be seen as  an  asymptotic limit of the free volume theory). \redlork{In this context, we note that the use of the free volume theory to rationalize experimental measurements of the self-diffusion for fluids in zeolites was reported in a seminal paper by \citeauthor{karger_self-diffusion_1980}.~\cite{karger_self-diffusion_1980}}
In the free volume theory, the effect of the confined fluid density, \ie.\ the adsorbed amount, is included in the exponential term through the
change in the free volume $V_\mathrm{f}$ as adsorption occurs ($V_\mathrm{f}^0$
is the porous volume which corresponds to the free volume when no molecule is
adsorbed in the porosity).
As shown in \cref{EQ4.52}, $D_\mathrm{s}(v^*)$ is the self-diffusivity at
infinite dilution, \ie.\ $N \rightarrow 0$, which can be estimated from 
molecular dynamics simulations for an isolated molecule. $D_\mathrm{s}(v^*)$ is
linked to the so-called mobility $\mu k_\textrm{B}T$ at infinite dilution.
For a single molecule at equilibrium in the nanoporosity of a host porous material,
the drag force responsible for diffusion $F_u = u/\mu$, balances the friction
force $F_f = - \xi u$ (where $u$ is the molecule velocity and $\xi$ is the
friction coefficient). In molecular dynamics simulations, it can be easily
checked that this important force balance condition is verified as the
confined fluid is at rest (\ie.\ no flow condition). In order to
illustrate the applicability of the free volume theory to diffusion in nanoporous
materials, \cref{fig:4_8} compares its predictions with molecular simulation data
for different alkane mixtures in a disordered  nanoporous
carbon.
Even for binary and ternary mixtures confined in such a heterogeneous
material, the free volume  model accurately predicts the self-diffusivity
$D_\mathrm{s}$ of each mixture component. The change in the accessible volume
$V_\mathrm{f}$ as  adsorption occurs scales with the adsorbed amount $\Gamma$,
$V_\mathrm{f}/V_\mathrm{f}^0 = 1 - \beta
\Gamma$ where $\beta = 0.37$ is the packing fraction of the confined molecules.
Inserting this expression into \cref{EQ4.52}, the self-diffusivity can be
rewritten as a function of $\Gamma$:
\begin{equation}
D_\mathrm{s} = D_\mathrm{s}(v^*) \exp \left\lbrack - \frac{\gamma \beta
\Gamma}{1- \beta
	\Gamma} \right\rbrack
\label{EQ4.53}
\end{equation}
In agreement with experimental and molecular simulation data at large confined
fluid loadings, \cref{EQ4.53} shows  that the self-diffusivity
decreases upon increasing the adsorbed amount $\Gamma$.
By taking the limit of the model at very low densities, $\beta \Gamma \ll 1$
(\ie.\ $V_\mathrm{f} \ll V_\mathrm{f}^0$),
$D_\mathrm{s}/D_\mathrm{s}(v^*) \sim 1 - \gamma \beta \Gamma$.
In Ref.~\citen{obliger_free_2016}, $\gamma = 2.76$ and $\beta = 0.37$ was found
for alkanes in a disordered nanoporous material.
Because $\gamma \beta \sim 1$ with these numbers,
this leads to  $D_\mathrm{s}/D_\mathrm{s}(v^*) \sim 1 - \Gamma$ which is
identical to the
surface diffusion model introduced previously. This suggests that
the free volume theory is very general and robust as its asymptotic limit
encompasses the surface diffusion model. \Cref{fig:4_8} compares the free volume
theory, the surface diffusion model, and the molecular simulation data. It is
found that the free volume theory describes more accurately the
self-diffusivity while the two models merge at low fluid densities
($\Gamma\ll 1$).

%%% Chapter 4: Diffusion at the porous network scale
%!TeX spellcheck = en_US
%!TeX encoding = utf8 
%!TeX program = pdflatex
%!TeX root = ../manuscript.tex

\section{Self and tracer diffusion in porous networks}
\label{chap:diffusion_network}

\subsection{Effective diffusion}

\subsubsection{Tortuosity}

The tortuosity, usually denoted by  $\tau$, is an important 
concept related to both diffusion and transport in porous 
materials~\cite{ghanbarian_tortuosity_2013,sobieski_analysis_2017}.
It characterizes the way in which the porous medium slows down transport
processes due to its geometrical complexity. While there have been attempts to
define $\tau$ based on a geometrical analysis
\cite{ghanbarian_tortuosity_2013,osullivan_perspective_2022}, it remains unclear
how these geometrical characterizations can be applied to the prediction of
transport properties. Therefore, in this review, we take the pragmatic point of
view that consists in defining $\tau$ as the ratio of a transport 
coefficient  for the bulk fluid, $L^0_{\alpha\beta}$, and for the confined 
fluid, $L_{\alpha\beta}$, \ie.\ 
$\tau = L^0_{\alpha\beta}/L_{\alpha\beta}$.
As will be seen in \cref{chap:transport_pore},  
any transport coefficient such as electrical conductivity $\sigma$, permeance $K$ 
or collective diffusivity $D_0 \sim K$, heat conductivity $\lambda_T$, 
\etc.\ can be considered for $L_{\alpha\beta}$. 
In the context of the present chapter, which is devoted to self-diffusivity 
in porous media, the self-diffusivity can be also used to define the tortuosity 
$\tau = D_\mathrm{s}^0/D_\mathrm{s}$. Usually, because of confinement and 
adsorption within a 
porous material, the tortuosity $\tau > 1$ since  $D_\mathrm{s}^0 > 
D_\mathrm{s}$. This 
definition is valid for a single pore but can be easily extended to porous 
materials made of different pores as: 
\begin{equation}
\tau = \frac{\phi D_\mathrm{s}^0}{ D_\mathrm{s}}  
\label{EQ4.54}
\end{equation}
In other words, the tortuosity as defined from the ratio of the bulk and 
\bluelork{As an important remark related to experiments on self-diffusion in nanoporous media, we note that the porosity in \cref{EQ4.54} is needed when tracer-permeation measurements are performed (to correct for the difference in the space explored upon transport). In contrast, such porosity correction is not needed when probing diffusion using PFG NMR.}
Confined transport coefficients must be normalized to the volume fraction in 
which transport occurs \ie.\ the porosity $\phi$. Typically, the latter 
definition ensures that $\tau \rightarrow 1$ for $\phi \rightarrow 1$ (since 
$D_\mathrm{s} = D_\mathrm{s}^0$ when the porosity $\phi = 
1$)~\cite{ghanbarian_tortuosity_2013}. 
\redlork{As an extension of \cref{EQ4.54}, the tortuosity can be defined as any transport coefficient normalized to its bulk value under the same thermodynamic conditions (self-diffusivity, collective diffusivity, electrical conductivity, thermal conductivity, \etc.). While these different  definitions are acceptable, they tend to assume a very unlikely situation where all transport coefficients are defined by a unique tortuosity. Of course, this description ignores that underlying physical phenomena can drastically differ from one transport coefficient to another. In fact, we know that  the mechanisms behind individual dynamical processes (self or tracer diffusion) and collective dynamical processes (e.g. permeability, electrical conductivity) involve very different mechanisms so that using a single tortuosity for all transport types is necessarily very restrictive.}

Several models have been proposed in the literature to describe the tortuosity 
of fluids confined in porous materials. While some of the tortuosity 
expressions are empirical relationships that possess the right asymptotic 
behavior, other expressions are more grounded on robust physical concepts. The 
Weissberg expression belongs  to the former category as it  corresponds to an 
empirical expression between $\tau$ and 
$\phi$~\cite{weissberg_effective_1963}: 
\begin{equation}
\tau = 1-p \ln \phi
\label{EQ4.55}
\end{equation}
where $p$ is a geometrical factor that depends on the morphology (pore shape) 
and topology (pore connectivity) of the host porous 
material~\cite{kolitcheff_tortuosity_2017}. Typically, $p$ takes 
well defined values for materials with simple geometry  such as 
overlapping spheres, cylinders, etc.\ but in practice it is used as an 
empirical parameter fitted against available experimental data. Another 
empirical relationship between tortuosity $\tau$ and porosity $\phi$ is 
Archie’s law that was initially observed for electrolyte transport in porous 
rocks~\cite{archie1942electrical}:
\begin{equation}
\tau = \frac{1}{\phi^m}
\label{EQ4.56}
\end{equation}
where $m$ is  an exponent that usually falls between 
1.8-2~\cite{glover_generalized_2010}.

Besides empirical equations, different models have been proposed to 
describe the tortuosity in porous materials~\cite{sobieski_analysis_2017}.
Using Maxwell’s effective medium theory for the 
conductivity of very diluted spherical particles in an 
electrolyte~\cite{markel_introduction_2016}, it is possible to obtain an approximate analytical expression for the 
tortuosity (defined from the ratio of the bulk and confined electrical 
conductivity but the derivation can be extended to any transport coefficient). 
While a formal derivation will be given in the next subsection on effective 
models (combination rules and effective medium approximation), in the Maxwell 
model (also known as Maxwell--Garnett model), the tortuosity $\tau$ is defined 
from the effective conductivity $\overline{\sigma}$ of an electrolyte having an 
intrinsic conductivity $\sigma_0$ and containing spherical particles with
conductivity $\sigma_p$: 
\begin{equation}
\tau = \frac{\phi \overline{\sigma}}{\sigma_0} = \phi \frac{2\sigma_0 + 
\sigma_p - 2(1-\phi_\mathrm{ext}) (\sigma_0 - \sigma_p)}{2\sigma_0 + \sigma_p + 
(1-\phi_\mathrm{ext})(\sigma_0-\sigma_p)}
\label{EQ4.57}
\end{equation}
where $\phi_\mathrm{ext}$ is the interparticle porosity (1 - 
$\phi_\mathrm{ext}$ is therefore 
the volume fraction of the spherical particles) and $\phi$ is the total 
porosity, \ie.\ the sum of the interparticle porosity and of the 
particle porosity $\phi_p$ (again, a formal derivation of this expression will 
be given in the next subsection). As shown by 
Barrande~\etal.~\cite{barrande_tortuosity_2007}, 
for non porous, \ie.\ non conducting particles ($\sigma_p = 0$ 
and $\phi_\mathrm{ext} = \phi$), \cref{EQ4.57} leads to $\tau = 1 + 0.5 (1 - 
\phi) = (3-\phi)/2$. Note that   Weissberg equation given in \cref{EQ4.55} in the asymptotic limit of  
large porosities $\phi \rightarrow 1$ is compatible with  Maxwell’s 
expression: 
$\tau = 1-p \ln \phi %= 1-p \ln (1-(1-\phi)) 
\sim  1 + p(1-\phi)$.
Maxwell’s effective model for spherical objects corresponds to the choice  $p = 0.5$.
%so that the latter expression leads to $\tau = 
%3/2-1/2 \phi$ as directly derived from Maxwell’s model.
Extensions of Maxwell’s effective medium theory have been proposed by 
Torquato~\cite{torquato_random_2005} or 
Landauer~\cite{landauer_electrical_1952}.
For futher reading on the different models available 
to describe tortuosity in porous materials, the reader is invited to consult 
references~\citen{barrande_tortuosity_2007,kolitcheff_tortuosity_2017,
	sobieski_analysis_2017,duda_hydraulic_2011} and the review 
paper by Sahimi and coworkers~\cite{ghanbarian_tortuosity_2013}.\

\subsubsection{Serial/parallel models}

\bluelork{Some important considerations in this section deal with the application of thermodynamic gradients. We recall that self-diffusion and tracer diffusion, which are at the heart of this section, occur both in systems at equilibrium (i.e.\ without any thermodynamic gradient) and in non-equilibrium systems (i.e.\ with thermodynamic gradients). Therefore, while some important derivations below involve thermodynamic gradients, they allow describing situations that correspond to self-diffusion or tracer diffusion where no such driving forces are applied.}
Many attempts have been made to establish simple models of diffusion in porous 
media based on its analogy with electrical transport in resistance networks. In 
what follows, we review the four models that can be established based on this 
analogy: resistance in series, resistance in parallel, Maxwell's equation, and 
the effective medium theory. All these models rely on simple transport 
equations in which a net flow is induced by a thermodynamic (or electrical 
gradient) $\Delta \chi$. However, thanks to the 
linear response theory, predictions from transport models are also relevant to 
diffusion at equilibrium (\ie.\ with no net flow) since the limit of a 
vanishing driving force $\Delta \chi \rightarrow 0$ can be taken.

\begin{figure}[htbp]
	\centering
	\includegraphics[width=0.95\linewidth]{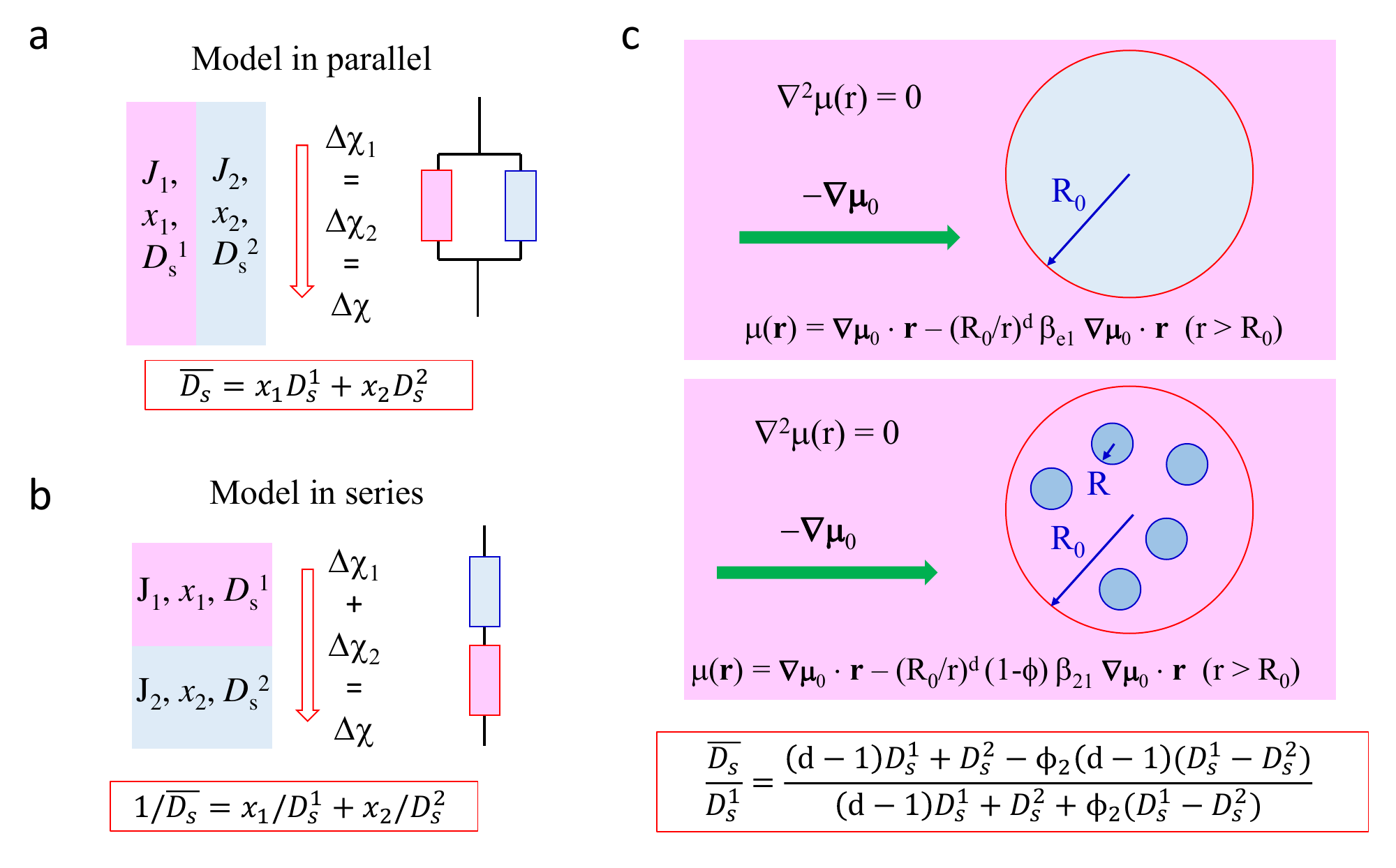}
	\caption{\textbf{Effective medium models.} (a) For a porous medium made of 
		domains aligned parallel to the thermodynamic gradient inducing 
		transport, 
		the total flux is proportional to the fluxes in each domain $J = J_1 + 
		J_2$. In this case, the effective diffusivity $\overline{D_\mathrm{s}}$ 
		is 
		the sum 
		of the self-diffusivity $D_\mathrm{s}^\alpha$ in each domain type 
		$\alpha$ 
		weighted by the fraction $x_{\alpha}$ of molecules located in this 
		domain 
		type. (b) For a porous medium made of domains aligned in series 
		with respect to the direction of the thermodynamic gradient inducing 
		transport, the total flux is equal to the flux in each domain $J = J_1 
		= 
		J_2$. In this case, the reciprocal of the effective diffusivity 
		$\overline{D_\mathrm{s}}$ is the sum of the reciprocal of the 
		self-diffusivity 
		$D_\mathrm{s}^\alpha$ in each domain type $\alpha$ weighted by the 
		fraction 
		$x_{\alpha}$ of molecules located in this domain type. (c) In  
		general, when a porous medium cannot be assumed to be made of domains 
		aligned in parallel or in series with respect to the direction of the 
		thermodynamic gradient inducing transport, the flow can be predicted 
		using 
		Maxwell's effective model. The physical solution is obtained by writing 
		that the chemical potential field verifies Laplace 
		equation, \ie.\ $\nabla^2 \mu = 0 \hspace{1mm} \forall r$,
		since there is 	no source or sink term that modifies the local chemical 
		potential $\mu(\textbf{r})$.} 
	\label{fig:5_1}
\end{figure}

Let us consider a network of domains 
in which a fluid is transported according to different transport coefficients. 
The derivation below is done for the collective diffusivity $D_0$ 
\ie.\ the permeance $K \sim D_0$ but the self-diffusivity $D_\mathrm{s}$ will 
be readily obtained by considering the limit for an infinitely diluted medium 
where $D_\mathrm{s} \sim D_0$. 
The collective diffusivity  in each domain type $\alpha$  is $D_{0}^\alpha$ and 
the system is subjected to a chemical potential gradient  $\nabla \mu = \Delta 
\mu/L$ where $L$ is the thickness of the sample. Let us first consider that the 
domains are organized in parallel as depicted in \cref{fig:5_1}(a) --- for the 
sake of simplicity, only two domain types $\alpha = 1, 2$ are shown but the 
derivation below can be generalized to any number of domains. The total flux 
$J$ across the material having a section area $A$ is the sum of the fluxes in 
each domain weighted by its section area 
$A_\alpha$, $J \times A = \sum_{\alpha} J_\alpha \times A_\alpha$. 
In the framework of the linear response theory, the flux $J$ 
induced by the chemical potential gradient is 
$J = - \rho \overline{D_0} \Delta \mu/L k_\textrm{B}T$
where $\overline{D_0}$ is the effective permeance of the 
composite material. As the number of molecules $N$ in the material is 
$N = \rho A L$,
$J \times A 
	= - N/L^2 \times \overline{D_0} \Delta \mu/k_\textrm{B}T$. 
Similarly, because the domains are subjected to the same driving force 
$\Delta \mu$ when set in parallel, the flow in each domain is 
$J_\alpha \times A_\alpha 
	= - N_\alpha/L^2 \times D_{0}^\alpha \Delta \mu/\left(k_\textrm{B}T\right)$ 
where $N_\alpha$ is the number of molecules in each domain $\alpha$. 
With these relations, by writing $J \times A = \sum_{\alpha} J_\alpha \times
A_\alpha$ and taking the limit $D_\mathrm{s} \sim D_0$, we obtain: 
\begin{equation}
\overline{D_\mathrm{s}} = \sum_{\alpha} x_\alpha D_\mathrm{s}^\alpha
\label{EQ4.58}
\end{equation}
where $x_\alpha = \rho_\alpha/\rho = N_\alpha/N$ is the mole fraction of 
molecules in the domain type $\alpha$. As mentioned above, while this equation 
was derived for the  collective diffusivity, \ie.\ permeance, it also 
applies to any other transport coefficient such as the electrical conductivity 
and the self-diffusivity.

Let us now consider a porous network in which the domains are organized in 
series as depicted in \cref{fig:5_1}(b). Again, in the framework of the linear 
response theory, the flow $J \times A$ induced by the chemical potential 
gradient is 
$J \times A 
	= - \rho A \times \overline{D_0} \Delta \mu/Lk_\textrm{B}T 
	= - N/L^2 \times \overline{D_0} \Delta \mu/k_\textrm{B}T$
where $\overline{D_0}$ is the effective collective diffusivity of the composite 
material and where $N = \rho A L$ is the total number of fluid molecules. 
Similarly, let us define the flow in each domain $\alpha$ as 
$J_\alpha \times A 
	= - \rho_\alpha A \times \overline{D_0} \Delta \mu/L_\alpha k_\textrm{B}T
	= - N_\alpha/L_\alpha^2 \times \overline{D_0} \Delta \mu/k_\textrm{B}T$
($N_\alpha = \rho_\alpha A L_\alpha$ and $L_\alpha$ 
are the number of molecules and the 
thickness for this domain $\alpha$, respectively). Because of mass conservation 
upon transport in the composite material, the flow $J_\alpha$ in each domain is 
equal to the overall flow $J$, \ie.\ $J_\alpha = J\,\forall \alpha$. 
Moreover, the difference $\Delta \mu_\alpha$ in the thermodynamic variable 
across the domain $\alpha$ must be such that 
$\Delta \mu_\alpha = \sum_\alpha \Delta \mu$.
With these conditions, it is straightforward to show that 
$JL^2/N\overline{D_0} = \sum_\alpha J L_\alpha^2/N_\alpha D_0^\alpha$. 
Finally, by noting that $N_\alpha \propto L_\alpha$ and taking the limit 
$D_\mathrm{s} \sim D_0$, we obtain:  
\begin{equation}
\frac{1}{\overline{D_\mathrm{s}}} = \sum_{\alpha} 
\frac{x_\alpha}{D_\mathrm{s}^\alpha}
\label{EQ4.59}
\end{equation}
where $x_\alpha = \rho_\alpha/\rho = N_\alpha/N$ is the mole fraction of molecules in the 
domain type $\alpha$.

\subsubsection{Maxwell model} 

The combination rules for domains in series and in 
parallel rely on a very simplified representation of real systems.
A more rigorous treatment of the problem was first derived by Maxwell in an 
attempt to describe the conductivity of an electrolyte containing conducting 
spheres at infinite dilution~\cite{torquato_random_2005}.
As shown in the following paragraph, this model can be used to describe the 
effective diffusivity at constant temperature --- $\nabla T = 0$ ---
of a heterogeneous medium made of spheres having a 
different self-diffusivity than the medium in which they are included. 
%More in  detail, let us consider the isothermal  situation 
%depicted in \cref{fig:5_1}(c). 
%where a spherical inclusion of a radius $R_0$ is 
%immersed in a medium subjected to a homogeneous chemical potential field
%$\mathbf{\nabla \mu}_0$  
Let us consider in \cref{fig:5_1}(c, bottom) a medium 1 with a spherical inclusion of
radius $R_0$ made up of medium 1 and small spherical domains of medium 2.
The whole system is subjected to a chemical potential gradient $\nabla \mu_0$ in
the far field.
Maxwell's model allows  modeling this system by replacing the heterogeneous
inclusion by an effective medium $e$ having the same radius $R_0$ as shown in
\cref{fig:5_1}(c, top).
In what follows, the subscripts $1$ and $2$ refer to the main medium and to the 
small spherical inclusions in the large inclusion, respectively.
The subscript $e$ refers to the large inclusion treated as an effective medium.
From a very general viewpoint, $\mathbf{\nabla \mu}_0$  can be replaced by any
thermodynamic gradient $\nabla \chi$ such as a pressure or temperature gradient
or an electrical field (so that the treatment below can be seen as a very
general solution to such a broad class of complex problems).
Because there is no molecule source/sink, the solution $\mu(\textbf{r})$ is
given by the  Laplacian equation $\Delta \mu = 0$ with  the following boundary
conditions:
(1) chemical potential continuity at $r = R_0$ i.e.  $\mu(r = R^+) = \mu(r = R^-)$, 
(2) if $\rho_1 D_0^1$ and $\overline{\rho} \overline{D_0}$ are the 
conductivity --- \ie.\ the collective diffusivity multiplied by the fluid density ---
in the medium 1 and in the effective medium $e$, the flux balance from 1 to $e$
and from $e$ to 1 at $r = R_0$ imposes that $\rho_1 D_0^1
\textbf{n} \cdot \mathbf{\nabla \mu})_+ = \overline{\rho} \overline{D_0} 
\textbf{n} \cdot \mathbf{\nabla \mu})_-$, and (3) in the far field, i.e.
$r \rightarrow \infty$, $\mu(r)  \rightarrow \mathbf{\nabla \mu}_0 \cdot \textbf{r}$.
The general solution of this system in a space of dimension $d = 3$ is:
\begin{align}
\mu(\textbf{r}) &= \mathbf{\nabla \mu}_0 \cdot \textbf{r} 
	- \beta_{e1} \hspace{1mm} \Big (\frac{R_0}{r} \Big )^3 
	\mathbf{\nabla \mu}_0 \cdot \textbf{r} 
\qquad & (r \geq R_0)  \nonumber \\
\mu(\textbf{r}) &= \mathbf{\nabla \mu}_0 \cdot \textbf{r} - \beta_{e1} \,
\mathbf{\nabla \mu}_0 \cdot \textbf{r} 
\qquad &(r \leq R_0) 
\label{EQ4.60}
\end{align}
where $\beta_{e1} =  [\overline{\rho} \overline{D_0} - \rho_1 D_0^1]/[\overline{\rho} \overline{D_0} + (d-1)\rho_1 D_0^1]$ with $\overline{\rho}$ and 
$\overline{D_0}$ the effective density and collective diffusivity in the 
large spherical inclusion.

Let us now treat explicitly the spherical inclusion in \cref{fig:5_1}(c) as made
up of medium 1 but with many small spherical inclusions of a radius $R$.
First, considering a single domain of medium 2 having a radius $R$, the solution
in the far field $r > R_0$ can be derived like in the previous treatment:  
$\mu(\textbf{r}) = \mathbf{\nabla \mu}_0 \cdot \textbf{r} - \beta_{21} (R/r)^3 \hspace{1mm} 
\mathbf{\nabla \mu}_0 \cdot \textbf{r}$ for $r \geq R_0$.\footnote{ 
When deriving this equation, the far field solution for $r > R_0$ applies
\textit{a fortiori} to $r > R$ since $R_0 \gg R$.
}
We now consider the more physical case where there are $N$ domains of medium 2 in the inclusion.
Within the infinite dilution regime, we can assume that all perturbations
created by the domains of medium 2 add up in the far field region $r \gg R_0$.
Since we have $N$ cavities, the  solution is simply obtained by multiplying by
$N$ the last equation: 
$\mu(\textbf{r}) = \mathbf{\nabla \mu}_0 \cdot \textbf{r} - \beta_{21} \hspace{1mm} 
N (R/r)^3 \mathbf{\nabla \mu}_0 \cdot \textbf{r}  
= \mathbf{\nabla \mu}_0 \cdot \textbf{r} - \beta_{21} \hspace{1mm} v (R_0/r)^3
\mathbf{\nabla \mu}_0 \cdot \textbf{r}$
[the last equality was obtained by introducing the volume fraction of the
domains 2 within the spherical inclusion, $v = N(R/R_0)^3$]. 
By comparing this last equation with \cref{EQ4.60}, we obtain that $\beta_{e1} =
v\beta_{21}$ or equivalently: 
\begin{equation}
\frac{\overline{\rho} \overline{D_0} -\rho_1 D_0^1}{\overline{\rho} 
\overline{D_0} + (d - 1) \rho_1 D_0^1}
= v \left \lbrack \frac{\rho_2 D_0^2-\rho_1 D_0^1}{\rho_2 D_0^2 + (d - 1) 
\rho_1 D_0^1}  \right \rbrack	
\label{EQ4.61}
\end{equation}
By noting that the conductivity $\sigma = \rho D_0$, 
and using  the volume fraction $v = 1 - \phi_\mathrm{ext}$ where
$\phi_\mathrm{ext}$ is the porosity of the large 
spherical inclusion and $d = 3$, it is straightforward to check that 
\cref{EQ4.61} is equivalent to Maxwell's formula given in \cref{EQ4.57} for the 
effective conductivity of an electrolyte containing spherical particles having 
a different conductivity than its bulk counterpart. Finally, taking the limit 
of an infinitely diluted medium \ie.\ by replacing the chemical 
potential gradient $\nabla \mu$  by $\nabla \rho/\rho$, the treatment above 
leads to the same equation as \cref{EQ4.61} but with the effective 
collective-diffusivity $\overline{D_0}$ replaced by the effective 
self-diffusivity $\overline{D_\mathrm{s}}$.

\subsubsection{Effective medium theories}

Generalization of Maxwell's formula given in \cref{EQ4.61} to any 
heterogeneous medium made of domains $\alpha = \lbrack 1, M \rbrack$ occupying 
a volume fraction $v^\alpha$ writes: 
\begin{equation}
\frac{\overline{\rho} \overline{D_\mathrm{s}} - 
\rho_1 D_0^1}{\overline{\rho} \overline{D_\mathrm{s}} + (d - 1) \rho_1 D_0^1} = 
\sum_{\alpha = 1,M} v_\alpha \left \lbrack \frac{\rho_\alpha 
D_0^\alpha-\rho_1 D_0^1}{\rho_\alpha D_0^\alpha + (d - 1) \rho_1 
D_0^1}  \right \rbrack
\label{EQ4.62}
\end{equation}
where we recall that $d$ is the dimension of the medium. Further development of Maxwell’s model was proposed by 
Bruggeman~\cite{bruggeman_berechnung_1935}
and Landauer~\cite{landauer_electrical_1978} who introduced in 
\cref{EQ4.62} a self-consistent approximation.
As shown in \cref{fig:5_1}(c), 
Maxwell’s formulation relies on the idea that the inclusion 
acts as a perturbation to the uniform field $\mu(\textbf{r})$ applied outside 
the inclusion. In the self-consistent scheme, the different perturbations 
induced by all the domains $\alpha = \lbrack 1,M \rbrack$ to the homogeneous 
field average to zero, $\langle \Delta \mu(\textbf{r}) \rangle = 0$.
In terms of conductivity, this condition implies that: 
\begin{equation}
\sum_{\alpha = 1,M} v_\alpha \frac{\rho_\alpha D_{0,\alpha} - \overline{\rho} 
\overline{D_0}}{\rho_\alpha D_{0,\alpha} - (d - 1) \overline{\rho} 
\overline{D_0}} = 0
\label{EQ4.63}
\end{equation}
Interestingly, as shown by Kirkpatrick~\cite{kirkpatrick_percolation_1973}, 
this formula can be retrieved by applying the percolation theory 
to an array of electrical conductors of different resistivities. In the case of 
heterogeneous materials made of two domain types, \cref{EQ4.63} can be 
easily solved to predict the diffusivity in porous media from the known 
diffusivities in the different domains. While the self-consistent approximation 
to Maxwell’s formula provides a very general framework to discuss diffusion and 
transport in complex porous media, its application is mostly limited to 
semi-quantitative prediction as it relies on crude assumptions which neglect 
issues such as interfacial transport limitations. For an example of the 
application of Bruggeman’s equation --- \ie.\ \cref{EQ4.63} --- to 
self-diffusion in heterogeneous media, the reader is referred to 
Refs.~\citen{bonilla_understanding_2014,galarneau_probing_2016}.

\subsection{Hierarchical and mesoscopic approaches}

\subsubsection{General considerations}

Statistical physics is a very robust and general framework to describe the 
complex problem of diffusion in heterogeneous media such as porous materials. A 
detailed description of the field of statistical mechanics and its application 
to diffusion in complex systems is out of the scope of the present context. However, we note that important statistical physics derivations can be found in 
Refs.~\citen{hansen_theory_2013,barrat_basic_2003} for general aspects and  
in Ref.~\citen{bouchaud_anomalous_1990} for diffusion in disordered media. In a previous section, the physics of the 
Intermittent Brownian Motion was introduced to describe the problem of 
diffusion in a single pore in which the confined fluid subdivides into an 
adsorbed phase at the pore surface and a bulk-like phase in the pore center. 
Here, we move to a larger scale by considering molecular diffusion in a 
heterogeneous medium such as a fluid confined in a disordered porous solid made 
of different domains. This problem can be treated in a rigorous fashion using 
theoretical frameworks such as Langevin and Fokker--Planck 
equations~\cite{roosen-runge_analytical_2016}.
As shown below, this formal approach allows 
predicting the dynamics in such complex media using intermediate functions 
which are readily obtained using experiments
(typically, quasi-elastic and inelastic neutron scattering).

Let us consider the situation depicted in \cref{fig:5_2}(a). A host solid 
is made of different porous domains --- a, b, c, d, e, and f --- in which a 
molecule diffuses. The molecule diffusion is illustrated using the 
trajectory shown as the black line between a set of initial and final 
positions corresponding to the grey circles.
In order to provide a concrete analogy,  
\cref{fig:5_2}(a) shows electron tomography data of a  zeolite crystal in 
which one distinguishes the microporous domains (`mi', the corresponding 
porosity is invisible because the experimental resolution is larger than the 
corresponding pore size), the mesoporous domains (`me'), and macroporous 
domains (`ma', located outside the zeolite crystal). 
In \Cref{sec:surface_versus_volume_diffusion}, we have 
already introduced the self-correlation function $G_\mathrm{s}(\textbf{r},t)$ 
defined in \cref{EQ4.25}. This function describes the probability that a 
molecule diffuses by a quantity $\textbf{r}$ over a time $t$. We have also 
introduced its spatial Fourier transform $I_\mathrm{s}(\textbf{q},t) = \int 
G_\mathrm{s}(\textbf{r},t) 
\textrm{e}^{i\textbf{q}\cdot\textbf{r}} \textrm{d}\textbf{r}$.
As already discussed in \Cref{sec:surface_versus_volume_diffusion}, in the 
limit of very slow switching rate between domains, 
\ie.\ $\tau_C^\alpha \gg \left(D_\mathrm{s}^\alpha q^2\right)^{-1}$ where 
$(D_\mathrm{s}^\alpha q^2)^{-1}$ is the time needed to diffuse through a 
typical domain $\alpha$ having a size $q \sim 2\pi/r$ while $\tau_C^\alpha$ is 
the typical crossing time estimated as the mean first-passage time, molecule 
exchange between the different domains is limited. In this limit, the 
correlation function $I_\mathrm{s}(\textbf{q},t)$, which is dominated by 
diffusion in 
the different domains, 
is given by the sum of all intra-domain diffusion contributions,  
$I_\mathrm{s}(\textbf{q},t) = \sum_\alpha x_\alpha \exp(-D_\mathrm{s}^\alpha 
q^2 t)$.

\subsubsection{Diffusion in heterogeneous media}

\begin{figure}[htbp]
	\centering
	\includegraphics[width=0.95\linewidth]{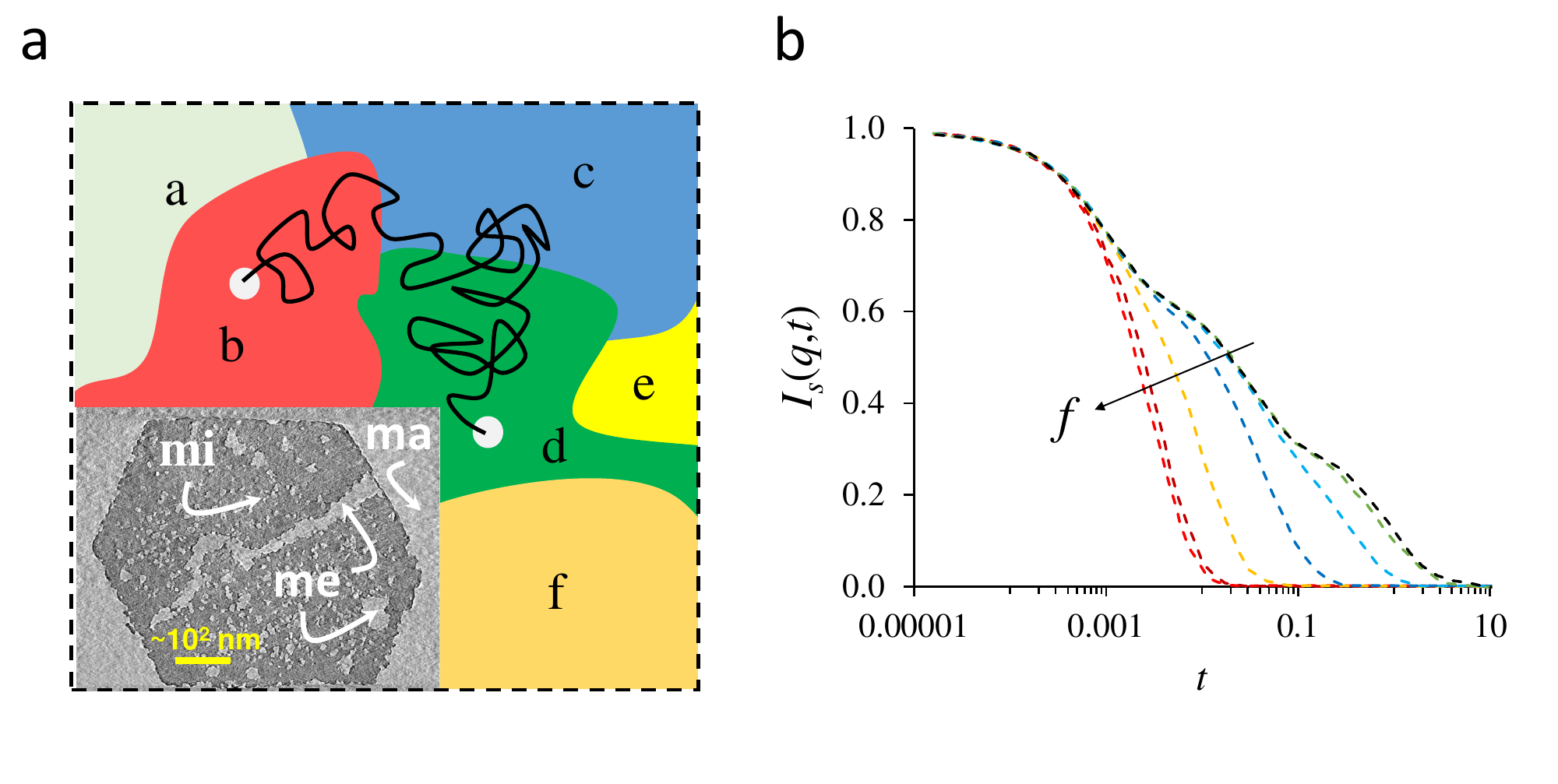}
	\caption{\textbf{Self-diffusion in heterogeneous media.} (a) Principle of 
	self-diffusion in heterogeneous media made of different domains --- here, 
	6 domains denoted with different colors and letters a, b, c, d, 
	e, and f. A molecule follows a single trajectory which is made of segments 
	in these different domains 
	b $\rightarrow$ c $\rightarrow$ d $\rightarrow$ c $\rightarrow$ d.
	As an example of such domain decomposition, electron tomography data of a hexagonal 
	zeolite crystal are shown in which micropores (`mi') and mesopores (`me') 
	domains are shown.
	Modified with permission from Ref.~\protect\citen{de_jong_zeolite_2010}.
	Copyright 2010 Wiley-VCH Verlag GmbH \& Co.\ KGaA.
	Typical self-diffusion in such hierarchical systems include trajectories
	segments in these micropores and mesopores but also in the macropores (`ma')
	which are located outside the zeolite grains. 
	(b) Incoherent scattering function $I_\mathrm{s}(q,t)$ determined for a
	heterogeneous medium made of 
	3 domain types (a, b, c) in which molecules display  different 
	self-diffusion coefficients: $D_\mathrm{s}^c \sim 33 D_\mathrm{s}^b \sim 
	1000 D_\mathrm{s}^a$.  
	The different color lines correspond to increasing switching rates $f$ 
	which vary from 0, 0.1, 1, 10, 100, 1000, 10000. Figure adapted with permission from 
	Roosen-Runge \etal.~\cite{roosen-runge_analytical_2016}.
	Copyright 2016 AIP Publishing.} 
	\label{fig:5_2}
\end{figure}

In order to derive a general framework to analyze diffusion in heterogeneous 
media, Roosen-Runge \etal.~\cite{roosen-runge_analytical_2016} used a 
path integral approach in which the self-correlation function 
$G_\mathrm{s}(\textbf{r},t)$ is written by considering every possible path that 
leads 
from a point $\textbf{r} = 0$ to a point $\textbf{r}$ over a time $t$. 
In more detail, upon assuming that the trajectory segment within a given domain 
is independent of that within the previous domain visited, 
$G_\mathrm{s}(\textbf{r},t)$ 
can be expressed as: 
\begin{align}
G_\mathrm{s}(\textbf{r},t) = \sum_{m = 0}^\infty \sum_{i_1, \ldots, 
\hspace{0.5mm} i_m}
&\int_V \textrm{d}\mathbf{r_1} \int_0^\infty \textrm{d} t_1 
P_{i_1}(\mathbf{r_1},t_1\lvert\mathbf{0},0) \hspace{0.2cm} \times
\nonumber \\
& \int_V \textrm{d}\mathbf{r_2} \int_{t_1}^\infty \textrm{d} t_2 
P_{i_2}(\mathbf{r_2},t_2\lvert\mathbf{r_1},t_1) \hspace{0.2cm} \times 
\hspace{0.2cm} \dots
\hspace{0.2cm} \times 
\nonumber \\
&\int_V \textrm{d}\mathbf{r_m} \int_{t_m-1}^\infty \textrm{d} t_m 
P_{i_m}(\mathbf{r_m},t_m\lvert\mathbf{r_{m-1}},t_{m-1})  \hspace{0.2cm} \times
%\nonumber \\
Q_{i_{m+1}}(\mathbf{r},t\lvert\mathbf{r_{m}},t_{m})
\label{EQ4.64}
\end{align}
In this equation, $P_{i_m}(\mathbf{r_m},t_m\lvert\mathbf{r_{m-1}},t_{m-1})$ 
denotes the probability that the molecule leaves a domain $m$ at time $t_m$ and 
position $\mathbf{r_{m}}$ while initially entering the same domain at time 
$t_{m-1}$ and at position $\mathbf{r_{m-1}}$ (this definition implies that 
$\mathbf{r_{m-1}}$ and $t_{m-1}$ are the position and time at which the 
molecule leaves the domain $m-1$ to enter the domain $m$).  
$Q_{i_{m+1}}(\mathbf{r},t\lvert\mathbf{r_{m}},t_{m})$ is the probability that a 
molecule is located at a position $\mathbf{r}$ in a domain $m+1$ at time $t$ 
after entering the same domain at a position $\mathbf{r_{m}}$ and time $t_{m}$. 
The subscripts $i_m$ and $i_{m+1}$ in $P_{i_m}$ and $Q_{i_{m+1}}$ indicate the 
domain types corresponding to the $m^\mathrm{th}$ and $m+1^\mathrm{th}$ visited 
domains, respectively (e.g. type a, b, c, d, e, or f in \cref{fig:5_2}). A few 
remarks are in order to fully understand the above equation.
\begin{itemize}
\item{In \cref{EQ4.64}, the first discrete sum over $m$  from 0 to 
$\infty$ accounts for the fact that the distance $\mathbf{r}$ traveled over a 
time $t$ can correspond to different sets of $m$ visited domains before ending 
up in the domain $m+1$ (with $m$ taking any integer number of trajectory 
segments in different domains). Typically, taken as an example, the 
individual trajectory shown in \cref{fig:5_2} corresponds to a traveled 
path made of $m + 1 = 5$ domain visits: b $\rightarrow$ c $\rightarrow$ d 
$\rightarrow$ c $\rightarrow$ d. The term $m = 0$ corresponds to a molecule 
that diffuses by a distance $\mathbf{r}$ over a time $t$ while staying within 
the same domain. On the other hand, the terms $m \neq 0$ corresponds to a 
trajectory over a distance $\mathbf{r}$ traveling through $m + 1$ different 
domains (even though the same domain can be visited several times as 
illustrated in \cref{fig:5_2}).}
\item{For a given $m$, there are $2m$ integrals contributing to 
$G_\mathrm{s}(\mathbf{r},t)$ in \cref{EQ4.64}; each set of two integrals is of 
the form $\int_V \textrm{d}\mathbf{r_k} \int_{t_{k-1}}^\infty \textrm{d}t_{k} 
P_{i_k}(\mathbf{r_k},t_k\lvert\mathbf{r_{k-1}},t_{k-1})$. Each of these terms 
accounts for the contribution from the trajectory segment within the 
$k^\mathrm{th}$ visited domain to the total self-correlation function 
$G_\mathrm{s}(\mathbf{r},t)$ (we 
recall that the subscript $i_k$ indicates the domain type corresponding to the 
$k^\mathrm{th}$ visited domain). In more detail, for each term, the probability 
$P_{i_k}(\mathbf{r_k},t_k\lvert\mathbf{r_{k-1},t_{k-1}})$  that the molecule 
leaves this domain at the position $\mathbf{r_{k}}$ at a time $t_{k}$ while 
entering at the position $\mathbf{r_{k-1}}$ at a time $t_{k-1}$ is integrated 
over the time $t_{k}$ (with $t_k$ going from $t_{k-1}$ to $\infty$) and over 
the position $\mathbf{r_k}$ (with $\mathbf{r_k}$ taking any possible value over 
the entire volume of the porous medium).} 
\item{The term $Q_{i_{m+1}}(\mathbf{r},t\lvert\mathbf{r_{m},t_{m}})$ accounts 
for the fact that after entering the last visited domain, \ie.\ domain 
$m + 1$, the particle remains located in this domain and reaches the position 
$\mathbf{r}$ at time $t$.}
\item{The second discrete sum in \cref{EQ4.64} corresponds to every 
combination of possible path sequence $\lbrack i_1, \dots, i_k, \dots, i_m 
\rbrack$ for the $m$ visited domains before ending in the final $m + 1$ domain 
--- where $i_k$ corresponds to the domain type (a, b, c, d, e, or f in 
\cref{fig:5_2}) --- before ending up in the domain $m+1$.  The sets of values 
taken by $\lbrack i_1, \dots, i_k, \dots, i_m \rbrack$  are restricted to  
integers with different consecutive values. As noted by Roosen-Runge 
\etal.~\cite{roosen-runge_analytical_2016}, the formalism above only relies on 
the assumption that trajectory segments within one domain are uncorrelated with 
that in the previous/next domains.
The heterogeneous dynamics is all included in the 
values taken by the passage positions and times, \ie.\ $\mathbf{r_m}$ 
and $t_m$, and in the domain connectivity which is accounted for in the values 
taken by the functions $P_{i_k}$ (two unconnected domain types $i_k$ and 
$i_{k+1}$ will not contribute to $G_\mathrm{s}(\mathbf{r},t)$ as no trajectory 
segment 
within a domain of type $i_k$  can lead to the border of a domain of type 
$i_{k+1}$).
} 
\end{itemize}

Assuming invariance under time and space translation of the different functions $P_{i_k}$ 
and $Q_{i_k}$, the integral expression in  \cref{EQ4.64} can be used to 
predict the heterogeneous dynamics in such complex environments. In practice, 
one calculates the following Fourier--Laplace transform which writes: 
\begin{align}
\tilde{I}(\mathbf{q},s) &= \int \textrm{d} \mathbf{r} \hspace{1mm} 
e^{i\mathbf{q} \cdot 
\mathbf{r}} \int \textrm{d}t \hspace{1mm} e^{-st} G_\mathrm{s}(\mathbf{r},t) 
\nonumber \\
&= \sum_{m = 0}^{\infty} \sum_{i_1, \ldots, \hspace{0.5mm} i_m} 
\tilde{P}_{i_1}(\mathbf{q},s) 
\tilde{P}_{i_2}(\mathbf{q},s) \ldots \tilde{P}_{i_m}(\mathbf{q},s) 
\tilde{Q}_{i_{m+1}}(\mathbf{q},s)
\label{EQ4.65}
\end{align}
with  $\forall k$:
\begin{align}
\tilde{P}_{i_k}(\mathbf{q},s) &= \int \textrm{d} \mathbf{r} \hspace{1mm} 
e^{i\mathbf{q} \cdot 
\mathbf{r}} \int \textrm{d}t \hspace{1mm} e^{-st} P_{i_k}(\mathbf{r},t) 
\nonumber \\
\tilde{Q}_{i_k}(\mathbf{q},s) &= \int \textrm{d} \mathbf{r} \hspace{1mm} 
e^{i\mathbf{q} \cdot 
\mathbf{r}} \int \textrm{d}t \hspace{1mm} e^{-st} Q_{i_k}(\mathbf{r},t) 
\label{EQ4.66}
\end{align}
Let us consider a porous medium made of two domain types $a$ and $b$. 
Typically, in the context of the present review, these two domains can 
correspond to domains within a porous particle and outside the porous particle 
or to different porosity types within the same porous particles. As treated in 
detail in Ref.~\citen{roosen-runge_analytical_2016}, for such two domain 
systems, \cref{EQ4.65} leads to: 
\begin{align}
\tilde{I}(q,s) &= \sum_{k = 0}^\infty 
(\tilde{P}_a \tilde{P}_b)^k \times 
\lbrack
c (\tilde{Q}_a + \tilde{P}_a \tilde{Q}_b) + (1 - c) (\tilde{Q}_b + \tilde{P}_b 
\tilde{Q}_a)
\rbrack
\nonumber \\
&= \frac{c (\tilde{Q}_a + \tilde{P}_a \tilde{Q}_b) + (1 - c) (\tilde{Q}_b + 
\tilde{P}_b)}{1 - \tilde{P}_a \tilde{P}_b}
\label{EQ4.67}
\end{align}
where $c = P_a(q,t=0)$ is the initial concentration in domain $a$.
Note that we omit in the above equation  to explicitly write the function 
dependence, \ie.\ $P_{a,b} = P_{a,b}(q,s)$. 

To predict the heterogeneous dynamics using the formalism above, one needs to 
provide simple expressions or determine using molecular simulation the function 
$P_{a,b}(r,t)$ and $Q_{a,b}(r,t)$. As far as formal expressions are concerned, 
a simple treatment consists of assuming that the switching probability 
$\psi_{a,b}(t)$ --- \ie.\ the probability that a molecule switches 
after a time $t$ --- and the diffusion propagator $g_{a,b}(r,t)$ --- 
\ie.\ the probability that a molecule diffuses by a distance $r$ over 
a time $t$ --- are independent of each other.
With these approximations, the probability distributions 
$P_{a,b}(r,t)$ are given by the probability that the molecule diffuses by $r$ 
over a time $t$ multiplied by the probability that the molecule switches to a 
different domain at time $t$. Similarly, the function $Q_{a,b}(r,t)$ is defined 
as the probability that the molecule diffuses by $r$ over $t$ multiplied by the 
probability that the molecule remains within the same domain
$\phi_{a,b}(t) = 1-\int_0^t \psi_{a,b}(t') \textrm{d}t'$.
These choices lead to the following expressions: 
\begin{align}
P_{a,b}(r,t) &= \psi_{a,b}(t) g_{a,b}(r,t)
\nonumber \\
Q_{a,b}(r,t) &= \phi_{a,b}(t) g_{a,b}(r,t)
\label{EQ4.68}
\end{align}
All the complex underlying phenomena involved in the heterogeneous dynamics are 
included in the functions  $\psi_{a,b}(t)$ and $g_{a,b}(r,t)$. In the most 
general situations, the switching functions $\psi_{a,b}(t)$ depend on the 
geometry and structure of the porous medium with some known rigorous 
mathematical forms in some specific cases~\cite{roosen-runge_analytical_2016}. 
Similarly, the propagator $g_{a,b}(r,t)$ can take various   forms depending on 
the type of diffusion involved. A reasonable approximation that 
applies to many situations is to take for  $g_{a,b}(r,t)$ a Gaussian 
propagator defined in normal diffusion with its Fourier transform given by 
$g_{a,b}(q,t) \sim \exp(-D_\mathrm{s}^{a,b} q^2t)$ where $D_\mathrm{s}^{a,b}$ 
is the self-diffusivity in domains $a$ and $b$, respectively. With these 
choices, the Fourier--Laplace transforms of \cref{EQ4.68} write: 
\begin{align}
\tilde{P}_{a,b}(q,s) &= \tilde{\psi}_{a,b}(s+D_\mathrm{s}^{a,b}q^2) 
\nonumber \\
\tilde{Q}_{a,b}(r,t) &= \tilde{\phi}_{a,b}(s+D_\mathrm{s}^{a,b}q^2) = \frac{1 - 
\tilde{P}_{a,b}(q,s)}{s + D_\mathrm{s}^{a,b}q^2}
\label{EQ4.69}
\end{align}
which, upon insertion in \cref{EQ4.67}, leads to: 
\begin{multline}
\tilde{I}(q,s) = \frac{c}{s+D_\mathrm{s}^{a}q^2} + 
\frac{1-c}{s+D_\mathrm{s}^{b}q^2} + (D_\mathrm{s}^a - 
D_\mathrm{s}^b)q^2 \times \\
\frac{\tilde{\psi}_{a}(s+D_\mathrm{s}^{a}q^2)\tilde{\psi}_{b}(s+D_\mathrm{s}^{b}q^2)
\lbrack 1-2c \rbrack + \tilde{\psi}_{a}(s+D_\mathrm{s}^{a}q^2)c - 
\tilde{\psi}_{b}(s+D_\mathrm{s}^{b}q^2) \lbrack 1-c \rbrack}{\lbrack 
1-\tilde{\psi}_{a}(s+D_\mathrm{s}^{a}q^2) 
\tilde{\psi}_{b}(s+D_\mathrm{s}^{b}q^2)\rbrack \lbrack 
s+D_\mathrm{s}^a q^2\rbrack \lbrack s+D_\mathrm{s}^b q^2 \rbrack}
\label{EQ4.70}
\end{multline}
In this equation, the third term carries all the dependency on the switching 
functions $\tilde{\psi}_{a,b}(q,s)$. On the other hand, the first two terms can 
be gathered as $\tilde{I}_{slow}(q,s)$ as they correspond to the limit defined 
above for the slow-switching limit which is independent of the switching 
distributions: 
\begin{align}
I_{slow}(q,t) &= c \exp{(-D_\mathrm{s}^a q^2 t)} + (1 - c) 
\exp{(-D_\mathrm{s}^b q^2 t)}
\nonumber \\
S_{slow}(q,\omega) &= \frac{c}{\pi} \frac{D_\mathrm{s}^a 
q^2}{\omega^2+D_\mathrm{s}^{a2} q^4} + 
\frac{1-c}{\pi} \frac{D_\mathrm{s}^b q^2}{\omega^2+D_\mathrm{s}^{b2} q^4}
\label{EQ4.71}
\end{align}

\Cref{EQ4.70} is very powerful as it allows describing the dynamics in 
complex porous media from simple functions that can be estimated using 
molecular simulation or for which known analytical solutions exist. This 
formalism can be extended using a transition matrix approach to more complex 
media that contain, for instance, more than two domain 
types~\cite{roosen-runge_analytical_2016}.
\Cref{fig:5_2} illustrates the scattering function 
$I_\mathrm{s}(q,t)$ expected for a medium containing three domain types (a, b, 
c) as a 
function of the switching rate $f$. As expected based on the approach  
described above, for a vanishing switching rate, \ie.\
$f \rightarrow 0$, $I_\mathrm{s}(q,t)$ follows the slow-switching limit where 
$I_\mathrm{s}(q,t)$ 
is the sum of three independent contributions with characteristic times given 
by $(D_\mathrm{s}^{a,b,c} q^2)^{-1}$.
In contrast, as the switching rate $f$ increases, the third term in 
\cref{EQ4.70} becomes important and a single characteristic time that 
depends on the switching rate as well as on the switching functions is 
observed.

\subsubsection{Random walk}
\label{sec:diffusionpore_hierarchical_random}

Random walk approaches constitute a powerful class of techniques that can be 
used to investigate the self-diffusion of molecules in heterogeneous media such 
as porous materials. 
\bluelork{In this context, monitoring of particles using “single particle
tracking” experiments has become a powerful tool for diffusion measurements.\cite{maris_single-molecule_2021}}
In fact, as will be discussed below, random walk methods 
allow one to go beyond the framework of self-diffusion as they can be applied to 
situations where transport is induced by a thermodynamic gradient 
(\ie.\ with a non zero net flow condition).
From a very general viewpoint, diffusion and 
transport in porous materials, and in any heterogeneous medium in general, can 
be described using the Fokker--Planck 
equation~\cite{pottier_nonequilibrium_2010,risken_fokker-planck_1996}. 
More precisely, we start with the Smoluchowski equation which 
 applies to the probability 
$P(\mathbf{r},t)$ to find a molecule at a position $\mathbf{r}$ at time 
$t$~\cite{barrat_basic_2003}:

\begin{equation}
\frac{\partial P(\mathbf{r},t)}{\partial t} = \nabla \cdot
\Big \lbrack 
D_\mathrm{s}  \frac{\nabla F(\mathbf{r},t)}{k_{\textrm{B}}T} + 
\nabla \left(D_\mathrm{s} P(\mathbf{r},t)\right)
\Big \rbrack
\label{EQ4.72}
\end{equation}

This equation  is analogous to the advection-diffusion equation introduced in 
\cref{chap:fundamentals} which describes the variations in space and time of 
the concentration in a heterogeneous medium. However, in contrast to the 
advection-diffusion equation, the Smoluchovski equation refers  to the 
molecule distribution $P(\mathbf{r},t)$, \ie.\ the propagator, so that it 
relies on a statistical mechanics approach rather than a deterministic picture 
of the problem. 
Interestingly, the Smoluchovski equation encompasses very general situations 
such as Brownian Dynamics as described by the  Langevin 
equation~\cite{barrat_basic_2003} but it is more general as it does not assume 
any stochastic dynamics \textit{a priori}. 
In \cref{EQ4.72}, the left hand side term 
describes the time evolution of the molecule distribution $P(\mathbf{r},t)$. 
Such an evolution arises from two contributions described in the right hand 
side of \cref{EQ4.72}.
The first term, \ie.\
$\nabla \cdot D_\mathrm{s} \nabla F(\mathbf{r},t)/k_{\textrm{B}}T$,
is an advection contribution where the molecule distribution $P(\mathbf{r},t)$ 
is transported through a velocity field
$v \sim \nabla F(\mathbf{r},t)/k_{\textrm{B}}T$
(in practice, $F(\mathbf{r},t)$ is a free 
energy gradient that can correspond to any thermodynamic quantity like 
pressure, temperature, etc.). The second term, \ie.\ $\nabla \cdot \nabla 
[D_\mathrm{s} P(\mathbf{r},t)]$, describes Fick’s diffusion where the 
particle distribution evolves because of fluctuations which are explicitly 
written as dependent on both time $t$ and position $\mathbf{r}$. 
\Cref{EQ4.72} is a very general expression that allows describing transport, 
\ie.\ diffusion and advection, in heterogeneous media under any 
flow conditions. 
In particular, as will be discussed below, the 
formalism briefly described above gives rise to a number of numerical 
approaches which include random walk strategies for transport in porous materials 
with different schemes available (continuous time random walk, time domain 
random walk, \etc.).~\cite{noetinger_random_2016}

The Smoluchowski equation is at the heart of random walk approaches to 
probe self-diffusion in heterogeneous media under no flow conditions 
(in the absence of an external force, \ie.\ no thermodynamic field
$F(\mathbf{r},t)$ is applied). 
In that case, \cref{EQ4.72} simplifies to yield the diffusion equation with a 
heterogeneous diffusion coefficient $D_\mathrm{s}$: 
\begin{equation}
\frac{\partial P(\mathbf{r},t)}{\partial t} =
\nabla \cdot \left[ D_\mathrm{s}(\mathbf{r},t)
\nabla  P(\mathbf{r},t)\right]
%D_\mathrm{s} \nabla^2  P(\mathbf{r},t)
\label{EQ4.73}
\end{equation}
\Cref{EQ4.73} can be interpreted in terms of the density of an ensemble of independent particles whose trajectories are determined by a stochastic differential equation, the Langevin equation. In the general case where $D_\mathrm{s}(\mathbf{r},t)$ depends on position and/or time, this interpretation is delicate, and requires a careful description of the calculus rule used for integrating the stochastic equation (the so-called Itô-Stratonovich rules discussed  in detail in specific 
textbooks, \eg.\ Ref.~\citen{risken_fokker-planck_1996}). If one restricts the approach to uniform, time independent $D_s$, 
the equation 
$\partial P(\mathbf{r},t)/\partial t \allowbreak= D_\mathrm{s} \nabla^2 
P(\mathbf{r},t))$
is equivalent to the Langevin equation:
\begin{equation}
m \frac{\textrm{d}\mathbf{v}}{\partial t} = - \xi m \mathbf{v} + \delta 
\mathbf{F}(t)
\label{EQ4.74}
\end{equation}
where $m$ and $\mathbf{v}$ are the molecule mass and velocity, respectively. In 
the Langevin equation, which relates to Newton's equation of motion, the left 
hand side term is the Brownian molecule acceleration while the right hand side 
corresponds to the different forces exerted on the molecule. The  contribution 
$-\xi v$ is the average friction term arising upon the motion of the Brownian molecule 
in the solvent while the contribution $\delta \mathbf{F}(t)$ is a random, 
fluctuating force describing the deviation from the average. 
The random force is zero on 
average, \ie.\
$\left\langle \delta F_\alpha(t) \right\rangle  = 0$ 
with $\alpha = x$, $y$, $z$.
Moreover, because the collisions with the solvent 
occur on a time scale much shorter than the characteristic time scale 
corresponding to the change in velocity described in \cref{EQ4.74}, the 
random force $\delta \mathbf{F}(t)$ must verify that 
$\left\langle \delta F_\alpha(t) \delta F_\beta(t') \right\rangle
 = 2 \delta F_0 \delta_{\alpha\beta} \delta(t-t')$ where
$\delta F_0$ is the amplitude of the fluctuating force, 
$\delta_{\alpha\beta}$ the Kronecker symbol and $\delta(t-t')$ the Dirac 
function. With these 
considerations, the solution of the Langevin equation is known (see Ref.~\citen{barrat_basic_2003} for more details):
\begin{equation}
\mathbf{v}(t) = \mathbf{v}(0) \exp(-\xi t) 
+ \frac{1}{m} \int_0^t \textrm{d}t' \exp\left\lbrack -\xi (t-t') \right\rbrack 
\delta \mathbf{F}(t')
\label{EQ4.75}
\end{equation}
While $\left\langle v(t) \right\rangle = 0$ in the absence of external 
driving force (\ie.\ thermodynamic field) to induce a velocity drift, 
$\left\langle v(t)^2 \right\rangle$ is non zero with 
a value obtained by squaring \cref{EQ4.75}:\footnote{
	To obtain this equation, we notice that the cross terms
	$\left\langle \mathbf{v}(0) \times \int_0^t 
	\textrm{d}t' \exp \left[- \xi (t-t') \right] \delta \mathbf{F(t')} 
	\right\rangle$ are equal to zero because the velocity at time $t = 0$ 
	only depends on the force exerted at times $t < 0$ and therefore not on the 
	force exerted at times $t \geq 0$. 
}
\begin{multline}
\left\langle v(t)^2 \right\rangle = \left\langle v(0)^2 \right\rangle 
\exp(-2\xi t) \\
+ \frac{1}{m^2} \int_0^t \int_0^t \textrm{d}t' \textrm{d}t''
\exp\left\lbrack -\xi (2t-t'-t'') \right\rbrack 
\times \left\langle \delta \mathbf{F}(t') \delta 
\mathbf{F}(t'')\right\rangle
\label{EQ4.76}
\end{multline}
The long time limit, $t \rightarrow \infty$, of this equation, \ie.\  
$\left\langle v^2\right\rangle \sim 3 \delta F_0/m^2\xi$, must be equal to its 
value defined via the equipartition energy theorem, 
$\left\langle v^2\right\rangle \sim 3k_\textrm{B}T/m$.
This leads to the important following relationship, which is an expression of 
the fluctuation-dissipation theorem: $m \xi k_\textrm{B}T = \delta F_0$.

To predict the trajectory of Brownian molecules, one can determine the 
velocity autocorrelation function from \cref{EQ4.75}, \ie.\
$\left\langle v(t)v(t')\right\rangle = 
3k_{\textrm{B}}T/m \times \exp(-\xi \lvert t-t' \lvert)$, which can be 
transformed as described in \cref{sec:selfColTransportD} into:
\begin{equation}
\left\langle \left\lvert \mathbf{r}(t) - 
\mathbf{r}(t^\prime)\right\lvert^2\right\rangle = \frac{6k_{\textrm{B}}T}{m\xi}
\left( \left\lvert t-t' \right\lvert + \frac{1}{\xi} \left(\exp \left\lbrack 
-\xi\left\lvert t-t' \right\lvert \right\rbrack - 1\right) \right)
\label{EQ4.77}
\end{equation}
This equation shows that the dynamics obeys normal diffusion for times 
longer than the characteristic time $\xi^{-1}$, \ie.\ 
$t \gg \xi^{-1}$. 
In this case, called the ``overdamped limit", the mean square displacement obeys normal diffusion, 
$\Delta r(t)^2 = 6 D_\mathrm{s} t$ with a self-diffusion coefficient given by 
$D_\mathrm{s} = k_{\textrm{B}}T/m\xi$, and the evolution of $P(r,t)$ is described by the Smoluchovski equation.

We have already introduced the concept of random walks when discussing Fickian 
diffusion at the pore scale in \Cref{sec:molecular_diffusion}.
In short, with this model, normal diffusion is probed by considering a 
molecule, the walker, that jumps from a site to a random neighboring site on a 
lattice of lattice constant $a$ within a time $\delta t$ (\cref{fig:4_3}).
The average mean-square displacement of the molecule at a time $t$ is given by 
$\langle r^2(t) \rangle \sim 2d D_\mathrm{s} t$ where $d$ is the dimensionality 
of the lattice (1D, 2D, \etc.) and $D_\mathrm{s} = k a^2$ is the 
self-diffusivity with $k$ the hopping rate (all details can be found in 
\Cref{sec:molecular_diffusion}).
The random walk model is very useful in the specific context of diffusion in 
porous media.
While the fundamentals are quite simple and general, different random walk 
approaches are available to investigate the self-diffusion in porous solids. In 
what follows, the two following approaches are presented: (1) on-lattice random 
walk and (2) random walk particle tracking. Other methods such as the 
correlated random walk 
theory~\cite{bhatia_stochastic_1986,bonilla_multicomponent_2012}
and biased-diffusion calculations for the determination of transport 
coefficients~\cite{schwartz_transport_1989,epicoco_mesoscopic_2013}
can also be used.

\begin{figure}[htbp]
	\centering
	\includegraphics[width=0.95\linewidth]{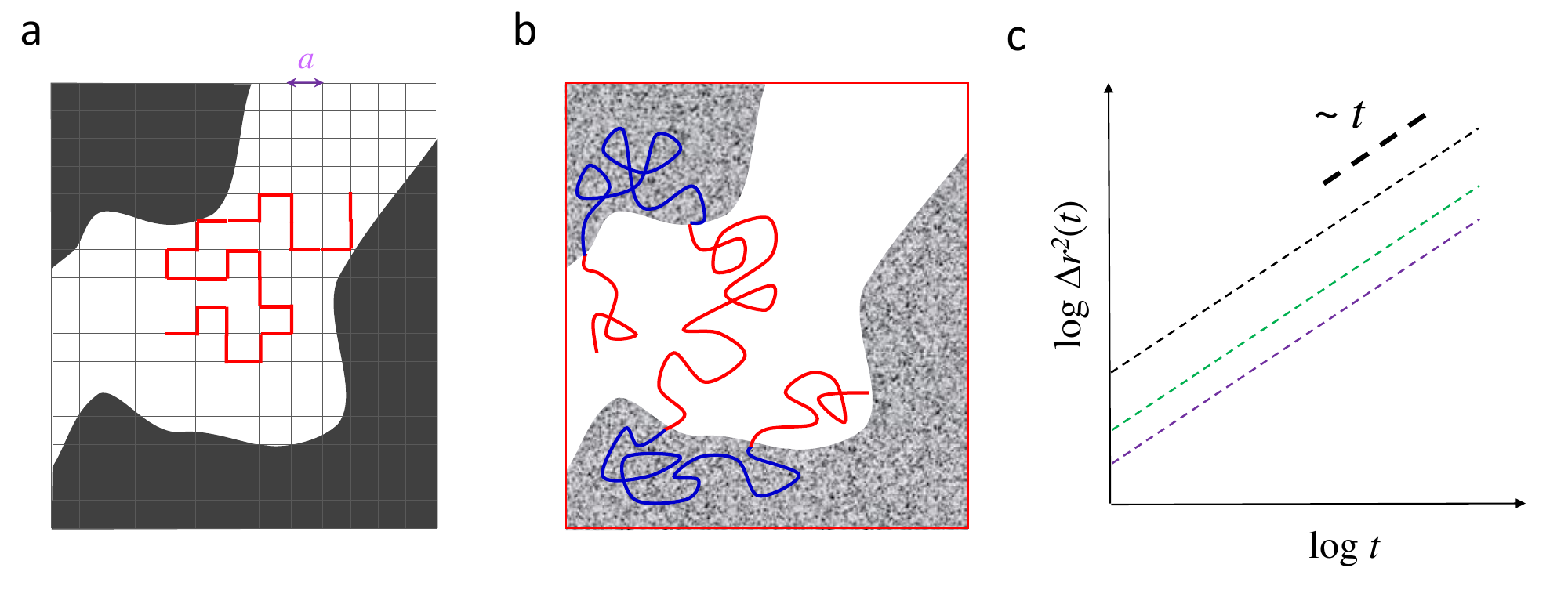}
	\caption{\textbf{Random Walk models in confined environments.} (a) Random 
	walk on a lattice model mapped onto structural data for a porous material 
	(with dark regions corresponding to non-porous solid domains and white 
	regions corresponding to the porous domains). The red segments correspond to the 
	molecule trajectory as described in this coarse-grained lattice random 
	walk approach. (b) In a constant time random walk (CTRW), a molecule 
	follows an off-lattice trajectory with consists of a succession of 
	trajectory segments within domains of different porosities (grey and white 
	regions in this example). In each domain, the molecule displays a different 
	self-diffusivity $D_\mathrm{s}^\alpha$ so that the distance between two 
	points 	separated by a constant time step $\Delta t$ depends on 
	$D_\mathrm{s}^\alpha$ (typically, $\vert\Delta r\vert \sim 
	\sqrt{D_\mathrm{s}^\alpha\Delta t}$). 
	(c) Regardless 	of the random walk approach used, the molecule mean square 
	displacement $\Delta r^2(t)$ scales linearly with time as expected for a 
	molecule obeying Brownian motion.} 
	\label{fig:5_3}
\end{figure}

\noindent \textbf{On-lattice random walk.} The most straightforward application of the 
random walk model to diffusion in porous materials consists of measuring the 
effective diffusivity. In particular, this method can be used to determine the 
tortuosity as the diffusivity factor which corresponds to the ratio of the bulk 
diffusivity to the effective diffusivity within the porous material, $\tau \sim 
D_\mathrm{s}^0/D_\mathrm{s}$~\cite{botan_bottom-up_2015,han_deviation_2009}.
In practice, with this approach, one starts with some structural data in 
real space of a porous material such as tomography data, transmission electron 
microscopy data, focused ion beam scanning electron microscopy data, 
\etc.
A 2D or 3D lattice is mapped onto such data with each node being assigned a 
domain type --- porous void or solid domain --- as shown in \cref{fig:5_3}(a).
The effective self-diffusivity $D_\mathrm{s}$ is then determined by simulating 
the random walk of a molecule onto such a lattice. More in detail, starting 
from an accessible node chosen randomly, the random walk is simulated by moving 
the molecule to an adjacent node chosen randomly while incrementing the time by 
a time step $\delta t$. If the chosen node is inaccessible, \ie.\ belongs  to 
the solid domains, the move is rejected and the molecule remains at the same 
site until the next move. The effective self-diffusivity $D_\mathrm{s}$ is 
determined from the slope of the mean-square displacement $\Delta r^2(t)$ as a 
function of time $t$, $D_\mathrm{s} = \Delta r^2(t)/6t$ as shown in 
\cref{fig:5_3}(c). An average over many initial points chosen randomly is 
needed to obtain data that are representative of the whole porous material --- 
\ie.\ independent of any particular initial condition. To obtain the 
tortuosity $\tau = D_\mathrm{s}^0/D_\mathrm{s}$, the effective diffusivity 
$D_\mathrm{s}$ must be compared to the bulk diffusivity $D_\mathrm{s}^0$ which 
is equal to $a^2/6\delta t$ (see \Cref{sec:molecular_diffusion}). The 
simple method above provides an estimate for the diffusion tortuosity that only 
accounts for morphological (pore shape) and topological (network connectivity) 
effects. However, such a geometrical description of a given porous material and 
its diffusion properties can be completed by including the description of 
adsorption effects as follows. In a simulation of random walk trajectories in a 
porous material, when the chosen adjacent node is inaccessible, the move is 
rejected and a residence time corresponding to a number $n$ of time step 
$\delta$ is added to the current time $t$. By increasing the current time $t$ 
to $t + n\delta t$, adsorption effects can be taken into account.
While this approach remains mostly empirical, it is efficient at capturing and 
disentangling adsorption and geometry effects on the diffusion in complex 
porous media. 
As a possible refinement, the effects of adsorption can be captured in a more 
realistic, \ie.\ semi-quantitative, fashion by choosing the residence 
time according to the time distribution probability  $\psi_A(t)$ which 
corresponds to the probability that an adsorption step lasts a time
$t$\cite{levitz_molecular_2013, bousige_bridging_2021} (this distribution was
defined in the Intermittent Brownian Motion model introduced 
in \Cref{sec:intermittent_brownian_motion}).

\noindent \textbf{Random walk particle tracking.} 
The on-lattice random walk approach described above is limited to binary 
structures made of an accessible (porous voids) and an inaccessible (solid 
domains) phases. In contrast, it is not suitable for more complex structures 
where the accessible phase can be made of domains with different diffusion 
properties. 
Examples of such heterogeneous media include complex, multiscale porous 
materials exhibiting domains with different pore sizes. For each domain, the 
confined fluid will diffuse according to a given self-diffusivity that depends 
on factors such as pore size and porosity. The random walk particle tracking 
method (RWPT)~\cite{salles_taylor_1993} is an efficient 
off-lattice method which allows probing the diffusion in such complex 
porous solids~\cite{hlushkou_pore_2010,tallarek_multiscale_2019}.
Like for the on-lattice method described above, such numerical simulations 
require to have available structural data which can be used to investigate 
their self-diffusion property according to the random walk model. Such real 
space data must be of a resolution good enough to allow distinguishing the 
different diffusing domains. Typically, for hierarchical porous adsorbents such 
as mesoporous zeolites shown in \cref{fig:5_2}(a), a segmentation 
procedure must be performed to assign each pixel to a domain type: microporous, 
mesoporous, and macroporous domain. When impermeable/inaccessible domains 
exist such as non-porous patches, an additional domain type must be considered 
with a corresponding self-diffusivity set to zero. A self-diffusivity value 
$D_\mathrm{s}^\alpha$ is then assigned to each domain type $\alpha$. Such 
values can be 
estimated from experimental data available for single-phase materials (zeolite 
crystal, mesoporous material, etc.) while the diffusivity in macroporous 
domains can be taken as the value for the bulk phase under the same 
thermodynamic conditions (pressure, temperature, chemical potential, etc.).

Once the structural data and domain diffusion properties have been set up, a 
RWPT simulation is carried out as follows. A molecule is placed randomly within 
the material at a position $\mathbf{r}$ belonging to a domain type $\alpha$. As 
illustrated in \cref{fig:5_3}(b), the random walk of this molecule is then 
simulated by moving the molecule by a random vector $\delta \mathbf{r}$ so that 
its new position is $\mathbf{r'} = \mathbf{r} + \delta \mathbf{r}$. The vector 
$\delta \mathbf{r}$ has a random orientation but a norm $\lvert \delta 
\mathbf{r} \lvert$ equal to $\sqrt{6D_\mathrm{s}\alpha \delta t}$ where $\delta 
t$ is 
the time step used to integrate the random walk trajectory. By virtue of the 
Brownian motion, this definition of the norm $\lvert \delta \mathbf{r} \lvert$  
for the random walk displacement ensures that the self-diffusivity of the 
particle is equal to $D_\mathrm{s}^\alpha$ if the particle remains in a domain 
of type 
$\alpha$.
Like with any discretized numerical approach, the time step $\delta t$ must be 
chosen small enough to ensure that the integration of the random walk 
trajectory is stable and accurate. If the new position at $t + \delta t$ 
belongs to the same domain as the position at time $t$,  the move is accepted 
and the time is increased from $t$ to $t + \delta t$. However, if the new 
position belongs to a different domain, a mass balance condition must be 
verified to impose that the flow from the domain $\alpha$ to the domain 
$\alpha'$ is equal to the flow from the domain $\alpha'$ to the domain 
$\alpha$. This condition is very important to impose the ergodicity 
condition --- \ie.\ the fact that the fraction of time spent by the 
molecule in domains of type $\alpha$ is equal to the fraction of molecules 
occupying these domains. Such a mass balance condition 
writes~\cite{hlushkou_morphologytransport_2011}: 
\begin{equation}
p_{\alpha\alpha'} c_\alpha \phi_\alpha D_\mathrm{s}^\alpha = 
p_{\alpha'\alpha} 
c_{\alpha'} \phi_{\alpha'} D_\mathrm{s}^{\alpha'}
\label{EQ4.78}
\end{equation}
In this equation, $c_k$, $\phi_k$, and $D_\mathrm{s}^k$ are the molecule 
concentration, porosity, and the self-diffusivity in domain $k$ ($k = \alpha$ 
or $\alpha'$). $p_{\alpha\alpha'}$ is the probability to accept the random move 
from a position in a domain $\alpha$ to a position in a domain $\alpha'$ (and 
\textit{vice-versa} for $p_{\alpha'\alpha}$).
The condition given in \cref{EQ4.78} 
can be understood as follows. The left hand side describes in a probabilistic 
fashion the molecular flow from a domain $\alpha$ to a domain $\alpha'$. More 
in detail, in this contribution,  $c_\alpha \phi_\alpha$ is proportional to 
the number of molecules in a domain $\alpha$ so that $D_\mathrm{s}^\alpha 
c_\alpha \phi_\alpha $ is proportional to the number of molecules attempting to 
leave the domain per unit of time. $p_{\alpha\alpha'}$ is the fraction of 
molecules that eventually succeeds in leaving the domain $\alpha$ to enter the 
domain $\alpha'$ (the same reasoning applies for the right hand side of the 
above equation but for the transfer from molecules in a domain $\alpha'$ to a 
domain $\alpha$).

With the algorithm above, the dynamics in heterogeneous media can be modeled 
while properly taking into account the ergodicity principle through the 
probability to accept or reject at each step the proposed 
random walk move (whenever it involves switching from one domain to another). 
In particular, the random walk particle tracking method is very general as it 
encompasses the first random walk technique described in this section. Indeed, 
by considering that solid domains are domains in the RWPT method but with a 
porosity set to zero, the acceptance probability to have a random walk move 
leading the molecule from the porosity to the impermeable solid skeleton is 
zero. In practice, as illustrated in Figs. \ref{fig:5_3}(b) and (c), once the 
rules defined above have been set up, the RWPT simulation is performed and 
analyzed in a very similar fashion as regular random walk simulations. Like 
with any random walk approach, the effective self-diffusivity $D_\mathrm{s}$ is 
determined from the slope of the mean-square displacement $\Delta r^2(t)$ as a 
function of time $t$, $D_\mathrm{s} = \Delta r^2/6t$ as shown in 
\cref{fig:5_3}(c).

\noindent \textbf{Other random walk approaches.} 
The random walk technique was  
introduced above in the limited context of diffusion (\ie.\ in the 
absence of any advective flow).
Here, we intend to describe its generalization to transport 
situations where both diffusion and advection are at play. Reviews 
such as Ref.~\citen{noetinger_random_2016} are available for a general 
presentation of random walk approaches to diffusion and transport in porous 
materials. In what follows, we only report some key aspects with a focus on the 
extension of these techniques to transport with flow conditions.
 
The particle tracking random walk (PTRW) can be used to treat the general 
problem of combined diffusive and advective transport.
Typically, this is achieved by integrating 
with a constant time step $\delta t$ the particle position; 
$\mathbf{r}(t)$ at a time $t$ becomes at a time $t + \delta t$: 
$\mathbf{r}(t+\delta t) = \mathbf{r}(t) + \mathbf{v}(\mathbf{r}) \delta t + 
\delta \mathbf{r} $ where $\mathbf{v}(\mathbf{r})$ is the local Stokes velocity 
field while $\delta \mathbf{r}$ is a vector having a random orientation. While 
the velocity field $\mathbf{v}$ can be pre-calculated using Lattice Boltzmann 
simulations for instance, the random vector $\delta \mathbf{r}$ must have a 
norm governed by the self-diffusion coefficient $[6D_\mathrm{s} \delta 
t]^{1/2}$ as already discussed above.
In heterogeneous materials, the constant time approach used in most random walk 
approaches can raise important issues --- especially when the coexisting 
diffusion domains have very different associated self-diffusivity coefficients 
(typically, diffusion will involve either very large displacements in the fast 
domains or very small displacements in the slow domains). Different strategies 
are available to overcome such limitations inherent to the constant time 
approach: namely, the continuous time random walk (CTRW) and the time domain 
random walk (TDRW). 
In CTRW, as discussed in Ref.~\citen{noetinger_random_2016} and 
references therein, the particle position $\mathbf{r}(t)$ and time $t$ are both 
implemented with position and time increments that depend on the position 
$\mathbf{r}$. Moreover, these increments are coupled through the probability 
distribution function  describing the probability that a molecule at a given 
position $\mathbf{r}$ moves  by a quantity $\delta \mathbf{r}$ over a time 
increment $\delta t$. Such a scheme allows transforming both the molecule 
position and time into continuous variables despite the use of a discrete 
trajectory integration frame. Such methods can prove to be very powerful to 
investigate transport and diffusion problems  in complex porous media such as  
hierarchical structures and fractured solids~\cite{yang_ubiquity_2019}.
In particular, owing to their simplicity and versatility, they can be used to 
model problems such as molecule (solute) dispersion under liquid flow in any porous 
structure type.
In the spirit of the CTRW, albeit different, the TDRW can also be employed to 
investigate the diffusion and transport of molecules in realistic numerical 
reconstructions of real porous media. In contrast to CTRW, the TDRW technique 
assumes --- regardless of the particle position $\mathbf{r}$ --- a 
constant position increment $\lvert \delta \mathbf{r} \lvert$ and calculates 
the time increment needed to achieve such a translation.

\subsubsection{Free energy and hierarchical simulations}

Molecular dynamics is a technique of choice to probe the self-diffusivity of 
adsorbates in porous media such as nanoporous catalysts. However, in many 
situations, this technique, which is limited to times of $\sim 
100\,\mathrm{ns}$ up to a few $\mu\mathrm{s}$ depending on the 
system size, proves to be inefficient at describing self-diffusivity. Such 
situations are often encountered for bulky molecules in nanoporous materials 
with very small pores.
Even in case of small molecules, probing diffusivity at low 
temperatures can become a complex task. Let us consider a nanoporous material 
with a heterogeneity length $L$. This length is defined as the typical 
length which must be explored by the diffusing molecules to obtain data that 
are representative of the overall macroscopic diffusivity (typically, in the 
case of zeolites, $L$ is of the order of the unit cell parameter). In order to 
efficiently probe diffusivity using molecular dynamics, we can assume that the 
molecules must diffuse over a distance that is at least twice the lengthscale 
$L$. The time $\tau$ required to meet this condition is given by $\tau \sim 
4L^2/D_\mathrm{s}$ where $D_\mathrm{s}$ is the self-diffusivity. For a bulky 
molecule such as 
cyclohexane, with a diffusivity of the order of $10^{-12}$ m$^2$/s in 
nanoconfinement~\cite{mehlhorn_probing_2014}, $\tau \sim 4 \,\mu\mathrm{s}$, 
\ie.\ a value beyond the typical time that can be achieved with molecular 
dynamics.
When regular molecular dynamics does not allow probing 
diffusivity in nanoporous media, other strategies such as hierarchical 
simulations are 
available~\cite{june_transition-state_1991,maginn_dynamics_1996}.
Such simulations, which consist of determining the diffusivity of 
an adsorbed molecule using a methodology based on the transition state theory, 
are very powerful at predicting the slow macroscopic dynamics in complex 
heterogeneous materials~\cite{schuring_entropic_2002,dubbeldam_molecular_2005,
abouelnasr_diffusion_2012,bai_understanding_2016}.
In practice, such molecular simulations rely on a two-step approach. First, one 
calculates a free energy map to identify all adsorption sites within the host 
porous solid. Second, using the transition state theory, a kinetic Monte Carlo 
algorithm or Brownian dynamics simulation is used to determine the 
macroscopic diffusivity through the motion on this free energy map. Different 
versions of these hierarchical simulations can be found in the literature. In 
what follows, we focus mostly on the original version as proposed by Theodorou 
and coworkers for monoatomic~\cite{june_transition-state_1991} and 
more complex fluids~\cite{maginn_dynamics_1996}. However, it should 
be noted that improvements have been proposed in more recent years such as an 
extension of the transition state theory to adsorbed loadings where collective 
effects strongly affect the dynamic correction factor $\kappa$ (see 
below)~\cite{beerdsen_understanding_2006,dubbeldam_molecular_2005}.

In hierarchical simulations, diffusion within the porous material is assumed to 
occur through uncorrelated jumps from a local free energy  minimum $i$ to 
another local free energy minimum $j$. In more
detail, as described in \cref{sec:molecular_diffusion}, the transition rate 
$k_{ij}$ between the two local minima is obtained from their free energies 
using \cref{EQ4.11}: 
\begin{equation}
k_{ij} = \kappa \frac{k_{\textrm{B}}T}{2\pi m} \times
\frac{\int_{S_{ij}} \exp(- U/ k_{\textrm{B}}T) \textrm{d}\mathbf{r}^2}{\int_{V 
_{i}} \exp(-U/ k_{\textrm{B}}T) \textrm{d}\mathbf{r}^3}
\label{EQ4.79}
\end{equation}
where $U$ is the system internal energy and $m$ the mass of the adsorbed 
atom. This expression is valid for a monoatomic fluid where there is no 
configurational entropy contribution. For a molecular fluid, $U$ in the 
expression above must be replaced by the free energy $F$ which is probed by 
considering at a given position $\mathbf{r}$ the following average over $N$ 
configurations $k$ picked randomly 
$F(\mathbf{r}) = - k_\textrm{B}T \ln \left[\sum_k^N 
	\exp(-U_k/ k_\textrm{B}T)/N\right]$
($U_k$ is the internal energy of the $k^{th}$ 
molecular configuration of the fluid molecule)~\cite{kim_large-scale_2013}. The integral 
in the numerator in \cref{EQ4.79} is evaluated at the surface boundary between 
sites $i$ and $j$ while the integral in the denominator is evaluated over the 
volume around the site $i$ (typically, a cutoff in energy $U$ can be used as 
positions around site $i$ with strongly repulsive energies do not contribute 
significantly to the integral in the denominator).
The term $\kappa$ in the above equation is the 
dynamic correction factor which was already introduced in 
\Cref{sec:molecular_diffusion}. This factor, which is 
comprised between 0 and 1, accounts for the fact that only a fraction $\kappa$ 
of molecules attempting to jump from site $i$ to site $j$ eventually succeeds 
in doing so. Such a contribution can be estimated using the Bennett--Chandler 
approach~\cite{chandler_statistical_1978} from 
Molecular Dynamics simulations in which molecules are positioned at the top of 
the energy barrier with an initial velocity $v_i(0)$ at time $t = 0$ selected 
randomly according to a Maxwell--Boltzmann distribution at temperature $T$. 
With these simulations, $\kappa$ is determined by counting the number of 
molecules that ends up in state $j$ at a time $t$ later, $\kappa = \sqrt{2\pi 
m/k_{\textrm{B}}T} \left\langle v_i(0) \theta_k(t) \right\rangle$
(where $\theta_k(t) = 1$ if the molecule ends up in site $j$ and 0 otherwise 
and $\left\langle \cdots \right\rangle$ denotes average 
over many realizations in the microcanonical NVE ensemble). As discussed by Smit 
and coworkers~\cite{beerdsen_understanding_2006,dubbeldam_molecular_2005},
the impact of loading, \ie.\ collective effects, in the 
determination of $\kappa$ and therefore in the macroscopic diffusivity in 
nanoconfinement is an important issue that is usually not accounted for.
These authors have proposed an extended version of the transition state theory 
in which the density dependence of $\kappa$ is measured to account for such 
dynamic corrections.
In practice, such an extension consists of determining the corrections 
$\kappa_{ij}$ for the transition between state $i$ and $j$ as a function of 
loading while taking into account the necessary mass balance condition between 
these two sites.

As discussed in \Cref{sec:molecular_diffusion}, for a homogeneous medium, 
\ie.\ with a constant jump rate $k$, the self-diffusivity is readily 
obtained as $D_\mathrm{s} = 1/6 k a^2$ where $a$ is the distance between 
adjacent sites. 
However, for a heterogeneous medium such as in nanoporous materials where 
adsorption sites are separated by free energy barriers, the jump rate is not 
constant and an additional numerical strategy is needed to simulate the 
molecular dynamics in the underlying free energy landscape.
 In practice, this is usually achieved using a kinetic Monte 
Carlo approach as described in detail in 
Refs.~\citen{june_transition-state_1991,abouelnasr_diffusion_2012}.
In this approach, the probability 
$\pi_{ij}(t)$ that a molecule jumps from a site $i$ to a site $j$ between the 
times $t$ and $t + \textrm{d}t$ is assumed to follow a Poisson 
distribution:\footnote{
	This condition is equivalent to assuming that two 
	subsequent jumps are uncorrelated as usually assumed in any random walk 
	approach.}
$\pi_{ij}(t) = \rho_{ij} \exp (-\rho_{ij} t) \textrm{d}t$. In this 
expression, $\rho_{ij}$ is the transition rate which is given by the jump rate 
$k_{ij}$ multiplied by the number of molecules in state $i$. As discussed by 
June \etal.~\cite{june_transition-state_1991}, with the underlying Poisson 
distribution, the average time between two jumps between states $i$ and $j$ is 
$\tau_{ij} = 1/\rho_{ij}$. When $N$ different events, \ie.\ moves,
are possible from a state $i$, the Poisson distribution defined above imposes 
that the overall rate parameter (frequency) is given by $\rho = \sum_k^N 
\rho_k$ where $\rho_k$ is the rate parameter of the move of type $k$. 
Moreover, the probability that a move $k'$ occurs first is given by: $P_{k'} = 
\rho_{k'}/\sum_k^N \rho_k$. 
With these theoretical considerations, the molecular dynamics in the 
pre-calculated free energy landscape is simulated as follows. First, 
considering the overall rate parameter $\rho$ discussed above, the time 
interval of the occurrence of the next move is chosen according to the Poisson 
distribution $\pi(t) = \rho \exp(-\rho t)$. In practice, using the technique 
known as the inverse transform sampling, $\tau$ is sampled as follows. $\Pi(t) 
= \int_0^{t'} \pi(t) dt$ is a cumulative function that describes the 
probability that a move occurs between a time $t = 0$ and a time $t$. 
A random number $\epsilon \in \lbrack 0, 1 \rbrack$ is chosen such that 
$\epsilon = \Pi(\tau)$. 
In the specific case of a Poisson distribution $\pi(t)$, one obtains
$\Pi(t) = 1-\exp(-\rho t)$ and $\tau = -\ln(1-\epsilon)/\rho$.\footnote{
	In more detail, $\pi(t) = \rho \exp(-\rho t)$ leads to $\Pi(t) = 
	\int_0^{t} \pi(t') \textrm{d}t' = \lbrack -\exp(-\rho t') \rbrack_0^t  = 
	1-\exp(-\rho t)$. 
	In turn, the condition $\epsilon = \Pi(\tau)$ leads to
	$\tau = -\ln(1-\epsilon)/\rho$
}
Then, in a second step, a particular move $k'$ is picked by comparing a random 
number $\xi \in \lbrack 0, 1 \rbrack$ with the probability $P = \sum_k^N P_k$. 
Once a given move $k'$ and a time interval $\tau$ have been selected, the 
trajectory of the molecule is updated. As usual, the self-diffusivity is 
readily obtained from the mean square displacements using the Einstein 
equation, $D_\mathrm{s} =  \lim_{t \rightarrow \infty} \Delta r^2(t)/6t$.

%%%%%%%%%%%%%%%%%%%%%%%%%%%%%%%%%%%%%%%%%%%%%%%%%%%%%%%%%%%%%%%%%%%%%%%%%%%%%%%%%%%%%%%%

% Molecular and Mesoscopic Dynamics in Nanoporous Materials

%!TeX spellcheck = en_US
%!TeX encoding = utf8 
%!TeX program = pdflatex
%!TeX root = ../manuscript.tex

%%% Chapter 6: Diffusion in nanoporous catalysts
%!TeX spellcheck = en_US
%!TeX encoding = utf8 
%!TeX program = pdflatex
%!TeX root = ../manuscript.tex

\section{Gradient-driven transport in nanoporous materials}
\label{chap:transport_pore}

\subsection{Flow mechanisms} 

\subsubsection{Convection, advection, diffusion} 

\begin{figure}[htbp]
	\centering
	\includegraphics[width=0.95\linewidth]{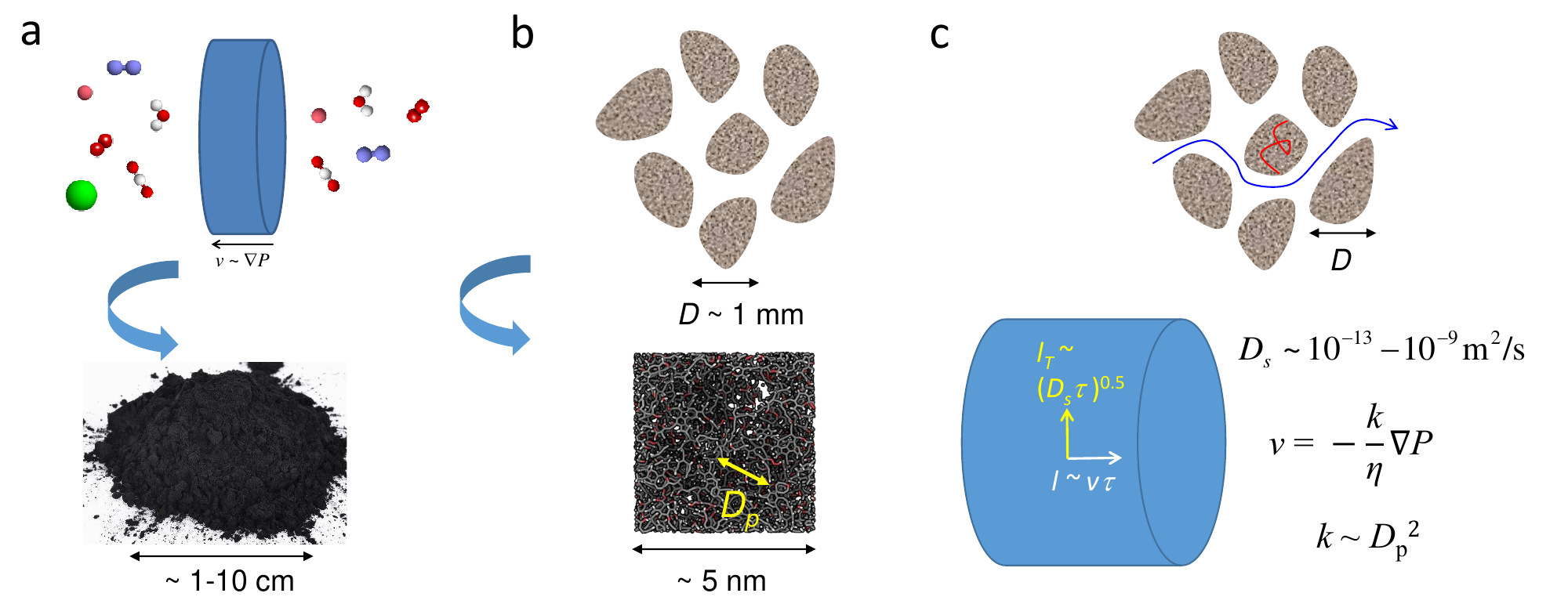}
	\caption{\textbf{Transport in nanoporous catalysts}. (a) Schematic 
	representation of gas and liquid transport across a pellet or a column 
	packed with grains of a nanoporous material (blue cylinder). The flow 
	across this system having macroscopic dimensions ($\sim$ 1 - 10 cm) is 
	typically induced by a pressure gradient $\nabla P$ which leads to a flow 
	rate $v$ proportional to $-\nabla P$ (in the linear approximation regime). 
	(b) At smaller scales, the pellet or packed column is made up of grains 
	having a size $D$ of the order of mm or even smaller. Each grain is made of a 
	nanoporous material having very small pores of a size $D_p$ down to the molecular scale 
	(here, a disordered porous carbon is shown for illustration). \bluelork{We note that $D$ refers to the grain size in the powder material while $D_p$ refers to the size of the pores forming the porosity in the grain.} (c) Transport 
	in such complex materials --- \ie.\ assemblies of packed 
	grains --- involves different flow regimes such as advection and possibly 
	convection in large voids between grains and diffusion and/or advection in 
	the very small porosity within the grains. The competition between 
	advection and diffusion at the smallest porosity scale depends on the 
	self-diffusion coefficient $D_\mathrm{s}$ and the permeability $k$. 
 	The image of the activated carbon powder is taken from
	Ref.~\protect\citen{wiki_2006_carbon} (creative commons license).}
	\label{fig:6_1}
\end{figure}

Let us now consider transport phenomena in materials containing nanopores while 
more complex structures including different porosity scales will be considered 
in \Cref{chap:transport_network}.
Concerning transport, we here refer to situations where a net 
fluid flow is induced by a driving force corresponding to a thermodynamic 
gradient such as a chemical potential, pressure or temperature gradient. In 
contrast, the diffusion aspects, which were covered in 
\Cref{chap:diffusion_pore,chap:diffusion_network}, correspond to a no net flow 
condition. The typical transport experiment considered in the present and 
subsequent chapters is depicted in \cref{fig:6_1}(a). A pellet or a column 
packed with grains made up of a nanoporous solid is set in contact with a fluid 
phase which is then subjected to a pressure gradient. The fluid phase, which 
flows under the action of the pressure gradient, is either a single component 
phase or, more generally, a mixture and can exist within the porosity
as a gas or a liquid. Provided the pressure gradient $\nabla P$ is not too 
large, the fluid response to the driving force is linear so that the flow rate 
$v$ is directly proportional to $\nabla P$. While transport occurs at the 
macroscopic scale typically over lengths in the range $10^{-2}$ -- $10^{-1}$ m, the solid 
sample at a smaller scale is made of small particles having a size $D \sim 
10^{-6}$ -- $10^{-3}$ m (roughly, for a granular medium, the intergranular 
spacing is of the same order of magnitude as $D$).
In turn, as illustrated in \cref{fig:6_1}(b), 
these particles are made up of a nanoporous material with pore sizes in the 
molecular range $D_\textrm{p} \sim $ 0.1 -- 100 nm. \bluelork{Generally speaking, as already discussed in Section 2, transport occurs through convection, advection, and/or diffusion~\cite{ho_gas_2006}. Here, to avoid ambiguity as conventions differ from one scientific community to another, we  recall that we refer to the following definitions throughout the present review: (1) convection corresponds to inertial effects (the non-linear part of the Navier--Stokes equation) and (2) advection corresponds to the transport carried by the flow parallel to the applied driving thermodynamic gradient  (linear part of Navier--Stokes equation). For more details, the reader is referred to the corresponding discussion in Section 2.1.} 
\bluelork{From a very general viewpoint, as discussed below, the exact nature of the flow between the different mechanisms above depends on many parameters such as the molecular size of the flowing fluid, the pore size and morphology/topology of the porous network, and the thermodynamic conditions that apply to the system. As an important illustration in the field of membrane and separation science, the flow of water and organic solvents in very narrow pores for reverse osmosis and nanofiltration was found to obey a pore flow regime (i.e.\ advection) rather than a solution-diffusion regime.~\cite{heiranian_mechanisms_2023,wang_water_2023, fan_physical_2024,
he_molecular_2024, fan_solution-diffusion_2024}}
%On the one hand, for a single component,  \ie. a pure gas or liquid phase, transport induced by a thermodynamic  gradient typically corresponds to convection and/or advection (diffusion as induced by a concentration gradient does  not apply in most situations for a single-component fluid since the  upstream and downstream phases do not differ in concentration). On the other  hand, for a fluid mixture, in addition to advection and convection, transport in porous materials can also involve a diffusion component if the downstream and  upstream concentrations differ.

Specific situations such as that considered in \cref{fig:6_1} will be addressed 
in the next subsection by discussing the influence of pore and particle sizes 
on transport but also of thermodynamic conditions (corresponding either to 
fluid in a gas or liquid state). However, generally speaking, regardless of the 
porous solid and flowing fluid considered, the competition between convection, 
advection, and diffusion can be assessed using the two following dimensionless 
numbers: the Reynolds number Re and the Peclet number Pe. As discussed in 
\Cref{sec:fundamentals_frameworks_hydrodynamics}, $\textrm{Re} = \rho v L / 
\eta$ characterizes the ratio of the convective to the advective contributions 
in the Navier--Stokes equation given in \cref{EQ3.1} ($v$ and $L$ are the flow 
velocity and characteristic lengthscale while $\eta$ and $\rho$ are the fluid 
dynamical viscosity and density). 
For practical situations as illustrated in \cref{fig:6_1}(c), two Re numbers 
can be defined as transport involves intragranular [$L \sim D_\textrm{p}$] and 
intergranular [$L \sim D$] flows. In practice, Re depends on different 
parameters, such as the pore/particle size and pressure gradient, which affect 
the flow rate. However, convection is usually negligible unless large 
$D_\textrm{p}$ or $D$ are considered.
In fact, while intergranular transport can involve a 
convective contribution (non-negligible Re because of the large intergranular 
size $\sim D$), intragranular transport does not include any convective 
contribution when nanoporous grains are considered (Re $\ll 1$ because 
$D_\textrm{p} \sim$ 0.1 -- 10 nm). In addition to Re, the Peclet number is a 
very useful parameter as it describes the balance between the diffusive and 
advective contributions to transport in porous materials: Pe $= v L/D_\mathrm{s}$ 
where $v$ is the fluid velocity, $D_\mathrm{s}$ the fluid self-diffusion 
coefficient, 
and $L$ the characteristic lengthscale. Like for Re, two Pe numbers can be 
defined as overall transport involves intragranular [$L \sim D_\textrm{p}$] and 
intergranular [$L \sim D$] flows. Again, the corresponding Pe values for a 
specific problem depend on parameters such as the pressure gradient and 
particle/pore sizes but, usually, intergranular transport is found to be 
dominated by the flow,  as the  velocity is large (large Pe). On the other 
hand, intragranular transport can be advective or diffusive depending 
on the pore size $D_\textrm{p}$ and the corresponding self-diffusivity 
$D_\mathrm{s}$. 
%Typically, considering the geometry in \cref{fig:6_1}~(c), for small pores  
%$D_\textrm{p}$, within a characteristic time $\tau$, the fluid in the 
%intergranular voids is transported by an advective flow over a length $l \sim 
%v\tau$ while the fluid diffuses in the transverse direction over a length $l_T 
%\sim (D_\mathrm{s} \tau)^{0.5}$.

\subsubsection{Gas and liquid flows}
\label{sec:transportPore_flow_gasAndLiquid}

In what precedes, the specific conditions to observe convection, advection, and 
diffusion upon fluid transport in porous media were presented. Using the generic 
dimensionless numbers Re and Pe, the different contributions and flow 
mechanisms can be predicted as a function of the following fluid, solid, and 
thermodynamic parameters. The relevant fluid parameters are the viscosity 
$\eta$ and density $\rho$ at the experimental temperature $T$ while those 
related to the porous solid are the particle size $D$, pore size 
$D_\textrm{p}$, and column/sample size $L_\mathrm{s}$. As for the thermodynamic 
conditions, the temperature $T$ and the pressure gradient $\nabla P = \Delta 
P/L_\mathrm{s}$ are the key parameters affecting the flow mechanisms. For a 
given system and thermodynamic conditions, the resulting flow mechanisms will 
be driven by the fluid self-diffusivity $D_\mathrm{s}$ and flow rate $v \sim 
\nabla P$ 
in the different porosity scales. 
%In order to discuss some specific yet representative examples, we consider the following porous solid and thermodynamic conditions: $D \sim 0.1 - 1$ mm, $D_\textrm{p} \sim 1 - 100$ nm, $L = 10$ cm, $\Delta P = 10$ bar, and $T = 300$ K. 
In what follows, liquid flow is first considered before discussing gas or 
supercritical fluid flow. This dichotomy fluid/liquid, which can appear as 
arbitrary at first, is important because of the following. On the one hand, 
liquids are incompressible or weakly compressible so that the transport 
parameters $D_\mathrm{s}$, $D_\textrm{p}$, $\eta$, $v$, \etc.\ at a 
given temperature can be treated as independent of the average pressure and 
pressure gradient. In other words, for most practical 
situations, unless very large pressure drops $\Delta P$ are used to induce 
transport, a single relevant flow type will be observed. On the other hand, 
owing to their non-negligible compressibility, even if other parameters are 
maintained constant, gas or supercritical fluid flow can pertain to very 
different mechanisms depending on the density $\rho$, and, therefore, pressure 
$P$.

\noindent \textbf{Liquids.} Let us consider as an illustrative example the case 
of a liquid flowing in a complex porous medium as depicted in 
\cref{fig:6_1}(c). 
The mean free path $\lambda \sim 1/\rho \sigma^2$ for a liquid remains nearly 
constant since $\rho$ does not vary or only very little with pressure ($\sigma$ 
is the molecular diameter of the liquid molecule). In practice, considering 
typical liquid densities, $\lambda$ is very small or of the order of magnitude 
of the pore size $D_\textrm{p}$. As a result, as discussed in 
\Cref{chap:fundamentals}, even if the density changes because of 
compressibility effects when the liquid is subjected to a pressure gradient, 
the liquid flow remains in the same regime with no change in the underlying 
transport mechanisms.
As discussed in \Cref{sec:fundamentals_frameworks_hydrodynamics},
for water at room temperature, the flow 
in large pores such as in the intergranular pores is advective or convective 
while the diffusive flow remains negligible. The nature of the flow --- being 
either advective or convective --- depends on the pressure drop (or flow rate) 
imposed as defined in the expression for Re. In contrast, in small pores such 
as in the intragranular space in \cref{fig:6_1}, Re is very small so that the 
liquid flow is either advective or diffusive. On the one hand, the advective 
flow rate $v$ is related to the pressure gradient $\nabla P$ through the 
permeability $k$.
As will be discussed further below, the permeability scales with the 
squared pore size, $k \sim D_\textrm{p}^2$. On the other hand, the 
self-diffusivity decreases with decreasing the pore size $D_\textrm{p}$ (see 
\cref{chap:fundamentals}). Therefore, for water at room temperature, when pore 
sizes are in the range $D_\textrm{p} \sim$ 1 - 100 nm, the self-diffusivity can 
vary from $10^{-13}$ to $10^{-9}$ m$^2/s$ (the latter value being close to the 
bulk diffusivity as no confinement effect is expected for large pores 
$D_\textrm{p} \sim$ 100 nm). Going back to the situation in \cref{fig:6_1}, the 
competition between advection and diffusion within the nanoporous particles 
depends on the flow rate $v$ and pore size $D_\textrm{p}$. To estimate the 
trade-off between these two mechanisms, it is instructive to consider the 
typical liquid displacement over a characteristic time $\tau$. In the direction 
parallel to the pressure gradient, advection induced by the pressure gradient 
$\nabla P$ in the large porosity displaces the liquid by a distance $l_T \sim v 
\tau$ with $v \sim - k/\eta \nabla P$. In the direction perpendicular to the 
pressure gradient, the liquid diffuses by a quantity $l \sim \sqrt{D_\mathrm{s} 
\tau}$ over the same characteristic time $\tau$. The ratio between these two 
characteristic lengths allows defining the efficiency of a given sample, by imposing  that $l_T= v\tau$ is the sample size. For 
large $l_T/l$, the advective transport is very fast so that diffusion in the 
nanoporous particles is limited --- in other words, most of the flowing liquid 
does not explore the entire granular particles because it moves across the 
sample in a very short time. On the other hand, for small $l_T/l$, the fluid 
flow rate in the sample is small and its diffusion through the intragranular 
space is very efficient.

\noindent \textbf{Gases and supercritical fluids.} In contrast to liquids, 
which are incompressible or weakly compressible, the transport of gases or 
supercritical fluids in porous materials is more complex as the density can vary 
drastically upon changing the applied thermodynamic conditions. Even when 
constant thermodynamic boundary conditions are applied, the gas flow induced by 
a pressure gradient implies that the fluid density along the flow varies with 
the local pressure. Moreover, because of possible adsorption effects, even when 
low fluid densities/pressures are considered, the fluid density within the 
porous material can be large. In order to rationalize gas or fluid transport in 
nanoporous materials, it is instructive to consider the mean free path $\lambda 
\sim 1/\rho \sigma^2$ as a function of the applied gas pressure. For gases or 
supercritical fluids, $\lambda$ varies drastically depending on the exact 
pressure conditions considered. $\lambda$ covers a broad range 
from values close to the pore size $D_\textrm{p}$ (``tight gas’’ conditions) to 
values very large compared to $D_\textrm{p}$ (diluted gases). As discussed in 
\Cref{chap:fundamentals}, for a given pore size $D_\textrm{p}$, the relevant 
dimensionless number to describe the change in flow regime with $\lambda$ is 
the Knudsen number Kn $= \lambda/D_\textrm{p}$~\cite{ziarani_knudsens_2012}.
In what follows, the different regimes that can be encountered depending on Kn 
are introduced (for a deeper presentation, the reader is also referred to 
Refs.~\citen{ziarani_knudsens_2012,bear_modeling_2018}).

\begin{itemize}
\item{\textbf{Viscous flow, Kn $\mathbf{< 0.01}$.} In this asymptotic regime, 
the flow is advective as described using Darcy law which will be introduced in 
\Cref{sec:transportPore_viscousFlow_darcy}. The flow is assumed to be laminar 
so that it only applies to situations corresponding to low Re.}

\item{\textbf{Slip flow, Kn $\mathbf{\sim [0.01-0.1]}$.} 
As the Knudsen number increases above 
0.01, the gas flows according to a corrected Darcy's law. Such a correction, 
known as Klinkenberg effect, will be also discussed in 
\Cref{sec:transportPore_viscousFlow_darcy}. This correction, which leads to 
faster flow rate compared to that predicted from the regular Darcy's law, 
corresponds to gas slippage at the pore surface because of compressibility 
effects.} 

\item{\textbf{Transition flow, Kn $\mathbf{\sim [0.1-10]}$.} Transition flow 
corresponds to a regime where transport occurs both through advection and 
diffusion. As noted in Ref.~\citen{karger_diffusion_2012}, in such conditions, 
the advective and diffusive fluxes sum up as the two contributions act in 
parallel in terms of transport. Indeed, while the flow rate associated with 
advection induces a displacement of the fluid  center of mass, diffusion adds 
up a mean-square displacement to the overall fluid molecular motion. In this 
transition regime, as the Knudsen number increases, a continuous transition is
observed from a pure advective flow towards a pure diffusive flow.} 

\item{\textbf{Knudsen diffusion, Kn $\mathbf{> 10}$.} In this asymptotic 
regime, the fluid mean free path is much larger than the pore size so that 
transport occurs through Knudsen diffusion (predominant collisions of the 
molecules with the pore surface). This regime was discussed at length in 
\cref{chap:fundamentals}.} 
\end{itemize}

\subsection{Viscous flow} 

\subsubsection{Poiseuille law} 

\begin{figure}[htbp]
	\centering
	\includegraphics[width=0.95\linewidth]{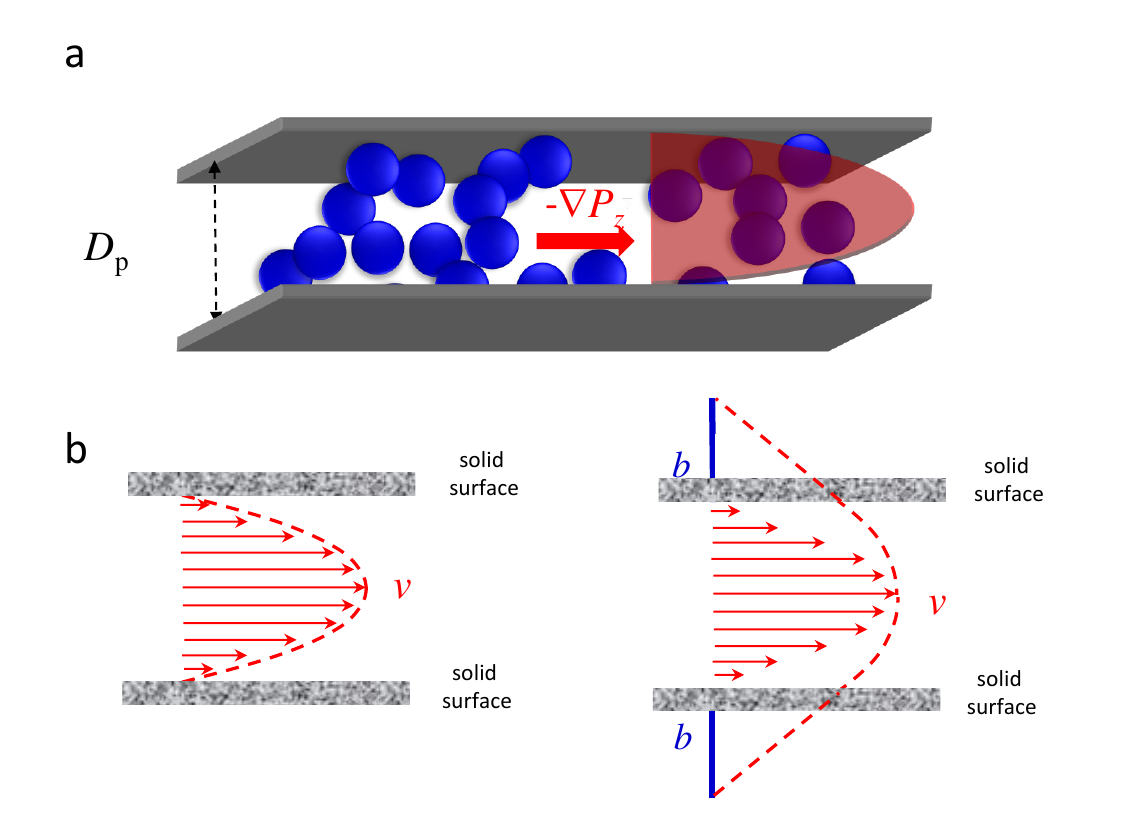}
	\caption{\textbf{Poiseuille flow with and without slippage}. (a) Poiseuille 
	flow for fluid molecules (blue spheres) confined in a slit nanopore having 
	a pore size $D_{\textrm{p}}$ (grey walls). A parabolic velocity profile 
	(red shaded area) is induced by a pressure gradient $\nabla P$ parallel to 
	the pore walls. 
	$x$ and $z$ are the directions perpendicular and parallel to the pore surface, respectively.
	(b) Depending on the boundary condition at the pore surface, the velocity 
	parabolic profile $v_z(x)$ tends to zero at the pore surface --- stick 
	or no-slip boundary condition (left panel) --- or to a finite value --- 
	non-stick or slip boundary condition (right panel).  In the latter 
	situation, slippage occurs with a slip length $b$ defined as the distance 
	where the extrapolated velocity profile reaches zero: 
	$- b \partial v_z/\partial x = v_z$.}
	\label{fig:6_2}
\end{figure}

\textbf{Incompressible flow.} Let us consider an incompressible liquid confined 
in a slit pore of width $D_\mathrm{p}$ and length $L$ as depicted in \cref{fig:6_2}. 
The liquid flows when subjected to a pressure gradient $\nabla P$ in the 
direction $z$ parallel to the pore surface ($x$ is the direction normal to the 
pore surface). Physically, even for pores as small as a few nm in width, the 
liquid flow in such a configuration is well captured by the Navier--Stokes 
relation for the momentun conservation given in \cref{EQ3.1}.\cite{kavokine_fluids_2021,
schlaich_hydration_2017,gravelle_transport_2022} 
In the laminar regime, \ie.\ for small Re, the inertial term is negligible, 
$\mathbf{v} \cdot \mathbf{\nabla v}\sim 0$ (where $\mathbf{v}$ is the fluid 
velocity field), so that the momentum conservation 
writes~\cite{bird_transport_2002}: 
\begin{equation}
\rho \frac{\partial \mathbf{v}}{\partial t} = - \nabla P + \eta \nabla^2 \mathbf{v}
\label{EQ6.1}
\end{equation}
where the gravity term $\mathbf{f}$ was omitted as this contribution is
negligible for liquids confined in nanopores. 
The stationary solution of \cref{EQ6.1} is obtained by imposing $\partial 
\mathbf{v}/\partial t = 0$, which leads to the differential equation $\nabla P 
= \eta \nabla^2 \mathbf{v}$. Considering the symmetry of the configuration 
shown in \cref{fig:6_2}, the velocity field is parallel to the $z$ direction 
with an amplitude that only depends on the position along $x$: $\mathbf{v} = 
v_z(x) \mathbf{k_z}$ where $\mathbf{k_z}$ is a unit vector in the $z$ 
direction. In practice, the pressure gradient is obtained by imposing a 
pressure difference $\Delta P > 0$ over the length of the pore $L$ so that 
$\nabla P = - \Delta P/L$.  A first integration with respect to $x$ of the
latter differential equation leads to $\partial v_z/\partial x = - x \Delta 
P/L\eta + C_1$. The constant $C_1$ is necessarily equal to zero as the velocity 
field is symmetrical in $x  = 0$ ($\partial v_z/\partial x = 0$ for $x = 0$). 
This allows defining the shear stress $\sigma_{xz} = - \eta \partial 
v_z/\partial x$ exerted by a liquid layer located in $x$ on the adjacent layer 
in the $x$ direction, $\sigma_{xz} = x\Delta P/L$.
This equation shows that the shear stress is 
maximum at the pore surface
$x = D_\textrm{p}/2$ with a value
$\sigma_{xz}^s = D_\textrm{p} \Delta P/2L$.

A second integration with respect to $x$ of the differential equation above 
leads to $v_z(x) = - \Delta P x^2 /2L\eta + C_2$ where the constant $C_2$ is is 
obtained by considering the velocity boundary condition at the pore surface. At 
the macroscopic scale, the boundary condition for a viscous fluid at the solid 
surface is ``no slip'', \ie.\ $v_z(x = D_\textrm{p}/2) = 0$, so that $C_2 = 
\Delta P/8L\eta \times D_\textrm{p}^2$.  This leads to 
the following parabolic profile for the velocity field: 
\begin{equation}
v_z(x) = \frac{\Delta P D_\textrm{p}^2}{8L\eta} \bigg[1 - \bigg(\frac{2x}{D_\textrm{p}}\bigg)^2\bigg]
\label{EQ6.2}
\end{equation}
The derivation above performed for a cylindrical pore of diameter $D_\textrm{p} = 2R_\textrm{p}$ leads to the following equation: 
\begin{equation}
v_z(r) = \frac{\Delta P R_\textrm{p}^2}{4L\eta} \bigg[1 - \bigg(\frac{r}{R_\textrm{p}}\bigg)^2\bigg]
\label{EQ6.3}
\end{equation}
Regardless of the pore geometry, the flow $J$ is obtained by multiplying the 
velocity profile by the fluid density $\rho$ and integrating over the pore 
section area (or, equivalently, by multiplying the constant fluid density by 
the average flow rate $\overline{v}$).

At the microscopic scale, as discussed in Ref.~\citen{barrat_large_1999}, the 
velocity boundary condition at the pore surface depends on the fluid wetting 
properties towards the solid. For perfectly wetting fluids, the strong 
interactions between the fluid and solid phases impose that the liquid velocity 
field vanishes at the fixed solid boundary. This situation, which is 
illustrated in the left panel of \cref{fig:6_2}(b), is described by 
\cref{EQ6.2}. On the other hand, for a partially wetting liquid, the fluid 
molecules at the pore surface slip so that their velocity does not vanish as 
illustrated in the right panel of \cref{fig:6_2}(b).
In this case, one defines the so-called slip 
length $b$ which corresponds to the distance at which the extrapolated velocity 
profile reaches zero, $- b\partial v_z/\partial x = v_z$. With such a
boundary condition for the slit pore geometry $C_2 = \Delta 
P D_\textrm{p}^2/8L\eta \times [1 + 4b/D_\textrm{p}]$. In turn, this constant 
also leads to a parabolic velocity profile but with a shifted boundary 
condition: 
\begin{equation}
v_z(x) = \frac{\Delta P D_\textrm{p}^2}{8L\eta} \bigg[
1+\frac{4b}{D_\textrm{p}}
- \bigg(\frac{2x}{D_\textrm{p}}\bigg)^2
\bigg]
\label{EQ6.4}
\end{equation}

\noindent \textbf{Compressible flow.} Regardless of the velocity boundary 
condition considered, the description above relies on the assumption that the 
flowing liquid is incompressible. This raises the question of the relevance of 
the Poiseuille flow to low density fluids such as gases as they are 
compressible. With this respect, as already discussed in 
\Cref{chap:fundamentals}, the Mach number is a very important dimensionless 
number defined as the ratio of the flow velocity $v$ to the sound 
velocity $c$, $\textrm{Ma} = u/c$. This dimensionless number indicates whether 
or not the fluid flow can be treated as incompressible.
In practice, for $\textrm{Ma} < 0.3$, the flow 
velocity is smaller than the sound velocity so that the fluid can be regarded 
as incompressible because the compression wave generated by its displacement 
does not interfere with the flow. In this case, the derivation provided above 
for the Poiseuille flow applies even if the flowing fluid is a gaseous phase. 
In contrast, for $\textrm{Ma} > 0.3$, the fluid compressibility must be taken 
into account because the compression wave induced by the fluid motion affects 
the flow.  In this case, the derivation provided above no longer applies as the 
change in the fluid density along the pressure gradient imposed to the flowing 
fluid leads to a different velocity profile.
For a compressible flowing fluid, the molecules slip at the 
pore surface but it is important to emphasize that such slippage is of 
different nature as that described for the Poiseuille liquid flow.
Indeed, while slippage in the Poiseuille liquid flow occurs from non-wetting 
surface interactions, it corresponds to a compressibility effect in the case of 
compressible flows.

In practice, as discussed in Ref.~\citen{bird_transport_2002}, the compressible 
flowing fluid behaves with a slip length $b$ defined like the slip boundary 
condition introduced earlier, $v_z = - b \partial v_z/\partial x$. Because of 
the analogy with the slip length defined in the context of the Poiseuille flow, 
the flow for a compressible fluid under a pressure gradient $\Delta P/L$ leads 
to the same mathematical expression as \cref{EQ6.4}. Experimentally, it is 
observed that the slip length $b$ scales with the reciprocal of the pressure, 
$b = b_0/\overline{P}$ in which $b_0$ is a constant and $\overline{P}$ is the 
average pressure between the upstream and downstream pressures, $\overline{P} = 
(P_\uparrow + P_\downarrow)/2$. This leads to the following equation for the 
velocity profile of a compressible flow in a slit pore of a width 
$D_\textrm{p}$:
\begin{equation}
v_z(x) = \frac{\Delta P D_\textrm{p}^2}{8L\eta} \bigg[
1+\frac{4b_0}{D_\textrm{p} \overline{P}}
- \bigg(\frac{2x}{D_\textrm{p}}\bigg)^2
\bigg]
\label{EQ6.5}
\end{equation}
This equation is an important result for compressible fluids such as gases and 
supercritical fluids. It is the essence of the Klinkenberg correction which was 
mentioned for confined gases as an intermediate regime between Knudsen 
diffusion and viscous flow. While this correction will be discussed in the next 
section dealing with Darcy's law, a few remarks are in order here. On the one 
hand, for large $\overline{P}$, the slip correction $\sim 
b_0/\overline{P}$ becomes negligible and \cref{EQ6.5} is equivalent to 
Poiseuille flow with no-slip condition. This result is an expected asymptotic 
regime as fluids become incompressible in the limit of large pressures. On the 
other hand, for small $\overline{P}$, the slip correction known as the 
Klinkenberg correction in the context of compressible flows, becomes very 
important as  it scales as $\sim 1/\overline{P}$.

\begin{figure}[htbp]
	\centering
	\includegraphics[width=0.95\linewidth]{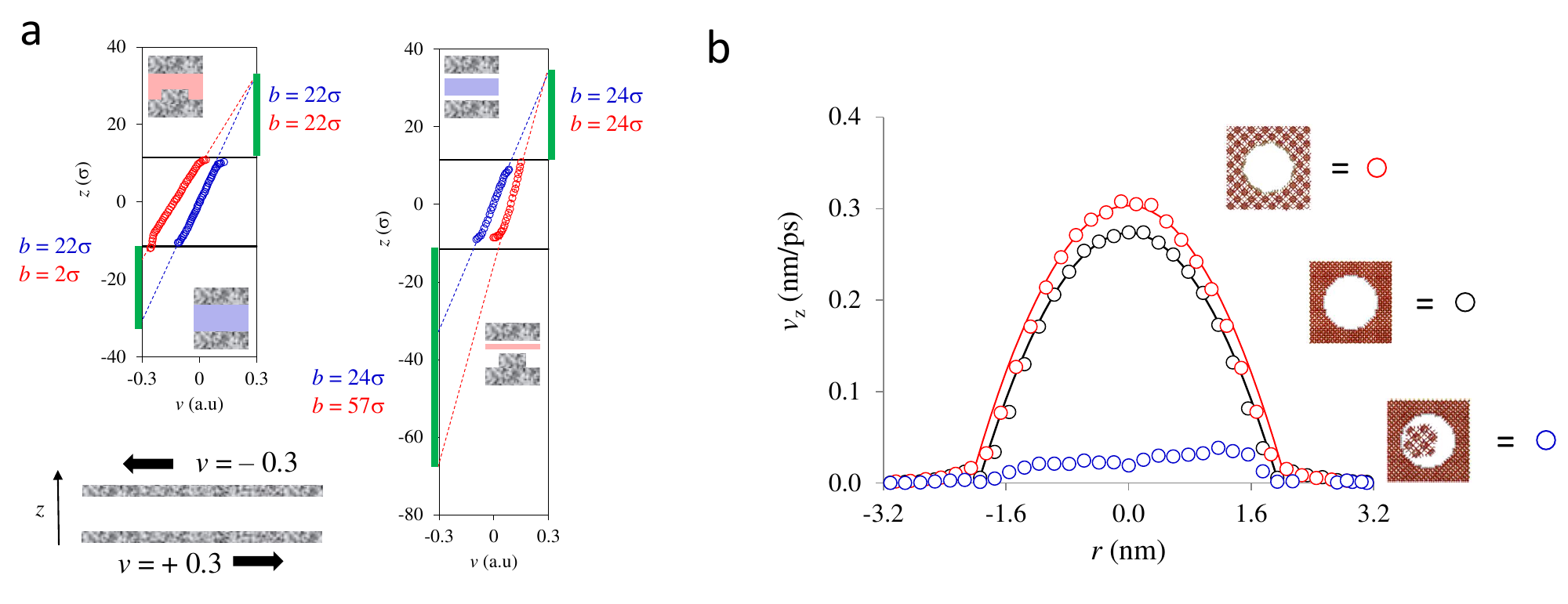}
	\caption{\textbf{Poiseuille flow at molecular surfaces}. (a) velocity 
		profile $v(z)$ for a wetting Lennard--Jones fluid confined in a slit pore 
		subjected to shear flow. While the upper surface is moved at a constant 
		velocity $v = -0.3$, the other surface is moved in the other direction 
		at a constant velocity $v = +0.3$ (all quantities are reduced with 
		respect to Lennard--Jones units). In the left panel, the velocity 
		profiles for a wetting fluid confined in a slit pore formed by two 
		smooth surfaces (blue data) and in a slit pore formed by a smooth 
		surface and nanopatterned surface (red data) are shown. The slip 
		lengths $b$ at each wall (upper and lower surfaces) are denoted by the 
		green thick lines and their numerical values are provided. Such slip 
		lengths are determined as the $z$-positions at which the  extrapolated 
		velocity profiles reach $v(z) = 0$. The solid black lines indicate the 
		positions of the upper and lower surfaces. In the right panel, the same 
		data are shown but for a non-wetting fluid. In this case, the confined 
		fluid does not wet the smooth surfaces (as illustrated by the ``vacuum 
		layers'' between the confined fluid and the surfaces) and 
		the confined fluid does not occupy the troughs at the nanopatterned 
		surface. Adapted with permission from Ref.~\protect\citen{cottin-bizonne_low-friction_2003}.
		Copyright 2003 Springer Nature.
		(b) Velocity profile for N$_2$ confined in a cylindrical silica pore (black 
		circles), a cylindrical silica pore with walls made up of zeolite (red 
		circles) and a cylindrical silica pore containing a zeolite 
		nanoparticle (blue circles). The abscissa $r$ denotes the radial 
		position with respect to the pore center.  In each case, the 
		cylindrical cavity has a diameter $D_{\textrm{p}} \sim 4.2$ nm. For the 
		data corresponding to the regular cylindrical pore and the cylindrical 
		pore with walls made up of zeolite, the lines are fits against the 
		Poiseuille flow law. Adapted with permisssion from Ref.~\protect\citen{coasne_molecular_2013}.
		Copyright 2013 American Chemical Society.}
	\label{fig:6_3}
\end{figure}

\noindent \textbf{Structured, patterned surfaces.} Most real materials exhibit 
rough surfaces which depart from the smooth, \ie.\ structurless, pore 
wall assumption used to derive the Poiseuille's law above. Surface roughness can 
refer either to atomistic surfaces or patterned surfaces such as 
crenelated walls with troughs as shown in \cref{fig:6_3}(a). In the context of 
the present work, surface roughness also includes situations illustrated in 
\cref{fig:6_3}(b) with mesoporous materials ($D_\textrm{p} \sim 
2-50\,\mathrm{nm}$) made of microporous walls ($D_\textrm{p} \leq 2$ nm).
These different situations raise the question of surface roughness and, more 
generally, of the role played by surface disorder on viscous flow in porous 
materials. 
In the case of patterned surfaces, this question was addressed using molecular 
simulation by Cottin-Bizonne \etal.\ who considered the hydrodynamics  
of a fluid confined in smooth and rough 
nanopores~\cite{cottin-bizonne_low-friction_2003}. More specifically, as shown in 
\cref{fig:6_3}(a), the flowing properties of the confined fluid were
probed by imposing a shear stress. To do so, the lower and upper walls of the 
slit nanopore were moved at a constant positive and negative velocity, 
respectively (parallel Couette flow geometry). This numerical set-up allows determining 
the liquid flow but also the hydrodynamic boundary conditions at each pore 
surface by monitoring the fluid velocity profile.

As shown in \cref{fig:6_3}(a), the authors in 
Ref.~\citen{cottin-bizonne_low-friction_2003} found different behaviors 
depending on the wetting properties of the confined fluid with respect to the 
solid phase. For smooth pore surfaces, the fluid response in terms of velocity 
profile conforms the classical picture provided by Poiseuille's law.
The velocity boundary condition at 
the pore surface depends on the strength of the fluid/wall interactions with a 
slip length $b$ that increases upon decreasing the wetting of the fluid phase. 
Typically, $\sigma$ being the fluid molecule size, $b$ varies from $22\sigma$ 
to $24\sigma$ as the thermodynamic condition switches from wetting to partial 
dewetting (in the latter case, as schematically illustrated in the right panel 
of \cref{fig:6_3}(a), a gas phase forms between the solid surface and the 
confined liquid). In the case of the patterned surfaces, the situation is 
more complex as the fluid flow properties drastically depend on the wetting 
scenario. 
Under wetting conditions, the fluid occupies the whole porosity as it also wets 
the troughs formed by the patterned surface.  As expected, the velocity 
boundary condition, \ie.\ slip length at the smooth pore surface, is 
identical to that obtained for the regular pore with two smooth surfaces ($b = 
22\sigma)$. 
However, at the rough surface, under wetting conditions, the disorder 
drastically suppresses the slippage as $b = 2\sigma$. This result is due to the 
fact that, when the whole porosity is filled by the liquid, the troughs hinders 
transport by imposing a boundary condition which slows down the liquid flow in 
the pore center. Under partial dewetting conditions, an opposite trend is 
observed. At the smooth pore surface, like under wetting conditions, the slip 
length is identical to that observed for the regular smooth slit pore ($b = 
24\sigma$). However, at the rough surface, because the liquid does not wet the 
troughs under dewetting conditions, the velocity boundary condition corresponds 
to a very large slip length $b = 57\sigma$. This result can be rationalized by 
noting that the gas layer formed under partial dewetting reduces drastically 
the friction at the solid surface. The results above are important as they 
show, in agreement with available experimental data, that the liquid/solid 
friction in flowing liquids can be either increased or decreased depending on 
the surface wetting properties by tuning surface disorder.

As another important illustration of the effect of surface roughness on liquid 
flow in porous materials, \cref{fig:6_3}(b) shows the parabolic profiles 
obtained for liquid $\mathrm{N}_2$ in cylindrical silica 
mesopores~\cite{coasne_molecular_2013}. 
Several porous materials are considered as they pertain to different physical 
situations encountered in real applications (in all cases, the cylindrical 
mesopore has a diameter $D_\textrm{p} = 4.2$ nm). First, as a reference system, 
a single cylindrical mesopore with dense, impermeable walls is carved out of an 
amorphous silica block. The silica surface is modeled in a realistic fashion by 
saturating all O dangling bonds with H atoms in order to form silanol groups. 
Second, the same silica pore is obstructed by adding into its mesoporosity a 
supported zeolite nanoparticle having a diameter of about 2 nm. Such a zeolite 
nanoparticle was prepared from a faujasite zeolite crystal having a silicon to 
aluminum ratio Si/Al = 15 (the compensating cation is Na$^+$). Third, a 
cylindrical mesopore of a diameter $D_\textrm{p} = 4.2\,\mathrm{nm}$ was carved 
out of a faujasite zeolite crystal (in order to compare the different samples,
the starting faujasite  material is identical to that used to prepare the second system). 
These three realistic materials allow probing the effect of surface disorder 
and patterning on liquid flow under conditions relevant to practical 
applications in catalysis, filtration, \etc.
In practice, using such atomistic structures, the flow and collective dynamics 
of the confined liquid is simulated by inducing transport through an external 
force $f$ acting on each molecule. This force is an equivalent description of a
chemical potential gradient $f = -\nabla \mu$ or, correspondingly, a pressure
gradient for an incompressible liquid (since the Gibbs--Duhem relation states
that $\rho \textrm{d}\mu = \textrm{d}P$).
Using such non equilibrium Molecular Dynamics simulations, the velocity profile
$v(r)$ can be determined in the course  of simulation runs [\cref{fig:6_3}(b)].
As expected, the velocity profile for the 
single silica mesopore obeys Poiseuille's law. In particular, the velocity 
profile shown for this system in \cref{fig:6_3}(b) is quantitatively described 
by a parabolic function as predicted with Poiseuille's law (the fitted viscosity 
was found to be in reasonable agreement with the experimental value). As 
expected, as a result of the strong fluid/wall interactions, the velocity of 
the molecules at the pore surface in the single mesopore is zero (no-slip 
boundary condition). The velocity profile for the cylindrical mesopore with 
porous walls also obeys Poiseuille's law. However, compared to the mesopore with 
impermeable walls, the flow for the model with porous walls is larger due to 
the fluid molecule transfer between the wall porosity and the mesopore. More 
precisely, molecules at the mesopore surface have a non zero velocity which 
contributes to enhance the flow within the mesopore. The corresponding slip 
length, $b \sim 0.2 -0.4\,\mathrm{ nm}$, is about the size of the fluid 
molecule ($\sigma_{\textrm{N}2} \sim 0.36$ nm). Considering that N$_2$ fully 
wets the silica surface, such a slip length induced by the surface 
microporosity is fully consistent with the results by Cottin-Bizonne \etal.\ 
reported above.\cite{cottin-bizonne_low-friction_2003} 
Indeed, in the case of wetting fluids, the slip length at the rough surface was 
found to be of the order of the trough depth. In the case the porous wall made 
of zeolite, the typical pore size is about  0.8 nm --- a value which 
is indeed roughly similar to the observed slip length.  While such a small slip 
length seems very small at first, it leads to a large flow enhancement 
because the flux corresponds to the fluid average velocity multiplied by the 
pore cross-section area and the fluid density (typically, the slip length 
observed in \cref{fig:6_3}(b) leads to a 20\% flow increase with respect to 
the no-slip condition). As for the third model shown in \cref{fig:6_3}(b), the 
velocity profile shows that the nanoparticle insertion into the single mesopore 
drastically reduces the overall fluid flow as it obstructs the porosity.

\subsubsection{Darcy's law and its extensions}
\label{sec:transportPore_viscousFlow_darcy}

\noindent \textbf{Darcy law.} In the previous section, the constitutive 
equations for Poiseuille flow,  with or without surface slippage, were obtained 
for pores of a simple geometry (planar or cylindrical pore). The use of a 
well-defined geometry is required to write a rigorous flow boundary condition 
--- typically the fluid velocity at the pore surface. However, when dealing with
most real man-made or natural porous media, the porosity does not consist of 
pores of a simple geometry. Yet, under the same assumptions as those used to 
derive the Poiseuille's law --- namely the linear response regime, the fluid 
incompressibility and the pure advective nature of the flow --- it is possible 
to extend its applicability to any porous material. This is the essence of 
Darcy's law which can be seen as a linear response model of an incompressible 
liquid submitted to a pressure gradient~\cite{ho_gas_2006,bear_modeling_2018}. 
More specifically, Darcy's law states that the average flow rate $\overline{v}$ in a 
given porous material is directly proportional to the pressure gradient 
inducing transport with a proportionality constant that depends on the fluid 
viscosity $\eta$ and the material permeability $k$: 
\begin{equation}
\overline{v} = - \frac{k}{\eta}  \nabla P
\label{EQ6.6}
\end{equation}
From this equation, the flow $J = \rho \overline{v}$ is readily obtained by 
multiplying the average flow velocity $\overline{v}$ by the fluid density $\rho$
(with the latter being constant since the fluid is assumed to be 
incompressible). In general, $k$ depends on the pore size but also the pore 
shape. However, in the case of simple geometries such as for planar and 
cylindrical pores, the permeability $k$ must be consistent with the flow 
predicted using Poiseuille flow. 
Typically, by averaging over the pore section area the velocity profile 
determined in \cref{EQ6.2,EQ6.3}, we obtain $\overline{v} = \Delta P 
D_\textrm{p}^2/12 L \eta$ for the planar geometry and $\overline{v} = \Delta P 
D_\textrm{p}^2/32 L \eta =  \Delta P R_\textrm{p}^2/8 L \eta$ for the 
cylindrical geometry. This implies that the permeability is $k = 
D_\textrm{p}^2/12$ and $k = R_\textrm{p}^2/8$  for the slit and cylindrical 
pores, respectively.

\Cref{EQ6.6} can be easily generalized to a disordered porous medium with pores 
of a distorted shape that are strongly connected. In practice, let us consider 
a porous solid having a porosity $\phi$ (defined from the free volume 
available to the flowing fluid) and pores with a non-regular shape and a 
tortuosity $\tau$. As discussed in \cref{chap:diffusion_network}, the 
tortuosity is defined as the average in-pore fluid flow path divided by the 
pore length.
For a disordered porous solid, with strongly interconnected pores, the 
tortuosity is defined as the average path $\langle L\rangle$ followed by the 
fluid molecules divided by the sample height $h$. Such a tortuosity implies 
that the physical pressure drop $\Delta P$ imposed across the sample does not 
exert a physical pressure gradient 
$\nabla P = \Delta P/h$ but $\nabla P = \Delta P/\tau h$ because the 
corresponding force applies to a physical length $\langle L\rangle = \tau h$
instead of $h$. 
As for the flux, $J = \overline{\rho} \overline{v}$,
it is obtained by multiplying the effective fluid density 
$\overline{\rho}=\phi\rho$ (where $\rho$ is the physical fluid density and 
$\phi$ the porosity) by the mean flow rate $\overline{v}$.
With these considerations, the flux in a given porous material can be defined 
as:
\begin{equation}
J = - \frac{\rho \phi k}{\eta}  \frac{\Delta P}{\tau h}
\label{EQ6.7}
\end{equation}
In the spirit of this last equation, the Kozeny equation is a 
generalized Darcy expression which was derived for an assembly of tubes of any 
arbitrary shape. Such a relation expresses the permeability $k$ 
as~\cite{torquato_random_2005,bear_modeling_2018}: 
\begin{equation}
k = \frac{\phi^3}{c_0 s^2}  
\label{EQ6.8}
\end{equation}
where $\phi$ is the solid porosity, $c_0$ is a numerical factor that depends on 
the  shape of the tube section, and $s$ is the pore surface area per unit of 
volume. 
As discussed in Ref.~\citen{torquato_random_2005}, $c = 2$ for a circular pore 
section,  $c = 5/3$ for a regular triangular pore section, and $c = 1.78$ for a 
regular square section. In fact, for real isotropic porous materials, $c$ is an 
adjustable parameter but $c = 5$ is often found to work well. In particular, 
this latter value is found to apply to the specific case of packed beds of 
solid particles. For such systems, the Kozeny--Carman relation 
reads~\cite{bear_modeling_2018}:  
\begin{equation}
k = \frac{1}{5 s^2} \times  \frac{\phi^3}{1 - \phi^2} \sim \frac{D^2}{180}   \times \frac{\phi^3}{1 - \phi^2}
\label{EQ6.9}
\end{equation}
where the second equation applies to spherical particles of a diameter $D$ for
which the surface area per unit of volume is $s = 6/D$ (so that
$5s^2 = 180/D^2$).

When using Darcy's equation, regardless of the pore morphological and 
topological disorders, it is often assumed that the two physical constants in 
the proportionality factor in \cref{EQ6.6} are decoupled. In more detail, while 
$\eta$ only depends on the fluid nature and temperature,  $k$ is assumed to be 
an intrinsic material parameter (\ie.\ a fluid independent quantity). 
This assumption is of practical use to engineering approaches as it allows predicting the 
permeability $k$ from a simple in-laboratory transport experiment. However, 
from a more physical standpoint, there is no fundamental reason to expect the 
permeability $k$ to be fluid independent. In fact, as explained above, 
\cref{EQ6.6} is a simple linear response theory in which the fluid flow rate is 
assumed to scale with the pressure gradient inducing transport. As will be 
shown in \Cref{sec:transportPore_viscousFlow_beyond}, the viscous approximation 
breakdown for small pores at the nm scale leads to effective permeabilities 
that are no longer fluid independent even though the linear response assumption 
still holds \cite{osullivan_perspective_2022}.

\noindent \textbf{Klinkenberg effect.} As discussed earlier, the Darcy 
equation --- which can be seen as a generalization of Poiseuille flow to any 
porous medium --- relies on the hypothesis that the liquid sticks to the pore 
surface (no slip boundary condition). However, as discussed in the section 
devoted to Poiseuille flow, compressibility effects such as with low 
density gases lead to slippage at the pore surface. Such slippage 
effects, inherent to compressible systems, are referred to in the context of 
Darcy's equation as Klinkenberg correction. 
Such slippage only manifests itself in pores of a diameter $D_\textrm{p}$ close 
to the gas mean free path $\lambda$ and leads to the following corrected 
permeability:  
\begin{equation}
k_c = k_\infty \Bigg (1 + 8 a \frac{\lambda}{D_\textrm{p}} \Bigg )
\label{EQ6.10}
\end{equation}
where $k_\infty$ is the permeability obtained for the liquid phase, 
\ie.\ under no-slip condition, while $a$ is a proportionality factor 
close to 1 in practice. Because the mean free path scales with the reciprocal 
of the gas pressure, $\lambda \sim P^{-1}$, the latter equation can be recast 
as: 
\begin{equation}
k_c = k_\infty \Bigg (1 + \frac{b}{P} \Bigg )
\label{EQ6.11}
\end{equation}
where $b$ is a constant that depends on the pore size but also the nature of 
the solid/gas combination. As discussed in 
\Cref{sec:transportPore_flow_gasAndLiquid}, the Knudsen number Kn 
$= \lambda/D_\textrm{p}$ is the appropriate dimensionless number to determine 
whether or not the Klinkenberg correction should be applied.

\noindent \textbf{Darcy--Forchheimer law.} The advective nature of the fluid 
flow is an important assumption at the root of Darcy's equation. As for the 
linear response hypothesis between the pressure gradient and the flow rate, it 
only applies to low Reynolds number --- typically, Re $< 10$. For larger 
Re, such an approximation breaks down and higher order terms in 
the driving force/flow rate relationship must be included:  
\begin{equation}
- \nabla P \sim a v + b v^2 + \mathcal{O}(v^3)
\label{EQ6.12}
\end{equation}
In this equation, known as Forchheimer equation, $\mathcal{O}(v^3)$ encompasses 
all higher order terms which are assumed to be negligible. Physically, 
\cref{EQ6.11} describes the fact that the flow resistance becomes non linear as 
the flow becomes turbulent. In order to account for such effects, using 
\cref{EQ6.11}, a corrected Darcy equation --- the so-called Darcy--Forchheimer 
law --- can be proposed: 
\begin{equation}
- \frac{k}{\eta} \nabla P  = v(1 + a' v)  = v(1 + a'' \textrm{Re})
\label{EQ6.13}
\end{equation}
where the second equality is obtained by noting that
$Re = \rho v D_\textrm{p}/\eta$.
As expected, for low Re, the second order correction becomes
negligible and one recovers the conventional Darcy equation.

\subsubsection{Beyond viscous flows}
\label{sec:transportPore_viscousFlow_beyond}

\noindent \textbf{De Gennes narrowing.} As discussed in 
\Cref{sec:selfColTransportD}, the 
permeability $k$ that characterizes the viscous flow of a confined fluid 
induced by a pressure gradient $\nabla P$ is strictly equivalent to the 
collective diffusivity $D_0$. More precisely, by writing the flow response of 
an incompressible liquid to a chemical potential gradient $\nabla \mu$, it is 
straightforward to show that the permeance $K = k/\eta = D_0/\rho k_\textrm{B}T$.
From a fundamental viewpoint, in contrast to the self-diffusivity, which can be 
described using simple models such as a free volume theory or a surface diffusion model (see 
\Cref{chap:diffusion_network}), modeling the collective diffusivity $D_0$ and,
therefore, the permeance $K$ of a confined fluid proves to be a complex task. 
This is due to the fact that, as shown in \cref{EQ3.40}, the time 
autocorrelation function of the fluid center of mass velocity, which defines 
the collective diffusivity in the Green--Kubo formalism, involves cross-terms 
between different molecules (in contrast to the self-diffusivity that only 
involves time correlations of the velocity of individual molecules). Such 
cross-correlations reflect the molecular interactions within the flowing 
liquid and are, therefore, crucial to calculate the permeance $K$. In the 
case of ultra-confined liquids or in the limit of very low loadings, such  
cross-correlations are often found to be negligible so that the collective 
diffusivity can be approximated by the self-diffusivity, $D_\mathrm{s} \sim 
D_0$ \cite{falk_subcontinuum_2015}. In this particular situation, the collective diffusivity can be modeled 
in a simple fashion using the different techniques cited above to describe and 
predict the self-diffusivity.
However, in most practical situations, this approximation does not hold and 
modeling the collective diffusivity or, equivalently, the permeance remains 
often empirical.

In fact, besides the Green--Kubo formalism which offers a robust framework to 
describe the collective diffusivity from measurements such as coherent neutron 
scattering, there is only one model that allows predicting in a simple way the 
collective diffusivity: De Gennes narrowing~\cite{de_gennes_liquid_1959}. 
This model, which was derived to describe the wave-vector $q$ dependence of 
neutron scattering in liquids, relates the collective diffusivity of the fluid 
to the structure factor $S(q)$ that characterizes its structural 
ordering~\cite{hohenberg_theory_1977}. Recently, it was shown that the concept
of De Gennes narrowing is a robust formalism to describe the flow of fluids in
nanoporous materials.\cite{kellouai_gennes_2024}
In detail, De Gennes narrowing model constitutes more
than a predictive tool for the collective diffusivity in flowing liquids
as it describes the wave-vector dependence of the collective diffusion,
$D_0(q)$, \ie.\ the collective dynamical response of the fluid to a density 
fluctuation over a length $l \sim 1/q$. To introduce the concept of De 
Gennes narrowing, let us consider a confined liquid whose local density at a 
time $t$ is given by the distribution $\rho(\mathbf{r},t)$ (see Chapter 12 in 
Ref.~\citen{barrat_basic_2003}). The free energy per unit of volume 
$\mathcal{F}$, which only depends on the distribution of molecules,
\ie.\ being a time independent quantity, is a functional form defined as 
$\mathcal{F}[\rho(\mathbf{r})]$. The local chemical potential $\mu(\mathbf{r})$ 
can be defined as the derivative of the local free energy with respect to the 
local density, \ie.\ $\mu(\mathbf{r}) = \delta \mathcal{F}/\delta 
\rho(\mathbf{r})$ where $\delta$ denotes the functional derivative. Locally, 
the mass conservation imposes that $\partial \rho(\mathbf{r}, t) / \partial t + 
\nabla \cdot J(\mathbf{r}, t) = 0$ where the flux $J$ is given by the following 
linear response relation $J(\mathbf{r}, t) = - \alpha \nabla \mu(\mathbf{r}, t)$. With these considerations, mass conservation leads to the following 
relationship: 
\begin{equation}
\frac{\partial \rho(\mathbf{r},t)}{\partial t} = \alpha  \nabla^2 \frac{\delta \mathcal{F}(\mathbf{r},t)}{\delta \rho(\mathbf{r},t)}
\label{EQ6.14}
\end{equation}
By describing the free energy functional and density distribution using Fourier
components, the above equation can be solved in Fourier space as: 
\begin{equation}
\frac{\partial \rho(\mathbf{q},t)}{\partial t} = - \alpha  q^2  \frac{\delta \mathcal{F}(\mathbf{q},t)}{\delta \rho(\mathbf{q},t)}
\label{EQ6.15}
\end{equation}
The Fourier transform of the free
energy can be written as a sum of uncoupled modes with a quadratic dependence,
$\mathcal{F}(\mathbf{q},t) = \sum_q \frac{1}{2} A(\mathbf{q})\vert \rho(\mathbf{q},t)\vert^2$.
In more detail, this expression assumes that, locally, the free energy profile 
is composed of harmonic oscillators in the local density $\rho(\mathbf{q},t)$. Inserting this free energy expression into \cref{EQ6.14} leads to the following differential equation, 
$\partial \rho(\mathbf{q},t)/\partial t = - \alpha  q^2  A(\mathbf{q}) \rho(\mathbf{q},t)$
for all $\mathbf{q}$, which admits the following solution: 
\begin{equation}
\rho(\mathbf{q},t) = \rho(\mathbf{q},0) \exp \left[- D_0(\mathbf{q})  q^2  t \right]
\label{EQ6.16}
\end{equation}
where $D_0(\mathbf{q}) = \alpha A(\mathbf{q})$ is the wave-vector dependent 
collective diffusivity. As a result  of the energy equipartition theorem, each 
thermodynamically accessible mode $\mathbf{q}$ involves a free energy contribution $\sim 
\kbT$, \ie.\ $A(\mathbf{q}) \rho(\mathbf{q})^2 \sim \kbT$,
which can be rewritten as: 
\begin{equation}
S(\mathbf{q}) = \langle \rho(\mathbf{q}) \rho(-\mathbf{q})  \rangle = 
\frac{k_\textrm{B}T}{A(\mathbf{q})}
\label{EQ6.17}
\end{equation}
where $S(\mathbf{q}) = \langle \rho(\mathbf{q}) \rho(-\mathbf{q})  \rangle$
is the structure factor and $\langle \cdots \rangle$ 
denotes ensemble average. Combining \cref{EQ6.17} with $D_0(q) = \alpha A(q)$ 
leads to: 
\begin{equation}
D_0(\mathbf{q}) = \frac{\alpha k_\textrm{B}T}{ S(\mathbf{q})} 
\label{EQ6.18}
\end{equation}
This is the essence of De Gennes narrowing which can be interpreted as follows.
Density fluctuations over a characteristic length $\sim 1/q$ are frequent 
[large $S(q)$] if the corresponding free energy cost is low [small $A(q)$]. In 
turn, because such fluctuations are thermodynamically favorable, the relaxation 
towards equilibrium is slow [$D_0(q)$ small]. On the other hand, unfavorable 
density fluctuations over a characteristic length $\sim 1/q$ [small $A(q)$] 
involve a large free energy cost so that they relax rapidly towards 
equilibrium, \ie.\ $D_0(q)$ is large.

A more refined treatment allows introducing collective interactions. As 
discussed in Ref.~\citen{poulos_communications_2010}, at infinite dilution, the 
constant $\alpha \sim D_\mathrm{s}/k_\textrm{B}T$ is proportional to the 
self-diffusivity as collective effects are negligible.
However, in finite dilution conditions, 
the constant $\alpha$ can be replaced by $D_\mathrm{s} H(q)/k_\textrm{B}T$ 
where $H(q)$ are the corrections that need to be taken into account because of 
collective interactions. 
This leads to a modified De Gennes narrowing expression: 
\begin{equation}
D_0(q) = \frac{D_\mathrm{s} \times H(q)}{S(q)} 
\label{EQ6.19}
\end{equation}
In the case of bulk molecular liquids or colloidal suspensions,  De Gennes 
narrowing is often invoked to provide a rational description of $q$-dependent 
neutron scattering 
data~\cite{alba-simionesco_gennes_2001,poulos_communications_2010}.
In particular, several attempts have been made over the years to provide a 
microscopic picture of De Gennes  narrowing 
effects~\cite{kleban_toward_1974,wu_atomic_2018}.
In contrast, less attention has been devoted to the use of De Gennes narrowing 
in the specific context of fluids confined in porous media. In this respect, it 
is interesting to note that Nygard \etal.\ have shown 
that dense confined fluids possess a wave-vector dependent 
collective diffusion consistent with De Gennes narrowing. 
Moreover, in contrast to their bulk counterpart, the bulk structure 
factor and de Gennes narrowing in confinement were found to be 
anisotropic~\cite{nygard_anisotropic_2016}. In the case of a nanoconfined fluid,
the fluid collective dynamics is shown to be accurately described through De
Gennes narrowing.\cite{kellouai_gennes_2024}

\begin{figure}[htb]
	\centering
	\includegraphics[width=0.95\linewidth]{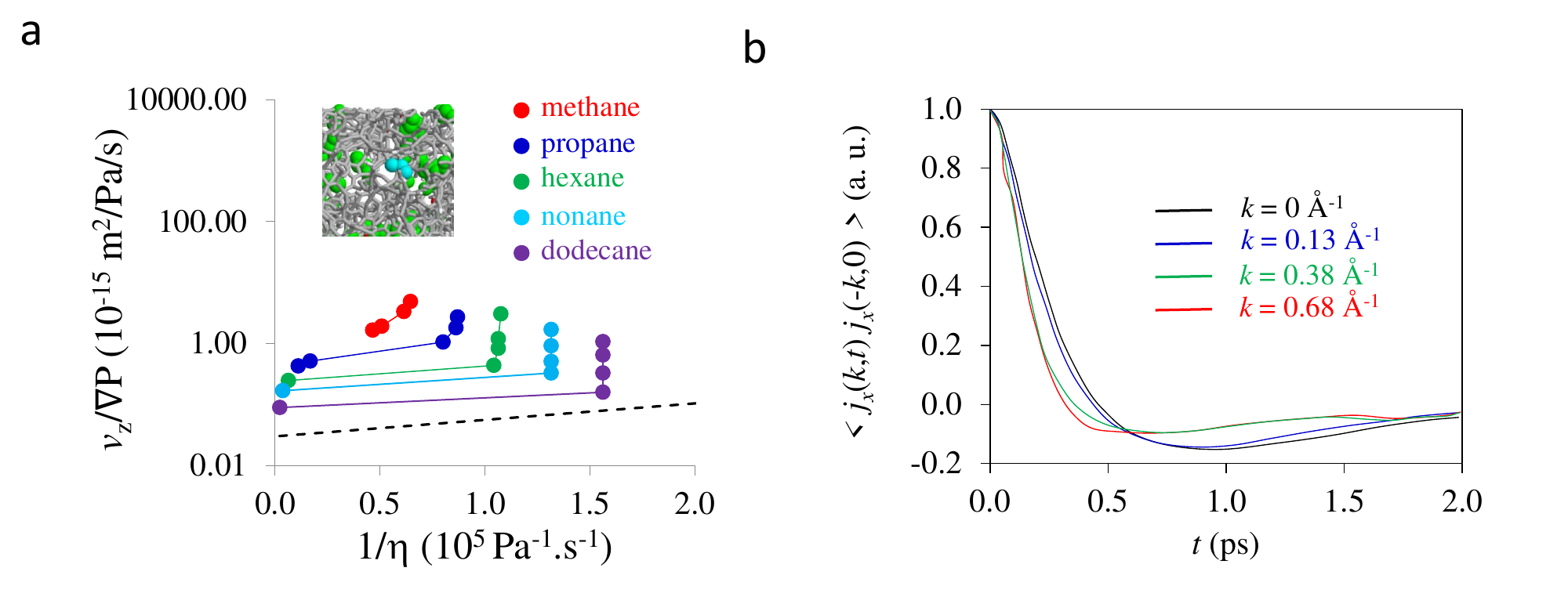}
	\caption{\textbf{Permeability and viscous flow in nanopores}. (a) Flow rate 
	normalized to the pressure gradient, $v_z/\nabla P$, as a function of 
	viscosity $\eta$ for different alkanes in a disordered nanoporous carbon: 
	methane (red circles), propane (blue circles), hexane (green circles), 
	nonane (cyan circles), and dodecane (purple circles). The dashed line 
	corresponds to Darcy's law as predicted from the typical pore size $\sim$ 1~nm. 
	The inset shows a typical molecular configuration for propane in a 
	disordered porous carbon (the blue and green spheres are the CH$_x$ groups 
	of propane while the grey segments are bonds between carbon atoms in the 
	host material). (b) Time autocorrelation function of the transverse current 
	for methane dodecane. Each color corresponds to a given $k$-value: $k$ = 0, 
	0.13, 0.38, and 0.68 \AA$^{-1}$. These transverse current autocorrelation 
	functions do not follow the expected exponential decay for a simple viscous 
	fluid. Adapted with permission from Ref.~\protect\citen{falk_subcontinuum_2015}.
	Copyright 2015  Macmillan Publishers Limited under Creative Commons
	Attribution 4.0 International License
	(http://creativecommons.org/licenses/by/4.0/.}
	\label{fig:6_4}
\end{figure}

\noindent \textbf{Deviations from Viscous flow.} Coming back to Darcy's law given in 
\cref{EQ6.6}, as already stated, it is usually assumed that the permeability 
$k$ is an intrinsic property of the host porous material and $\eta$ equals
the bulk value  (\ie.\ interfacial effects are neglected).
However, when dealing 
with nanoporous materials, many experimental and theoretical works report 
important failure of this hypothesis at the heart of Darcy's approach. As an 
illustration, \cref{fig:6_4} shows molecular simulation data of the flow 
observed for different alkanes --- methane, propane,
hexane, nonane, and dodecane --- in the disordered porosity of a host nanoporous 
carbon~\cite{falk_subcontinuum_2015}. These data show the flow rate
normalized to the pressure gradient, $v_z/\nabla P$, as a function of the fluid 
viscosity $\eta$ for different alkanes at various densities.  At constant 
temperature, for the different pressure gradients considered in this study, the 
confined alkanes were found to flow with a flux obeying the linear response 
regime \ie.\ $J \sim - K \nabla P$. As expected from the 
fluctuation-dissipation theorem, the proportionality factor $K$, known as the permeance, 
was found to be consistent with the collective diffusivity $D_0 = K \rho 
k_\textrm{B}T$ measured using the same confined liquid but under 
equilibrium (\ie.\ no flow condition).  However, in contrast to the 
expected Darcy behavior, the permeability $k$ depends on the flowing molecule 
as well as on its fluid density as shown in \cref{fig:6_4}.

For reasons that will become clearer below, the non-Darcy behavior observed in
\cref{fig:6_4} has its roots in the strong adsorption and confinement effects in
nanoporous solids. Such effects inherent to fluids confined within nanosized
cavities lead to a breakdown of the viscous flow hypothesis by introducing
memory effects into the time autocorrelation of the fluid velocity. To identify
the origin of this breakdown of viscous flow, it is instructive to probe the
transverse component of the fluid momentum fluctuations in Fourier space \cite{barrat_basic_2003}: 
\begin{equation}
j_x(k_z,t)=\sum_l mv^{(l)}_x(t)\exp \left[-\textnormal{i}k_z z^{(l)}(t)\right]
\label{EQ6.20}
\end{equation}
where $x$ and $z$ are two orthogonal directions and $m$ is the fluid molecule 
mass. The sum runs over all molecules $l$ of the fluid. As discussed in 
Ref.~\citen{barrat_basic_2003}, for a bulk viscous fluid, the  
correlation  time for momentum must be consistent with the momentum 
conservation law,  $\rho \partial \textbf{v}/\partial t = - \nabla P + \eta  
\nabla^2 \textbf{v}$. Solving this equation in Fourier space and in a 
direction $x$ normal to the fluid flow (component $\nabla P_x$ = 0)  leads to the 
following differential equation, $\partial j_x(\mathbf{k},t)/\partial t = - 
\eta k^2/\rho j_x(\mathbf{k},t)$. In turn, this equation implies that the time 
correlation function of the momentum Fourier components is of the  form: 
$\langle j_x(k,t) j_x(-k,0)\rangle \sim \exp (-\eta k^2 /\rho t)$ (where 
$\langle \cdots \rangle$ denotes statistical ensemble average). For a fluid 
confined inside a porous solid, the momentum conservation relation in the 
laminar flow regime, which leads to Darcy's equation, reads:
\begin{equation}
\rho \frac{\partial \textbf{v}}{\partial t} = 
- \nabla P + \eta  \nabla^2 \textbf{v} - \xi \rho \textbf{v}
\label{EQ6.21}
\end{equation}
where $\xi$ is the friction parameter at the solid/fluid interface.
Solving this equation in the Fourier space and in a direction $x$ normal to the
flow leads to the differential equation $\partial j_x(\mathbf{k},t)/\partial t =
-(\eta/\rho)/k^2 + \xi j_x(\mathbf{k},t)$ which admits the following solution: 
\begin{equation}
\left \langle j_x(k,t) j_x(-k,0)\right \rangle \sim \exp \left[ -\left(\frac{\eta}{\rho} k^2 + \xi\right) t \right]
\label{EQ6.22}
\end{equation}
As shown in \cref{fig:6_4}, 
it was found that the transverse momentum fluctuations of the confined fluids 
in the disordered nanoporous solid do not follow the behavior expected from 
\cref{EQ6.22}. This demonstrates that the viscous flow approximation at the 
root of Darcy's equation does not hold in such extreme confinements. This 
suggests that  memory effects such as  Mori--Zwanzig functions defined in 
generalized hydrodynamics should be included (see Chapter 9 in 
Ref.~\citen{hansen_theory_2013}).

From a fundamental viewpoint, the breakdown of the viscous flow approximation 
can be rationalized as follows~\cite{bocquet_nanofluidics_2010}. The use of the 
continuum conservation law in \cref{EQ6.21} implicitly assumes that there is a 
time scale separation between the molecular relaxation time $\tau_R$ of the
fluid molecules and the fluid momentum transfer time  $\tau$. On the one hand, 
the relaxation time within the fluid is given by the characteristic 
decay time in the stress tensor since $\eta = 1/Vk_\textrm{B}T \int 
\sigma_{\alpha\beta}(t) \sigma_{\alpha\beta}(0) \textrm{d}t$ with $\alpha,\beta = x,y,z$.
In practice, $\tau_R$ depends on the nature of the  fluid  but also on thermodynamic 
conditions such as temperature and density,  but it is typically of the order of 
$10^{-12}\,\mathrm{s}$.
On the other hand, the characteristic time for  momentum transfer is given by the 
decay time of the correlation function defined in \cref{EQ6.22}:  
$\langle j_x(k,t) j_x(-k,0)\rangle \sim \exp [-t/\tau]$ with $\tau = \left[(\eta k^2 
+ \xi)/\rho\right]^{-1}$. In the hydrodynamic limit, the time scale separation, 
which is inherent to viscous flows, is valid since $k \rightarrow 0$, \ie.\ 
$\tau_R \ll \tau$. 
However, for a nanoconfined fluid, momentum transfers occur with 
characteristic wave vectors $k \sim 1/L$ where $L$ is typically of the order of 
the pore size $\sim$ 1~nm. As shown in Ref~\citen{bocquet_nanofluidics_2010}, for 
such ultra-confined fluids,  $\tau_R\sim \tau$ and the time scale separation 
leading to the viscous flow description is no longer valid.

\subsection{Adsorption/desorption kinetics}

\subsubsection{Liquid imbibition} 
\label{sec:transportPore_kinetics_imbibition}

\begin{figure}[htb]
	\centering
	\includegraphics[width=0.95\linewidth]{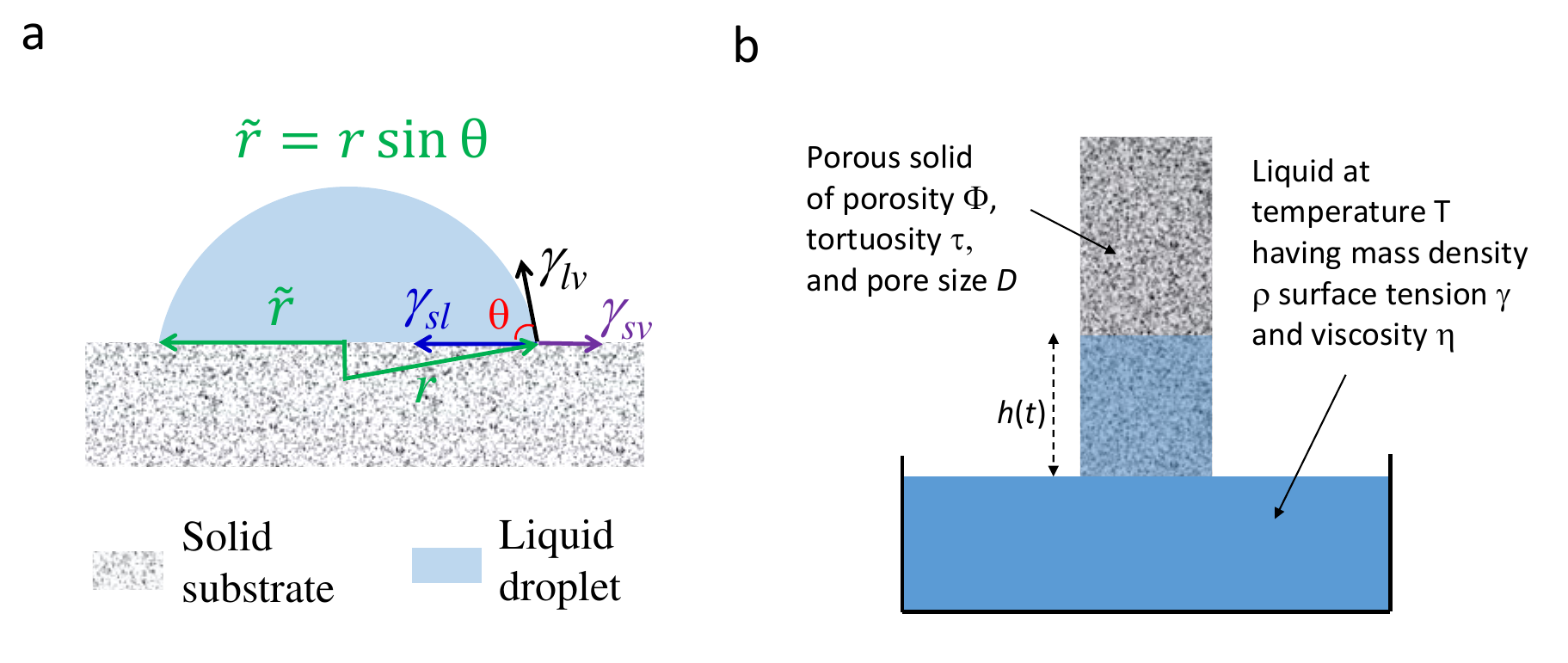}
	\caption{\textbf{Imbibition in nanoporous materials}.
	(a) Schematic representation of a liquid droplet of a radius $r$ at the surface of a solid substrate.
	    The angle $\theta$ formed by the droplet with the substrate depends on
	    the liquid-vapor ($\gamma_{lv}$), solid-liquid ($\gamma_{sl}$), and
	    solid-vapor ($\gamma_{sv}$) surface tensions through Young equation. 
	(b) A porous solid (grey, stone-like aspect) with pores of a diameter $D$ is
		immersed partially into a wetting fluid (blue) at a temperature $T$.
		The porous solid has a porosity $\phi$ and tortuosity $\tau$ while the
		fluid has a density $\rho$, surface tension $\gamma$ and a viscosity
		$\eta$.
		Owing to the capillary force, which overbalances the friction force, the
		fluid rises in the porosity of the porous solid with an imbibed height
		at a time $t$ that is proportional to the square-root of time, $h(t)
		\sim \sqrt t$.}
	\label{fig:6_5}
\end{figure}

We now consider non stationary problems such as imbibition at a temperature $T$ 
of a liquid having a density $\rho$, a surface tension $\gamma_{lv}$ and a 
viscosity $\eta$ into a porous solid having a porosity $\phi$, a tortuosity 
$\tau$ and an average pore radius $R_\mathrm{p}$ (\cref{fig:6_5}).
When the porous solid is brought to contact with the liquid, imbibition occurs
due to capillary forces.\cite{gruener_spontaneous_2009, gruener_anomalous_2012,
huber_soft_2015}
Let us first assume that the porous solid can be
described as an assembly of cylindrical pores having a pore radius
$R_\mathrm{p}$. In order to determine the height $h(t)$ of the
liquid that has penetrated the solid at a time $t$, one has to solve the
following equation of motion: 
\begin{equation}
\frac{\textrm{d}(m v)}{\textrm{d}t} = - mg + 2 \pi R_\mathrm{p} \gamma_{lv} \cos{\theta} - 8\pi \eta h v    
\label{EQ6.23}
\end{equation}
where $m$ is the mass of fluid inside the porous material, $\theta$ is the 
contact angle that is characteristic of the liquid/solid couple, and $v$ is the 
imbibition velocity. The left hand side term in the above equation is 
the rate of change of the momentum of the  liquid contained in the pore.
The first term on the right hand side is the gravity term which limits 
imbibition while the second term is the capillary term (\ie.\ the driving force 
for imbibition). 
As shown in \cref{fig:6_5}(a), the contact angle $\theta$ depends on the
liquid/vapor, liquid/solid and gas/solid surface tensions through Young
equation.
The last term in the above equation is the viscous (friction) 
term which is described using Poiseuille--Hagen flow. 
By noting that the mass $m = \rho \pi R_\mathrm{p}^2 h$ is related to the height
$h$, the above equation can be rewritten as:
\begin{equation}
\rho \frac{\textrm{d}(h v)}{\textrm{d}t} = - \rho g h + \frac{2  \gamma_{lv}\cos{\theta}}{R_\mathrm{p}} - \frac{8\eta h v}{R_\mathrm{p}^2}
\label{EQ6.24}
\end{equation}
This equation can be obtained starting from Poiseuille law for the flow rate 
$v$ induced by a pressure gradient $\nabla{P}$, $v = - k/\eta \nabla{P}$ where 
$k$ is the permeability. For a cylindrical pore of radius $R_\mathrm{p}$ the 
latter is $k = R_\mathrm{p}^2/8$ so that Poiseuille equation can be rewritten 
as  $v = - R^2/(8\eta) \nabla{P}$.
Considering that $\nabla{P} = \Delta P/h$ for an imbibition height $h$, the 
latter expression can be inverted to arrive at 
$\Delta P = - 8 \eta hv/ R^2$. By noting that the viscous force $F_v$ acting on 
the fluid corresponds to the pressure difference multiplied by \bluelork{the pore cross-section 
area}, we can write $F_v = \Delta P \times \pi R^2 = - 8 \pi \eta h v$ which is 
identical to the expression used in \cref{EQ6.23}. \Cref{EQ6.24} was derived 
for a cylindrical pore having a regular, straight section. Extension of 
\cref{EQ6.24} to a cylindrical pore having a tortuous shape can be done by 
introducing the tortuosity $\tau$ in the permeability $k = 
R_\mathrm{p}^2/8\tau$ where the permeability for a regular cylindrical geometry 
is recovered for $\tau = 1$~\cite{tsunazawa_experimental_2016}.
Using this definition, the tortuosity describes the fact that the friction 
force (between the liquid and the porous material) is proportional to the 
physical length of the pore and not its height. We note that this tortuosity 
differs from another common definition via the diffusion factor (in the 
latter, the tortuosity is defined as the ratio of the bulk diffusivity to the 
diffusivity of the confined liquid taken under the same thermodynamic 
conditions). After introducing the tortuosity $\tau$, \cref{EQ6.24}
simply reads: 
\begin{equation}
\rho \frac{\textrm{d}(h v)}{\textrm{d}t} = - \rho g h + \frac{2  \gamma_{lv}\cos{\theta}}{R_\mathrm{p}} - \frac{8\eta \tau h v}{R_\mathrm{p}^2}
\label{EQ6.25}
\end{equation}
The inertial term $\rho \textrm{d}(h v)/\textrm{d}t$ only manifests itself in 
the short-time range~\cite{quere_inertial_1997}. After a rapid transient 
regime, liquid imbibition becomes stationary and inertia becomes negligible so 
that \cref{EQ6.25} can be recast as: 
\begin{equation}
- \rho g h + \frac{2  \gamma_{lv}\cos{\theta}}{R_\mathrm{p}} - \frac{8\eta \tau h v}{R_\mathrm{p}^2} = 0
\label{EQ6.26}
\end{equation}
For liquid imbibition in a horizontal setup, the gravity term is to be 
discarded and one arrives at the simple, known solution for \cref{EQ6.26}:
\begin{equation}
h(t) = \sqrt{\frac{R \gamma_{lv} \cos{\theta} \times t}{2 \eta \tau}} 
\label{EQ6.27}
\end{equation}
where we used that $v = \dot{h}$ and $\int{2h \dot{h}\textrm{d}t} = h^2$ (the dot in these notations indicate time derivative). 
The same derivation can be extended in case the liquid slips at the solid 
surface as done in Ref.~\citen{joly_capillary_2011}
In case gravity cannot be neglected (vertical experiment), \cref{EQ6.25} can be recast as: 
\begin{equation}
\dot{h} = \frac{a}{h} - b 
\label{EQ6.28}
\end{equation}
with $a = 2 \gamma_{lv} \cos{\theta} R_\mathrm{p}/8 \tau \eta$ and $b = \rho g R_\mathrm{p}^2/8 \tau 
\eta$. The solution of \cref{EQ6.28} is obtained by noting that it can be 
expressed as: $\textrm{d}t/\textrm{d}h = h/(a - bh)$
which leads to the solution: 
$t(h) = - a/b^2 \times \ln{(1 - bh/a)} - h/b$. \Cref{EQ6.28} can also be solved using Lambert function [$x = W(x) \exp{W(x)}$]
to obtain directly $h(t)$~\cite{fries_analytic_2008}. However, for practical 
purposes, experimental data can be fitted  to estimate $a$ 
and $b$ from which $\tau$ can be inferred since $a = 2 \gamma_{lv} \cos{\theta} 
R/8 \tau \eta$ and $b = \rho g R^2/8 \tau \eta$.    
Moreover, usually, one measures the mass uptake $m(t)$ rather than the liquid height $h(t)$.
However, these two values can be linked in a straightforward way since $m(t) =
\rho A \phi h(t)$ where $\phi$ is the solid porosity (assumed to be constant
throughout the sample), $A$ the cross-section area of the solid face through
which imbibition occurs, and $\rho$ the liquid density.

\subsubsection{Mass uptake: surface, internal and external resistances} 
\label{sec:transportPore_kinetics_massuptake}

Let us now consider adsorption kinetics which corresponds to the filling of a porous solid by a fluid phase having a density $\rho$.
Initially, the solid porosity is either empty or filled with the fluid but at a lower density. In what follows, we present a simple yet representative picture of sorption kinetics into porous solids.
For a deeper discussion of the problems at play, the reader is referred 
to the book by Kärger, Ruthven and Theodorou~\cite{karger_diffusion_2012}.
\bluelork{From a very general viewpoint, looking at the schematic picture  given in \cref{fig:6_1}(b), the dynamical process leading to adsorption  involves three mechanisms: transport within the bulk external phase, transport across the interface between the porous solid and external phase, and transport within the porous solid. Identifying the relative importance of these  underlying dynamics is a key problem as it allows one to determine the overall adsorption kinetics and the limiting rate. In particular, we emphasize here that the observed mass transfer for a given fluid/material couple results from a competition between the driving force that induces the motion within the material and any resistances arising from surface barriers at the interface or from internal and external diffusion limitations.} 

\redlork{From this general viewpoint, the overall resulting transfer occurs by involving subsequent steps in series of the three mechanisms above (transport within the solid porosity, transport through the external phase and transfer across the interface). 
In the different subsections below, we  consider asymptotic limits where one or two of these mechanisms constitute the limiting step. This is for instance the essence of the Linear Driving Force model, where one assumes that external diffusion and in-pore diffusion are very rapid so that surface transfer is the limiting step.~\cite{sircar_why_2000,brandani_exact_2021}
With these assumptions, the fluid density in the gas phase and within the pores are predicted to be homogeneous and the interface is the bottleneck for gas transfer. Describing how the different regimes combine when an asymptotic limit cannot be taken is a central question in this field. In that context, the effective medium approaches such as those presented in Section 4 are very important. In these strategies, one considers a far field chemical potential gradient (which corresponds to a concentration or pressure gradient applied at the macroscopic level) and then solves the problem of combining external and in-pore diffusion. From a formal viewpoint,  the problem can also be solved by using the work by Roosen-Runge et al.\ which was presented in Section 4.2.2. Finally, engineering-based approaches such as the random walk models described in Section 4.3.2 also offer a relevant numerical strategy. With this approach, one performs numerical simulations of a random walker on domains distributed on a lattice (each domain can be a pore, a wall, or the external phase). The approach is rendered stochastic to include surface transfer limitations by adding mass balance conditions across domains (this is the adsorption isotherm that  can be  used to ensure that at equilibrium random walkers are distributed according to the free energy landscape) and a probability to overcome the free energy barrier at interfaces.}

From a general standpoint, predicting the overall transport in multiscale porous materials combining surface transfer with external and internal transport is a complex issue as the three dynamical mechanisms depend on temperature, fluid density in the external and confined phases, \etc.\ Also, for given thermodynamic conditions, the impact of each dynamical step depends on the partition distribution between the bulk fluid, fluid at the  solid/fluid interface, and confined fluid. Typically, for a large bulk crystal, \ie.\ having a small surface to volume ratio, adsorption kinetics will be 
mostly governed by the diffusion within the solid phase. On the other hand, the 
adsorption kinetics for a very finely divided porous solid (grains at the 
$\sim$  nm scale) can be largely controlled by surface 
barriers~\cite{remi_role_2016,fasano_interplay_2016,
	gulin-gonzalez_influence_2006}.

On top of this intrinsic complexity, the situation can be very puzzling 
depending on the physical chemistry phenomena occurring at the solid external 
surface and within its porosity. Typically, while physical adsorption is always 
exothermal, chemisorption and/or chemical reactions within the porosity of the 
host solid can be either exothermal  or endothermal. Moreover, in the latter 
situation, the existence of different species, which can convert from and into 
each other, raises the question of additional transport driving forces arising 
from local concentration gradients. In any case, the existence of exothermal or 
endothermal  processes makes the general solution of adsorption  kinetics very 
complex. Because a general presentation of this problem is beyond the scope of 
this review, we invite those interested in more details to read reference 
documents such as Chapter 6 of Ref.~\citen{karger_diffusion_2012}. In what 
follows, we provide the main elements by considering the following situation. 
We assume that the timescale for heat dissipation within the porous solid and  
adsorbed phase is short compared to that involved in adsorption kinetics so 
that the problem can be assumed to be isothermal. Moreover, by restricting the 
problem to low loadings and small concentration changes inducing sorption 
kinetics, the diffusivity can be assumed to be constant (\ie.\ no 
loading dependence of the transport coefficients).
In the rest of this subsection, we present in the following 
order: diffusion at and across the interface between the porous solid and the 
fluid external phase, diffusion within the host porous solid, and diffusion in 
the external fluid phase.
The coupling between these different mechanisms can be also treated formally 
with different possible combinations (namely, combination of surface diffusion 
and in-pore diffusion or combination of in-pore diffusion and diffusion in the 
external phase). For the sake of clarity, we do not treat these coupled regimes 
in what follows as specialized documents are available on the topic. However, 
the specific example of combined surface and in pore diffusion is discussed in 
\Cref{sec:transportPore_kinetics_barriers} where available molecular 
simulation techniques to probe transport mechanisms and diffusion barriers are 
presented.

\noindent \textbf{Transfer at the interface.} In what follows, $\rho$ denotes the fluid density or concentration in the bulk external 
phase while $n$ refers to the density (concentration) of the adsorbed phase in 
the porous solid. Initially, the porous solid is filled with an adsorbed phase 
having a density $n_0$ in equilibrium with the external fluid at a density 
$\rho_0$. At a time $t = 0$, the density in the external fluid is 
increased (or decreased) from $\rho_0$ to $\rho$. Such a density change 
is induced by a chemical potential change, $\Delta \mu$. Fick’s first law, 
which was introduced in \cref{chap:fundamentals}, allows writing: 
\begin{equation}
\frac{\textrm{d}n}{\textrm{d}t} =  \frac{\rho k_s}{k_\textrm{B}T} \frac{\Delta \mu}{a} 
\sim  \rho k_s \frac{\Delta \rho}{a} 
\label{EQ6.31}
\end{equation}
where the last expression was obtained by considering the low density regime 
where $\mu \sim k_\textrm{B}T \ln \rho$. In the equation above, $k_s$ is 
the mass transfer coefficient --- related to the surface diffusivity --- while 
$a$ is the thickness of the interface across which diffusion occurs. Typically, 
$a$ depends on the nature of the surface diffusion process which can be of 
thermodynamic or geometrical nature. On the one hand, diffusion at the solid 
external surface can be limited by the so-called fluid external film 
resistance, \ie.\ the diffusion of adsorbing molecules through the 
laminar flow surrounding the solid particles. In this case, $a$ is simply the 
volume to surface ratio of the solid particles with $a = l$ for platelet 
particles of a width $2l$ and $a = R_\textrm{p}/3$ for a spherical particle of 
radius  $R_\textrm{p}$.
On the other hand, if diffusion is limited by geometrical 
constraints such as pore partial or complete obstruction, $a \sim \delta$ 
where $\delta$ is the width over which these defects are present at the solid 
external surface.

Generally speaking, surface diffusion can involve diffusion limitations 
because of external fluid resistance and geometrical resistance. 
Because these two resistances act in series, the corresponding mass transfer 
coefficients, $k_f$ and $k_g$, should be combined into an 
effective mass transfer coefficient $\overline{k}_s$: 
\begin{equation}
\frac{1}{\overline{k_s}} =  \frac{1}{k_f} +  \frac{1}{k_g}   
\label{EQ6.32}
\end{equation}
This equation was derived in \Cref{chap:diffusion_network} when considering the 
effective diffusivity in composite materials with domains aligned in series 
with respect to the flow direction. Henry’s regime is a special situation where 
\cref{EQ6.31} can be simplified~\cite{coasne_gas_2019}. In this regime, valid 
in the limit of very small $\rho$, the density of the adsorbed 
phase is proportional to its bulk counterpart \ie.\ $n = K_\mathrm{H} \rho$ 
where $K_\mathrm{H}$ is the so-called Henry constant. Using this relationship 
into \cref{EQ6.31}, we obtain $\textrm{d}n/\textrm{d}t = k_s/K_\mathrm{H} 
\times \Delta n /a$. 
If we use the initial conditions $\rho = 0$ for $t \leq 0$, this 
differential equation leads to the following solution: 
\begin{equation}
n(t) = K_\mathrm{H} \rho \left(1 - \exp \left[ - \frac{k_s t}{a} \right] \right)
\label{EQ6.33}
\end{equation}

\noindent \textbf{In-pore diffusion.}
In-pore diffusion, often termed as intra-particle diffusion, corresponds to the
diffusion of the adsorbed molecules within the porosity of the host solid.
The solution of the diffusion equation leads to different equations depending on
the geometry of the porous solid particle.
For spherical particles, the diffusion equation can be written in spherical
coordinates as: 
\begin{equation}
\frac{\partial n}{\partial t} 
= \frac{1}{r^2} \frac{\partial}{\partial r} 
\left( r^2 D_\mathrm{T} \frac{\partial n}{\partial r} \right) 
\sim  D_\mathrm{s}\left( \frac{\partial^2 n}{\partial r^2} + \frac{2}{r} \frac{\partial 
n}{\partial r} \right) 
\label{EQ6.34}
\end{equation}
where $D_\mathrm{T}=D_0(\partial \ln f / \partial \ln \rho)_T$ is the transport 
diffusivity as introduced in \cref{EQ3.44}. In the second equality, the 
self-diffusivity was assumed to be constant upon changing the density $n$. The 
latter is in general valid in the infinite dilution limit where, as discussed 
in \cref{sec:selfColTransportD}, the Darken factor $(\partial \ln f / \partial 
\ln \rho)_T=1$ is constant and $D_0=D_\mathrm{s}$. As a result, upon transient 
sorption kinetics, $D_\mathrm{T}=D_0=D_\mathrm{s}$ does not depend on the 
location $r$ within the porosity and can therefore be taken out of the 
differential equation. Starting from a configuration filled with an adsorbed 
phase $n_0$ in equilibrium with the bulk fluid at a density $\rho_0$, a sudden 
change from $\rho_0$ to $\rho$ at a time $t = 0$ induces a change in the 
adsorbed  density $n(t)$ described by \cref{EQ6.34}.
In many cases, the sorption starts from $\rho_0 \sim 0$ in which case the 
infinite dilution limit on the right hand side of \cref{EQ6.34} is justified. 
Using constant $D_\mathrm{s}$, the solution can be obtained 
as~\cite{jaeger1959conduction}:
\begin{equation}
\frac{n(t) - n_0}{n(\infty) - n_0} =  
1 - \frac{6}{\pi^2} \sum_{m = 1}^{\infty}  \frac{1}{m^2} \exp 
\left[   - \frac{m^2 \pi^2 D_\mathrm{s} t}{R^2}  \right]
\label{EQ6.35}
\end{equation}
where $R$ is the radius of the spherical particle and $n(\infty)$ is the 
solution at equilibrium ($t \rightarrow \infty$). We recall that this solution 
is only valid for solid particles having a spherical shape. In the short time 
limit $t \rightarrow 0$, \cref{EQ6.35} leads to the solution $n(t)-n_0 \sim 
\left[n(\infty)-n_0\right] \times  6\sqrt{D_\mathrm{s} t/\pi R^2}$ with a 
typical dependence in $t^{1/2}$ as expected for diffusion. In contrast, in the 
long time limit $t \rightarrow \infty$, \cref{EQ6.35} can be approximated by 
keeping only the leading term $m = 1$ in the series. This leads to $n(t)-n_0 
\sim \left[n(\infty)-n_0\right] \times  (1-6/\pi^2 \exp [-\pi^2 D_\mathrm{s} 
t/R^2])$ where the exponentially decaying term describes the fact that the 
adsorption kinetics slows down as the system reaches asymptotically equilibrium 
(leading to a vanishing driving force $\Delta n$ for adsorption).

\begin{figure}[htb]
	\centering
	\includegraphics[width=0.95\linewidth]{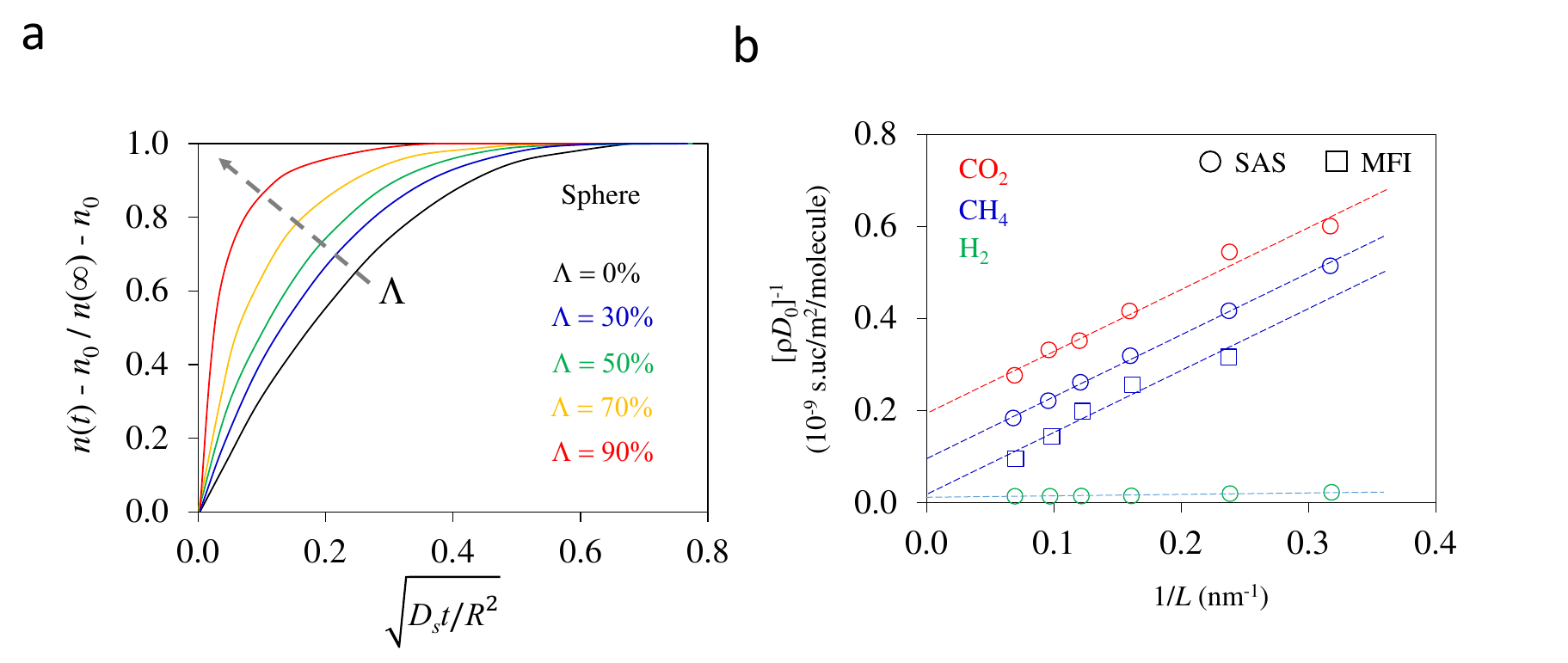}
	\caption{\textbf{Mass uptake and surface resistance}. (a) Theoretical mass 
	uptake showing the change in the adsorbed density $n(t)$ as a function of 
	time normalized to the change in the adsorbed density reached at 
	equilibrium ($t \rightarrow \infty$). These data are predicted by solving 
	for different $\Lambda$ the diffusion equation for a porous particle having 
	a spherical geometry. The parameter $\Lambda$ corresponds to the 
	concentration change in the bulk fluid between the initial and final states 
	(\textit{see text}). Adapted from Ref.~\protect\citen{karger_diffusion_2012}.
	Copyright 2012 Wiley-VCH.
	(b) Effective collective diffusivity $D_0$ (proportional to the permeance $K$) as a 
	function of the membrane thickness $L$ for gas transport across a zeolite 
	membrane. The data show molecular simulation results for different zeolites 
	(circles for SAS and squares for MFI) and gases (red for carbon dioxide, 
	blue for methane, and green for hydrogen). The surface resistance to 
	transport is obtained from such data (\textit{see text}).
	The density is given in molecules/unit cell (uc).
	Adapted with permission from Ref.~\protect\citen{dutta_interfacial_2018}.
	Copyright 2018 Royal Society of Chemistry.}
	\label{fig:6_6}
\end{figure}

In the above derivation, it was assumed that the concentration $\rho$ in the 
external phase remains constant. In practice, if the porous solid has a 
large adsorption capacity, this implies that the volume of the external phase 
has to be large enough to verify this condition of constant external density. 
However, as discussed in Ref.~\citen{karger_diffusion_2012}, if this condition 
is not verified, the general solution of \cref{EQ6.34} is given by: 
\begin{equation}
\frac{n(t) - n_0}{n(\infty) - n_0} =  
1 - 6 \sum_{m = 1}^{\infty}  \frac{\exp[-D_\mathrm{s} p^2_m t/R^2]}{9\Lambda/(1 - \Lambda) + (1 - \Lambda)p_m^2} 
\label{EQ6.36}
\end{equation}
where $\Lambda= (\rho-\rho_\infty)/(\rho-\rho_0)$ and $\rho $ is the 
density imposed in the bulk fluid at $t = 0$ while $\rho_\infty$ is the 
residual bulk density when the adsorption process is complete. In practice, in 
the previous solution where the bulk density  was assumed to remain constant, 
$\Lambda = 0$. In fact, $\Lambda$ can be seen as the fraction of fluid 
molecules added upon the density change during adsorption in the porous solid 
when equilibrium is approached. As for the constants $p_m$, they correspond to 
the non zero solutions of the equation $\tan p_m = 3 p_m/[3 + (1/\Lambda-1) 
p_m^2]$. \Cref{fig:6_6}(a) shows the result of \cref{EQ6.36} for different 
values of $\Lambda$. As expected, the short and long time results show the 
expected asymptotic solutions with $\sim t^{1/2}$ for $t 
\rightarrow 0$ and $\exp(-t)$ for $t \rightarrow \infty$.

\noindent \textbf{External diffusion.} If the diffusion is limited by the  diffusion in the external fluid phase, a different mass balance equation must 
be considered. \redlork{In this context, we stress that external diffusion includes diffusion in the system's macroporosity -- in fact, external diffusion refers to all molecules that can be treated as equivalent to those in the bulk external phase (therefore, this definition applies to any molecules located in regions where the contribution from surface adsorption is negligible.} Instead of \cref{EQ6.34}, the overall mass conservation for the 
adsorbed and external phases writes: 
\begin{equation}
 \epsilon_\textrm{p} \frac{\partial \rho}{\partial t}
= \epsilon_\textrm{p} D_\mathrm{s}^0 \left[\frac{\partial \rho^2}{\partial r^2} 
+ 
\frac{2}{r}  \frac{\partial \rho}{\partial r} \right]   -(1 -\epsilon_\textrm{p}) \frac{\partial n}{\partial t}  
\label{EQ6.37}
\end{equation}
where $D_\mathrm{s}^0$ is the self-diffusivity of the external phase (which is 
assumed to 
be density independent). $\epsilon_\textrm{p}$ is the external porosity which is 
proportional to the volume accessible to the external phase. The last term in Eq. (\ref{EQ6.37}) is a sink term that describes the loss of molecules in the fluid due to mass uptake of the solid fraction. For a system in 
the low  density region, \ie.\ in the Henry regime, the adsorbed 
density is $n = K_\mathrm{H}\rho$ where $K_\mathrm{H}$ is the Henry constant. Using this 
relationship, \cref{EQ6.37} can be recast as: 
\begin{equation}
\frac{\partial \rho}{\partial t} = 
\frac{\epsilon_\textrm{p} D_\mathrm{s}^0}{\epsilon_\textrm{p} + (1 
-\epsilon_\textrm{p})K} 
\left( 
\frac{\partial \rho^2}{\partial r^2} + \frac{2}{r}  \frac{\partial 
\rho}{\partial r}
\right)
\label{EQ6.38}
\end{equation}
Such an equation leads to the same expression as in \cref{EQ6.35} but with the 
self-diffusivity of the confined phase, $D_\mathrm{s}$, replaced by an 
effective bulk diffusivity, 
$\overline{D_\textrm{p}^0} = \epsilon_\textrm{p}/[\epsilon_\textrm{p} + 
(1-\epsilon_\textrm{p})K_\mathrm{H}]$. Because of the equivalence between 
\cref{EQ6.38,EQ6.34}, the analytical solution given in \cref{EQ6.35} 
remains valid but $D_\mathrm{s}/R^2$ must be replaced by 
$\overline{D_\textrm{p}^0}/R_\textrm{p}^2$.

\subsubsection{Transport barriers}
\label{sec:transportPore_kinetics_barriers}

Like with experimental set-ups, there are different molecular simulation 
strategies that can be undertaken to identify barriers in transport of confined 
fluids. In experiments, possible surface barriers can be probed either under 
stationary flow conditions or under transient kinetic regimes. In more detail, 
while transport barriers in sorption kinetics are probed using fluid uptake 
measurements as described in the previous section, those involved in stationary 
regimes will manifest themselves in experiments at equilibrium such as 
pulsed-field gradient NMR or neutron scattering (at the mesoscopic $\sim \mu$m 
lengthscale for the former and at the $\sim $ nm scale for the latter). 
Similarly, in molecular simulation, transport barriers will be identified under 
different conditions but always at the microscopic level because of the 
intrinsic molecular scale involved in these approaches. 
As will be discussed in this section, such techniques include at 
equilibrium~\cite{combariza_influence_2011,crabtree_simulation_2013,
	sastre_molecular_2018} and 
non equilibrium~\cite{zimmerman_selection_2011,dutta_interfacial_2018,
liu_interfacial_2016} molecular dynamics simulations but also Monte 
Carlo~\cite{zimmermann_surface_2012} and free energy~\cite{lee_activated_2016}
approaches.
Finally, while this is not treated here for the sake of simplicity, we also 
emphasize that some authors have investigated the problem of thermal effects 
on adsorption/desorption~\cite{inzoli_thermal_2007,simon_adsorption_2014}.
In what follows, molecular simulation strategies that can be 
used to probe surface barriers in stationary regimes are first considered 
before treating transient regimes. The end of this subsection is devoted to the 
important problem of activated transport where surface barriers lead to 
phenomena over timescales beyond those accessible with molecular simulation. 
Other molecular simulation aspects that will not be covered below include the 
investigation of geometrical surface effects in biomimetic 
systems~\cite{gravelle_optimizing_2013} and sub- or super-additive effects when 
considering pore array systems~\cite{gadaleta_sub-additive_2014}. 
Only a few representative works have been selected rather than a complete
overview of the molecular simulation literature available on the topic and we
acknowledge that other significant contributions have been reported on phenomena
involving surface barriers in transport of nanoconfined fluids.

\noindent \textbf{Transport barrier in stationary flow.} Using molecular 
simulation approaches, Bhatia and coworkers have considered in detail the 
surface barriers involved in the transport of fluids through various porous 
media including carbon nanotube~\cite{liu_interfacial_2016,liu_effects_2018} 
and zeolite~\cite{dutta_interfacial_2018} membranes.
In their study on CO$_2$ and CH$_4$ transport across a zeolite 
membrane, these authors investigated in depth the existence of surface barriers 
for various zeolite structures and temperatures~\cite{dutta_interfacial_2018}. 
The strategy consists of simulating the stationary 
flow across a zeolite membrane of thickness $L$ as induced by a chemical 
potential gradient $\nabla \mu$ at constant temperature $T$. In practice, as 
already discussed earlier, the chemical potential gradient is simulated by 
applying to each molecule the corresponding force, $f = -\nabla \mu$. In the 
framework of Onsager theory of transport, provided the driving force remains 
small (typically $\Delta \mu < k_\textrm{B}T$), the flux should be proportional 
to the force inducing transport,
$J = - \rho D_0/k_\textrm{B}T \times \nabla \mu$, where $D_0$ is the 
collective diffusivity and $\rho$ is the density of the confined fluid (we 
recall that $K = D_0/\rho k_\textrm{B}T$ is the permeance as defined in 
flow experiments). For an infinite zeolite crystal, such molecular 
simulations allow determining the intrinsic collective diffusivity $D_0$ which 
can be compared to its experimental part obtained using coherent neutron 
scattering. However, for a system with an interface as considered in 
Ref.~\citen{dutta_interfacial_2018}, because the surface 
barrier --- if any --- acts in series with respect to transport within the 
zeolite membrane, one expects the effective collective diffusivity, 
$\overline{D_0}$, to obey the following combination rule (see 
derivation in \Cref{chap:diffusion_network}):
\begin{equation}
\frac{L}{\rho A \overline{D_0}} = \frac{L}{\rho A D_0 } + R_s
\label{EQ6.39}
\end{equation}
where $L$ is the membrane thickness and $A$ the membrane section area through 
which fluid transport occurs. $R_s$ is the surface resistance barrier which 
subdivides into an entrance surface resistance and an exit surface resistance, 
$R_s = R_{in} + R_{out}$ (these two resistances add up because they are in 
series). Using the simple formalism above, by probing the effective collective 
diffusivity as a function of the membrane thickness, Dutta and Bhatia 
determined the surface resistance for different gases, temperatures, 
\etc. In more detail, by rewriting \cref{EQ6.39} as $(\rho 
\overline{D_0})^{-1} = (\rho D_0)^{-1}  + AR_s/L$, the surface resistance $R_s$ 
can be readily obtained from the data shown in \cref{fig:6_6}(b).
As expected, significant differences can be observed between different 
gases/zeolites with small gases such as H$_2$ being unlikely prone to 
surface barriers at room temperature owing to its small size and weak molecular interactions.

\begin{figure}[htb]
	\centering
	\includegraphics[width=0.95\linewidth]{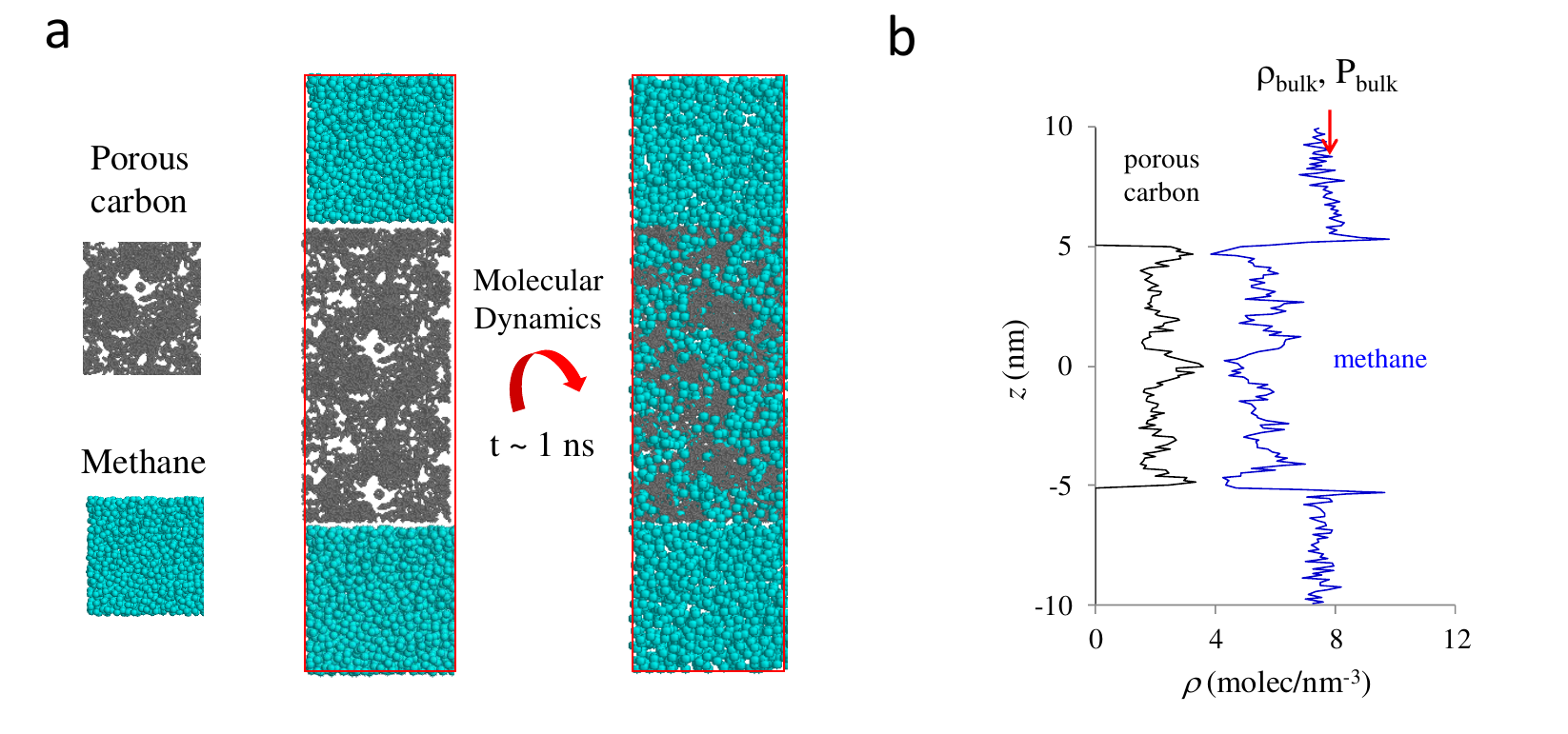}
	\caption{\textbf{Mass transfer}. (a) Mass transfer techniques can be used 
	to investigate the adsorption kinetics while accounting for 
	inaccessibility issues (\ie.\ the fact that part of the porosity 
	might be closed and, therefore, inaccessible to an external fluid 
	phase). 
	In practice, a reservoir of fluid molecules at an initial pressure $P_0$ 
	and temperature $T$ is set in contact with a porous solid. A molecular 
	dynamics simulation is then performed so that the fluid molecules diffuse 
	through the fluid/solid interface to fill the porosity --- except for 
	porous regions 	inaccessible to the fluid phase on a timescale 
	corresponding to the typical simulation run (\SI{1}{\nano\second} -- 
	\SI{1}{\micro\second}).
	(b) By monitoring the fluid density profile within the porous solid, one 
	can measure the mass uptake $N(t)$ as a function of time $t$ (which is 
	obtained by integrating the fluid density profile between boundaries 
	corresponding to the external surfaces of the host porous solid). 
	Eventually, in the long time limit, the adsorbed amount at equilibrium is 
	readily obtained by plotting $N(t \rightarrow \infty)$ as a function of the 
	final pressure $P_{\textrm{bulk}}$ reached in the fluid reservoir (the 
	latter can be determined from the density $\rho_{\textrm{bulk}}$ using the 
	bulk equation of state at the temperature $T$).}
	\label{fig:6_7}
\end{figure}

\noindent \textbf{Transport barrier in transient regime.}  In the previous 
example, a stationary flow was considered as would be involved in 
chromatography experiments for instance. However, in many relevant experimental 
situations, probing transport barriers involved in adsorption kinetics 
is very important. A rather straightforward molecular simulation method, which 
can be easily implemented, is illustrated in \cref{fig:6_7}(a) (see 
Refs.~\citen{combariza_influence_2011,falk_effect_2015,sastre_molecular_2018} 
for examples where this technique was successfully applied). 
To determine the fluid adsorption kinetics in a given porous solid 
(here, a disordered porous carbon), one prepares a porous solid film of a given 
thickness $L$ in contact with two fluid reservoirs under given thermodynamic 
conditions $T$, $P_0$ (the fluid density is $\rho_0$). First, in the framework 
of molecular dynamics simulations at constant $T$, reflective boundary 
conditions between the fluid reservoirs and the porous solid are used so that 
the fluid molecules cannot invade the solid porosity. At a time $t = 0$, these 
boundary conditions are removed so that the fluid invades the 
porosity --- because Molecular Dynamics simulations at constant temperature $T$ 
are used, the temperature in the bulk and confined fluids remain constant at 
all times. 
On the other hand, the pressure $P(t)$ in the bulk fluid decreases with time as 
the number of molecules $N(t)$ within the porosity increases. By monitoring the 
density profile in the porous solid as shown in \cref{fig:6_7}(b), $N(t)$ as a 
function of $t$ can be determined. This allows probing the adsorption kinetics 
under conditions similar to experimental situations. Indeed, in many cases, the 
experimental sorption kinetics is determined using finite reservoir conditions 
by measuring the decrease in the bulk external pressure (or molecule 
concentration for liquid phase adsorption). In particular, for such final 
reservoir conditions, as shown in Ref.~\citen{karger_diffusion_2012}, the 
solution of the diffusion equation with or without surface transport barriers 
is known so that adsorption kinetics data can be used to quantitatively assess 
the different transport mechanisms at play.

As an important consistency check, in the long time regime, the value of $N(t 
\rightarrow \infty)$ as a function of the residual pressure in the reservoir 
$P(t \rightarrow \infty)$ must fall onto the adsorption isotherm $N(P)$ (which 
can be determined using standard molecular simulations such as Monte Carlo 
simulations in the Grand Canonical ensemble). In addition to providing useful 
information regarding the adsorption kinetics for a given fluid/solid system, 
the simulation strategy described above has another important merit. It allows 
one to probe accessibility issues which can be severe and lead to much lower fluid 
concentrations within the porous solid compared to those obtained using methods 
that ignore such limitations. Indeed, as shown in 
Ref.~\citen{bousige_realistic_2016}, in case of complex disordered porous 
materials, many pores can  be accessible through very narrow necks. In 
practice, depending on the timescale probed in experiments, these cavities can 
be considered inaccessible. In that sense, the simple simulation method 
proposed above allows accounting for such inaccessibility issues. However, when 
using this method for such purpose, it should be kept in mind that only 
barriers inaccessible on very short time scales are probed --- \ie.\ the 
typical time scale accessible to molecular simulation, roughly 
\SI{1}{\nano\second} -- \SI{1}{\micro\second}. In the context of the 
transition state theory described in \Cref{chap:diffusion_pore}, if a molecular 
dynamics simulation is performed over a time $t_\mathrm{run}$ at a temperature 
$T$, this means that only cavities accessible through a free energy barrier 
$\Delta F^\ast$ such that $t_\mathrm{run} > \tau_0 \exp[\Delta F^\ast /k_\textrm{B}T]$ 
will be visited on average (we recall that $\tau_0$ is the characteristic 
molecular time of the order of $10^{-12}$ s).
This is an important limitation of the method proposed above when probing pore 
inaccessibility issues in porous solids. In the next subsection, we address 
this problem by considering free energy techniques which are specifically 
designed to treat the complex problem of such free energy barriers.

 \begin{figure}[htbp]
	\centering
	\includegraphics[width=0.95\linewidth]{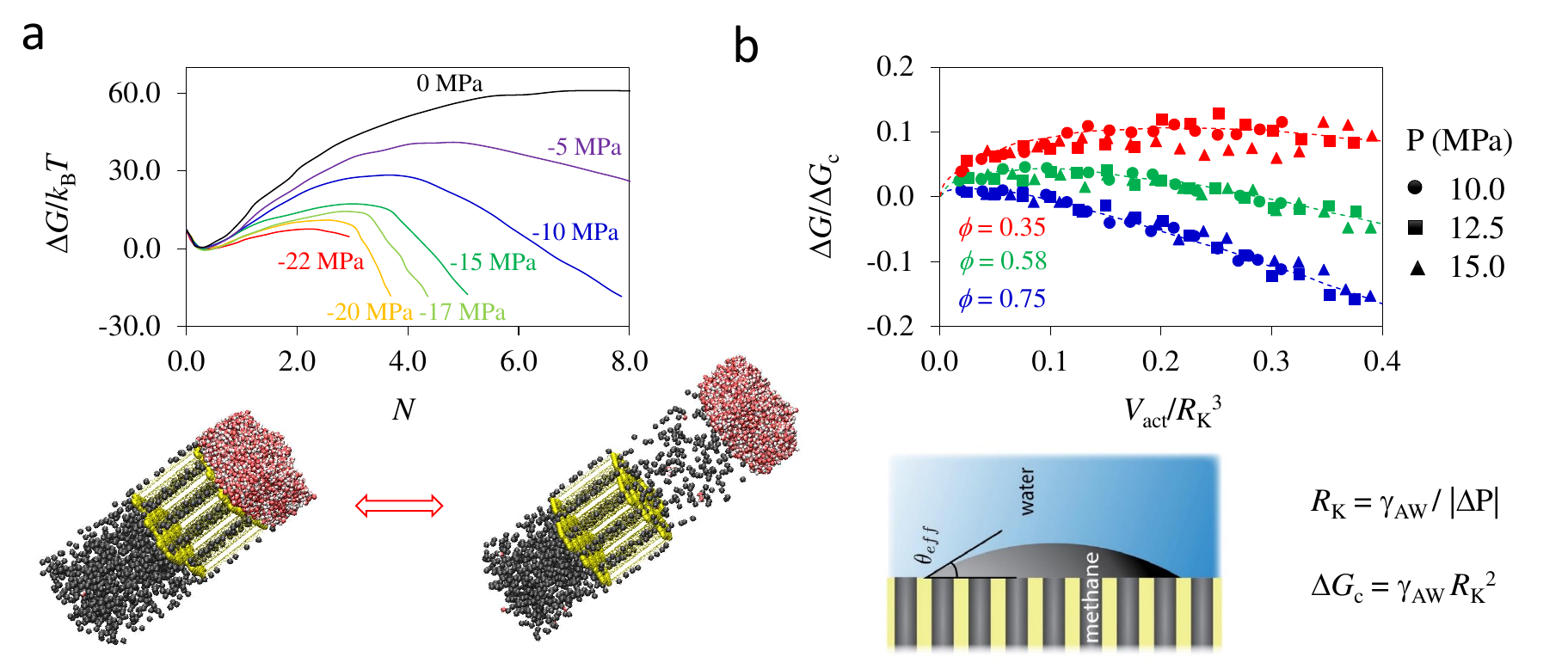}
	\caption{\textbf{Activated desorption kinetics}. (a) Free energy $\Delta 
	G/k_{\textrm{B}}T$ as determined using molecular simulation combined with 
	umbrella sampling as a function of the amount of methane $N$ extracted per 
	unit of surface area. These data are obtained for different pressure 
	differences $\Delta P$ as indicated in the graph. The free energy is 
	shifted with respect to its value $N \sim 0$ when most methane is trapped 
	within the porous material. The initial ($N \sim 0$, before desorption) and 
	final ($N \gg 0$, after desorption) states are illustrated in the typical 
	molecular configurations provided below the graph. The porous carbon 
	hosting initially the methane molecules is a nanoporous membrane made up of 
	carbon nanotubes having a pore radius $r \sim 0.59$ nm and separated from 
	each other by a pore spacing $D_\mathrm{p} \sim 1.7$ nm. (b) Free energy $\Delta G$ 
	for the methane nucleus formed at the surface of the carbon nanotube 
	membrane covered by water as a function of its volume $V_{\textrm{act}}$. 
	The free energy $\Delta G$ is normalized to the value $\Delta G_c = 
	\gamma_{\textrm{AW}}R_\textrm{K}^2$ while the volume $V_{\textrm{act}}$ is 
	normalized by $R_\textrm{K}^3$ ($R_\textrm{K}$ is the Kelvin radius 
	$R_\textrm{K} = \gamma_{\textrm{AW}}/\lvert \Delta P \lvert$). These data 
	were obtained by minimizing the surface energy for several pressure 
	differences as indicated in the graph ($\Delta P =$ 10, 12.5 and 15 MPa 
	which correspond to different symbols). The different colors correspond to 
	different membrane porosities $\phi$, \ie.\ pore spacings as indicated in 
	the graph. For a given porosity, all data corresponding to various pressure 
	drops collapse onto a single curve. The solid lines are the free 
	energy predicted for a spherical cap geometry with an effective contact 
	angle $\theta_\mathrm{eff}$. The geometry of the contact angle formed by 
	the methane spherical cap (grey phase) at the interface between the carbon 
	nanotube membrane (yellow) and water (blue) is schematically illustrated 
	below the graph. Adapted with permission from Ref.~\protect\citen{lee_activated_2016}.
	Copyright 2016 Macmillan Publishers Limited under Creative Commons
	Attribution 4.0 International License
	(http://creativecommons.org/licenses/by/4.0/).
	}
	\label{fig:6_8}
\end{figure}

\noindent \textbf{Activated transport.}
The molecular simulation methods proposed above implicitly assume that 
transport limitations induced by surface barriers occur over timescales shorter 
than those probed (typically, \SI{1}{\nano\second} -- \SI{1}{\micro\second} 
with molecular dynamics). While this constraint is often verified for many 
examples of fluid transport across porous membranes, there is a number of 
situations where this criterion is not met. In what follows, we provide an 
example of such complex timescale issues taken from 
Ref.~\cite{lee_activated_2016}. While 
the situation below might appear as a  specific situation at first, it 
pertains to the general physics of fluid immiscibility in confined 
environments which is relevant to both basic and applied sciences. Moreover, it 
should be emphasized that the general strategy below, which relies on free 
energy techniques such as umbrella sampling, can be applied to 
any transport situation where a non negligible free energy barrier leads to 
activated transport. As shown in Ref.~\citen{lee_activated_2016}, such activated 
transport can lead to very intriguing phenomena which span several orders of 
magnitude in time and length scales.

Let us consider a porous membrane --- made of an array of carbon nanotubes for 
the sake of simplicity --- filled with methane at a given temperature $T$ and 
pressure $P$ as shown at the bottom of \cref{fig:6_8}(a). On the left side of 
the membrane, a piston is pressurizing a methane bulk reservoir in contact with 
confined methane to maintain the pressure $P_\uparrow$ and therefore the 
chemical potential constant in the system. On the right side of the system, the 
membrane is fully wetted by water which is maintained at constant pressure 
through a second piston at a pressure $P_\downarrow$.  At equilibrium, when the 
two pressures are equal, \ie.\ $P_\uparrow = P_\downarrow$, the 
confined fluid is at rest.
Because of the very low solubility of methane in water, the former remains 
confined within the porous membrane while water remains outside the membrane.
Using Molecular Dynamics simulations at constant temperature, methane 
desorption induced by a pressure drop was simulated by lowering $P_\downarrow$ 
to a value inferior to $P_\uparrow$. Even for large pressure drops, the amount 
of desorbed methane as a function of time was found to remain constant for a 
long time before extraction. Moreover, by repeating the same simulation from 
different initial configurations, the retention time $\tau_{\textrm{act}}$ 
prior to extraction was found to be non reproducible. However, on average, 
$\tau_{\textrm{act}}$ was found to scale as $\tau_{\textrm{act}} \sim 
\exp{(v^\star\Delta P/k_\mathrm{B}T)}$ where  $v^\star \sim 1.2$ nm$^3$ is a characteristic
molecular volume. This scaling relation suggests that methane desorption 
through the wet external surface corresponds to an activated 
process. Even for large pressure drops, $\Delta P = P_\uparrow - P_\downarrow 
\sim 10 - 25$ MPa, $\tau_{\textrm{act}}$ is of the order of $1$ -- 
\SI{100}{\nano\second}. 
Clearly, this timescale falls in the range of Molecular Dynamics simulation 
times routinely available. As discussed above, this raises the question of the 
relevance of short Molecular Dynamics runs to probe such activated processes.

For such activated transport, being either relevant to adsorption or 
desorption, free energy techniques such as umbrella sampling offer a mean to 
probe transport mechanisms at play even though they involve timescales 
beyond the microscopic and mesoscopic times. In practice, such simulations 
consist of biasing the simulation algorithm to force the system to explore 
large free energy regions of the phase space (for details, the reader is 
referred to specialized textbooks such as 
Ref.~\citen{frenkel_understanding_2001}).  
\cref{fig:6_8}(a) shows the free energy $\Delta G/k_\textrm{B}T$ as a function 
of the amount of desorbed methane $N$ for different pressure drops $\Delta P$. 
As expected, for $\Delta P = 0$, the stable solution corresponds to $N = 0$ (no 
desorbed methane). In contrast, for large  $\Delta P$, the stable solution 
corresponds to $N \neq 0$ as the strong driving force leads to methane 
desorption. By probing different surface areas, pore diameters, \etc., 
it was shown that the free energy barrier involved in the activated desorption 
process illustrated in \cref{fig:6_8}(a) corresponds to the free energy 
cost of replacing the initial membrane/water interface [left configuration at 
the bottom of \cref{fig:6_8}(a)] by a membrane/methane interface and 
methane/water interfaces [right configuration at the bottom of
\cref{fig:6_8}(a)].
Using a mesoscopic model, it was shown that the critical nucleus, which allows 
triggering the desorption of methane through the wet external surface, 
corresponds to the hemispherical cap shown in \cref{fig:6_8}(b). As expected 
in the classical nucleation theory, which allows describing activated 
processes, this nucleus shape was obtained by minimizing the surface energy at 
constant volume.  The free energy $\Delta G$ of the methane nucleus is shown in 
\cref{fig:6_8}(b) as a function of its volume $V_{\textrm{act}}$ for different 
membrane porosities $\phi$ and pressures drops 
$\Delta P$. For each system, the solution corresponds to a hemisphical cap of 
methane at the pore surface as illustrated in the bottom panel of 
\cref{fig:6_8}(b). The effective contact angle of this critical nucleus, 
$\theta_\mathrm{eff}$, is close to the solultion provided by the Cassie--Baxter 
equation  (this model describes the contact angle on the porous surface as a 
linear combination of the contact angles on the different regions 
\ie.\ porous and non porous surfaces).
As a confirmation that this model rigorously captures 
the physics of the activated desorption observed in the molecular simulations, the 
contact angle $\theta_\mathrm{eff}$ was found to lead to free energies that are 
consistent with those determined using the complete numerical calculations as 
shown in \cref{fig:6_8}(b).

%%% Chapter 7: Diffusion at the porous network scale
%!TeX spellcheck = en_US
%!TeX encoding = utf8 
%!TeX program = pdflatex
%!TeX root = ../manuscript.tex

\section{Gradient-driven transport in porous networks} 
\label{chap:transport_network}

\subsection{Coarse-grained models} 

\subsubsection{Lattice Boltzmann method}

\noindent \textbf{Boltzmann equation.} 
While Onsager theory of transport provides a macroscopic framework to 
describe transport in porous media by defining transport, or response
coefficients more in general, it does not rely on a microscopic description of
the underlying dynamics.
In contrast, statistical physics provides a set of descriptions --- typically
based on the Langevin, Boltzmann, and Fokker--Planck equations --- which adopt a
microscopic point of view by relying on the probability density
$f(\mathbf{r},\mathbf{v},t)$. 
Accordingly,
$f(\mathbf{r},\mathbf{v},t) \textrm{d}\mathbf{r}^3 \textrm{d}\mathbf{v}^3$ is 
the probability that, at a time $t$, a molecule is located in a small volume 
element $\textrm{d}\mathbf{r}^3$ around $\mathbf{r}$ with a velocity in an 
element $\textrm{d}\mathbf{v}^3$ around $\mathbf{v}$.
In the following, we focus on the Boltzmann equation, which was the first
equation in which the dynamics of dilute gas molecules was described using the
distribution function $f(\mathbf{r}, \mathbf{v}, t)$
long before the  derivation of the Fokker--Planck equation.
The Boltzmann equation relates the density distribution current $\partial 
f/\partial t$ to the forces exerted on the fluid as well as to the collisions 
between the fluid molecules~\cite{risken_fokker-planck_1996}: 
\begin{equation}
\left(\frac{\partial}{\partial t}  + \mathbf{v} \cdot \nabla + 
\frac{\mathbf{F}(\mathbf{r},t)}{m} \nabla_\mathbf{v} \right) 
f(\mathbf{r}, \mathbf{v}, t) = \left( 
\frac{\partial f}{\partial t} 
\right)_\textrm{coll.}
\label{EQ7.13}
\end{equation}   
where $\nabla$ and $\nabla_v$ are the gradient operators with respect to 
position and velocity, and the second and third terms on the left hand side
refer to the diffusive and force contributions, i.e.\
$\mathbf{F}(\mathbf{r},t)$ is the force acting on the fluid at position
$\mathbf{r}$ and time $t$. 
The right hand side term in \cref{EQ7.13} corresponds to the change in the
distribution function $f$ because of collisions between fluid molecules.
The complexity of the Boltzmann equation, which can be shown to be equivalent to
the Fokker--Planck equation, lies in the collision operator
$(\partial f/\partial t)_\textrm{coll.}$.
Rigorously speaking, this operator can be expressed using 
the scattering cross section 
$\sigma\left(\mathbf{v}_1,\mathbf{v}_2 \lvert \mathbf{v'}_1,\mathbf{v'}_2\right)$
of two molecules colliding at a position $\mathbf{r}$ with velocities 
$\mathbf{v}_1$ and $\mathbf{v}_2$ before the collision and $\mathbf{v'}_1$ and 
$\mathbf{v'}_2$ after the collision:
\begin{multline}
\left( \frac{\partial f}{\partial t} \right)_\textrm{coll.} = 
\int \textrm{d}\mathbf{v}_1 \int \textrm{d} \Theta \left| \mathbf{v}_2 - \mathbf{v}_1 \right| 
\sigma\left(\mathbf{v}_1,\mathbf{v}_2 \lvert \mathbf{v'}_1,\mathbf{v'}_2\right) \times\\
\left[
f(\mathbf{r},\mathbf{v'}_2,t) f(\mathbf{r},\mathbf{v'}_1,t) - 
f(\mathbf{r},\mathbf{v}_2,t) f(\mathbf{r},\mathbf{v}_1,t)  
\right]
\label{EQ7.14}
\end{multline}   
where $\Theta$ is the angle formed by the two vectors $\mathbf{v'}_2 - 
\mathbf{v'}_1$ and  $\mathbf{v}_2 - \mathbf{v}_1$. The two integrals in 
\cref{EQ7.14} denote that all initial velocities $\mathbf{v}_1$ and
$\mathbf{v}_2$ (given through the angle $\Theta$) must be considered.
For  a system at local equilibrium, 
$(\partial f/\partial t)_\textrm{coll.} = 0$, the distribution function must 
follow the Maxwell--Boltzmann distribution $f^\textrm{eq.}$: 
\begin{equation}
f(\mathbf{r},\mathbf{v},t)^\textrm{eq.} =  \frac{\rho}{(2\pi v_T^2)^{d/2}} 
\exp \left[ 
-	\frac{(\mathbf{v} - \mathbf{u} )}{2v_T^2}
\right]
\label{EQ7.15}
\end{equation}   
where $d$ = 1, 2, 3 is the dimension of space, $\rho$ the fluid density, 
$\mathbf{u}$ the local fluid velocity and $v_T = \sqrt{\kbT/m}$ the thermal 
velocity.

\begin{figure}[htbp]
	\centering
	\includegraphics[width=0.95\linewidth]{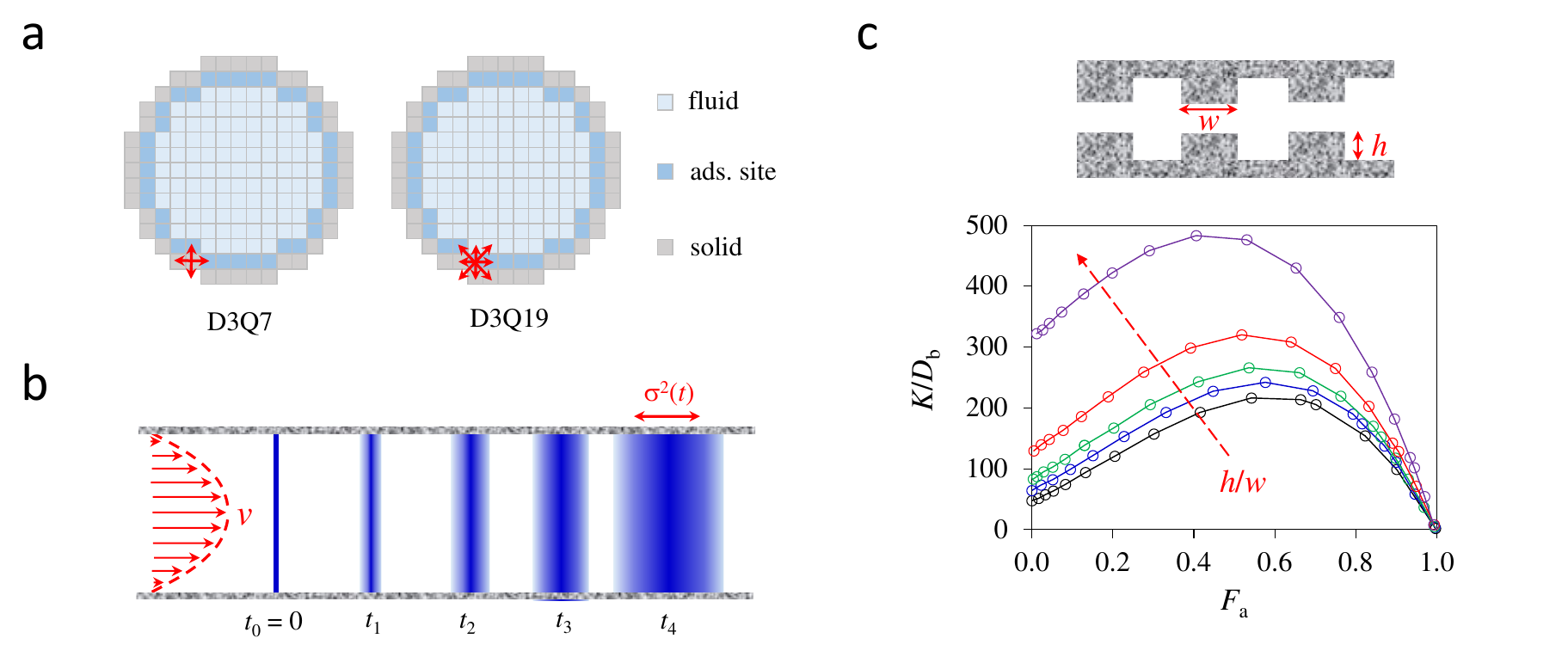}
	\caption{\textbf{Lattice Boltzmann method}. (a) Grid mapping onto a 
	cylindrical pore in Lattice Boltzmann calculations. The grey sites are 
	solid domains which are not accessible to the fluid while the blue sites 
	are accessible to the fluid. For the latter, the dark blue sites are 
	adsorption sites as they are located in the vicinity of the solid surface 
	while the light blue sites are non-adsorption sites (bulk-like, \ie.\ free 
	fluid). Adapted from Ref.~\protect\citen{vanson_interdependence_2016}. In Lattice 
	Boltzmann simulations, different choices can be made to discretize the 
	velocities as indicated by the codes `D3Q7' and `D3Q19' (see text). (b) 
	Taylor dispersion for adsorbing particles dispersed in a liquid flowing 
	into a cylindrical pore. As expected, the liquid flow obeys a parabolic 
	(\ie.\ Poiseuille) velocity profile $v$ with a velocity maximum at the pore 
	center. The concentration profile of adsorbing particles are denoted by the 
	blue code color (with a concentration that increases from light blue to 
	dark blue). At time $t_0 = 0$, a concentration profile of adsorbing 
	particles, where all particles are located in an infinitely narrow stripe 
	in the pore center, is injected in $x = 0$ ($x$ is the position along the 
	pore axis). At a time $t$ later, the concentration profile broadens with a 
	variance $\sigma^2(t)$ that scales linearly with time. Adapted from 
	Ref.~\protect\citen{zaafouri_modelisation_2020}. (c) Dispersion coefficient, $K = 
	\lim_{t \rightarrow \infty} \sigma^2(t)/2t$, normalized to the bulk 
	self-diffusivity of the adsorbing particles, $D_\mathrm{b}$, as a function 
	of the fraction of adsorbed molecules $F_\textrm{a}$. These data are 
	obtained for a slit pore having patterned surfaces as shown in the 
	subfigure above the graph. The pattern corresponds to a crenelated pore and 
	the data shown in the graph are obtained for different aspect ratios $h/w 
	=$ 0, 1, 2, 4, 10. Adapted with permission from Ref.~\protect\citen{vanson_unexpected_2015}.
	Copyright 2015 Royal Society of Chemistry.}
	\label{fig:7_1}
\end{figure}

\noindent \textbf{Discretized Boltzmann equation.} Despite its complexity, the 
Boltzmann equation can be transformed into a robust and efficient numerical 
method to describe the transport of fluids in different 
environments of arbitrary geometry~\cite{ladd_lattice-boltzmann_2001}. This technique, which is known 
as the Lattice Boltzmann method (LBM), relies on a discretized version of the 
Boltzmann equation given in \cref{EQ7.13}~\cite{verberg_accuracy_2001}.
In more detail, with this technique, both the space position 
$\mathbf{r}$ and velocity $\mathbf{v}$ are discretized. 
Spatial discretization is achieved by treating the fluid and solid on a lattice 
so that $\mathbf{r}$ can only take values corresponding to lattice nodes. As 
for the velocities, only a set of velocity vectors are treated by considering 
different propagation directions of the fluid onto the grid.\cite{kruger_lattice_2017}
As illustrated in \cref{fig:7_1}(a), the convention D$_n$Q$_m$ is a convenient meshing 
classification which contains all needed information on the space and velocity 
discretization scheme adopted. On the one hand, the value $n$ indicates the 
space dimension so that $n$ = 1, 2 or 3. On the  other hand, $m$ indicates the 
number of  propagation directions considered. There are only a few values that 
$m$ can take for a given $n$. For instance, when $n = 2$,  the smallest possible
value is $m=5$ (4 nearest neighbor directions and the direction $\mathbf{v} = 0$
for which the fluid density distribution does not propagate) while the
second-smallest value is $m=9$.
In the  D$_n$Q$_m$ convention, these two discretization schemes are referred to
as D2Q5 and D2Q9.
For $n = 3$, available discretization schemes are D3Q7, D3Q15, D3Q19, \etc. 
For a given space dimension $n$, the accuracy of the Lattice Boltzmann method
strongly depends on the discretization scheme adopted.
A trade-off between a large value of $m$ leading to high accuracy and a
reasonable computational time is usually sought for. 
Discretization allows replacing the probability distribution $f(\mathbf{r},
\mathbf{v}, t)$ by its  discretized version $f_i(\mathbf{r}, t)$ where $i$
corresponds to a propagation velocity $\mathbf{c}_i$ so that $i \in [0, m]$ and
$\mathbf{r}$ only takes values corresponding to node positions on the lattice.
With this discretized probability distribution, the Boltzmann equation defined
in \cref{EQ7.13} can be recast as: 
\begin{equation}
f_i(\mathbf{r} + \mathbf{c}_i \Delta t, t + \Delta t) = f_i(\mathbf{r}, t) + 
F_i(\mathbf{r}) + \Omega_i
\label{EQ7.16}
\end{equation}   
where $F_i$ corresponds to the external force acting on the fluid at the node
position $\mathbf{r}$ while $\Omega_i$ is the collision operator introduced
above.

As mentioned earlier, generally speaking, the collision operator $\Omega_i$ is 
of a complex form. However, it is possible to make some simplifications to 
render the  of \cref{EQ7.16} feasible and efficient. In the Lattice 
Boltzmann method, one typically approximates the non-linear collision operator
by the linear BGK operator that makes the local distribution function
$f_i(\mathbf{r}, t)$ relax towards its equilibrium value 
$f_i^{\textrm{eq.}}(\mathbf{r}, t)$ over a characteristic time $\tau$:\cite{kruger_lattice_2017}
\begin{equation}
\Omega_i = - \frac{1}{\tau}  \Big [ f_i(\mathbf{r},t) - f_i^{\textrm{eq.}}(\mathbf{r}, t)
\Big ]
\label{EQ7.17}
\end{equation}   
The BGK operator is the simplest possible collision operator that respects the
fundamental conservation laws of mass, momentum, and total and internal energy.
For small Mach numbers, upon introducing the sound velocity $c_s$, the 
discrete equilibrium distribution can be written as a power series in 
$c_i/c_s$: $f_i^{\textrm{eq.}} = w_i \rho [1 + 1/c_s^2 \times \mathbf{c}_i 
\cdot \mathbf{u} +1/2c_s^4 \times (\mathbf{c}_i \cdot \mathbf{u})^2 - 
u^2/c_s^2]$ ($w_i$ is a coefficient that accounts for the weight of the 
propagation direction $i$ in a given lattice geometry). 
In the Lattice Boltzmann method, the evolution of the 
probability distribution $f_i$ then is determined using \cref{EQ7.16,EQ7.17} which 
allow determining key quantities at each site $\mathbf{r}$ such as the mass 
density $\rho = \sum_i f_i$,  momentum density $\rho \mathbf{u} = \sum_i 
\mathbf{c}_i f_i$ and momentum-flux tensor   $\Pi_{\alpha\beta}  = \sum_i 
c_{i\alpha}c_{i\beta} f_i$. 
For more detailed information on the Lattice Boltzmann method the reader is
referred to the text book by Kr\"uger et al.\cite{kruger_lattice_2017}
The Lattice Boltzmann method is receiving increasing attention owing to its 
broad applicability to fluid flow in porous media or in the vicinity of solid 
particles. 
As a very classical example, \cref{fig:7_1}(b) illustrates the solution obtained
using the Lattice Boltzmann method to the problem of Taylor
dispersion~\cite{zaafouri_modelisation_2020}.
Let us consider a regular slit or cylindrical pore in 
which a liquid (solvent) is subjected to a pressure gradient. As discussed 
earlier, the solution of this simple problem corresponds to Poiseuille flow 
where the liquid velocity profile is parabolic with a maximum value in the pore 
center. At a time $t = 0$, solute particles are injected in the pore at a 
position $x = 0$ with a concentration profile given by a delta function, $c(x) 
= c_0 \delta(x-x^\prime)$ ($x$ is the direction parallel to the pressure gradient 
inducing the liquid flow). The solution of this problem, known as Taylor 
dispersion, is given by a concentration peak being transported at a velocity 
equal to the average liquid velocity while its variance $\sigma^2(t)$ increases 
linearly with time $t$.

\noindent \textbf{Adsorption and moment propagation scheme.}  The Lattice 
Boltzmann method has been extended to account for adsorption effects. For the 
sake of clarity, we here only present the key elements of the formal treatment 
of adsorption/desorption phenomena into the Lattice Boltzmann method (for more 
details, readers are referred to Ref.~\citen{levesque_accounting_2013}).
Such an extension to include adsorption effects relies on the moment 
propagation technique as developed by Lowe and 
Frenkel~\cite{lowe_super_1995,lowe_hydrodynamic_1996}. 
In this method, as discussed in Ref.~\citen{levesque_accounting_2013}, the
probability density $P(\mathbf{r},t)$ at node position $\mathbf{r}$ and time $t$
is propagated to/from adjacent lattice nodes while being modified by
adsorption/desorption.
The time evolution for $P^\star(\mathbf{r},t)$, defined in absence of any
adsorption effects, obeys the following equation (streaming step in the Lattice
Boltzmann method):\cite{kruger_lattice_2017}
\begin{equation}
P^\star(\mathbf{r},t + \Delta t)   = \sum_i  \left[ P(\mathbf{r} - 
\mathbf{c}_i\Delta t, t) p_i(\mathbf{r} - \mathbf{c}_i\Delta t, t)\right] + 
P(\mathbf{r}, t) \left[1 - \sum_i p_i(\mathbf{r}, t)\right]
\label{EQ7.18}
\end{equation}
where the sum runs over all velocity propagation directions $i$ that lead to
site $\mathbf{r}$ and $ p_i(\mathbf{r}, t)$ is the probability that the fluid
particle leaves site $\mathbf{r}$ along the velocity propagation direction $i$
at time $t$.
Physically, \cref{EQ7.18} describes that the quantity $P$ in $\mathbf{r}$ at
time $t+\Delta t$ is given by the sum of two terms: 
(1) The value of $P$ in site $\mathbf{r - \mathbf{c}_i \Delta t}$ at time $t$
multiplied by the probability that fluid particles leave site $\mathbf{r} -
\mathbf{c}_i \Delta t$ in the direction  of site $\mathbf{r}$ (first term).
(2) The value of $P$ minus the fraction of fluid particles that leave this site
(second term).

In order to account for adsorption and desorption, one introduces the probability density 
$P_\textrm{ads}$ associated with the adsorbed particles ($P_\textrm{ads}$ is
only non zero for sites in the vicinity of the solid surface). The evolution of
the quantity $P_\textrm{ads}$ is given by: 
\begin{equation}
P_\textrm{ads}(\mathbf{r},t + \Delta t)   =  P(\mathbf{r}, t) p_a - P_\textrm{ads}(\mathbf{r}, t)  p_d
\label{EQ7.19}
\end{equation}
where $p_a$ and $p_d$ are linked to the probability that fluid molecules get
adsorbed and desorbed, respectively.  The probability density $P$ associated
with free (\ie.\ not adsorbed) fluid molecules then obeys the following 
evolution equation: 
\begin{equation}
P(\mathbf{r},t + \Delta t)   =  P^\star(\mathbf{r}, t) + 
P_\textrm{ads}(\mathbf{r}, t)  p_d
- P(\mathbf{r},t) p_a
\label{EQ7.20}
\end{equation}
where the second term accounts for the adsorbed particles that get desorbed 
while the third term accounts for free molecules that get adsorbed.

The method above was used in different works to simulate the dynamics of fluids 
in porous materials in the presence of adsorption/desorption 
effects~\cite{levesque_accounting_2013,vanson_unexpected_2015}.
As an illustration, \cref{fig:7_1}(c) shows the dispersion coefficient, $K = 
\lim_{t \rightarrow \infty} \sigma^2(t)/2t$, as a function of the fraction of 
adsorbed molecules $F_\textrm{a}$ in  patterned slit pores [$\sigma^2(t)$ is the time-dependent variance of the molecule distribution function as indicated in \cref{fig:7_1}(c)].
Such an effective 
diffusivity displays an optimum in $F_\textrm{a}$ as the dispersion --- which 
reflects the variance of the molecule distribution --- is maximal when the 
fluid molecules subdivide into free molecules being transported by the 
advective flow and adsorbed molecules sticking to the surface (in contrast, for 
$F_\textrm{a} = 0$ or 1, all molecules are either transported with  the 
advective flow or adsorbed at the pore surface). As expected, for a given 
adsorbed molecule fraction $F_\textrm{a}$, the dispersion increases with 
increasing the pore surface area through an increase in the aspect ratio $h/w$ 
of the patterned surface~\cite{vanson_unexpected_2015}.

\subsubsection{Dynamic Monte Carlo}
\label{sec:dynamic_monte_carlo}

Dynamic or kinetic Monte Carlo methods (DMC) extend the classical Monte 
Carlo methods for equilibrium systems to non equilibrium problems such as 
chemical reactions and diffusion.
They are based on the time $\tau$ between jumps in 
the system phase space which corresponds to a hopping rate 
$k \sim \langle \tau \rangle^{-1}$, where the brackets denote the expectation value.
This time can be the chemical reaction rate or the 
typical adsorption/desorption time of a molecule on a surface.
The time between successive transitions $\tau$ is distributed exponentially
around its mean according to $P(\tau) \sim e^{-k\tau}$.
The latter is a direct result from a discrete-time Markov jump process.
\footnote{
	A Markov process is a stochastic model describing a sequence of possible 
	events in which the probability of each future event depends only on the 
	current state.
	Roughly speaking, this Markov property can be characterized as 
	``memorylessness".
	Prominent examples cover random walks, Brownian motion and Markov chain 
	Monte Carlo simulations.
}$^,$\cite{gagniuc_markov_2017}

The basic idea behind Dynamic Monte Carlo schemes is as follows.
A particle is randomly selected and the trial reaction is chosen in a random 
manner from all possible reaction sites (Random Selection Method, RSM).
The reaction will occur with a probability proportional to $k$ which is 
determined using another random number.
For each attempted jump, the time is incremented which corresponds to some unit 
of real time, $t/N$, where $N$ is the number of MC molecules.
In so doing, processes corresponding to surface diffusion, 
adsorption/desorption or chemical reaction can be implemented straightforwardly.
The drawback of this method, however, is that if the transition probabilities 
$1/k$ are small, there will be many unsuccessful transition attempts for each
update of the system configuration.
In the case of surface diffusion, for instance, many unsuccessful jump 
attempts will occur for every successful jump~\cite{kang_dynamic_1989}.

A more common approach, which significantly improves the efficiency of Dynamic 
Monte Carlo approaches, is to 
construct a list of all possible jumps, $W_{\alpha\beta}$, from the current 
configuration $\alpha$ to another configuration $\beta$.
A success probability weighted by the cumulative transition probability for 
the change $\alpha\to\alpha^\prime$ to take place is associated to each 
possible jump in this First Reaction Method (FRM): $f_{\alpha\alpha^\prime} = 
k_{\alpha\alpha^\prime}/\sum_{\beta} k_{\alpha\beta}$. The new time $t^\prime$ 
is then selected according to $t^\prime = t + \Delta t$ with
\begin{equation}
	\Delta t = -\frac{1}{\sum_{\beta} k_{\alpha\beta}} \ln r
\end{equation}
where $r$ is a uniform random number taken in the interval $]0,1]$.
After each update $\Delta t$, the transition $\alpha\to \alpha^\prime$ is 
selected with a probability $f_{\alpha\alpha^\prime}$ and the reaction list is 
updated.
This approach ensures that each iteration of the simulation
results in a change in the configuration.
Although the FRM and RSM yield the same results for a given model, the 
computational effort differs drastically depending on the system.
A detailed introduction to the the formal connection with the master equation 
and variations of the presented algorithm are given in 
Ref.~\citen{jansen_introduction_2003}.

A central point in the DMC methods lies in the fact that the time scale for the 
evolution of the simulated system is exact if the rates are correct and the 
underlying reaction is Markovian~\cite{serebrinsky_physical_2011}. 
Note that the Markovian assumption implies that reactions are uncorrelated --- 
a strong assumption for many systems. 
The main disadvantage with DMC is that all possible rates $k_{\alpha\beta}$
and configurational changes or reactions have to be known in advance.
Common ways to obtain $k_{\alpha\beta}$ from theoretical considerations are 
quantum chemical calculations or atomistic simulations. Yet, due to the 
exponential scaling of the hopping probability, $\exp(-k\tau)$, small errors 
can lead to enormous uncertainties.
Thus, for most practical applications, the rates are determined experimentally 
from adsorption/desorption isotherms or diffusion measurements.

If reaction paths and rates are well known, DMC can provide insights into simple 
and binary diffusion in strongly heterogeneous media~\cite{trout_diffusion_1997,
	coppens_dynamic_1999,paschek_diffusion_2001,liu_dynamic_2009} or 
reaction mechanisms in porous materials~\cite{schumacher_generation_2006}.
Today, DMC algorithms are incorporated into many open source codes as well as 
commercial software packages~\cite{noauthor_carlos_nodate}. Algorithmic 
development allows one to treat increasingly complex 
problems with DMC~\cite{rai_efficient_2006,dybeck_generalized_2017}.
However, structural confinement can significantly influence reaction 
kinetics~\cite{janda_effects_2016}, making quantitative predictions using DMC 
highly sensitive to the detailed input parameters.
Summarizing, the main advantage of the DMC method is that, if we assume a 
simple problem where for every state there is one fast pathway, the simulation 
time step $\Delta t$ scales with the free energy barrier (\ie.
inversely with the rate $k$) associated to this path. For rare events, this 
enables capturing simulation times of minutes to hours at room temperature. On 
the other hand, DMC is usually not exact as it requires the full rate list for 
all possible pathways in the system under study~\cite{voter_introduction_2007}.

\subsubsection{Dynamic Mean-Field DFT}

In the Dynamic Mean-Field Density Functional Theory (Dynamic Mean-Field DFT in 
short), one considers a lattice model of a fluid confined in a porous material 
interacting via nearest neighbor interactions only.
The configurational energy is given by:
\begin{equation}
	U = -\frac{\epsilon}{2} \sum_\mathbf{i} \sum_\mathbf{j=i+a} n_\mathbf{i} 
	n_\mathbf{j} + \sum_\mathbf{i} n_\mathbf{i} \phi_\mathbf{i}
	\label{eq:kineticmeanfieldH}
\end{equation}
where $\epsilon$ is the nearest neighbor interaction strength, $n_\mathbf{i}$ 
is the occupancy (0 or 1) of the lattice site at position $\mathbf{i}$, 
and $\mathbf{a}$ denotes the vector to the nearest neighbors.
The second term denotes the interaction of each lattice site $i$ with the 
external potential $\phi_\mathbf{i}$, which typically corresponds to the 
interaction between the fluid particles
and the pore wall or other solid particles.
Within the lattice gas framework, the self-diffusion constant $D_\mathrm{s}$ of 
a tracer molecule can be obtained by following its Kawasaki 
dynamics. Such a dynamical evolution is obtained by performing mass-conserving 
moves to neighboring sites in contact with a heat bath according to a 
Metropolis criterion~\cite{kawasaki_kinetics_1972}.
The latter approach is therefore somehow similar to the Dynamic Monte Carlo 
simulations discussed in the previous section, but without taking into account 
properly the energy barrier between two successive 
states~\cite{kang_dynamic_1989}.
This yields the advantage that, for a homogeneous lattice gas, $D_\mathrm{s}$ 
can be obtained analytically~\cite{godreche_anomalous_2003}.
Kawasaki dynamics simulations have been applied for example to the phase 
separation of confined fluid 
mixtures~\cite{monette_wetting_1992,chakrabarti_kinetics_1992} and the 
relaxation dynamics of capillary 
condensation/evaporation~\cite{luzar_dynamics_2000,
	woo_phase_2003,valiullin_exploration_2006}.

A significant reduction in the computational cost of the model outlined above 
can be achieved using a mean-field approximation. Within this assumption, one 
minimizes the following Helmholtz free energy:
\begin{equation}
	F = \kbT \sum_{\mathbf{i}} \left[ \rho_\mathbf{i} \ln \rho_\mathbf{i}
	+ (1-\rho_\mathbf{i}) \ln (1-\rho_\mathbf{i}) \right]
	- \frac{\epsilon}{2} \sum_\mathbf{i} \sum_\mathbf{j=i+a}
	\rho_\mathbf{i} \rho_\mathbf{j} + \sum_\mathbf{i} \rho_i \phi_i.
	\label{eq:meanfield_lattice}
\end{equation}
The first term in \cref{eq:meanfield_lattice} represents the system entropy, 
the second term the interaction energy between the nearest neighbors and the 
last term stems from the external field $\phi$
(corresponding to the interaction of the fluid with wall particles).
In the standard mean-field approach, \cref{eq:meanfield_lattice} is minimized 
with respect to the mean density at lattice site $\mathbf{i}$ under the 
constraint:
\begin{equation}
	\frac{\partial F}{\partial \rho_\mathbf{i}} - \mu = 0
	\label{eq:meanfield_mu}
\end{equation}
where $\mu$ is the chemical potential acting as a Lagrange multiplier in order 
to fix the total number of molecules $N \sim \sum_{\mathbf{i}} \rho_\mathbf{i}$.
Combination of \cref{eq:meanfield_lattice,eq:meanfield_mu} results in a set of 
coupled equations that can be solved numerically.
The basic idea of a dynamic mean-field theory is to follow the time 
evolution of the local density~\cite{gouyet_description_2003,monson_mean_2008},
\begin{equation}
	\frac{\partial\rho_\mathbf{i}}{\partial t} = - \sum_{\mathbf{j}} 
	J_\mathbf{ij}(t)
\end{equation}
where $J_\mathbf{ij}(t)$ is the net flux from site $\mathbf{i}$ to its 
neighboring site $\mathbf{j}$ at time $t$.
Within the mean-field approximation, 
$J_\mathbf{ij} = w_\mathbf{ij} \rho_\mathbf{i}(1-\rho_\mathbf{j}) - 
w_\mathbf{ji}\rho_\mathbf{j}(1-\rho_\mathbf{i})$
where the transition probabilities $w_\mathbf{ij}$ can be obtained using 
Kawasaki dynamics according to a Metropolis scheme,
\begin{equation}
	w_\mathbf{ij} = w_0 \exp(-\Delta E_\mathbf{ij})
	\label{eq:kineticmeanfieldfac}
\end{equation}
In the latter equation, $\Delta E_\mathbf{ij}$ is the energy difference between
lattice sites $\mathbf{i}$ and $\mathbf{j}$ and the prefactor $w_0$ essentially
takes care of relating the chemical potential difference between neighboring
sites to their difference in local densities.
Due to the use of the underlying mean-field approach, 
correlation effects in collective transport --- which we treated in detail in 
\cref{chap:diffusion_pore} --- are necessarily described in an approximated fashion.
With these limitations, dynamic mean-field DFT has been applied to study the 
dynamics of capillary condensation/evaporation of simple fluids and mixtures in 
porous materials~\cite{kierlik_spontaneous_2011,leoni_spontaneous_2011,
	edison_dynamic_2013,casselman_modelling_2015}.
Finally, we note that lattice-free dynamic DFT methods have been developed 
earlier to study the relaxation dynamics of fluids~\cite{marconi_dynamic_1999,
	archer_dynamical_2004} and hydrodynamic interactions 
can be incorporated on a continuum level, see \eg.\ 
Refs.~\citen{kikkinides_dynamic_2015,rathi_comparison_2018,
	rathi_nonequilibrium_2019} 
and Ref. \citen{wu_density-functional_2007} for a review.

\subsection{Network models}

\subsubsection{Pore network models}

Pore network models are a class of systems in which the transport properties of 
a given porous structure are solved on a lattice~\cite{obliger_pore_2014,
gjennestad_stable_2018}.
In this approach, the porous medium of a dimension $d$ = 1, 2, or 
3 is described as an assembly of pores  connected by channels. Each pore is 
located on a lattice node while the connecting channels correspond to segments 
between nodes. Upon applying external driving forces (gradients) at the scale 
of the whole lattice, the resulting transport properties are obtained by 
solving the corresponding conservation equations (\ie.\ the flux of 
conserved quantities must be the same at the entrance and exit of the lattice). 
Pore entrance/exit effects are assumed to be negligible so that transport corresponds to 
fluxes through the nodes. The volume, particle and heat flux through each node 
are typically assumed to be in the linear regime, \ie.\ directly proportional to 
the local gradient. Despite this fundamental linear response assumption, owing 
to the complexity of the node/channel distributions, the overall transport 
response to the external macroscopic gradients can be non linear.

Let us consider a porous medium filled with a fluid subjected to different 
gradients. As a very general example, we consider that pressure $\nabla P$, 
temperature $\nabla T$, and chemical potential $\nabla \mu$ gradients are 
applied, but note thar other driving forces could be applied 
(in fact, in many practical situations, only one driving force is used which
greatly simplifies the problem). 
By virtue of Onsager theory of transport, the transport properties 
of this system can be described using a matrix solution in which the flux of 
each conjugated variable of a given thermodynamic gradient is given 
by~\cite{yoshida_generic_2014}:
\begin{equation}
\begin{pmatrix}
J_E \\
J_N \\
J_V 
\end{pmatrix}
= - 
\begin{pmatrix}
L_{EE} & L_{EN} & L_{EV} \\
L_{NE} & L_{NN} & L_{NV} \\
L_{VE} & L_{VN} & L_{VV} 
\end{pmatrix}
\begin{pmatrix}
\nabla T \\
\nabla \mu \\
\nabla P 
\end{pmatrix}
\label{EQ7.27}
\end{equation}
where $J_E$, $J_N$ and $J_V$ are the energy, molecule, and volume fluxes. The 
values $L_{\alpha\beta}$ with $\alpha,\beta = E, N, V$ are the 
transport coefficients which characterize the system flux in a quantity 
$\alpha$ to a gradient in the variable $\beta$. To illustrate how the 
pore network model can be solved to determine the coefficients $L_{\alpha\beta}$, 
we discuss below an ideal case which consists of a filled porous medium 
submitted to a pressure gradient. This example is identical to the well-known 
problem of electrical transport in resistance networks as treated in detail by 
Kirkpatrick~\cite{kirkpatrick_percolation_1973}.

\begin{figure}[htb]
	\centering
	\includegraphics[width=0.95\linewidth]{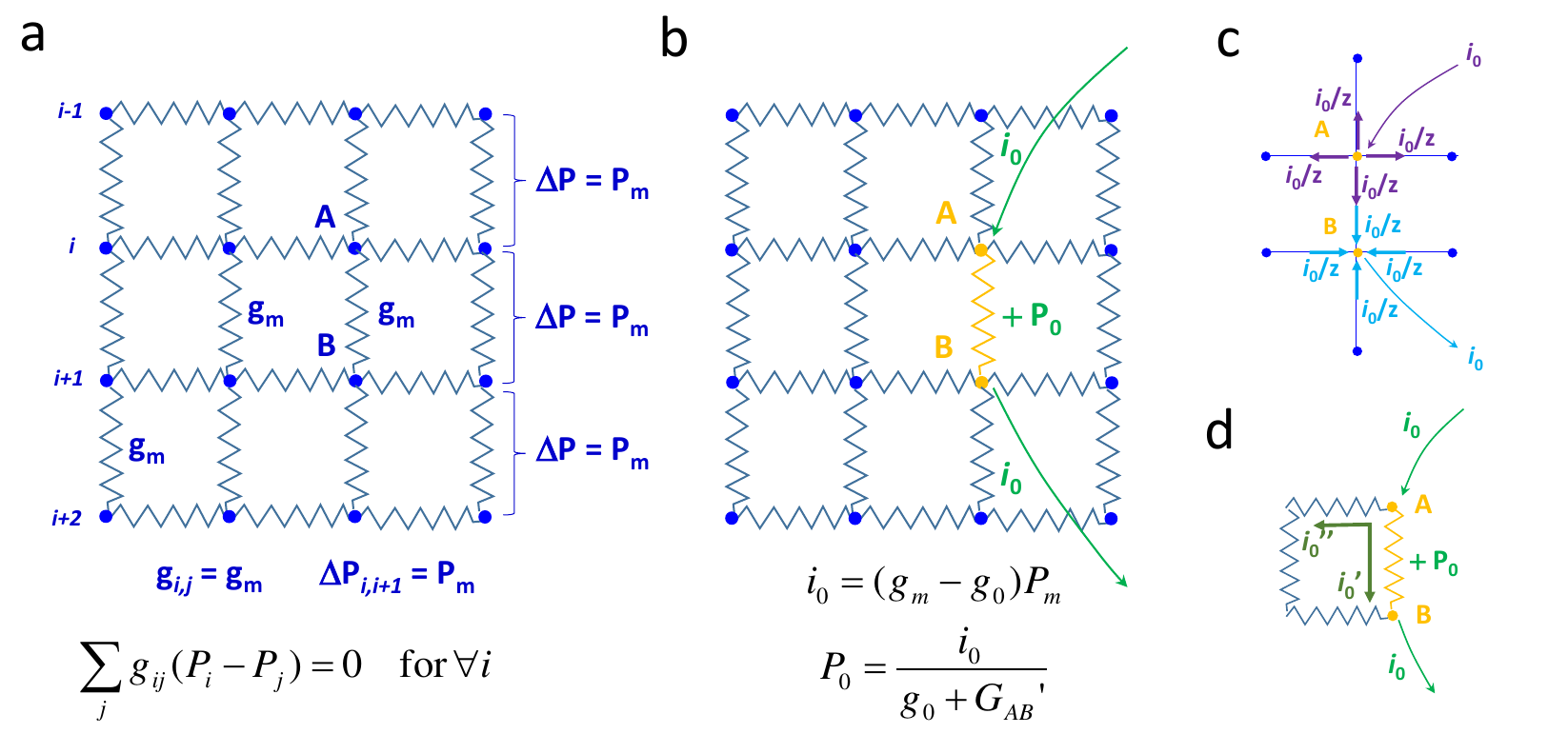}
	\caption{\textbf{Pore network model}. (a) Network model as defined in the
	Kirkpatrick percolation theory of electrical transport in resistor 
	networks. Here, by analogy with the Kirkpatrick model, a 3D network of 
	fluid 	conducting domains --- pores --- are subjected to a pressure drop 
	$\Delta P = P_m$ (the domain length is assumed to be equal to unity). 
	In the equivalent network, which allows determining the effective transport 
	properties such as the effective pore conductance $g_m$, the stationary 
	solution is given by the fluid mass conservation at each node $i$, $\sum_j 
	g_{ij} (P_i - P_j) = 0$. (b) In order to solve the problem stated in (a), 
	a perturbation is induced by varying the conductance of the domain 
	between nodes A and B. Such a perturbation induces a variation of $+P_0$ in 
	the local pressure difference $\Delta P$ as well as an additional fluid current $i_0$ between 
	nodes A and B. (c) and (d): $i_0$ can be measured by noticing that the 
	extra-current is injected in the domain network at node A and removed at 
	node B.}
	\label{fig:7_2}
\end{figure}

Let us consider the pore network model depicted in \cref{fig:7_2}(a). Each 
node $i$ corresponds to a pore which is connected to its neighbors $j$ through 
segments $ij$ corresponding to channels. The whole lattice is subjected to a 
pressure gradient $\Delta P/L$ where $L$ is the physical size of the connection 
between two pores (in total, the pressure drop across the lattice is $n\Delta 
P$ where $n$ is the number of nodes in the direction of the pressure-induced 
flow). Upon introducing the permeance $g_{ij}=K_{VV}(i\to j)$ of a channel $ij$ 
and the pressures $P_i$ and $P_j$  in pores $i$ and $j$, the transport 
properties of the porous medium mimicked using this pore network model can be 
determined by solving the following conservation equation at each lattice node 
$i$: 
\begin{equation}
\sum_j g_{ij} (P_i - P_j) = 0 \hspace{3mm} \forall i
\label{EQ7.28}
\end{equation}
where the sum runs over all nearest neighbors $j$ of the pore $i$. To solve 
this problem, one introduces the equivalent network where each 
segment has the same conductance $g_m$ and each node $i$ is at a pressure $P_i 
= P_\downarrow + P_m \times k(i)$ where $P_\downarrow$ is the downstream 
pressure at the entrance of the network and $k(i)$ is the position of the pore 
$i$ among the $n$ layers along the flow, \ie.\ $k(i) \in [1,n]$ (as shown in 
\cref{fig:7_2}(a), the pressure difference between two subsequent layers is 
$\Delta P = P_m$). 
Let us now consider a small perturbation to this equivalent network which is 
introduced by modifying the initial permeance $g_m$ of the segment A-B to a 
value $g_{AB} = g_0 \neq g_m$. Replacing $g_m$ by $g_0$ fails to satisfy the 
conservation condition given in \cref{EQ7.28} and, as shown  in 
\cref{fig:7_2}(b), an extra-current $i_0$ must be introduced in A and removed 
in B so that mass conservation is recovered with: 
\begin{equation}
i_0 =  (g_m - g_0)P_m
\label{EQ7.29}
\end{equation}
The generation of this extra-current requires a shift in the local pressure 
drop by a quantity $\Delta P_{AB} = P_0$. In order to determine the shift 
$P_0$, one can use the equivalent system shown in \cref{fig:7_2}(d) where the 
blue segments replace all the initial network connecting A and B except the 
segment AB itself. Mass conservation implies that the extra-current $i_0$ is 
the sum of the current running directly along A and B and the current running 
along the network in the absence of the segment AB: $i_0 = i_0' + i_0''$. Upon 
introducing the total conductance $G'_{\textrm{AB}}$ from A to B without 
including the conductance $g_0$ through the segment A-B, the latter current 
conservation relation leads to: 
\begin{equation}
P_0 = i_0/(g_0 + G'_{\textrm{AB}})
\label{EQ7.30}
\end{equation}

Let us now introduce the total conductance $G_{\textrm{AB}} = G'_{\textrm{AB}} 
+ g_m$ from A to B, which includes conduction through the domain A-B as well as 
through any other path connecting A and B. As illustrated in 
\cref{fig:7_2}(c), if $z$ is the network connectivity ($z$ = 6 for a cubic 
network), each current $i_0$ splits into a contribution $i_0/z$ from each 
segment. The total current along the segment A-B is therefore $2i_0/z$ (note 
that we recover the simple case $i_\textrm{AB} = i_0$ for a 1D system). This 
implies that $G_\textrm{AB} = z/2 g_m$ and $G'_\textrm{AB} = (z/2 - 1) g_m$ 
since  $G_{\textrm{AB}} = G'_{\textrm{AB}} + g_m$. Combining these relations 
with Eqs. (\ref{EQ7.29}) and (\ref{EQ7.30}) leads to: 
\begin{equation}
P_0 = \frac{P_m(g_m - g_0)}{g_0+ (z/2 - 1)g_m}
\label{EQ7.31}
\end{equation}
If we assume a continuous distribution $f(g)$ of $g_{ij}$, the average value
$\langle P_0 \rangle$ must be equal to zero because the homogeneous model is the
macroscopic solution by definition.
This implies: 
\begin{equation}
\int \frac{g_m - g}{g + (z/2 - 1)g_m} f(g) \textrm{d}g = 0
\label{EQ7.32}
\end{equation}
Similarly, using a discretized version where there are $M$ different (conducting) porous domains, one obtains: 
\begin{equation}
\sum_{\alpha = 1}^M v_\alpha \frac{k_\alpha - k_e}{k_\alpha + (z/2 - 1)k_e} = 0
\label{EQ7.33}
\end{equation}
where $k_\alpha$ is the conductance of a domain type $\alpha$ while $k_e$ is 
the effective conductance that is the solution of the equivalent problem. 
$v_\alpha$ are the volume fractions occupied by each phase $\alpha$. 
Interestingly, by defining the conductance as the the density $\rho$ multiplied 
by the transport coefficient $D$, \cref{EQ7.33} is strictly equivalent to 
\cref{EQ4.63} obtained using the effective medium theory presented in 
\Cref{chap:diffusion_network}.
While \cref{EQ7.32} and its discretized version \cref{EQ7.33} are refined
effective medium descriptions of transport in porous networks,
it is an important remark to note that they are not exact results.

\subsubsection{Other lattice models}

The mesoscopic methods proposed above allow one to gain insights into molecular transport at the porous network scale. However, a robust strategy in which nanoscale adsorption and transport phenomena are coupled is still lacking in 
the literature. Such approaches would allow accounting for complex adsorption 
(film formation, in-pore relocation, irreversible capillary condensation, 
\etc.)  and transport mechanisms (surface diffusion, slippage effects, viscous 
flow approximation breakdown, activated transport, \etc.)  
that are usually omitted in conventional techniques or included only in an effective 
way. As an illustration of possible developments, in what follows, we present 
the simple multiscale model that has been proposed by Botan \etal.\ to 
describe adsorption and transport in heterogeneous porous 
materials~\cite{botan_bottom-up_2015}.
This lattice model allows upscaling molecular simulation of adsorption and 
transport in porous materials at larger length and time scales. In particular, 
this bottom-up technique accounts for changes in adsorption and transport upon 
varying pore size, pressure, temperature, \etc, so that it does not 
require to assume {\textit{a priori} any adsorption or 
flow regime. In practice, this approach accounts for any adsorption effects 
and possible changes in the confined fluid state  upon transport by relating at 
each time step $t$  the local density $\rho(r)$ and chemical potential 
$\mu(r)$.

\begin{figure}[htbp]
	\centering
	\includegraphics[width=0.95\linewidth]{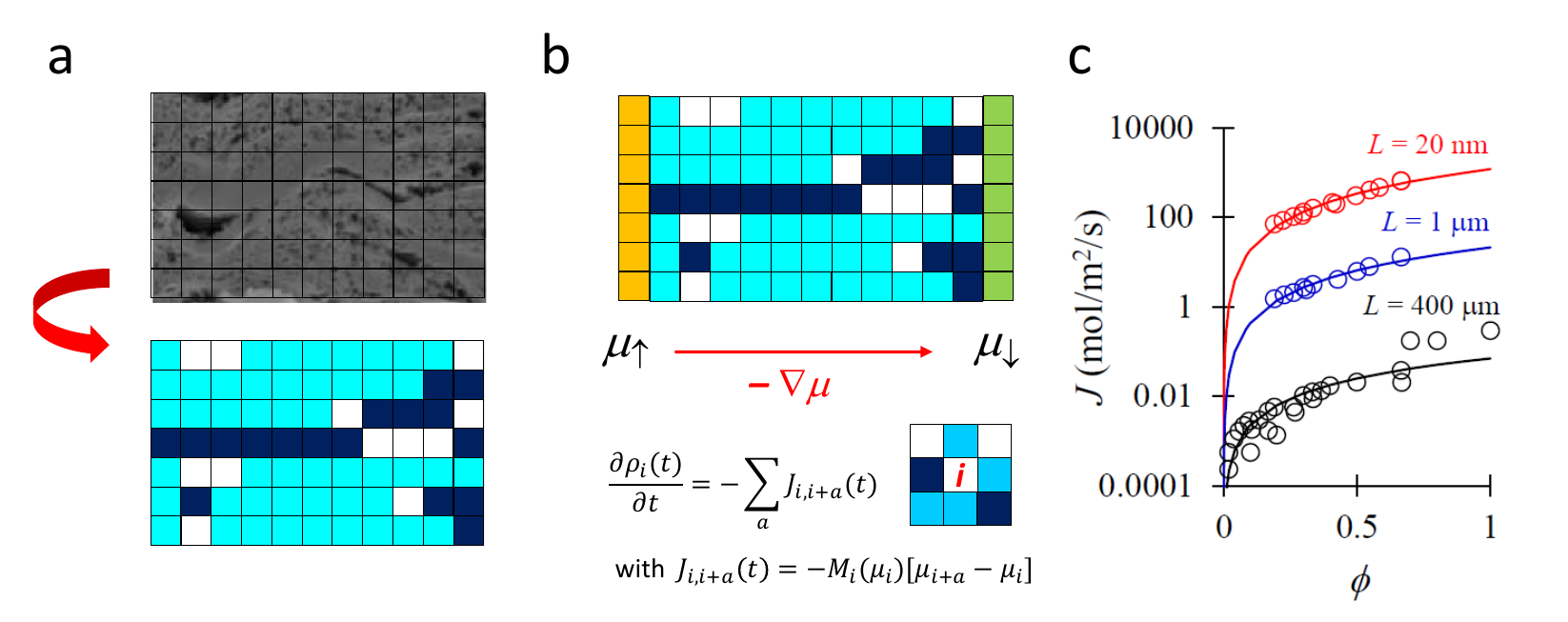}
	\caption{\textbf{Upscaling strategy using a lattice model}. (a) A lattice 
	model is mapped onto experimental structural data such as tomography or 
	FIB-SEM data. \bluelork{$L$ is defined as the size of the porous solid over which the fluid flow is considered as a response to a thermodynamic gradient.} Each domain type $i$ is assigned: (1) an adsorption isotherm 
	$\rho_i(\mu)$ where $\rho$ and $\mu$ are the fluid density and chemical 
	potential, respectively, and (2) a transport relation $J =  - M_i(\mu) 
	\nabla \mu$ where $J$ is the flux and $M_i$ the transport coefficient 
	associated to the driving force $\nabla \mu$. For each domain type, these 
	two sets of data can be determined from molecular simulation and/or from 
	available experimental data. In the situation depicted here, three domain 
	types are considered: white, cyan and blue. (b) Once the lattice model has 
	been defined, transport is simulated at large scale by placing the lattice 
	between two boundary conditions: the chemical potentials upstream and 
	downstream are constant and equal to $\mu_\uparrow$ and $\mu_\downarrow$, 
	respectively. Because of local mass conservation, the change in the local 
	density at site $i$ is simply given by the sum of the incoming and outgoing 
	fluxes from/towards neighboring sites. (c) Flux $J$ as a function of 
	porosity $\phi$ for an assembly of domains with different porosity scales. 
	The red and blue data are for lattice systems with different lengths 
	indicated in the graph. The black data, which are for a system having a 
	much larger size (\SI{400}{\micro\meter}), were determined by upscaling the 
	data for the lattice with \SI{1}{\micro\meter}.
	In all cases, the upstream and downstream 
	chemical potentials were set to values corresponding to pressures of 11 and 
	10 bar (the temperature is 423 K). The lines are fits against Archie's law, 
	$J \sim \phi^\alpha$. Adapted with permission from Ref.~\protect\citen{botan_bottom-up_2015}.
	Copyright 2015 American Physical Society.}
	\label{fig:7_3}
\end{figure}

\Cref{fig:7_3} schematically presents the multiscale strategy developed in 
Ref.~\citen{botan_bottom-up_2015}. First, 2D or 3D structural data as obtained 
for a real porous solid using tomography or microscopy experiments are used to 
map a lattice model. \bluelork{Let us introduce the  size $L$ of the porous solid over which the fluid flows as a response to a thermodynamic gradient (in Fig.  \ref{fig:7_3}, we consider as an example the fluid flow induced by a chemical potential gradient).} 
Using well-known segmentation techniques,  each node  of 
the lattice model can be assigned a porosity domain type (in this context, a 
type corresponds to a given  pore size and/or surface chemistry, 
\etc.). \bluelork{In this context, we define a domain type here as a solid region where the porosity scale is homogeneous so  that the physical properties of the fluid in this domain are equal in all points belonging to this domain. As illustrated in Ref.~\citen{botan_bottom-up_2015}, this domain can be made up of microporous, mesoporous, or macroporous regions  (and even non-porous regions if domains impermeable to the fluid are to be considered).}
For each domain type $m$, molecular simulation or available 
experimental data are used to provide the two following constitutive equations: 
(1) $\rho_m = \rho_m(\mu)$ and $J_m(\mu) = - M_m(\mu) \nabla \mu$. The first 
equation is a thermodynamic relation, known as adsorption isotherm, which 
relates the confined fluid density for a given domain type $m$ to the local 
chemical potential $\mu$. The second equation is a linear response 
relationship that relates the local flux in a given porous domain of type $m$ 
to the local gradient in the chemical potential $\nabla \mu$ (which can be
related e.g.\ to pressure or density gradients) via the transport coefficient $M$.
Once this mapping has been achieved, as shown in 
\cref{fig:7_3}(b), the lattice model is set in contact with an upstream and a 
downstream reservoir imposing chemical potentials equal to $\mu_\uparrow$ and 
$\mu_\downarrow$, respectively. Transport is then predicted at the porous 
network scale as a function of the fluid thermodynamic state and driving force imposed 
across the sample as follows. At each lattice node $i$, the following mass 
conservation equation must be fulfilled:
\begin{equation}
\frac{\partial \rho_i}{\partial t} =\sum_j J_{i \rightarrow j}(t) = \sum_j 
M_{ij} \Big( \frac{\mu_j - \mu_i}{l} \Big )   
\label{EQ7.34}
\end{equation}
where $\mu_i$ and $\rho_i$ are the local chemical potential and density at site 
$i$, respectively, while $l$ is the distance between two nearest neighbor 
sites. The sum runs over all nearest neighbor sites $j$ with $J_{i 
\rightarrow j}$ the flux between site $i$ and $j$ (which can be positive or 
negative depending on the sign of the local chemical potential difference 
$\mu_j - \mu_i$). In this equation, the local flux between two neighboring 
sites $i$ and $j$ is proportional to the local transport coefficient $M_{ij}$ 
which is defined as $M_{ij} = 2M_{ii}(\mu_i)M_{jj}(\mu_j)/[M_{ii}(\mu_i) + 
M_{jj}(\mu_j)]$ to ensure local mass conservation at each node.  As can be seen 
from \cref{EQ7.34}, local adsorption and transport effects at each node are 
taken into account as (1) the local chemical potential is directly related to 
the local density and (2) the transport coefficient depends on the local 
chemical potential. As shown in \cref{fig:7_3}(c), the approach above was 
capable to recover physical situations such as the known regime of Archie’s law 
in which the flux varies as a power law of the porosity, $J \sim \Phi^\alpha$,
(with $\alpha$ an exponent usually comprised between 1.8 and
2).\cite{archie1942electrical, glover_generalized_2010}
An interesting feature of such lattice models is that they can be upscaled in an
iterative fashion to reach the macroscopic scale while transferring at each step
all adsorption/transport information from the smaller lengthscale.
This is illustrated in \cref{fig:7_3}(c) where the flux versus porosity
relationship is shown for different lattice models and sizes (with sizes ranging
from 20 nm to 400 microns).

%%%%%%%%%%%%%%%%%%%%%%%%%%%%%%%%%%%%%%%%%%%%%%%%%%%%%%%%%%%%%%%%%%%%%%%%%%%%%%%%%%%%%%%%

% Conclusions

%!TeX spellcheck = en_US
%!TeX encoding = utf8 
%!TeX program = pdflatex
%!TeX root = ../manuscript.tex

\section{Discussion and Perspectives}
\label{chap:discussion}

In this review, we have addressed the complex issue of  diffusion and transport
of molecules in nanoporous materials under various conditions.
By considering phenomena occurring either at the molecular or  mesoscopic 
scales, we have discussed the  appropriate simulation techniques to tackle
questions relevant to each level of description.
While this review focuses on  small molecules in prototypical nanoporous
materials (e.g.  zeolites, active carbons, metal organic frameworks, mesoporous
oxides), we believe it applies to a very large number of molecule families. Yet, we acknowledge
that the behavior of more complicated molecules can display additional
complexity. 
In any case, for the sake of clarity and concision, we have omitted several
important topics relevant to this review on adsorption and transport in
nanoporous materials such as transport of multicomponent mixtures and transport
in electrical fields (and all electrokinetic effects in general). 
As for mixture systems, considering that such aspects have been covered in depth in several
already available reviews, we feel that a detailed treatment of the transport of
multicomponent mixtures was beyond the scope of our review. As for transport in
electrical fields, the dynamics of charged species in nanoconfined solvents is
interest of primary interest for the broad range of domains and applications
mentioned in our introduction. 
However, despite being a very interesting aspect of transport in nanoporous
materials, we feel that transport in electrical fields is also beyond the scope
of our review to be addressed in a reasonable fashion.
Moreover, outstanding reviews on adsorption and transport of charges in
uncharged and charged nanopores are already
available.\cite{schoch_transport_2008, fedorov_ionic_2014,
bocquet_nanofluidics_2010, kondrat_theory_2023}
In particular, all electrokinetic effects encompassing electrophoresis,
diffusion/osmosis, ionic conductivity in nanoconfined geometries and near solid
surfaces are the topic of very specific transport mechanisms that involve long
range electrostatic effects.

In what follows, we briefly discuss a non-exhaustive list of perspectives which
have been identified. While such additional topics go beyond the scope of this review, they are needed to tackle important challenges relevant to energy and environment applications. In particular, important efforts are currently devoted to conceiving and developing bio-inspired systems which allow reaching efficiency and targets only achieved by nature. Of utmost importance, different axes of development include central problems such as desalination and energy conversion,\cite{logan_membrane-based_2012, werber_materials_2016} CO$_2$ separation/capture \cite{reiner_carbon_2019} and the removal of pollutants including novel classes of molecules such as PFAS.\cite{gluge_overview_2020}

\noindent  \textbf{Chemo-mechanical effects.}
More and more reports are available in the literature on transport phenomena in
nanoporous environments in which the dynamics of confined fluids couples with
mechanical effects. In particular, when complex nanoporous solids such  as hybrid porous materials (e.g.\ metal organic frameworks) or disordered materials relevant to soft matter (polymers, cellulose, etc.) are considered, novel mechanical phenomena occur which include swelling or breathing effects as well as reorganization of the host material.\cite{coudert_responsive_2015, chen_role_2018}
While the interplay between these effects and the thermodynamics of confined
fluids has received significant attention,\cite{coussy_poromechanics_2004}
their impact in terms of fluid transport remains to be fully understood. 
\bluelork{In this context, we note that recent studies have considered
transport in polymer membranes with nanometric
pores.~\cite{heiranian_mechanisms_2023,wang_water_2023, fan_physical_2024,
he_molecular_2024, fan_solution-diffusion_2024}}
From a formal viewpoint, available approaches have been already proposed to address
these effects \cite{marbach_transport_2018}, but more research is needed to
fully unravel how such coupling proceeds.
Similarly, many interesting phenomena  relevant to such chemo-mechanical
effects are also reported for fluids in complex  media relevant to biophysics
(\eg.\ molecular and ion transport in biological membranes) and natural materials
(\eg.\ wood, concrete).\cite{wei_micro-scale_2019,
berthonneau_mesoscale_2018,
hoseini_effect_2009, zhang_effect_2018,
bai_coupled_2000}

\noindent  \textbf{Separation and phase transition at the nanoscale.}
The large majority of theoretical studies on fluid transport in nanoporous
materials deals with pure fluids.
However, many practical situations involve mixtures, which lead to strong
competitive adsorption effects with sometimes very large selectivities (defined
as the ratio of mole fractions in confinement with respect to the same ratio in
the bulk system).\cite{mitschke_confinement_2020}
In particular,  oversolubility refers to large solubility enhancements observed
when a solute is set in contact with a solvent nanoconfined within a nanoporous
material.\cite{ho_gas_2013, ho_solubility_2015, coasne_gas_2019}
While such thermodynamic effects are at the heart of  physical processes
relying on nanoporous materials, they remain to be better understood.
Of particular relevance to the present review, phase separation in nanoporous
materials is also observed when mixtures made up of fluids with different
physical interactions are involved.
From a fundamental viewpoint, such nanoscale separation is an important field as
it raises important questions. 
In particular, while thermodynamic frameworks are available to describe such
segregation effects, their impact on fluid transport at a large scale remains to
be fully explored. 
In this context, while more specific in nature, the case of conformers and
enantiomers separation is an important issue for both basic and applied
research.\cite{emmerling_olefin_2022}

\noindent \textbf{Solid/fluid coupling at interfaces.}
There is now a large body of research  unravelling the impact of the
solid dynamics of the host porous medium on the transport of fluids at their
surface.
This includes the impact of lattice vibrations on the diffusion or permeability
of the nanoconfined fluid.\cite{noh_phonon-fluid_2022}
Other interesting situations correspond to the coupling between a mechanical
solicitation imposed to the host solid and the fluid dynamical response.
In this context, Marbach \etal.\ have recently developed a theoretical framework
for studying the impact of fluctuations of the confining surfaces on diffusion.
These authors found that diffusion can be enhanced or suppressed depending on the
fluctuations.\cite{marbach_transport_2018}
Recent work has also highlighted the impact of pore compliance on the structure
and dynamics of the confined liquid water, revealing that the water density
profiles depend on the pore elastic properties, which in turn can drastically
enhance transport.\cite{schlaich_bridging_2024}
Other couplings between the host solid and confined fluid includes interesting
effects such as the concept of quantum
friction.\cite{kavokine_fluctuation-induced_2022}
With this concept, electronic relaxation within the solid phase set in contact
with the vicinal fluid phase leads to novel phenomena in the fluid thermodynamic
and transport properties.
Owing to the quantum mechanics nature of the solid response, new formalisms need
to be developed to rationalize and predict this physical coupling  at the
frontier between solid-state and liquid-state physics.\cite{coquinot_quantum_2023}

\textbf{Upscaling transport in reactive or phase transition conditions.}
Both stationary and transport conditions in nanoporous materials involve a large
variety of thermodynamic states.\cite{coasne_adsorption_2013}
This includes phase transition such as gas/liquid coexistence but also
separation mechanisms with coexisting states as described above.
These heterogeneous systems can be described using mesoscopic approaches
including Lattice Boltzmann simulations or computational fluid dynamics tools.
However, understanding the microscopic mechanisms at play in these coexisting
states remain an important challenge with many questions left unexplained.
In particular, as briefly discussed in this review, interfacial transport in the
sense of fluid dynamics at the interfaces between different thermodynamic phases
involves large free energy barriers with complex underlying dynamics.
While different in spirit, reactive transport also falls in the category of
complex molecular mechanisms with a strong impact on macroscopic transport and
overall process efficiency.
Like for transport in systems involving phase transition, taking into account
the ``reactivity'' of the system (evolution of interfaces upon phase transitions
or transformation of molecular species into others upon chemical reactions) is a
major challenge that needs to be addressed as it corresponds to a large number
of practical situations involving nanoporous materials.

%\change{
%    Summarizing, we conclude that on each individual length- and time-scale
%    elaborate approaches have been established that we have carefully, yet not
%    exhaustively presented in this work.
%    However, future developments need to tackle the problem of transferring
%    obtained microscopic insight consistently to mesoscale methods and
%    subsequently predict experimentally accessible observables.
%    To this end, data-integrated, multi-scale and multi-physics approaches for
%    diffusion and transport in nanoporous materials need to be established.
%}

%%%%%%%%%%%%%%%%%%%%%%%%%%%%%%%%%%%%%%%%%%%%%%%%%%%%%%%%%%%%%%%%%%%%%%%%%%%%%%%%%%%%%%%%

\section*{Biographies}

Alexander Schlaich studied physics at the University of Stuttgart and received
his PhD in 2017 from Freie Universität Berlin on water effects on the
interaction and friction between polar surfaces.
After postdoctoral assignments in Grenoble and Stuttgart he received a
fellowship to set up an independent junior research group within the clusster of
excellence "data-integrated sinulation science" in Stuttgart in 2021. 
Since 2024, he has held a chair in atomistic modeling of materials in aqueous
media at Hamburg University of Technology.
In his research, Alexander Schlaich combines different simulation and
theoretical approaches to bridge the description of nanoporous functional matter
from the fundamentals at the atomistic scale to engineering applications.

Jean-Louis Barrat received  a doctorate from the University of Paris in 1987,
working on statistical physics of liquids and phase transitions. He worked as a
postdoctoral researcher in Munich and Santa Barbara. After being a CNRS
researcher at the ENS in Lyon, he joined in 1994 the University of Lyon, where
he created a research group on modeling in materials science. In 2011, he joined
the University of Grenoble where he created the statistical physics group
within the Interdisciplinary Physics Lab. His research focuses on statistical
physics, condensed matter and materials physics.

Benoit Coasne obtained his PhD in Physics on capillary condensation in
nanoporous materials (Paris, 2003). After a postdoc in Raleigh, NC, USA
(2003-2005), Benoit Coasne was appointed CNRS researcher in Montpellier (2005)
and promoted CNRS Research Director (2015). During a 3 year visiting stay, he
was leading a fundamental research group on multiscale modeling of adsorption
and transport in the CNRS/MIT lab at MIT in Boston (2012/15). He is currently
affiliated CNRS Research Director in the Interdisciplinary Physics Lab in
Grenoble, France. He is also Permanent Affiliate of the Theory Group at the
Institute Laue Langevin in Grenoble. Benoit Coasne’s research consists of
studying by means of statistical mechanics and molecular simulation tools the
thermodynamics and dynamics of fluids confined in nanoporous media. 

\begin{acknowledgement}
The authors acknowledge financial support from EUROKIN which is a consortium of academic and industrial members (www.eurokin.org). We are also grateful to the French Agence Nationale de la Recherche for funding (Project TAMTAM ANR-15-CE08-0008-01 and Project  TWIST ANR-17-CE08- 0003). A.S. also acknowledges funding from the DFG under Germany's Excellence Strategy -- EXC 2075-390740016 and SFBs 1313/2-327154368 and 1333/2–358283783, and support by the Stuttgart Center for Simulation Science (SimTech). We wish to thank Sophie Marbach for useful comments and discussions. 
A preprint version of this review is available on arXiv \cite{schlaich_theory_2024}.
\end{acknowledgement}

\bibliography{eurokin,eurokin_new,eurokin_additional,nomanew2}

\begin{mcitethebibliography}{292}
\bibitem[Not()]{Note-2}
\label{footnoteMa} For a fluid, the sound velocity $c$ is given by the Newton--Laplace equation which involves the stiffness coefficient $K_s$ (\ie.\ the bulk elastic modulus for fluids) and the mass density $\rho$, $c = \sqrt{K_s/ \rho}$. In the case of an ideal gas at a given pressure $P$, $K_s = \gamma_a P$ so that $c = \sqrt{\gamma_a P/\rho}$ where $\gamma_a = C_p/C_v$, the adiabatic index, is equal to the ratio of the heat capacities at constant pressure and constant volume ($\gamma_a = 1.4$ for a diatomic ideal gas such as N$_2$). Using the ideal gas law, $P = \rho/M RT$ where $M$ is the molar mass, $R$ the ideal gas constant, and $T$ the temperature, we obtain $c = \sqrt{\gamma R T/M} \sim 350$ m/s for N$_2$ at room temperature.\relax
\mciteBstWouldAddEndPunctfalse
\mciteSetBstMidEndSepPunct{\mcitedefaultmidpunct}
{}{\mcitedefaultseppunct}\relax
\EndOfBibitem
\bibitem[Not()]{Note-3}
\label{maxentropy} Let us consider a thermodynamic system subdivided into two subsystems 1 and 2. For each conserved quantity $X$ such as the energy $E$ or number of molecules $N$, the sum of their contribution in each subsystem is constant, \ie.\ $X_1 + X_2 = 0$, which implies that $\partial X_1 = - \partial X_2$. Moreover, the equilibrium condition is given by the entropy maximization principle, \ie.\ $\partial S/\partial X_1 = \partial S_1/\partial X_1 - \partial S_2/\partial X_1 = 0$ which leads to $F_1 = F_2$ since $\partial S/\partial X = F$.\relax
\mciteBstWouldAddEndPunctfalse
\mciteSetBstMidEndSepPunct{\mcitedefaultmidpunct}
{}{\mcitedefaultseppunct}\relax
\EndOfBibitem
\bibitem[Not()]{Note-5}
Let $G(f)$ and $H(f)$ be the Fourier transform of two functions $g(t)$ and $h(t)$ in the time domain. The convolution theorem states that \begin{equation*} G(f)H(f) = \int\limits_{-\infty}^\infty g(\tau) h(t-\tau) \mathrm{d}\tau. \end{equation*} For real functions $g$ and $h$, we can use that $H(-f) = H^\star(f)$, where the $\star$ denotes the complex conjugate. This allows to write the correlation theorem, \begin{equation*} G(f) H^\star(f) = \int\limits_{-\infty}^\infty g(\tau+t)h(\tau) \mathrm{d}\tau, \end{equation*} which after backwards Fourier transform yields the correlation between $g$ and $h$. If $g$ and $h$ are discrete and finite ($t_\mathrm{max}<\infty$) observables defined only for $t\geq 0$ as in \cref{EQ3.20}, one can make use of the cyclic property of the discrete time Fourier transform and average $g(\tau+t)h(\tau)$ efficiently over all initial times $\tau$ using the Fast Fourier Transform (FFT), \begin{equation*} \left<g(\tau+t)h(\tau)\right>_\tau = \frac{1}{N_\tau(t)}\mathsf{iFFT}\left(\mathsf{FFT}(g) \cdot \left[\mathsf{FFT}(h)\right]^\star\right), \end{equation*} where the normalizing factor $N_\tau(t)$ counts the number of samples (\ie.\ the number of initial times $\tau$) that are considered in the data series for each lag time $t$. For the special case of an autocorrelation function, $g=h$, one obtains $G(f)\cdot G^\star(f)=\left|G(f)\right|^2$. Also note, that because of the cyclic property of the FFT one usually adds zeros to the signal for $-t_\mathrm{max}<t<0$ (zero padding). The advantage of employing the FFT for correlation functions is the scaling of $\mathcal{O}(N\log N)$ as compared to $\mathcal{O}(N^2)$ for a direct calculation.\relax
\mciteBstWouldAddEndPunctfalse
\mciteSetBstMidEndSepPunct{\mcitedefaultmidpunct}
{}{\mcitedefaultseppunct}\relax
\EndOfBibitem
\bibitem[Not()]{Note-8}
Henry’s law can be derived as follows by starting from the equality of the fugacity $f$ (or, equivalently, the gas chemical potential $\mu$) in the gas phase and in the liquid phase: $f_g = f_l$. Under the assumption that the gas behaves as an ideal gas, $f_g = P$. As for $f_l$, we can write $f_l = \gamma x f_l^{(0)}$ where $\gamma$ is the activity coefficient, $x$ is the mole fraction of gas molecules in the liquid, and $ f_l^0$ is the fugacity of the liquid phase for a pure fluid, \ie.\ $x = 1$. Using a polynomial expansion for $\ln \gamma \sim a_1 (1 - x) + a_2 (1 - x)^2 + a_3 (1 - x)^3 + \ldots$, one recovers the two important asymptotic limits: (a) $\ln \gamma = 0$ for $x = 1$ as expected for a pure liquid for which $\gamma \sim 1$ and (b) $\ln \gamma \sim$ constant for $x \ll 1$. As a result, for $x \ll 1$, and noting that $\rho \sim x$, one obtains Henry’s law with $K_\mathrm{H} \sim \left[ \gamma f_l^{(0)} \right]^{-1}$ as a good approximation that describes the fugacity equality between the gas and liquid phases. In practice, the simple scaling between gas concentration and pressure in Henry’s law holds provided that the gas partial pressure remains low enough and that the host solid or liquid does not chemically react with the solubilized gas.\relax
\mciteBstWouldAddEndPunctfalse
\mciteSetBstMidEndSepPunct{\mcitedefaultmidpunct}
{}{\mcitedefaultseppunct}\relax
\EndOfBibitem
\bibitem[Not()]{Note-9}
Let us consider a single fluid molecule diffusing through a medium containing a density $\rho$ of frozen objects with which the molecule can collide. Introducing the cross-section area for collision $s \sim \pi \sigma^2$, the mean free path can be written as $\lambda \sim 1/\rho \pi \sigma^2$. For a gas, since the objects are other gas molecules, they cannot be considered as immobile so that the relative velocity of the gas molecules must be taken into account which leads to a slightly different expression: $\lambda \sim 1/\sqrt{2} \rho \pi \sigma^2$. Assuming that the gas can be treated as ideal, \ie.\ $\rho = P/k_\textrm{B}T$, leads to: $\lambda = k_\textrm{B}T /\sqrt{2 } P \pi \sigma^2$.\relax
\mciteBstWouldAddEndPunctfalse
\mciteSetBstMidEndSepPunct{\mcitedefaultmidpunct}
{}{\mcitedefaultseppunct}\relax
\EndOfBibitem
\bibitem[Not()]{Note-10}
Tortuosity will be introduced later in detail --- here, it should be simply considered as the ratio of the pore physical length to the pore height.\relax
\mciteBstWouldAddEndPunctfalse
\mciteSetBstMidEndSepPunct{\mcitedefaultmidpunct}
{}{\mcitedefaultseppunct}\relax
\EndOfBibitem
\bibitem[Not()]{Note-11}
A formal derivation of this relationship is as follows. $L(t) \sim \int_{0}^{T} I(\tau)I(t+\tau) \textrm{d}\tau$ so that $L'(t) \sim \int_{0}^{T} I(\tau)I'(t+\tau) \textrm{d}\tau \sim \int_{0}^{T} I(\tau - t)I'(\tau) \textrm{d}\tau$ where the last equality is obtained by using the invariance of the function $L'(t)$ upon time shift \ie.\ $A(\tau)B(t+\tau) = A(\tau -t)B(\tau)$. Starting from the last expression, we derive a second time to obtain $L''(t) \sim - \int_{0}^{T} I'(\tau-t)I'(\tau) \textrm{d}\tau \sim - \int_{0}^{T} I'(\tau)I'(t+\tau) \textrm{d}\tau$.\relax
\mciteBstWouldAddEndPunctfalse
\mciteSetBstMidEndSepPunct{\mcitedefaultmidpunct}
{}{\mcitedefaultseppunct}\relax
\EndOfBibitem
\bibitem[Not()]{Note-12}
Here, we use that $\psi_A(t)$ and $\psi_B(t) = 0$ for $t < 0$ so that $\tilde{\psi}_A(\omega) = \int_{-\infty}^\infty \psi_A(t) \exp(i\omega t) \textrm{d}t = \int_{0}^\infty \psi_A(t) \exp(i\omega t) \textrm{d}t$ and $\tilde{\psi}_B(\omega) = \int_{-\infty}^\infty \psi_B(t) \exp(i\omega t) \textrm{d}t = \int_{0}^\infty \psi_B(t) \exp(i\omega t) \textrm{d}t$. As a result, upon calculating $\tilde{L}''(\omega)$, we use $\tilde{L}''(\omega) \sim -2 + \int_0^\infty \psi_A(t) \exp(i \omega t) \textrm{d}t + \int_{-\infty}^0 \psi_A(-t) \exp(i \omega t) \textrm{d}t + \cdots \sim -2 + \operatorname{Re} [2 \times \tilde{\psi}_A(\omega) + \cdots]$\relax
\mciteBstWouldAddEndPuncttrue
\mciteSetBstMidEndSepPunct{\mcitedefaultmidpunct}
{\mcitedefaultendpunct}{\mcitedefaultseppunct}\relax
\EndOfBibitem
\bibitem[Not()]{Note-13}
This simple treatment is important as it allows capturing the overall effect of fluid density and pressure on the effective diffusivity in nanopores in the low pressure range (see \cref{fig:4_5}). The above treatment can be recovered easily from the simple scaling behavior derived earlier $\overline{D_\mathrm{s}} = \sum_{\alpha = 1,2} x_\alpha D_\mathrm{s}^\alpha$ with $\alpha = 1$ and $\alpha = 2$ for the adsorbed phase, respectively. Within the approximation $\rho_g \sim 0$, such a linear combination rule leads to $\overline{D_\mathrm{s}} \sim D_\mathrm{s}^1$ since the fraction of molecules is $x_g \sim \rho_g \sim 0$. In other words, at temperatures low enough, molecular diffusion through the gas phase can be neglected so that, independently of the diffusion combination rules considered, diffusion only occurs through the adsorbed phase with a diffusion coefficient equal to the surface diffusion.\relax
\mciteBstWouldAddEndPunctfalse
\mciteSetBstMidEndSepPunct{\mcitedefaultmidpunct}
{}{\mcitedefaultseppunct}\relax
\EndOfBibitem
\bibitem[Not()]{Note-15}
To obtain this equation, we notice that the cross terms $\left\langle \mathbf{v}(0) \times \int_0^t \textrm{d}t' \exp \left[- \xi (t-t') \right] \delta \mathbf{F(t')} \right\rangle$ are equal to zero because the velocity at time $t = 0$ only depends on the force exerted at times $t < 0$ and therefore not on the force exerted at times $t \geq 0$.\relax
\mciteBstWouldAddEndPunctfalse
\mciteSetBstMidEndSepPunct{\mcitedefaultmidpunct}
{}{\mcitedefaultseppunct}\relax
\EndOfBibitem
\bibitem[Not()]{Note-16}
This condition is equivalent to assuming that two subsequent jumps are uncorrelated as usually assumed in any random walk approach.\relax
\mciteBstWouldAddEndPunctfalse
\mciteSetBstMidEndSepPunct{\mcitedefaultmidpunct}
{}{\mcitedefaultseppunct}\relax
\EndOfBibitem
\bibitem[Not()]{Note-18}
A Markov process is a stochastic model describing a sequence of possible events in which the probability of each future event depends only on the current state. Roughly speaking, this Markov property can be characterized as ``memorylessness". Prominent examples cover random walks, Brownian motion and Markov chain Monte Carlo simulations.\relax
\mciteBstWouldAddEndPunctfalse
\mciteSetBstMidEndSepPunct{\mcitedefaultmidpunct}
{}{\mcitedefaultseppunct}\relax
\EndOfBibitem
\end{mcitethebibliography}

%\newpage
%\begingroup
%\parindent 0pt
%\parskip 2ex
%\def\enotesize{\normalsize}
%\theendnotes
%\printendnotes           % <-- instead of \theendnotes
%\endgroup

\begin{tocentry}
	\includegraphics{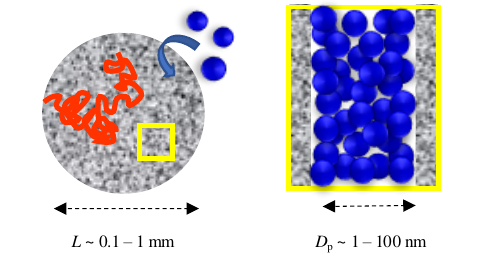}
\end{tocentry}

\end{document}